\documentclass[5p]{elsarticle}       
\usepackage[utf8]{inputenc}
\usepackage{amsfonts}
\usepackage{amsmath}
\usepackage{multicol}
\usepackage{amssymb}
\usepackage{graphicx}
\usepackage{mathtools}
\usepackage{xcolor}

\usepackage{bm} 
\usepackage{algorithm}
\usepackage[noend]{algpseudocode}
\usepackage{colortbl,booktabs}
\usepackage{xcolor}
\usepackage{etoolbox}
\let\bbordermatrix\bordermatrix
\patchcmd{\bbordermatrix}{8.75}{4.75}{}{}
\patchcmd{\bbordermatrix}{\left(}{\left[}{}{}
\patchcmd{\bbordermatrix}{\right)}{\right]}{}{}
\definecolor{mygray}{gray}{.9}
\usepackage{algorithm}
\usepackage{algorithmicx}
\usepackage{algpseudocode}

\usepackage{float}
\usepackage[labelformat=simple]{subcaption}

\usepackage{tikz}
\usepackage{tikz-qtree}

\usepackage{tikz}
\usetikzlibrary{calc}

\usepackage{hf-tikz}

\newcounter{theExample}
\newenvironment{Example}{
	\par\smallskip\small\refstepcounter{theExample}%
	\noindent\hspace*{-0em}\textbf{Example~\arabic{theExample}}:~%
	\leftskip0em\ignorespaces%
}{
	\par\smallskip
}
\newcounter{theRemark}
\newenvironment{Remark}{
	\par\smallskip\refstepcounter{theRemark}%
	\noindent%
	\textbf{Remark~\arabic{theRemark}}:~%
	\ignorespaces%
}{
	\par\smallskip
}

\usepackage{enumitem}
\usepackage{xr}
\usepackage{units}

\algnewcommand\algorithmicoutput{\textbf{Output:}} 
\algnewcommand\Output{\item[\algorithmicoutput]}
\algnewcommand\algorithmicinput{\textbf{Input:}} 
\algnewcommand\Input{\item[\algorithmicinput]}

\usepackage{xr}
\begin{document}
\title{Graph Signal Processing -- Part III: \\
	Machine Learning on Graphs, from Graph Topology to Applications}

\author[etf]{Ljubi\v{s}a Stankovi\'{c}}
\ead{ljubisa@ucg.ac.me}
\author[ic]{Danilo Mandic}
\ead{d.mandic@imperial.ac.uk}
\author[etf]{Milo\v{s} Dakovi\'{c}} 
\ead{milos@ucg.ac.me} 
\author[etf]{\\Milo\v{s} Brajovi\'{c}}
\ead{milosb@ucg.ac.me}
\author[ic]{Bruno Scalzo}
\ead{bruno.scalzo-dees12@imperial.ac.uk}
\author[ic]{Shengxi Li}
\ead{shengxi.li17@imperial.ac.uk}
\author[ic]{Anthony G. Constantinides}
\ead{a.constantinides@imperial.ac.uk}

\address[etf]{University of Montenegro, Podgorica, Montenegro }
\address[ic]{Imperial College London, London, United Kingdom }

\date{Received: date / Accepted: date}

\setcounter{tocdepth}{3}


\begin{abstract} 
Many modern data analytics applications on graphs operate on domains where graph topology is not known a priori, and hence its determination becomes part of the problem
definition, rather than serving as prior knowledge which aids the problem solution. Part III of this monograph starts by addressing ways to learn  graph topology, from the case where the physics of the problem already suggest a possible topology, through to most general cases  where the graph topology is learned from the data. A particular emphasis is on graph topology definition based on the correlation and precision matrices of the observed data, combined with additional prior knowledge and structural conditions, such as the smoothness or sparsity of graph connections. For learning sparse graphs (with small number of edges), the least absolute shrinkage and selection operator, known as LASSO is employed, along with its graph specific variant, graphical LASSO. For completeness, both variants of LASSO are derived in an intuitive way, and explained.  An in-depth elaboration of the graph topology learning paradigm is provided through several examples on physically well defined graphs, such as electric circuits, linear heat transfer, social and computer networks, and spring-mass systems. As many graph neural networks (GNN) and convolutional graph networks (GCN) are emerging, we have also reviewed the main trends in GNNs and GCNs, from the perspective of graph signal filtering. We have in particular studied the diffusion process over graphs and have shown that the trend of various improvements on GCNs can also be understood from the graph diffusion perspective. Given that the existing GCNs have been introduced largely in a heuristic manner, the definition of different diffusion processes can also serve as a basis for a new design of GCNs. Tensor representation of lattice-structured graphs is next considered, and it is  shown that tensors (multidimensional data arrays) are a special class of graph signals, whereby the graph vertices reside on a high-dimensional regular lattice structure. This part of monograph concludes with two emerging applications in financial data processing and underground transportation networks  modeling.  By means of portfolio cuts of an asset graph, we show how domain knowledge can be meaningfully incorporated into  investment analysis. In the underground traffic example, we demonstrate how graph theory can be used to identify the stations in the London underground network which have the greatest influence on the functionality of the traffic, and proceed, in an innovative way, to assess the impact of a station closure on service levels across the city.
\end{abstract}

\maketitle

\setcounter{tocdepth}{3}

\tableofcontents

\section{Introduction}

 Graph data analytics has already shown its enormous potential, as its flexibility in the choice of graph topologies (irregular data domains) and connections between the entities (vertices) allows for both a rigorous account of irregularly spaced information such as locations and social connections, and also for the incorporation of semantic and contextual cues, even for otherwise regular structures such as images.
 
In Part I and Part II of this monograph, it was assumed that the graph itself is already defined prior to analyzing data on graphs. The focus of Part I has been on defining graph properties through the mathematical formalism of linear algebra, while Part II introduced graph counterparts of several important standard data analytics algorithms, again for a given graph. However, in many modern applications,  graph topology is not known a priori \cite{friedman2008sparse,meinshausen2006high,pavez2016generalized,pourahmadi2011covariance,epskamp2018tutorial,das2017interpretation,dong2016learning,Dong,stankovic2017LLLvertex,stankovic2018reduced,stankovic2019intuitive,rabiei2019estimating,hamon2019transformation,cioacua2019graph}, and the focus of this part is therefore on simultaneous estimation of data on a graph and the underlying graph topology. Without loss of generality, it is convenient to  assume that the vertices are given, while the edges and their associated weights are part of the  solution to the problem considered and need to be estimated from the vertex geometry and/or the observed data   \cite{slawski2015estimation,ubaru2017fast,caetano2009learning,Thanou2014,camponogara2015models,zhao2012huge,yankelevsky2016,Zheng,segarra2016blind,stankovic2017vertex,stankovic2019vertexTEL,pasdeloup2019uncertainty,dal2019wavelet,tanaka2019oversampled,bohannon2019filtering,gu2019local,mao2019time}. 

Three scenarios for the estimation of graph edges from vertex geometry or data  are considered:

\begin{itemize}
	\item Based on the \textit{geometry of vertex positions}. In various sensing network setups (such as temperature, pressure, and transportation), the locations of the sensing positions (vertices) are known beforehand, while the vertex distances convey physical meaning about data dependence and  thus may be employed for edge/weight determination. 
	  \item 
	Based on \textit{data association and data similarity.} Various approaches are available to serve as data association metrics, with the covariance and precision matrices  most commonly used.  A strong correlation between data on two vertices would indicate a large weight associated with the corresponding edge. A small degree of correlation would indicate nonexistence of an edge (after thresholding).
 \item 
 \textit{ Physically well defined relations among the sensing positions}. Examples include electric circuits, power networks, linear heat transfer, social and computer networks, spring-mass systems, to mention but a few.  In these cases, edge weighting can be usually well defined based on the underlying physics of the considered problem.  
  
\end{itemize}
 
 Each of these scenarios has been considered in this part of the monograph. After a detailed elaboration of graph definition and graph topology learning techniques, a summary of graph learning from data using probabilistic generative models is given. In the sequel, graph neural networks (GNN) are reviewed, with a special attention to the convolutional graph networks (GCN). The analysis is considered from the perspective of graph signal filtering presented in Pat II. Graph  data analysis is further  generalized to the tensor representation of lattice-structured graphs, whereby the graph vertices reside on a high-dimensional tensor structure. At the end of this part of monograph, two applications of graph-based analysis are given: i) An example for domain knowledge being incorporated into a financial data analysis (the investment analysis), by means of portfolio cuts; (ii) the graph data processing framework is also applied to the underground traffic system. The later example demonstrates how graph theory can be used to identify the stations in the London underground network which have the greatest influence on the functionality of the traffic, and to assess the impact of a station closure on service levels across the city.  
 
\section{Geometrically Defined Graph Topologies} 

For a graph that  corresponds to a network with  geometrically distributed vertices, it is natural to relate the edge weights with the distance between vertices. Consider vertices $m$ and $n$ whose locations in space are defined by the position vectors (coordinates) $\mathbf{r}_m$ and $\mathbf{r}_n$. The Euclidean distance between these two vertices is then
$$r_{mn}=\text{distance}(m,n)=\left\Vert \mathbf{r}_m-\mathbf{r}_n \right\Vert_2.$$ 
 A common way to define the graph weights in such networks is through an exponentially decaying function of the distance, for example as
\begin{equation}
W_{mn}=\begin{cases}
e^{-r_{mn}^2/\tau ^ 2},  & \text{ for } r_{mn}\le \kappa\\
0, & \text{ for } r_{mn}> \kappa \text{ and } m=n,
\end{cases} \label{GaussWE}
\end{equation}
where $r_{mn}$ is the Euclidean distance between the vertices $m$ and $n$, and $\tau$ and $\kappa$ are chosen constants.  
This is also physically well justified, as based on $e^{-r_{mn}^2/\tau ^ 2 }$ the weights tend to $1$ for closely spaced vertices and diminish for distant vertices.

The rationale for this definition of edge weights is the assumption that the signal value measured at a vertex $n$ is similar to signal values measured at its neighboring vertices.
Then, the estimation of a signal at a vertex $n$ should also involve neighboring vertices  connected with  larger weights (close to $1$), while the signal values sensed at farther vertices would be less relevant, and are  associated with smaller weighting coefficients or are not included at all.

The Gaussian function, used in (\ref{GaussWE}), is appropriate in many applications, however, other forms to penalize data values associated with the vertices which are far from the considered vertex may also be used. Examples of such functions include various kernels, such as the kernel
\begin{equation}
W_{mn}=\begin{cases}
e^{-r_{mn}/\tau},  & \text{ for } r_{mn}\le \kappa\\
0, & \text{ for } r_{mn}> \kappa \text{ and } m=n
\end{cases} \label{GaussWEE}
\end{equation}
or the inverse Euclidean distance between vertices $m$ and $n$, given by
\begin{equation}
W_{mn}=\begin{cases}
\frac{1}{r_{mn}},  & \text{ for } r_{mn}\le \kappa\\
0, & \text{ for } r_{mn}> \kappa \text{ and } m=n. 
\end{cases} \label{InvLinWEE}
\end{equation}
Obviously, the simplest form for the edge weighting coefficients is a binary scheme
\begin{equation}
W_{mn}=A_{mn}=\begin{cases}
1,  & \text{ for } r_{mn}\le \kappa\\
0, & \text{ for } r_{mn}> \kappa \text{ and } m=n, 
\end{cases} \label{GaussWEEA}
\end{equation} 
which corresponds to an unweighted graph, with $\mathbf{W}=\mathbf{A}$.

\begin{Example}
We shall illustrate the geometry-based formation of graph structure on the well-known Swiss manifold as a domain for data acquisition This is a  three-dimensional surface with the space coordinates, $(x,y,z)$, defined as functions of two parameters, $u$ and $v$, in the following form 
\begin{gather}   
    x=\frac{1}{4\pi} v\cos(v) \nonumber \\
    y=u \nonumber \\
    z=\frac{1}{4\pi} v\sin(v). \label{coordSwiis}
\end{gather}
 The Swiss manifold shown in Fig. \ref{VF_graph3ab_Part3}(a) was created for the parameters, $u$ and $v$ within the intervals $-1 \le u \le 1$ and 
$\pi \le v \le 4\pi $. 
 
More specifically, we considered a graph with $N=100$ vertices, which were randomly placed on the Swiss roll surface, with the coordinates $(x_k,y_k,z_k)$, $k=1,2,\dots,N$, whereby
\begin{gather*} 
    u_k \text{ was unform random within } -1 \le u_k \le 1 \\
    v_k  \text{ was unform random within } \pi \le v_k \le 4\pi.  
\end{gather*}
The vertices were connected with edges whose weights are defined as in (\ref{GaussWE}), that is
$$W_{mn}=\exp(-r_{mn}^2/\tau^2),$$
for $r_{mn}>0.6$, with $W_{mn}=0$ for  $r_{mn} \le 0.6$ and $m=n$; $\tau=1/2$. The symbol $r_{mn}$ denotes the shortest geodesic distance between the vertices $m$ and $n$, measured along the Swiss roll manifold, in the following way
 \begin{gather*} r^2_{mn}=l^2_{mn}+(y_m-y_n)^2 \\
 l_{mn}=\frac{1}{4\pi}\int_{v_m}^{v_n}\sqrt{\Big(\frac{d(v\cos(v))}{dv}\Big)^2+\Big(\frac{d(v\sin(v))}{dv}\Big)^2}dv\\
 =\frac{1}{4\pi}\int_{v_m}^{v_n}\sqrt{1+v^2}dv \\
 =\Big(\frac{1}{2} v\sqrt{v^2 + 1}  + \frac{1}{2} \ln (\sqrt{v^2 + 1} + v) \Big) \Big|_{v_m}^{v_n}.
 \end{gather*} 
 Small weight values were hard-thresholded to zero, in order to reduce the number of edges associated with each vertex by keeping only a few  strongest ones.
 
 The so produced three-dimensional graph is shown  in Fig. \ref{VF_graph3ab_Part3}(b), and its two-dimensional presentation in Fig. \ref{VF_graph3ab_Part3}(c). The vertices were ordered so that the values of the Fiedler eigenvector, $u_1(n)$, were nondecreasing; the vertices were colored based on the two-dimensional and three-dimensional spectral vectors,  $\mathbf{q}_n=[u_1(n),u_2(n)]$ and $\mathbf{q}_n=[u_1(n),u_2(n),u_3(n)]$ of the Swiss role in Fig. \ref{VF_graph3ab_Part3}(d) and (e). This kind of vertex marking can also be used for clustering with, for example,  the $k$-means clustering presented in {\color{red}Part 1, Remark \ref{I-clusteringRem}}.
\begin{figure*}
	\centering
	
	\includegraphics[]{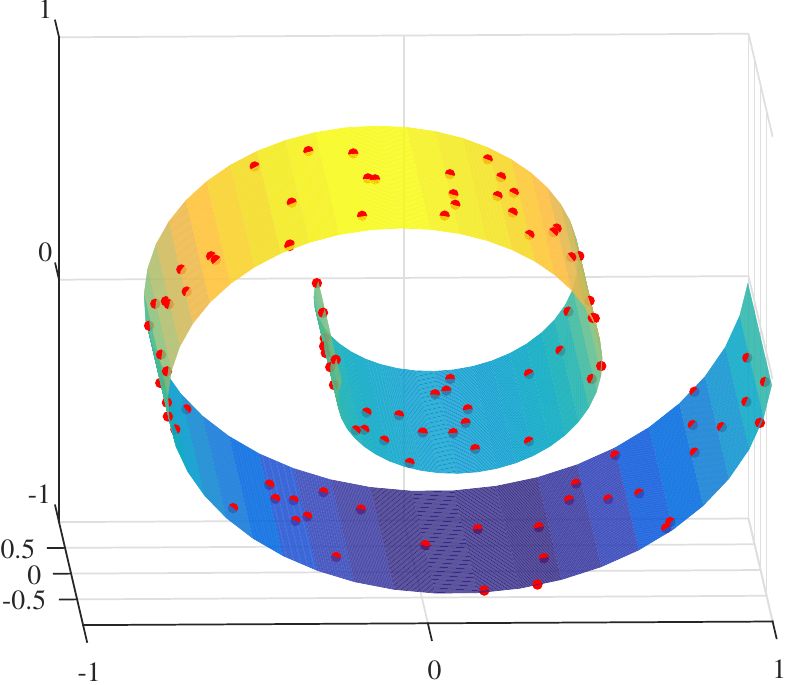}\hspace{2mm}(a)\hspace{5mm}
	\includegraphics[]{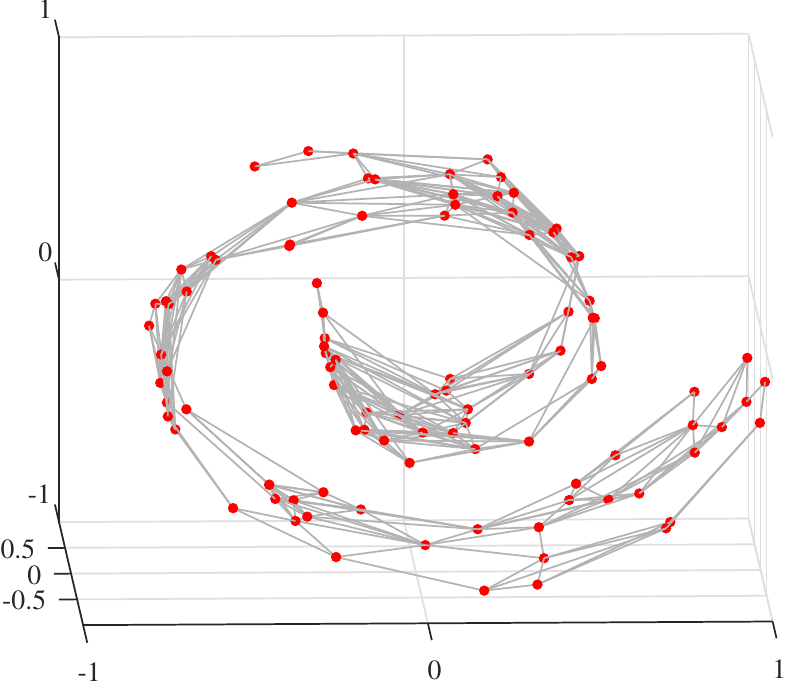}\hspace{2mm}(b)
	
	\vspace{10mm}
	
	\includegraphics[]{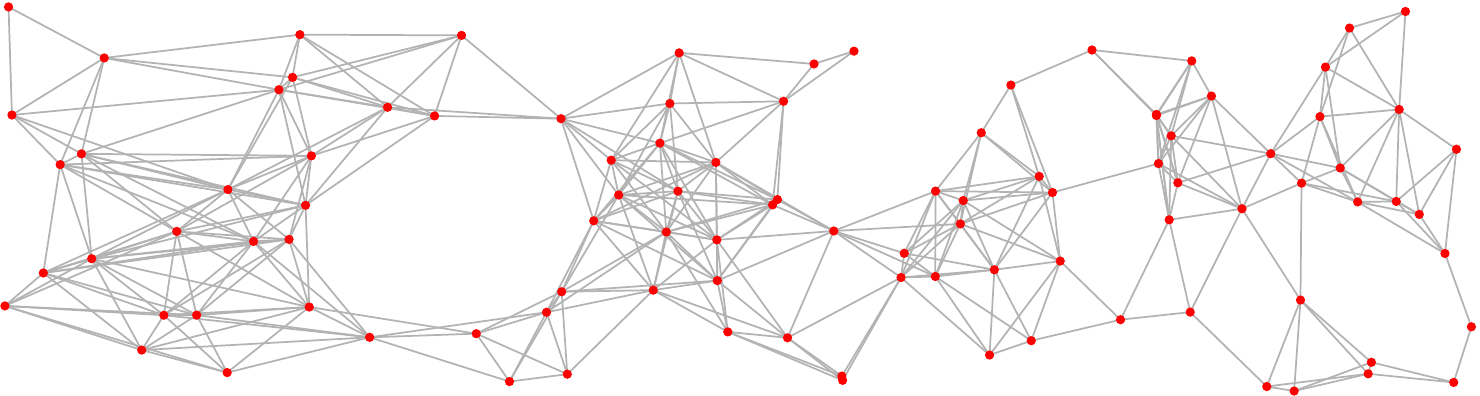}\hspace{7mm}(c)
	
	\vspace{10mm}
	
	\includegraphics[]{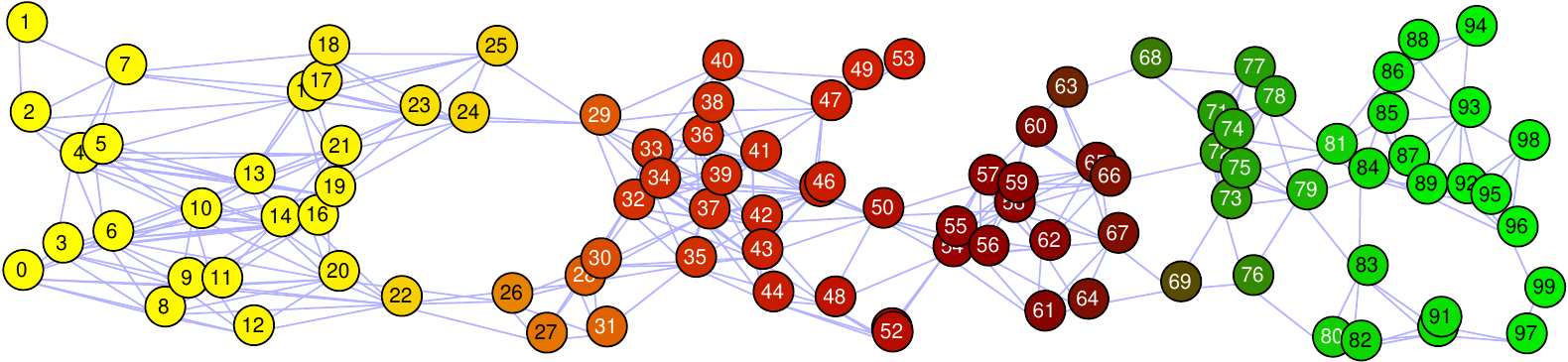}\hspace{7mm}(d)
	
	\vspace{10mm}
	
	\includegraphics[]{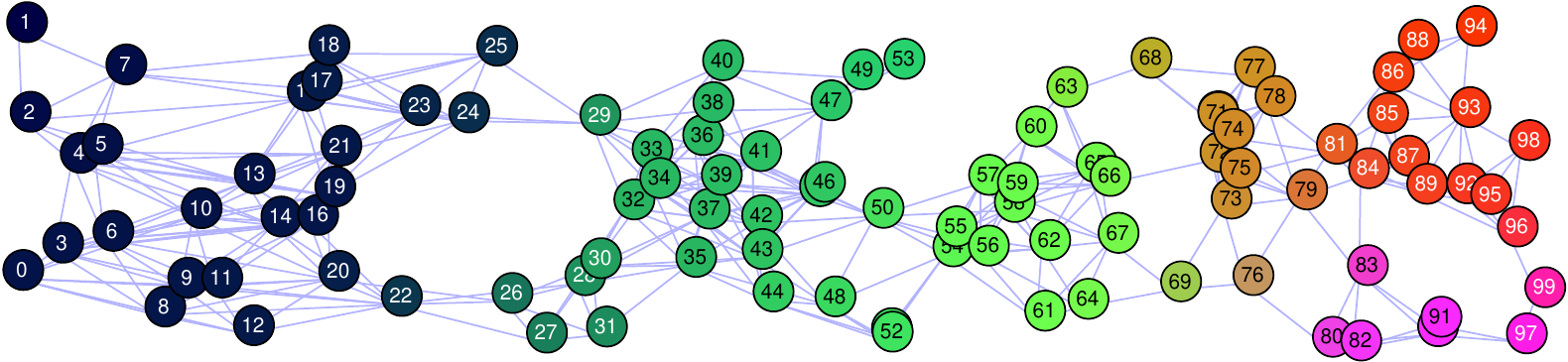}\hspace{7mm}(e)
	 
%
%
	
	\caption{Concept of graph definition based on problem geometry. (a) Vertices (points) on a three-dimensional  manifold called the Swiss roll surface. (b) A graph representation on the Swiss roll manifold. (c) Two-dimensional presentation of the three-dimensional graph from (b) obtained by unfolding the original 3D surface. (d) Vertices colored using the spectral vector, $\mathbf{q}_n=[u_1(n),u_2(n)]$, formed from the two smoothest \textit{generalized eigenvectors} of the graph Laplacian, $\mathbf{u}_1$ and $\mathbf{u}_2$. (e) Vertices colored using the spectral vector, $\mathbf{q}_n=[u_1(n),u_2(n),u_3(n)]$, formed from the three smoothest \textit{eigenvectors} of the graph Laplacian, $\mathbf{u}_1$, $\mathbf{u}_2$, and $\mathbf{u}_3$. The vertex indexing in (d) and (e) is performed based on the sorted values of the smoothest (Fiedler) eigenvector, $\mathbf{u}_1$.}
	\label{VF_graph3ab_Part3}
\end{figure*}
\end{Example}

\noindent\textbf{Classical Gaussian filter within graph topology formulation.} To illustrate this classical operation on the discrete-time domain data, assume that we desire to perform classical smoothing of a discrete-time domain signal, $x(n)$, at a vertex/instant $n$, through a moving average operation on data at neighboring vertices/instants, $x(m)$, using a truncated Gaussian weighting function given by 
	$$g(m,n)=e^{-(m-n)^2/\tau^2}$$
for $|m-n|\le \kappa$ and $g(m,n)=0$ for  $|m-n|> \kappa$.
The smoothed discrete-time domain signal, $y(n)$, can be expressed in classical data analysis as 
\begin{equation}y(n)=\sum_m e^{-\frac{(m-n)^2}{\tau^2}} x(m) \label{smoothineq}
\end{equation}
where the summation is performed for instants/vertices $m$ such that $|n-m| \le \kappa$. 

We shall now reformulate this classical data processing problem  within the graph topology framework.  The distance between the sampling instants/vertices, $\text{distance}(m,n)$, plays a crucial role in the smoothing, and is defined  as
$$\text{distance}(m,n)=r_{mn}=\left\Vert m-n \right\Vert_2=|m-n|.$$

The corresponding edge weights can be defined based on the Gaussian smoothing function, and are given by $W_{mn}=e^{-r_{mn}^2/\tau^2}$ for $r_{mn}\le \kappa$, and $W_{mn}=0$ for  $r_{mn}> \kappa$  and $m=n$.

 The classical smoothed signal, $y(n)$, defined in (\ref{smoothineq}) can now be expressed in the form appropriate for graph framework as 
$$y(n)=x(n)+\sum_m x(m)W_{mn}=x(n)+\sum_m e^{-\frac{(m-n)^2}{\tau^2}} x(m)$$
where the summation is performed for vertices $m$ such that $|m-n| \le \kappa$ and $m \ne n$. 
This operation can be defined within the graph analysis framework as a simple first order system on graph given by
$$
\mathbf{y}=\mathbf{W}^0\mathbf{x}+\mathbf{W}^1\mathbf{x}
$$
where the edge weights between the vertices $m$ and $n$ are defined by $W_{mn}$.

\begin{figure}[h]
	\centering
	\includegraphics[]{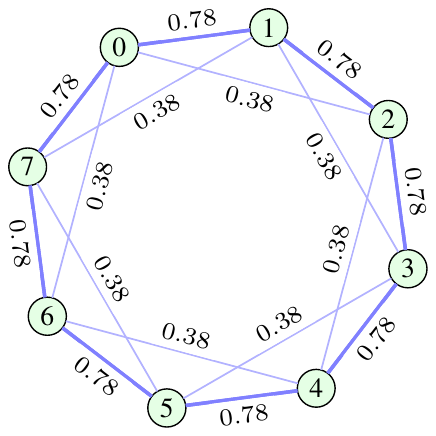}
	\caption{Graph which corresponds to the weighted moving average operator with Gaussian weights given in (\ref{smoothineq}).}
	\label{fig-spec-graph_Cl_Fil}
\end{figure}

For example, for $\tau=2$ and $\kappa=2$, the edge weights $W_{mn}$ are shown in Fig. \ref{fig-spec-graph_Cl_Fil} and this graph-based formulation is identical to the classical discrete-time domain weighted moving average  
\begin{align}
y(n)\!=\!x(n)+\!\sum_{m}W_{mn}x(m)\!=\!\!\sum_{m=n-2}^{n+2} \!\!e^{-\frac{(m-n)^2}{4}}x(m),
\end{align}
with the output signal samples, $y(n)$, being equal to the output of a first-order system on the graph given by
$$
\mathbf{y}=\mathbf{W}^0\mathbf{x}+\mathbf{W}^1\mathbf{x}=3.32\mathbf{L}^0\mathbf{x}-\mathbf{L}^1\mathbf{x}.
$$

For image input data, where the vertices correspond to the pixel positions and the Euclidean distance between pixels is used to model the image domain as a graph, the previous example would model a moving average filtered image, using a radial Gaussian window.

\begin{figure}[]
	\centering
	\includegraphics[]{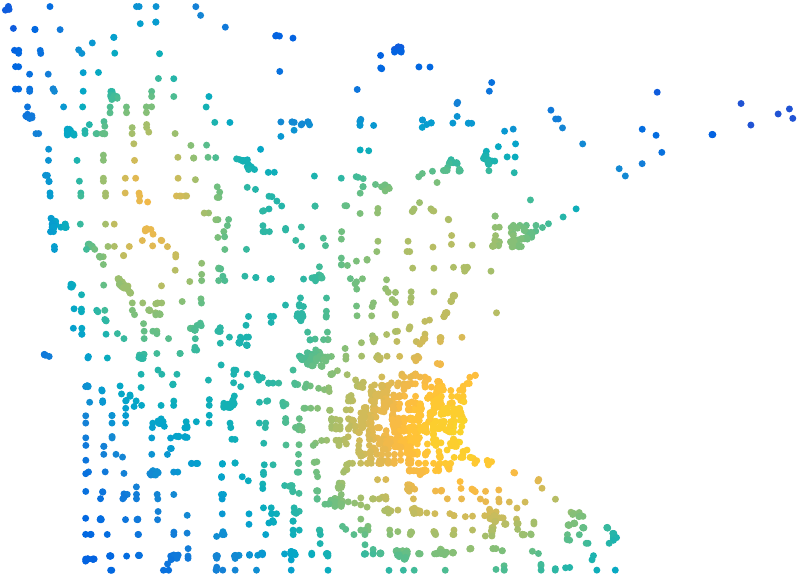}
	\\(a)
	\\ \vspace{5mm}
	\includegraphics[]{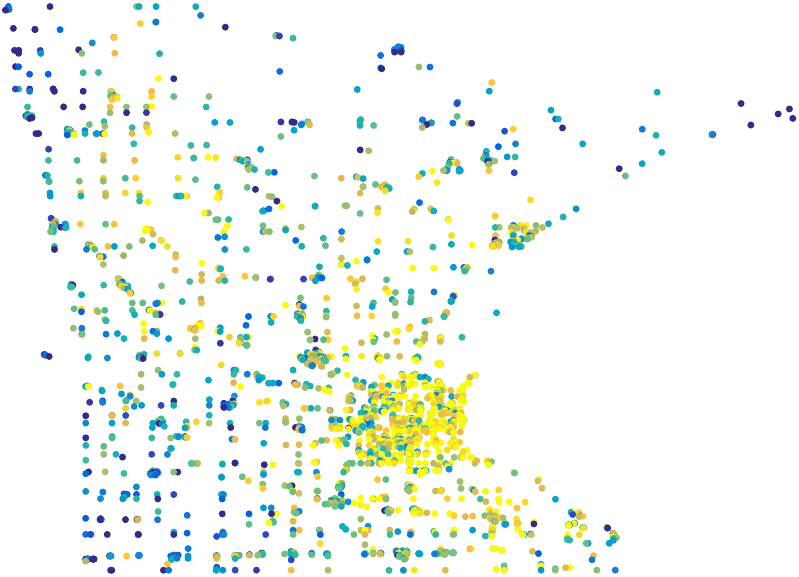}\\
	(b)
	\\ \vspace{5mm}
	\includegraphics[]{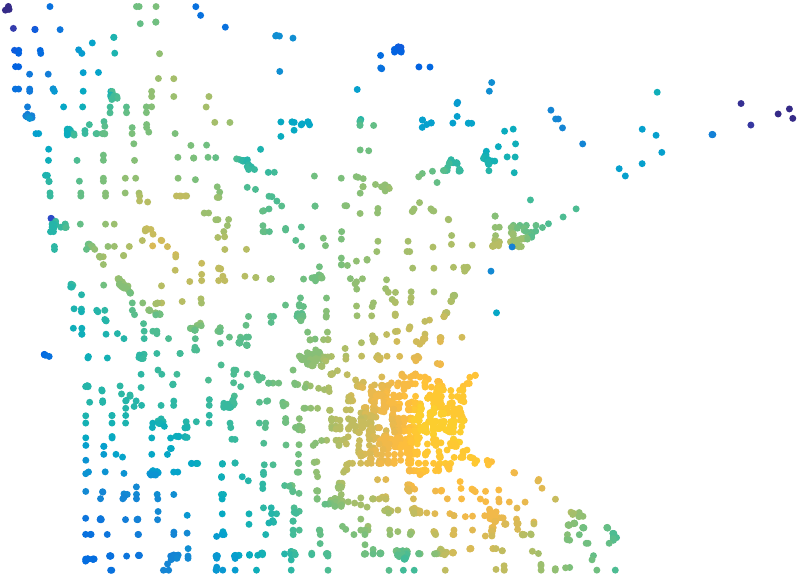}\hspace{5mm}\\
	(c)
	\caption{Temperatures simulated on the Minnesota roadmap graph. (a) Original synthetic temperature field signal. (b) Noisy temperature signal. (c) Low-pass filtered temperature signal from (b). The signal values are designated by the corresponding vertex color.}
	\label{minne_edges2}
\end{figure}

\begin{Example} Consider the benchmark Minnesota roadmap graph, for which the connectivity map (adjacency matrix) is designated by the road connections and the vertices are at the road crossings. The edges are defined by the adjacency matrix and were weighted according to their Eucledian distances using the weighting scheme in (\ref{GaussWEE}), with $\tau=25$km, to give
	$$W_{mn}=e^{-r_{mn}^2/\tau ^ 2},$$	
	where the threshold $\kappa$ was not used since the connectivity is already determined by the given adjacency matrix. 
	
	We considered a simulated temperature signal in the Minnesota area (normalized temperature filed) which was calculated as 
	\begin{gather*}x(n)=0.9\Big( 0.1+0.8e^{-(\frac{x-150}{100})^2-(\frac{y-400}{200})^2}, \\ + 0.5e^{-(\frac{x-450}{200})^2-(\frac{y-400}{100})^2} 
	+ e^{-(\frac{x-500}{250})^2-(\frac{y-150}{200})^2} \Big)+\nu(n)
	\end{gather*}
	where $\nu(n)$ is white Gaussian noise with standard deviation $\sigma_{\nu}=0.3$. The  noise-free and noisy version of this graph temperature signal are given respectively in Fig. \ref{minne_edges2} (a) and (b). The noisy signal was filtered in the vertex domain by a low-pass filter implemented using Taubin's $\alpha-\beta$ algorithm (presented in {\color{red} Part II, Section \ref{II-TaubSec}}) with $\alpha=0.15$ and $\beta=0.1$, and the so enhanced temperature signal is shown in Fig. \ref{minne_edges2} (c). For the input SNR of 9.35 dB the output SNR of 19.34 dB was achieved, a gain of 10 dB.
\end{Example}

 \section{Graph Topology Based on Signal Similarity}
 
 In the previous sections, graph weights were defined on the assumption that the geometric distance of vertices, where the signal is sensed, is a reliable indicator of  data similarity, or a more general data association. Indeed, this is the case with, for example, the measurements of atmospheric temperature and (barometric) pressure when the terrain configuration has no influence on the similarity of measured data. However, in general, the geometric distance between vertices may not be a good indicator of data similarity. 
 
One such example is in image processing, where the pixel color values themselves can be used as an indicator of  signal similarity; this can be achieved in combination with the distances between pixels, which play the role of vertices. If the  intensity values at pixels indexed by $m$ and $n$ are denoted by $x(m)$ and $x(n)$, then the difference of intensities is defined by  
 $$\text{Intensity distance}(m,n)=r_{mn}=|x(m)-x(n)|,$$
and the corresponding weights may be defined as 
 $$W_{mn}=\begin{cases}
 e^{-(x(m)-x(n))^2/\tau^2}, & \text{ for } r_{mn}\le \kappa\\
 0, & \text{ for } r_{mn}> \kappa \text{ and } m=n,
 \end{cases}$$
 where $r_{mn}$ is a geometric distance between the considered pixels/vertices and $\tau$ and $\kappa$ are chosen constants.

More reliable measures of data similarity can be defined when it is possible to collect more than one snapshot data for a given set of sensing points/vertices. Assume that at every vertex $n=0,1,\dots,N-1$ we have acquired $P$ signal values, denoted by $x_p(n),~p=1,2,\dots,P$. Such a dataset may be equally treated as multivariate data or signal measurements in a sequence. Then, an appropriate similarity measure function for a real-valued signal at vertices $m$ and $n$ may be
\begin{gather}
r^2_{mn}=
 \frac{\sum_{p=1}^P\big(x_p(m)-x_p(n)\big)^2
 }{
   \sum_{m=1}^{N-1}\sum_{n=1}^{N-1}\sum_{p=1}^P \Big(x_p(m)-
    x_p(n)\Big)^2
  } \label{eqcorrecoef}
\end{gather}
so that $\sum_{m=1}^{N-1}\sum_{n=1}^{N-1}r^2_{mn}=1$.
 
 The graph weights can again be defined using any of the previous forms, for example,
 $$W_{mn}=\begin{cases}
 e^{-r_{mn}^2/\tau^2}, & \text{ for } r_{mn}\le \kappa\\
 0, & \text{ for } r_{mn}> \kappa \text{ and } m=n,
 \end{cases}$$
or
 $$W_{mn}=\begin{cases}
 e^{-r_{mn}/\tau}, & \text{ for } r_{mn}\le \kappa\\
 0, & \text{ for } r_{mn}> \kappa \text{ and } m=n.
 \end{cases}$$
 
 \noindent\textbf{Random observations.}  When the signal values, $x_p(n)$, acquired over $P$ observations, $p=1,2,\dots,P$ at $N$ vertices $n=0,1,\dots,N-1$, are drawn from zero-mean random noise with equal variances $\sigma^2_x=1$, the similarity measure can be defined by   
 $$r^2_{m,n}=
 \frac{\sum_{p=1}^P\big(x_p(m)-x_p(n)\big)^2
 }{\sqrt{
   \sum_{p=1}^P x_p^2(m)
   \sum_{p=1}^P x_p^2(n)
  }  }=2\Big(1-R_x(m,n)\Big)$$
 where 
 $$R_x(m,n)=\frac{1}{P}\sum_{p=1}^Px_p(m)x_p(n)$$
 represents the normalized autocorrelation function and $\sigma^2_x=\frac{1}{P}\sum_{p=1}^P x_p^2(n)=1$ for sufficiently large $P$.
 
 \medskip
 
 \noindent\textbf{Similarity metrics for images.} The same structure can be used for other applications, such as image classification or handwriting recognition. In these cases, the \textit{distance between an image $m$ and  an image $n$} is equal to  
 \begin{equation}
 	r_{mn}=\text{Image distance}(m,n)=\Vert \mathbf{x}_m-\mathbf{x}_n \Vert_F, \label{imagedist}
 \end{equation}
 where  \begin{equation*}
 \Vert \mathbf{x} \Vert_F=\sqrt{\sum_{m}\sum_{n}|x(m,n)}|^2.
 \end{equation*} is the Frobenius norm  of an image  matrix $\mathbf{x}$ (that is, the square root of the sum of  squared image values over all pixels).
 
 \noindent\textbf{Block collaborative image processing.} A class of recent efficient image processing algorithms is based on detecting similar blocks within an image, followed by collaborative processing using those similar blocks. Image enhancement algorithms then assume that the basic images are also similar within these blocks, while the corresponding noise is not related and can be averaged out. The similarity between the image blocks, $\mathbf{x}_m$ and $\mathbf{x}_n$,  may then be defined similar to (\ref{imagedist}), using their distance given by
 $$r_{mn}=\text{Block distance}(m,n)=\Vert \mathbf{x}_m-\mathbf{x}_n \Vert_F.$$
 The similarity among the blocks in an image can be modeled by a graph, and such graph models may be used as bases for collaborative processing of image blocks. Recall that a block of $B \times B$ pixels is an example of a vertex in a $B^2$-dimensional space, since it is defined by $B \times B$ independent pixel values (vertex coordinates).
 
 \noindent\textbf{Generalized distance measure.} The Euclidean distance is typically used in the calculation of the distance between two blocks of data, $\mathbf{x}_m$ and $\mathbf{x}_n$. It may be generalized by introducing the inner product matrix, $\mathbf{H}$, into distance calculation to yield
 $$r^2_{mn}=(\mathbf{x}_m-\mathbf{x}_n)^T\mathbf{H}(\mathbf{x}_m-\mathbf{x}_n),$$ 
 where the data sets $\mathbf{x}_m$ and $\mathbf{x}_n$ are represented in the column vector form. When the inner product matrix, $\mathbf{H}$, is an identity matrix, $\mathbf{H}=\mathbf{I}$, the standard Euclidean distance is obtained. If we use, for example, $\mathbf{H}=\mathbf{U}_C\mathbf{U}^T_C$, where $\mathbf{U}_C$ is the matrix with cosine transform basis functions as its columns, we will arrive at         
\begin{gather*}r^2_{mn}=(\mathbf{x}_m-\mathbf{x}_n)^T\mathbf{U}_C\mathbf{U}^T_C(\mathbf{x}_m-\mathbf{x}_n) \\
=(\mathbf{C}_m-\mathbf{C}_n)^T(\mathbf{C}_m-\mathbf{C}_n)=\Vert \mathbf{C}_m -\mathbf{C}_n \Vert _2^2,
\end{gather*}
where $\mathbf{C}_n$ is the  2D discrete cosine transform (2D DCT) of $\mathbf{x}_n$, written in a vector column format. By virtue of this representation, problem dimensionality can straightforwardly be reduced using only the $K$ slowest-varying basis functions, $\mathbf{U}^{(K)}_C$, instead of the full 2D DCT transformation matrix (this operation corresponds to low-pass filtering of  $\mathbf{x}_n$ in the 2D DCT domain, by keeping the $K$ slowest varying coefficients). In this case, the distance, $r^2_{mn}$, is of the form 
\begin{gather*}r^2_{mn}=(\mathbf{x}_m-\mathbf{x}_n)^T\mathbf{U}^{(K)}_C \mathbf{U}^{(K)^T}_C (\mathbf{x}_m-\mathbf{x}_n) \\
=\Vert \mathbf{C}^{(K)}_m -\mathbf{C}^{(K)}_n \Vert _2^2,
\end{gather*}    
and is calculated based on the reduced original dimensionality of $\mathbf{x}_n$ or $\mathbf{C}_n$  to the dimensionality $K$ of $\mathbf{C}^{(K)}_n$.   

 Another interesting form of the inner product matrix is the inverse covariance matrix $\mathbf{H}=\boldsymbol{\Sigma}^{-1}$, which will be discussed later in Section \ref{secGLASSO} and Section \ref{GrafRanDSIG}.
  
\begin{Example}
	A noisy image with a designated set of 29 blocks of pixels is shown in Fig. \ref{Collarbor_graph_Image} (a). The similarity between any two of the blocks was defined based on the distance 
	$$r^2_{mn}=\frac{1}{B^2}||\mathbf{C}_m-\mathbf{C}_n||_F^2,$$
	where $\mathbf{C}_n$ represents the matrix form of the 2D DCT of the image block $\mathbf{x}_n$. 
	
	The 2D DCT was then hard-thresholded, with a threshold equal to $0.1\max|\mathbf{C}_n|$, to reduce the influence of noise (and problem dimensionality), that is, all 2D DCT coefficients bellow this threshold were set to zero
	$$C_n(k,l)=\begin{cases} C_n(k,l), & \text{ if } |C_n(k,l)|>0.1\max|\mathbf{C}_n| \\
	0, & \text{elsewhere}. \end{cases} $$
	
	The edge weights, $W_{mn}$, for a graph representation of the considered blocks (as vertices) were then calculated as
	$$W_{mn}=\exp(-r_{mn}^2B),$$ 
	for
	$r_{mn} \le 0.26$, and $W_{mn}=0$ for $r_{mn} > 0.26$, and $m=n$, with $B=16$.
	
	 The so obtained graph, which indicates block similarity, is given in Fig. \ref {Collarbor_graph_Image}(b). This graph representation is very convenient  for collaborative image processing, since the graph structure will ensure that the processing is performed independently on the sets of blocks which share relevant information (connected subgraphs). Notice that the blocks within each subgraph can be considered as a 3D signal of RGB components. Then, for example, a simple averaging over  similar blocks (within one subgraph), will not significantly degrade the image detail, while at the same time it will reduce the corresponding noise, as it is uncorrelated in different blocks. 
	 
	 This is precisely the principle of the Block-Matching and 3D filtering (BM3D) algorithm, where the noise and the image are estimated from the set of similar blocks (in our example, from the blocks within a subgraph). The estimation of the related set of blocks in the image and the estimation of noise power is then used to define the Wiener filter. Such Wiener filter is used to filter all related blocks (within the subgraph). The procedure is repeated for each set of similar blocks (subgraphs). Of course, in the case of the BM3D algorithm, for each considered (reference) block, $\mathbf{x}_n$, it is desirable to search over the whole image and to find as many similar blocks as possible in order to obtain the best possible Wiener filter and consequently achieve maximum possible noise reduction.   
	    
	    In this example, the blocks and the threshold for edge weights, $W_{mn}$, were selected so as to produce  disconnected graph components and a clear segmentation scheme. If this was not the case, vertex clustering and graph segmentation could be performed using the theory presented in Part 1. 
	 
	\begin{figure}[]
		\centering
		\includegraphics[scale=0.9]{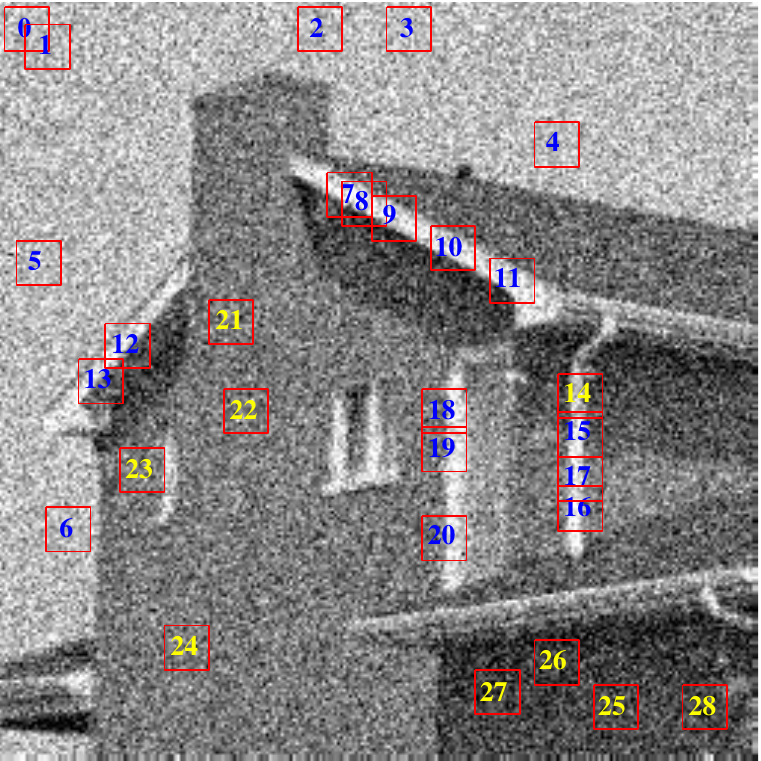}
		\\(a)
		\\ \vspace{3mm}
			\includegraphics[]{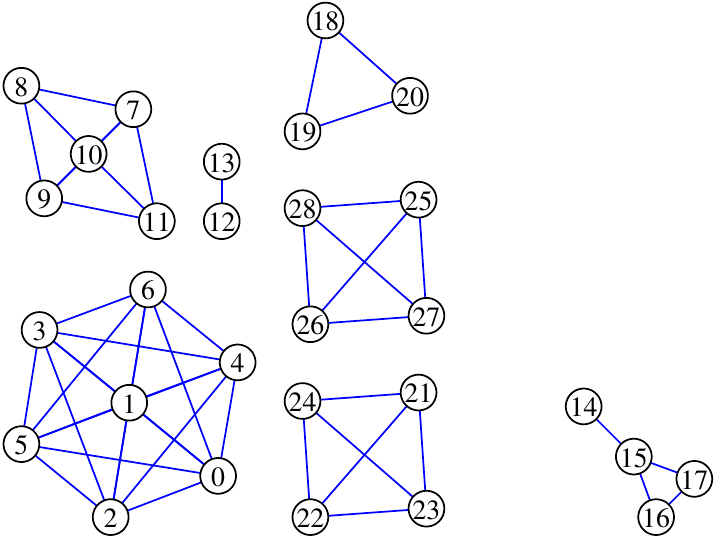}\\(b)
		\caption{Graph learning based  on the similarity of blocks of image data. (a) Original image with designated blocks of pixels. (b) The graph produced from the blocks in (a). Notice that the resulting graph consist of seven disconnected subgraphs, which correspond to the seven different groups of blocks.}
		\label{Collarbor_graph_Image}
	\end{figure}

	\end{Example}

 Recall that in {\color{red} Part I, Example \ref{I-Ex:images}} the \textit{structural similarity index (SSIM)}, was used instead of the simple difference/distance, to relate and cluster images.

	\begin{Example}
Eight images with the hand-written letter "b" were considered and the task was to create their graph representation.
The SSIM was calculated for each pair of images and the edge-weights were equal to the calculated SSIM values,  as shown in Fig. \ref{image-grpaph_Part_3}(a). For the graph  from Fig. \ref{image-grpaph_Part_3}(b), the generalized eigenvectors of the Laplacian were calculated and the vertices were colored using the smoothest (Fiedler) eigenvector, $\mathbf{u}_1$, and the smoothest two eigenvectors  $\mathbf{u}_1$ and $\mathbf{u}_2$,  as a basis for image clusterings, as respectively shown in Fig. \ref{image-grpaph_Part_3}(c) left and right.    
\begin{figure}
	\centering
	\includegraphics[scale=.55]{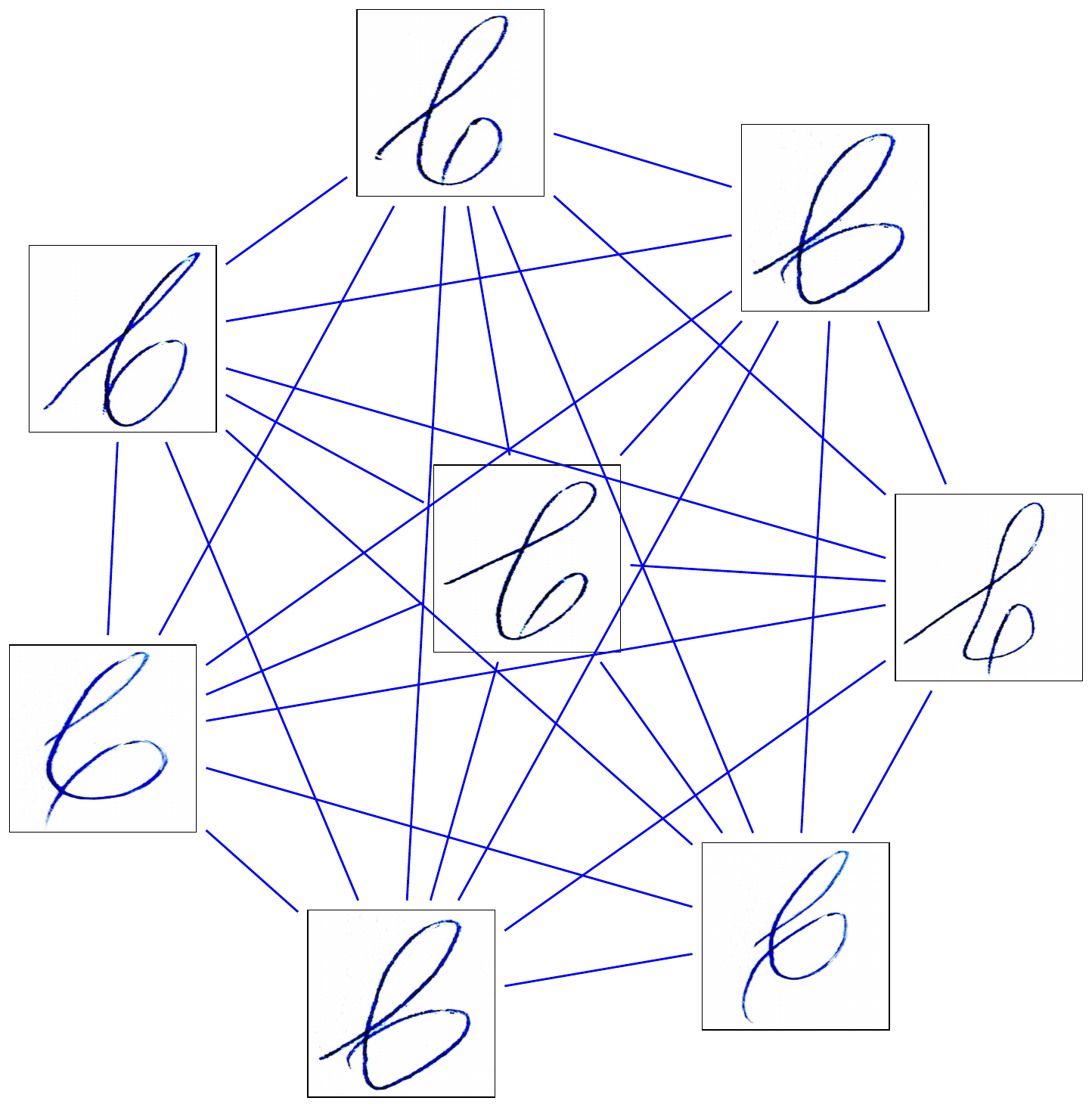} \\ (a)
	\\
	\vspace{2mm}
	\includegraphics[scale=0.9]{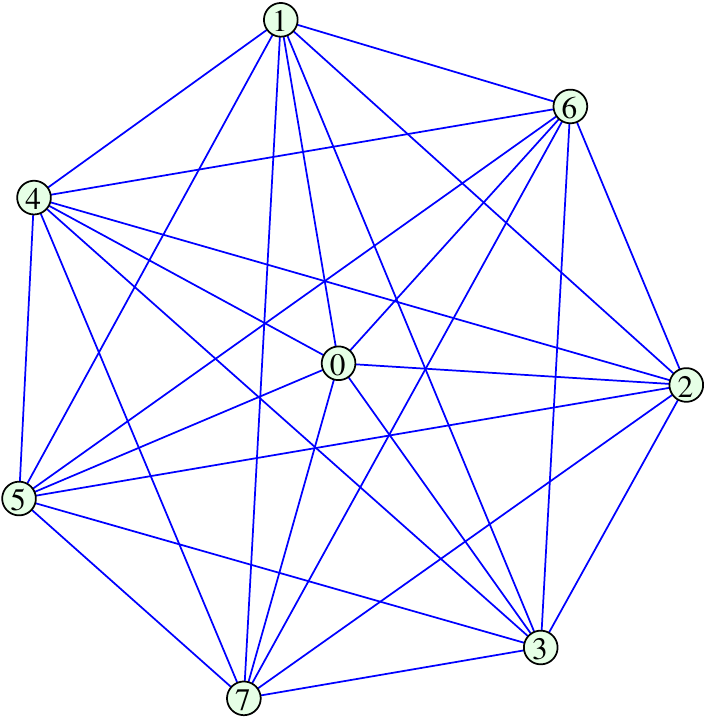}\\(b)
	\\
	\vspace{2mm}

	\includegraphics[scale=1]{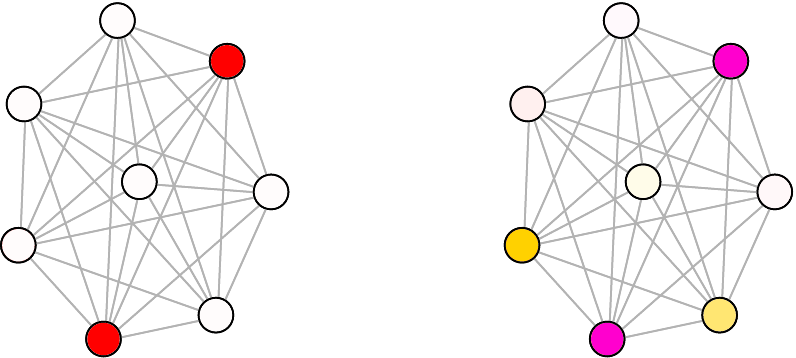} \\
	(c)
	\caption{Graph representation of a set of  hand-written images of the letter "b". The images serve as vertices, while the weight matrix for the edges is defined through the  structural similarity index (SSIM) between the images, with $W_{mn}=\mathrm{SSIM}(m,n)$.  The vertices are colored in (c) using the smoothest (Fiedler) eigenvector, $\mathbf{u}_1$, and the smoothest two eigenvectors,  $\mathbf{u}_1$ and $\mathbf{u}_2$,  of the generalized eigenvectors of the Laplacian (with spectral vectors $\mathbf{q}_n=[u_1(n)]$ and $\mathbf{q}_n=[u_1(n),u_2(n)]$) are respectively shown in Fig. \ref{image-grpaph_Part_3}(c) (left) and (right).}  
	\label{image-grpaph_Part_3}
\end{figure}
	\end{Example}

 \section{Learning of Graph Laplacian from Data}\label{CorrGraphLear}
 
 Consider a graph signal for which we have available $P$ independent observations. Denote the observed signal at a vertex, $n$, and for an observation, $p$, as $x_p(n)$. The column vector with graph signal samples from the $p$-th observation is denoted by $\mathbf{x}_p$. All observations from this graph signal can then be arranged into an $N\times P$ matrix, given by
 $$\mathbf{X}_P=
  \begin{bmatrix}
  \ \mathbf{x}_1, \  \ \mathbf{x}_2, \   \ \dots,   \ \mathbf{x}_P    
  \end{bmatrix} .
  $$
  Designate the $(n+1)$-th row of this matrix by a row vector, $\mathbf{y}_n$, which corresponds to the vertex $n$, that is
  \begin{equation}
  \mathbf{y}_n=\begin{bmatrix}
   \ x_1(n), \  \ x_2(n), \   \ \dots,   \ x_P(n)    
   \end{bmatrix}. \label{verttoR}
 \end{equation}
  Then, the matrix of observations can also be written as
  $$\mathbf{X}_P=
   \begin{bmatrix}
   \mathbf{y}_0 \\  \mathbf{y}_1 \\    \vdots   \\ \mathbf{y}_{N-1}    
   \end{bmatrix} .
   $$
  
  The correlation coefficient between vertices $m$ and $n$, estimated by averaging over the set of $P$ observations, is given by
  $$
 R_x(m,n)=\frac{1}{P}\sum_{p=1}^P x_p(m) x_p(n)= \frac{1}{P} \mathbf{y}_m \mathbf{y}_n^T
  $$
  or in a matrix form
 \begin{equation}
   \mathbf{R}_x=\frac{1}{P} \mathbf{X}_P \mathbf{X}_P^T. \label{CorrMatrixP}
  \end{equation}
 If the observations are not zero-mean, then we should use the covariance matrix, 
 \begin{equation}
    {\Sigma}_x(m,n)=\frac{1}{P}\sum_{p=1}^P \Big(x_p(m) -\mu (m)\Big) \Big(x_p(n)-\mu (n)\Big), \label{CorrMatrixPPP}
   \end{equation}
 where $\mu (n)$ is the mean of the observations at the vertex $n$. 
  
\begin{Remark} Since the correlation matrix in (\ref{CorrMatrixP}) includes contribution from signals at all vertices, it accumulates correlations obtained through all possible walks from the current vertex, $n$, to any other vertex, $m$. This also means that the correlation coefficient between two vertices will produce misleading results if there exists one or more other vertices, $q$, where the signal is strongly correlated with both of the considered vertices, $m$ and $n$. This is  why the naive use of correlation tends to overestimate the strength of direct vertex connections; this renders it a poor metric for establishing direct links (edges) between vertices. To resolve this issue, either additional conditions should be imposed on the correlation matrix, or other statistical parameters may be be used for edge weights estimation. 
\end{Remark}

\begin{Example}\label{SimCooEX}Consider four random graph signals observed at the vertices $n=0,1,2,3$, and given by 
\begin{gather}	
    x_p(0)=\nu_0(p) \nonumber \\
	x_p(1)=x_p(0)+\nu_1(p) \nonumber \\
	x_p(2)=x_p(1)+\nu_2(p) \nonumber \\
	x_p(3)=x_p(2)+\nu_3(p), \label{Corre-exsyst}
\end{gather}
where $\nu_0(p), \nu_1(p),\nu_2(p),\nu_3(p) $ are mutually uncorrelated, white  random variables with zero mean and unit variance.
The elements of the correlation matrix for the above signals can be calculated as, for example
$$R_x(0,1)=\mathrm{E}\{x_p(0)x_p(1)\}=\mathrm{E}\{x_p(0)(x_p(0)+\nu_1(p))\}=1$$
or
\begin{gather*}
R_x(0,2)=\mathrm{E}\{x_p(0)x_p(2)\}=\mathrm{E}\{x_p(0)(x_p(1)+\nu_2(p))\}\\
=\mathrm{E}\{x_p(0)(x_p(0)+\nu_2(p)+\nu_2(p))\}=1.
\end{gather*}
Observe from (\ref{Corre-exsyst}) that, although the signal value $x_p(2)$ is not directly related to $x_p(0)$, the correlation coefficient, $R_x(0,2)$, is nonzero  and even equal to $R_x(0,1)$, since there is an indirect link between these two signal values through $x_p(1)$. In practical applications, it is therefore desirable to avoid this indirect cumulative contributions to the correlation coefficient which results in an overestimated edge weight. 
	
All correlation coefficients for the above example can be written in a matrix form as
 \begin{equation}
 \mathbf{R}_x= 
 \begin{bmatrix}
 1  &  1  &  1  &  1 \\
 1  &  2  &  2   & 2 \\
 1  &  2  &  3   & 3 \\
 1  &  2  &  3   & 4 
 \end{bmatrix}, \label{RsimEX}
 \end{equation}
 with the inverse correlation matrix, called the \textit{precision matrix}   \begin{equation}\mathbf{C}=\mathbf{R}_x^{-1}= 
 \begin{bmatrix}
 \begin{array}{rrrr}
 2  & -1  &  0  &  0 \\
 -1 &  2 &  -1  & 0 \\
 0  & -1   & 2  &  -1 \\
 0 &  0  &  -1  & 1 \\
 \end{array}
 \end{bmatrix}. \label{CsimEX}
\end{equation}

\begin{Remark}
Observe that while the autocorrelation in (\ref{RsimEX}) overestimates the strength of edge links, the precision matrix in (\ref{CsimEX}) produces the desired results, since for example, $C(0,2)=0$, which indicates  that there is no direct relation between $x_p(0)$ and $x_p(2)$, although $x_p(2)$ is indirectly linked to $x_p(0)$ through $x_p(1)$. 
\end{Remark}
 
Similar to the normalized correlation, the normalized precision matrix, $\mathbf{C}^{(N)}$, is defined by $C^{(N)}_{mn}=C_{mn}/\sqrt{C_{mm}C_{nn}}$ to produce  
\begin{equation}
	\mathbf{C}^{(N)}= 
	\begin{bmatrix}
	\begin{array}{rrrr}
	1  & -0.5  &  0  &  0 \\
	-0.5 &  1 &  -0.5  & 0 \\
	0  & -0.5   & 1  &  -1/\sqrt{2} \\
	0 &  0  &  -1/\sqrt{2}  & 1 \\
	\end{array}
	\end{bmatrix}. \label{PrecisionNorm}
\end{equation}

\end{Example}	

\subsection{Imposing Sparsity on the Connection Metric}\label{SpartImpos}

The minimization of the \textit{sparsity of the weight matrix} keeps the number of its nonzero values to the minimum \cite{stankovic2001measure,stankovictutorial}, thus resulting in graphs with the smallest possible number of edges.   
 
 Consider the vertex $n=0$ and the graph signal observation vector as in (\ref{verttoR}), at this vertex. We can estimate the edge weights from this vertex to all other vertices,  $\beta_{0m}$, $m=1,2,3,\ldots,N-1,$ by minimizing
 \begin{equation}
 J_0=\Vert \mathbf{y}_0 - \sum_{m=1}^{N-1} \beta_{0m} \mathbf{y}_m \Vert_2^2 + \rho \sum_{m=1}^{N-1} | \beta_{0m} |.\label{cf1}
 \end{equation}
 Physically, the first term promotes the correlation between the observations $\mathbf{y}_0$ at the considered vertex (with $n=0$)  and the observations $\mathbf{y}_m$ at all other vertices, for $m=1,2,3,\ldots,N-1$; the second term promotes sparsity in the coefficient vector $\boldsymbol{\beta_{0}}$ (number of nonzero coefficients $\beta_{0m}$), while the parameter $\rho$ balances between these two conditions.

 The matrix form of the cost function (\ref{cf1}) is given by
 \begin{equation}
 J_0=\Vert \mathbf{y}_0^T - \mathbf{Y}_0^T \boldsymbol{\beta}_{0}^T  \Vert_2^2 + \rho  \Vert \boldsymbol{\beta}_{0} \Vert_1, \label{zalasso}
 \end{equation}
 where 
 $\mathbf{Y}_0$ is obtained from the matrix $\mathbf{X}_P$ after the first row is removed, and 
 $$\boldsymbol{\beta}_{0}=\begin{bmatrix}
   \ \beta_{01}, \  \ \beta_{02}, \   \ \dots,   \ \beta_{0N-1}    
   \end{bmatrix}.
 $$
\begin{Example}\label{ExLassoSim}
	For the correlation matrix from Example \ref{SimCooEX} and the observation vector, $\mathbf{y}_0$, at the vertex $n=0$, given by 
	$$\mathbf{y}_0=[\ x_1(0),  \ x_2(0),  \dots,  x_P(0)]=[\ \nu_0(1), \  \nu_0(2),  \dots,   \nu_0(P)],$$
	we can find the solution to (\ref{zalasso}) with $\rho=0$, which corresponds to the two-norm minimization of the error function, given by
	$$\frac{\partial J_0}{\partial \boldsymbol{\beta}_{0}^T  }=2\mathbf{Y}_0(\mathbf{y}_0^T - \mathbf{Y}_0^T \boldsymbol{\beta}_{0}^T )=\mathbf{0}$$
	or 
	$$\boldsymbol{\beta}_{0}^T =(\mathbf{Y}_0\mathbf{Y}_0^T)^{-1}\mathbf{Y}_0\mathbf{y}_0^T=\begin{bmatrix}
	2  &  2   & 2 \\
	2  &  3   & 3 \\
	2  &  3   & 4 \\
	\end{bmatrix}^{-1}\begin{bmatrix}
	1 \\
	1 \\
	1 \\
	\end{bmatrix}=\begin{bmatrix}
	0.5 \\
	0 \\
	0
	\end{bmatrix},$$
since $\mathbf{Y}_0\mathbf{Y}_0^T$ and $\mathbf{Y}_0\mathbf{y}_0^T$ are submatrices of correlation matrix $\mathbf{R}_x$, given in (\ref{RsimEX}).  

In the same way, the other three coefficient vectors, $\boldsymbol{\beta}_{1}^T$, $\boldsymbol{\beta}_{2}^T$, $\boldsymbol{\beta}_{3}^T$, were calculated to produce (with added zero-values (in red) at the diagonal) the coefficient matrix
\begin{equation} \boldsymbol{\beta}=\begin{bmatrix}
 \color{red}0  & 0.5  &  0   & 0 \\
 0.5  & \color{red}0 & 0.5   & 0 \\
 0  &  0.5 & \color{red}0  & 0.5 \\
 0  &  0  & 1 & \color{red}0\\
 \end{bmatrix}. \label{betadire}
 \end{equation}
 
 Since this procedure does not guarantee symmetry of $\beta_{nm}=\beta_{mn}$, the edge weights could have also been calculated through the geometric mean,
\begin{equation} W_{nm}=\sqrt{\beta_{nm} \beta_{mn}},
\label{betadiregm}
\end{equation}
 to produce
 \begin{equation} 
 \mathbf{W}=\begin{bmatrix}
 \color{red}0  & 0.5  &  0   & 0 \\
 0.5  & \color{red}0 & 0.5   & 0 \\
 0  &  0.5 & \color{red}0  & 1/\sqrt{2} \\
 0  &  0  & 1/\sqrt{2} & \color{red}0\\
 \end{bmatrix}.
 \label{betamatr}
 \end{equation}
This weight matrix is symmetric and corresponds to an undirected graph.  
 
The graph Laplacian, $\mathbf{L}=\mathbf{W}-\mathbf{D}$, is then obtained by changing the signs of the elements in $\mathbf{W}$ and adding appropriate diagonal elements, $\mathbf{D}$, such that the sum for each row or column is zero, that is
$$ \mathbf{L}=\begin{bmatrix}
\begin{array}{rrrr}
\color{red}0.5 & -0.5  &  0   & 0 \\
-0.5  & \color{red}1 &   -0.5   & 0 \\
0  &  -0.5 & \color{red}1.207  & -0.707 \\
0  &  0  & -0.707  & \color{red}0.707 \\
\end{array}
\end{bmatrix}.$$
Notice that the structure of nonzero off-diagonal elements in this matrix is the same as in the normalized precision matrix in (\ref{PrecisionNorm}), although the corresponding values were obtained through two quite different approaches to the estimation of the relations among graph data observed at different vertices.  
\end{Example}
 
 \noindent\textbf{LASSO approach.} In general, the problem in (\ref{zalasso}) can be solved using the well established least absolute shrinkage and selection operator  (LASSO) minimization, the regression analysis method that performs both variable selection and regularization, as
 $$
 \boldsymbol{\beta}_{0}=\mathrm{lasso}(\mathbf{Y}_0^T, \mathbf{y}_0^T, \rho). $$ 
 For more detail on the derivation and implementation of LASSO see Section \ref{SecLassGlasso} and Algorithm \ref{LassoAlg}. 
 
 \begin{algorithm}[htb]
 	\caption{\!\!\textbf{.} LASSO (ISTA variant), $\mathbf{B}$=lasso($\mathbf{Y},\mathbf{y},\rho$)}
 	\label{LassoAlg}
 	\begin{algorithmic}[1]
 		\Input
 		\Statex
 		\begin{itemize}
 			\item Observation column vector $\mathbf{y}$, $P\times 1$
 			\item Observation matrix $\mathbf{Y}$, $P\times N$
 			\item Sparsity promotion parameter $\rho$
 		\end{itemize}
 		\Statex
 		\State $\mathbf{B} \gets \mathbf{0}_{N\times 1}$
 		\State $\alpha \gets 2\max\{\mathrm{eig}(\mathbf{Y}^T\mathbf{Y})\}$
 		\Repeat
 		\State $\displaystyle \mathbf{s} \gets \frac{1}{\alpha}\mathbf{Y}^T (\mathbf{y}-\mathbf{Y}\mathbf{B})+\mathbf{B}$
 		\smallskip
 		\For{$k \gets 1$ to $N$}
 		\smallskip
 		\State $\displaystyle B(k) \gets 
 		\begin{cases} 
 		s(k)+\rho, & \text{for } s(k)<-\rho \\
 		0, & \text{for } |s(k)| \le \rho \\
 		s(k)-\rho, & \text{for } s(k)>\rho
 		\end{cases}
 		$
 		\EndFor
 		\Until{stopping criterion is satisfied}
 		\Statex
 		\Output
 		\Statex
 		\begin{itemize}
 			\item Reconstructed coefficients $\mathbf{B}$
 		\end{itemize}
 	\end{algorithmic}
 \end{algorithm}

 For the data from Example \ref{ExLassoSim} the LASSO approach yields
 $$\boldsymbol{\beta}_{0}=\mathrm{lasso}(\mathbf{Y}_0^T, \mathbf{y}_0^T, 0.01)=  [0.49,  \ 0, \ 0].$$
This result is almost the same as the first row (excluding the first element assumed to be zero) in the matrix $\boldsymbol{\beta}$ in (\ref{betadire}), as was expected since the solution in the first row in (\ref{betadire}) is already with maximum sparsity. 
Since in this setting the number of independent observations, $P$, could be significantly larger than the number of coefficients, $\beta_{0m}$, for this case the least squares estimation  is optimal and there are no additional degrees of freedom available to improve the sparsity of the solution (the solution, in this case is already with one nonzero element, that is, with minimum possible sparsity). On the other hand, ways to to promote sparsity would be necessary if the number of observations is smaller than the number of vertices (compressive sensing theory framework). 
 
 The minimization in (\ref{zalasso}) was performed for the vertex $n=0$, and should be repeated for all vertices  $n=1,2,\ldots,N-1$, through the cost function
 \begin{equation*}
 J_n=\Vert \mathbf{y}_n^T - \mathbf{Y}_n^T \boldsymbol{\beta}_{n}  \Vert_2^2 + \rho  \Vert \boldsymbol{\beta}_{n} \Vert_1,
 \end{equation*}
to obtain 
$$\boldsymbol{\beta}_{n}=\mathrm{lasso}(\mathbf{Y}_n^T, \mathbf{y}_n^T,\rho).$$ 

In general, if the resulting weight matrix, $\boldsymbol{\beta}$, is not symmetric then the edge weights could be calculated as 
$W_{nm}=\sqrt{\beta_{nm} \beta_{mn}},$ as mentioned in (\ref{betadiregm}).

 \begin{Example} As an example for graph learning from data using the LASSO algorithm, consider the graph from {\color{red} Fig. 2, Part I} and $P=3,000$ observations, which was simulated by assuming external white Gaussian  sources with zero-mean and variance $\sigma^2=1$, located at two randomly chosen vertices (see Section \ref{SecLassGlasso} and Fig. \ref{fig:ec2}). An $N\times P$ matrix of observed signal values, $\mathbf{X}_P$, was then formed, and from its rows the vector $\mathbf{y}_n$ and matrix $\mathbf{Y}_n$ were obtained. The  matrix of coefficients $\boldsymbol{\beta}=[\beta_{mn}]_{N\times N}$ follows from $\mathrm{lasso}(\mathbf{Y}_n^T, \mathbf{y}_n^T, \rho)$ with $n=0,1,2,3,4,5,6,7$ and $\rho=0.2,$ to yield
  \begin{equation*}
 \boldsymbol{\beta}=
 \left[
 \begin{array}
 {llllllll}
 0  &  0.0  &   0.75   &   0.16   &   0   &   0   &   0   &   0 \\
 0.03   &   0   &   0.35   &   0   &   0.19   &   0   &   0   &   0.18 \\
 0.75   &   0.35   &   0   &   0.10   &   0.11   &   0   &   0   &   0 \\
 0.16   &   0   &   0.10   &   0   &   0   &   0   &   0.45   &   0 \\
 0   &   0.19   &   0.11   &   0   &   0   &   0.74   &   0   &   0  \\
 0   &   0   &   0   &   0   &  0.74  &  0  &  0  &  0.19 \\
 0  &  0  &  0  &  0.45  &  0  &  0  &  0  &  0.58 \\
 0  &  0  &  0  &  0  &  0  &  0.19 &  0.58  &  0 \\
 \end{array}
 \right].
 \end{equation*}
 \begin{figure}
 	\centering
 	\hspace{7mm} Ground truth \hspace{20mm} LASSO with $\rho=0.2$
 	\includegraphics[]{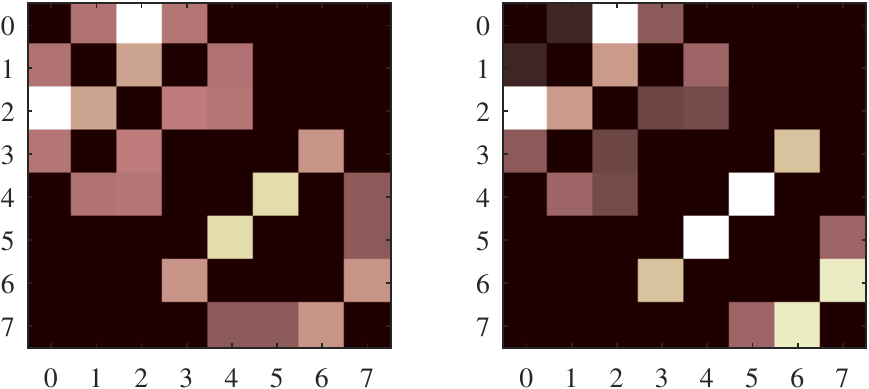} \\
 	(a) \hspace{40mm} (b) \\
 	
 	\vspace{5mm}
 	
 	\hspace{-1mm} LASSO with $\rho=0.05$ \hspace{15mm} LASSO with    $\rho=1$ 
 	
 	\includegraphics[]{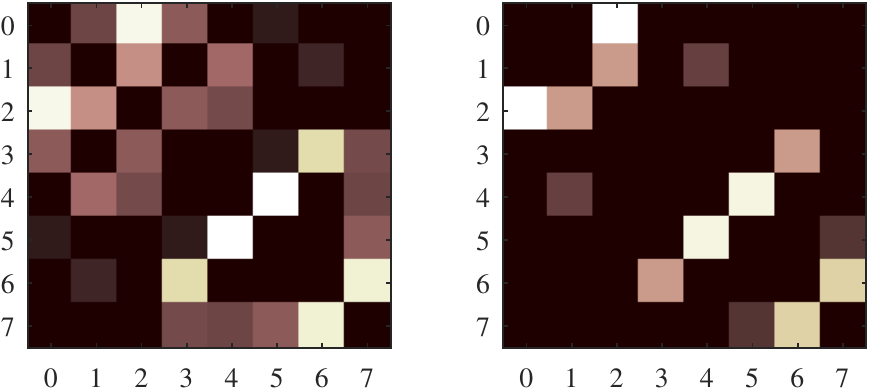}\\
 	(c) \hspace{40mm} (d) \\
 	
 	\caption{Estimation of the weight matrix for the graph from {\color{red} Fig. 2 in Part I} with color-coded element values.  (a) Ground truth weight matrix. (b) Estimated weight matrix with LASSO and $\rho=0.2$.  (c) Estimated weight matrix with LASSO and $\rho=0.05$.  (d) Estimated weight matrix with LASSO and $\rho=1$. }
 	\label{topo_el_circ_LASSO}
 \end{figure}
 
 \begin{figure}
 	\centering
 	\hspace{7mm} Ground truth \hspace{20mm} LASSO with $\rho=0.2$
 	\includegraphics[]{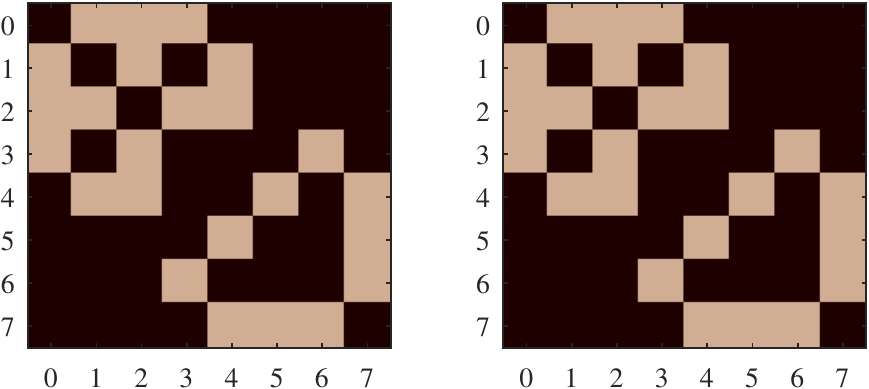} \\
 	(a) \hspace{40mm} (b) \\
 	\caption{Adjacency matrix for the unweighted graph from {\color{red} Fig. 1(a) in Part I}.  (a) Ground truth adjacency matrix. (b) Estimated adjacency matrix with LASSO and $\rho=0.2$. }
 	\label{topo_el_circ_LASSO2}
 \end{figure}

 The ground truth weights and the estimated weights through the LASSO are shown in Fig. \ref{topo_el_circ_LASSO} (a), (b). The estimation was repeated for the cases of (i) a smaller value of balance parameter $\rho=0.05$ (reducing the sparsity contribution and resulting in an increased number of nonzero weights, as in Fig. \ref{topo_el_circ_LASSO} (c)), and (ii) a larger  balance parameter $\rho=1$ (strengthening the sparsity  contribution and resulting in a reduced number of nonzero weights, as Fig. \ref{topo_el_circ_LASSO} (d)).
 
 The same experiment was next repeated for the unweighted graph from {\color{red} Fig. 1(a) in Part I}, and the result is shown in Fig. \ref{topo_el_circ_LASSO2}. In this case, the obtained values of $\boldsymbol{\beta}$ were used to decide whether $A_{mn}=1$ or $A_{mn}=0$.

\end{Example}
 
 \begin{Example}
 	The graph topology in the temperature estimation example in {\color{red}Part 2, Section \ref{II-sec2}} was determined based on the geometry and geographic distances of the locations/vertices where the temperature is sensed \cite{stankovic2019understanding}. Now, we will revisit this example by simulating the temperature field, $\mathbf{X}$, at the locations shown in  Fig. \ref{topo_LASSO_H}(a) and over a period of time with the aim to learn the graph topology from this data. The simulated temperature field  over $P=150$ days is shown in Fig. \ref{topo_LASSO_H}(b). 
 	The weight matrix calculated from the geographical positions of the vertices is denoted as the ground truth weight matrix, $\mathbf{W}$, and shown in Fig. \ref{topo_LASSO_H}(c). The  corresponding weight matrix, which is learned from data in Fig. \ref{topo_LASSO_H}(b) using the column LASSO with $\rho=0.2$, is given in Fig. \ref{topo_LASSO_H}(d). Before the calculation of the correlation matrices, the mean value of the sensed temperatures was removed from $x_p(n)$ for each observation $p$.  
 	
 \begin{figure}
 	\centering
 		\includegraphics[scale=0.95]{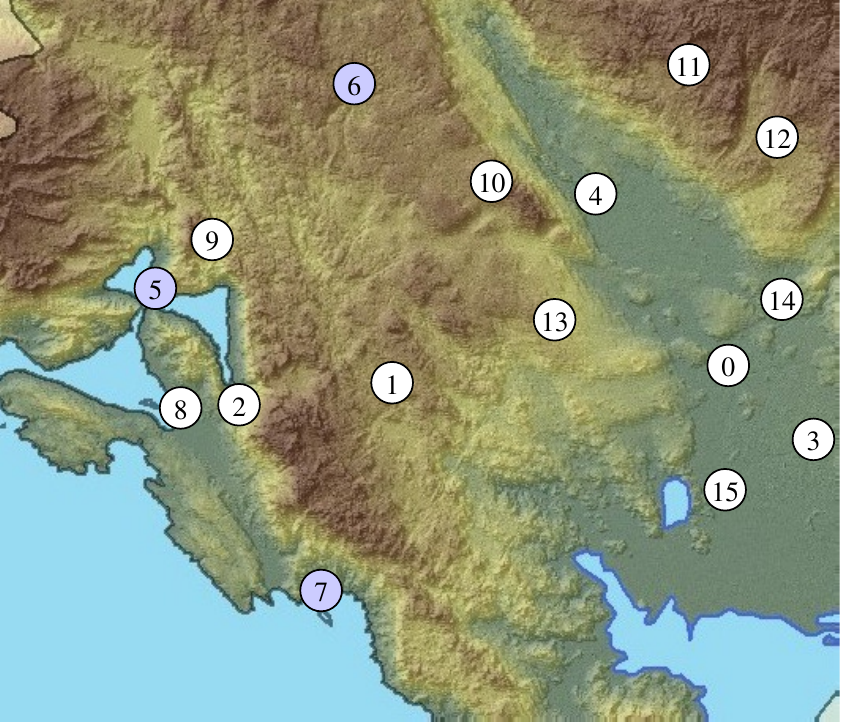} \ \ (a) \\
 	
 	\vspace{5mm}
 	
 	 \includegraphics[scale=0.95]{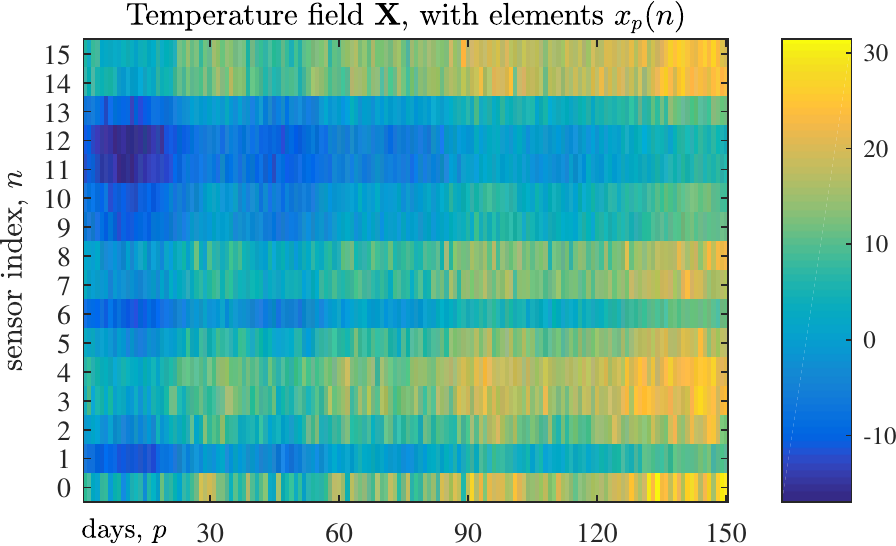} (b) \\
 	 
 	 	\vspace{5mm}
 	 	
 	  \includegraphics[]{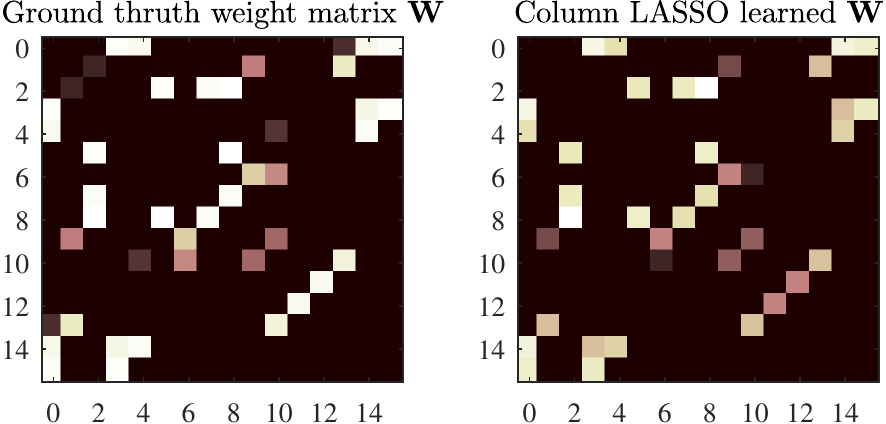} \\
 	(c) \hspace{40mm} (d) \\
 		\caption{Data-based learning of graph topology in the temperature sensing example from {\color{red}Part 2, Section \ref{II-sec2}}.
 		(a) Sensing locations in a geographic region along the Adriatic sea.  (b) Temperatures measured at $N=16$ sensing locations over $P=150$ days. (c) Ground truth weight matrix, $\mathbf{W}$, obtained through geographic properties of the sensing locations as in {\color{red}Part 2, Section \ref{II-sec2}}. (d) The weight matrix, $\mathbf{W}$, estimated solely based on the analysis of data from (b) and using the LASSO approach. }
 	\label{topo_LASSO_H}
 \end{figure}

\end{Example}
 
 \subsection{Smoothness Constrained Learning of Graph Laplacian}\label{SubSGC}
 
 Consider a set of noisy graph data, $x_p(n)$, measured over $P$ observations, $p=1,2,\dots,P$, at $N$ vertices $n=0,1\dots,N-1$, of an undirected graph. The aim is to learn the graph connectivity (its graph Laplacian) from the observed data. To this end, it is necessary to find a signal, $y_p(n)$,
 that is close to the observations, $x_p(n)$,  under the condition that $y_p(n)$ is  as  smooth as possible on a graph. This formulation is similar to that addressed in {\color{red} Part I}. 
 
 \begin{Remark}\label{smoothRemark}
 	\textit{The smoothness condition} may be imposed based on the physically meaningful expectation that the data at close and strongly related vertices should have similar values, that is, without abrupt changes in signal values from vertex to vertex. This requirement imposes  gradual  change of data over the graph domain, as is the case in many practical applications.
 	\end{Remark}

 The graph signal $y_p(n)$ can now be found by minimizing the cost function
   $$
   J_p=\frac{1}{2}\Vert \mathbf{y}_p-\mathbf{x}_p\Vert_2^2 + \alpha \mathbf{y}_p^T \mathbf{L} \mathbf{y}_p, \text{ for } p=1,2,\dots,P,
	   $$ 
whereby the first term aims at finding $\mathbf{y}_p$ which is as close as possible to $\mathbf{x}_p$, while the second term, $ \mathbf{y}_p^T \mathbf{L} \mathbf{y}_p$, promotes the smoothness of graph signal $\mathbf{y}_p$. 

	\begin{Remark} The difference in the problem considered here from the smoothing problem addressed in  {\color{red} Part I} is that \textit{the graph Laplacian  (graph edges  and their weights) is now unknown}. In other words, the graph Laplacian, $\mathbf{L}$, has to be determined along with the output signal $\mathbf{y}_p$, that is, the graph topology has to be learned from data. 
	\end{Remark} 
  
  Since we have available $P$ graph-wise observations, we can form the $N \times P$ matrices
 $$\mathbf{X}_P=
   \begin{bmatrix}
   \mathbf{x}_1,  \mathbf{x}_2, \dots,   \mathbf{x}_P    
   \end{bmatrix}$$
   and
   $$\mathbf{Y}_P=
   \begin{bmatrix}
    \mathbf{y}_1,  \ \mathbf{y}_2,  \dots,   \mathbf{y}_P    
   \end{bmatrix}.
   $$
   Notice that here the vectors $\mathbf{y}_n$ above have to be calculated, and they are not related to the rearranged signal vectors, defined with the same notation, in the previous section.  
   
   \subsection{Graph Topology Estimation with the Graph Laplacian Energy Condition}\label{energConL} In addition to the smoothness condition, it is very useful to introduce the energy of graph Laplacian as an optimization condition, since \textit{none of the above conditions is sensitive to the scaling of the graph Laplacian elements} and their possible large values. Such cost function is then of the following form 
   \begin{align}
   J&=\sum_{p=1}^{P}\Big[\frac{1}{2}\Vert \mathbf{y}_p-\mathbf{x}_p\Vert_2^2 + \alpha \mathbf{y}_p^T \mathbf{L} \mathbf{y}_p\Big]+\beta \Vert \mathbf{L}\Vert_F^2,\notag
   \end{align}
   where the penalty for the energy (squared Frobenius norm of a matrix) of the graph  Laplacian, given by
   $$\Vert \mathbf{L}\Vert_F^2=\sum_m\sum_nL_{mn}^2$$
  is involved in order to keep its values as low as possible.

   The cost function for the whole set of $P$ observations can now be  written in a compact form as
\begin{align}
 J= \frac{1}{2}\Vert \mathbf{Y}_P-\mathbf{X}_P\Vert_F^2 + \alpha \mathrm{Trace}\{\mathbf{Y}_P^T \mathbf{L} \mathbf{Y}_P\}+\beta \Vert \mathbf{L}\Vert_F^2,\label{convexpr}
\end{align}
 where $\mathrm{Trace}\{\mathbf{Y}_P^T \mathbf{L} \mathbf{Y}_P\}$ is a scalar which can be written as a matrix form of the term $\sum_{p=1}^{P}\mathbf{y}_p^T \mathbf{L} \mathbf{y}_p$, that is
   $$\sum_{p=1}^{P}\mathbf{y}_p^T \mathbf{L} \mathbf{y}_p=\mathrm{Trace}\{\mathbf{Y}_P^T \mathbf{L} \mathbf{Y}_P\}.$$  
 
 The above analysis assumes that the Laplacian has been first  normalized. In order to avoid trivial solutions, the condition 
\begin{equation} \mathrm{Trace}\{\mathbf{L}\}=N  \label{NormDiagL}
\end{equation}
 is also used (as the diagonal elements of the ground truth normalized graph Laplacian are $L_{nn}=1$), along with the condition that the off-diagonal elements are either zero or negative, that is
 \begin{equation}L_{mn}=L_{nm} \le 0 \text{ for $n\ne m$.}
 \label{NormDiagL2}
 \end{equation}
 As with any Laplacian matrix, the sum of the graph Laplacian elements over every row or column is zero, that is
 \begin{equation}
 \sum_{m=0}^{N-1}L_{nm}=0 \text{ and } \sum_{n=0}^{N-1}L_{nm}=0. \label{NormDiagL3}
 \end{equation}
 
 \begin{Remark}
 	The optimization problem in (\ref{convexpr}) aims to learn the graph topology from the graph data and by finding the graph Laplacian which is most likely, in the second order sense, to generate the observed graph data. This formulation in (\ref{convexpr})  obviously jointly convex with respect to both the observed signal and the Laplacian, and can be solved in an iterative two-step procedure, given in Algorithm \ref{ReconMetAlg}.
\end{Remark} 
  
 \begin{algorithm}[!h]
 	\caption{\!\!\textbf{.} Iterative procedure for solving the problem of graph learning from data, given in  (\ref{convexpr})}
 	\label{ReconMetAlg}
 	\begin{algorithmic}[1]
 	\State Assume that 
 	$$\mathbf{Y}_P=\mathbf{X}_P.$$
 	\State Estimate the graph Laplacian, $\mathbf{L}$, by minimizing 
 	$$
 	J_1= \alpha \mathrm{Trace}\{\mathbf{Y}_P^T \mathbf{L} \mathbf{Y}_P\}+\Vert \mathbf{L}\Vert_F^2
 	$$ 
 with the conditions given in (\ref{NormDiagL}), (\ref{NormDiagL2}), and (\ref{NormDiagL3}), for the normalized graph Laplacian form.
 	\State For the Laplacian obtained  in the Step 1, the signal $\mathbf{Y}_P$ is calculated by minimizing 
 	$$
 	J_2=\frac{1}{2}\Vert \mathbf{Y}_P-\mathbf{X}_P\Vert_F^2 + \alpha \mathrm{Trace}\{\mathbf{Y}_P^T \mathbf{L} \mathbf{Y}_P\}.
 	$$
 	\end{algorithmic}
 
  	Iteratively repeat Step 2 and Step 3. 
  	
  	Step 3 has a closed form solution explained in {\color{red} Part I}.
 \end{algorithm}

 \subsection{Learning of Generalized Laplacian - Graphical LASSO}\label{secGLASSO}
 
 The generalized Laplacian, $\mathbf{Q}$, is defined as 
 $$\mathbf{Q}= \alpha \mathbf{I}-  \mathbf{N},$$
 where  $\mathbf{N}$ is a nonnegative symmetric matrix and $\mathbf{Q}$ is a symmetric positive semidefinite matrix.  
 Any generalized Laplacian can be written as a sum of a standard Laplacian, $\mathbf{L}$, and a diagonal matrix, $\mathbf{P}$, that is 
 $$\mathbf{Q}= \mathbf{L}+ \mathbf{P}.$$
 
\begin{Remark} 
The generalized Laplacian allows for self-loops on the vertices; these self-loops are defined by matrix $\mathbf{P}$.
\end{Remark}

\begin{Example}For the data in Example \ref{SimCooEX}, the precision matrix is of the form
	$$\mathbf{C}=\mathbf{R}_x^{-1}= 
	\begin{bmatrix}
	\begin{array}{rrrr}
	2  & -1  &  0  &  0 \\
	-1 &  2 &  -1  & 0 \\
	0  & -1   & 2  &  -1 \\
	0 &  0  &  -1  & 1 \\
	\end{array}
	\end{bmatrix}.
	$$
 It  may be considered as a generalized graph Laplacian since
 \begin{gather*}
 \mathbf{R}_x^{-1}= 	\begin{bmatrix}
 	\begin{array}{rrrr}
 	2  & -1  &  0  &  0 \\
 	-1 &  2 &  -1  & 0 \\
 	0  & -1   & 2  &  -1 \\
 	0 &  0  &  -1  & 1 \\
 	\end{array}
 	\end{bmatrix}	\\
 	=\begin{bmatrix}
 	\begin{array}{rrrr}
 	1  & -1  &  0  &  0 \\
 	-1 &  2 &  -1  & 0 \\
 	0  & -1   & 2  &  -1 \\
 	0 &  0  &  -1  & 1 \\
 	\end{array}
 	\end{bmatrix}+	\begin{bmatrix}
 	\begin{array}{rrrr}
 	1  & 0  &  0  &  0 \\
 	0 &  0 &  0  & 0 \\
 	0  & 0   &  0  &  0 \\
 	0 &  0  &  0  & 0 \\
 	\end{array}
 	\end{bmatrix}\\
 	=\mathbf{L}+\mathbf{P}.
 \end{gather*}
This means that $\mathbf{R}_x^{-1}$ in this example may be interpreted as standard  graph Laplacian with a self-loop at the vertex $n=0$. 
 \end{Example}
 
 We will show next that owing to its physically relevant properties the precision matrix,  $\mathbf{C}=\mathbf{R}_x^{-1}$, can be used as an estimate of the generalized Laplacian, $\mathbf{Q}$.
 
\noindent\textbf{Estimation of graph Laplacian through precision matrix.} Consider a set of noisy signals $x_p(n)$ acquired over $P$ observations, $p=1,2,\dots,P$,  on $N$ vertices $n=0,1\dots,N-1$ of an undirected graph. Our aim is to learn the graph connectivity (its Laplacian) based on the condition that the observed graph signal in the $p$th realization, $\mathbf{x}_p$,  is as smooth as possible on the graph defined by a generalized Laplacian, $\mathbf{Q}$, as explained in Remark \ref{smoothRemark}. The cost function to achieve this goal can be conveniently defined by the signal smoothness function
 $$
 J_p= \mathbf{x}_p^T \mathbf{Q} \mathbf{x}_p, \text{ for } p=1,2,\dots,P.
 $$ 
 The cumulative smoothness for all data $\mathbf{x}_p$, $p=1,2,\dots,P$, is then expressed as  
 \begin{align}
 J= \frac{1}{P} \sum_{p=1}^P \mathbf{x}_p^T \mathbf{Q} \mathbf{x}_p,
 \end{align}
while the correlation matrix of the all considered observations can be written as 
 \begin{gather*}\mathbf{R}_x= \frac{1}{P}\sum_{p=1}^P\mathbf{x}_p\mathbf{x}_p^T\\
 =\frac{1}{P}\begin{bmatrix}
 \ \mathbf{x}_1,  \ \mathbf{x}_2,  \dots,  \mathbf{x}_P    
 \end{bmatrix}\begin{bmatrix}
 \ \mathbf{x}_1,  \ \mathbf{x}_2,  \dots,  \mathbf{x}_P    
 \end{bmatrix}^T
 =\frac{1}{P} \mathbf{X}_P \mathbf{X}_P^T.
 \end{gather*} 
 The smoothness index for all observations is now of the following form 
 \begin{align}
 J= \frac{1}{P} \sum_{p=1}^P \mathbf{x}_p^T \mathbf{Q} \mathbf{x}_p=\mathrm{Trace}\{\mathbf{R}_x \mathbf{Q}\}.\notag
 \end{align}
 since
 \begin{gather*}
J= \frac{1}{P}\sum_{p=1}^P \mathbf{x}_p^T \mathbf{Q}\mathbf{x}_p
 \\
 =\frac{1}{P}\mathrm{Trace}\{\begin{bmatrix}
 \ \mathbf{x}_1,  \ \mathbf{x}_2,  \dots,  \mathbf{x}_P    
 \end{bmatrix}^T\mathbf{Q}\begin{bmatrix}
 \ \mathbf{x}_1,  \ \mathbf{x}_2,  \dots,  \mathbf{x}_P    
 \end{bmatrix} \} \\
 = \frac{1}{P}\mathrm{Trace}\{\mathbf{X}^T_P \mathbf{Q}\mathbf{X}_P\}= \frac{1}{P}\mathrm{Trace}\{\mathbf{X}_P \mathbf{X}_P^T\mathbf{Q}\}\\
 =\mathrm{Trace}\{\mathbf{R}_x \mathbf{Q}\}.\notag
 \end{gather*} 
 
 To avoid a trivial solution, the conditions for the generalized Laplacian should be incorporated. For symmetric positive definite matrices, all eigenvalues are positive, and since for every matrix $\mathbf{Q}$ the product of its eigenvalues is equal to $\mathrm{det}(\mathbf{Q})$, this condition can be included by adding the term $\ln (\mathrm{det}(\mathbf{Q}))$ to the cost function, to give
\begin{equation}
 J= -\ln (\mathrm{det}(\mathbf{Q}))+\mathrm{Trace}\{\mathbf{R}_x \mathbf{Q}\}.\label{glcost}
\end{equation}
 
 \noindent\textbf{Maximum likelihood interpretation.} The interpretation of the cost function in (\ref{glcost})  within the theory of Gaussian random signal and maximum likelihood estimation is given in Section \ref{GrafRanDSIG}. If we assume that the graph data at $N$ vertices are $N$-dimensional random variables, with zero-mean and an unknown precision matrix $\mathbf{Q}$, then their $N$-dimensional probability density function is given by
 $$P(\mathbf{x}_p) =  \frac{1}{\sqrt{(2\pi)^p}}\sqrt{\textrm{det}(\mathbf{Q})} \exp{(-\frac{1}{2}\mathbf{x}^T_p\mathbf{Q}\mathbf{x}_p)}.$$
 Within the  maximum likelihood framework the goal is to find the unknown parameter (matrix)  $\mathbf{Q}$  so that the distribution fits the data in an optimal form. This optimal parameter matrix is obtained by differentiating the probability or its logarithm (log-likelihood) function,  
  \begin{gather}
  -\ln\{P(\mathbf{x}_p)\sqrt{(2\pi)^p}\} =  -\ln\{\sqrt{\textrm{det}(\mathbf{Q})} \exp{(-\frac{1}{2}\mathbf{x}^T_p\mathbf{Q}\mathbf{x}_p)}\}\nonumber \\
  =-\frac{1}{2}\ln\{\textrm{det}(\mathbf{Q})\} +\frac{1}{2}\mathbf{x}^T_p\mathbf{Q}\mathbf{x}_p, \label{LLfun}
  \end{gather}
  and setting to zero.
  \begin{Example}\label{ExamGaussMD1}
  	The concept of finding the best precision, $Q$, the reciprocal of the variance of Gaussian distribution, $Q=1/\sigma^2$, to fit the data will be now illustrated on a simple data setup. Assume that four observations of signal $x_p(n)$, $p=1,2,3,4$, at the vertex $n=0$ are available, and are given by $x_1(0)=0.2$, $x_2(0)=-0.3$, $x_3(0)=-0.4$, and  $x_4(0)=-0.5$. It is also known that the data are zero-mean. The goal is to find the precision, $Q=1/\sigma^2$, or variance, $\sigma^2$, of the Gaussian distribution of the observed data,
  	$$P(x_p(0)) =  \frac{1}{\sigma\sqrt{2\pi}} \exp{(-\frac{x^2_p(0)}{2\sigma^2}})= \sqrt{\frac{Q}{2\pi}} \exp(-\frac{1}{2}x_p(0)Qx_p(0))$$
  	which corresponds the best fit to the observed data.
  	The log-likelihood function of the joint distribution of these four observed data is then given by
   \begin{gather*} J=-\ln(P(x_1(0))P(x_2(0))P(x_3(0))P(x_4(0)))\\
   =-\ln( \frac{1}{4\pi^2}Q^2e^{-\frac{1}{2}0.2^2Q} e^{-\frac{1}{2}0.3^2Q}e^{-\frac{1}{2}0.4^2Q}e^{-\frac{1}{2}0.5^2Q})\\
   =2\ln(2\pi)-2\ln(Q) +\frac{1}{2}(0.2^2+0.3^2+0.4^2+0.5^2)Q \\
   =2\ln(2\pi)-2\ln(Q) + \frac{1}{2} 0.54 Q.
   \end{gather*}
  	The differentiation of this expression with respect to $Q=1/\sigma^2$  produces $-2/Q+\frac{1}{2}0.54=0$ or $Q=4/0.54=7.4$ and $$\sigma=\sqrt{1/Q}=0.36.$$ The same value would have been produced by a simple standard deviation estimator $\sigma=\sqrt{(0.2^2+0.3^2+0.4^2+0.5^2)/4}$.
\end{Example}

  \begin{Example}\label{ExamGaussMD2}	
  	Similar analysis, as in the previous example, can be performed for $P$ observations at two vertices, $n=0$ and $n=1$, $[x_p(0), x_p(1)]^T$. The goal is to estimate the parameters of precision matrix
  	$$\mathbf{Q}=\begin{bmatrix}
  	\begin{array}{rrrr}
  	Q_{11}  & Q_{12} \\
  	Q_{21}  & Q_{22} 
  	\end{array}
  	\end{bmatrix}$$ 
  	of the joint Gaussian distribution of $[x_p(0), x_p(1)]^T$, defined as
  		\begin{gather}P([x_p(0),x_p(1)]^T) =  \frac{\sqrt{\det(\mathbf{Q})}}{2\pi} e^{-\frac{1}{2}[x_p(0), x_p(1)]\mathbf{Q}[x_p(0), x_p(1)]^T}\\
  		=\frac{\sqrt{Q_{11}Q_{22}-Q_{12}Q_{21}}}{2\pi} e^{-(Q_{11}x^2_p(0)+(Q_{12}+Q_{21})x_p(0)x_p(1)+Q_{22}x^2_p(1))} \label{norm2Vp}
  		\end{gather}
  	Using $P$ available realizations, $$[x_1(0),x_1(1)],[x_2(0),x_2(1)],\dots,[x_P(0),x_P(1)]$$ and the corresponding $P$-variate normal distribution  of two variables as a product of $P$ distributions as in (\ref{norm2Vp}), we can find parameters $Q_{11},Q_{12},Q_{21},Q_{22}$ which produce the best fitted distribution using the partial derivatives of the log-likelihood function.
  	
  	 For example, a partial derivative of the log-likelihood function with respect to $Q_{11}$ would produce
  	$$-\frac{P}{2}\frac{Q_{22}}{\sqrt{Q_{11}Q_{22}-Q_{12}Q_{21}}}+\frac{1}{2}\Big(x^2_1(0)+x^2_2(0)+\cdots +x^2_P(0)\Big)=0.$$
  	Observe that the term $$\frac{Q_{22}}{\sqrt{Q_{11}Q_{22}-Q_{12}Q_{21}}}=\frac{Q_{22}}{\sqrt{\det(Q)}}$$
  	 is just the first element of the inverse of matrix $\mathbf{Q}$, while the term $(x^2_1(0)+x^2_2(0)+\cdots +x^2_P(0))$ is the first element of the correlation matrix $\mathbf{R}_x$, multiplied by $P$. In a similar way, the derivations over $Q_{12}$, $Q_{21}$, and $Q_{22}$, will produce the remaining elements of the inverse of matrix $\mathbf{Q}$ and the correlation matrix $\mathbf{R}_x$. In the matrix notation,  the solution to the so obtained system of four equation is given by 	
  	$$\mathbf{Q}^{-1}=\frac{1}{P}\begin{bmatrix}
  	\begin{array}{ll}
  	\sum_{p=1}^{P}x^2_p(0)  &\sum_{p=1}^{P}x_p(0)x_p(1) \\
  	\\
  	\sum_{p=1}^{P}x_p(1)x_p(0)   & \sum_{p=1}^{P}x^2_p(1) 
  	\end{array}
  	\end{bmatrix}=\mathbf{R}_x.$$ 
  	Notice that at least $P=2$ independent observations, $P\ge N$, are needed, since for $P=1$ observation,  $P< N$, and the rank of correlation matrix, $\mathbf{R}_x,$ would be $1$, which is lower than its dimension. In that case, the correlation matrix would not be invertible.     
\end{Example}

   The cost function in (\ref{glcost}) minimizes the logarithm of the joint probability density function of a graph signal $\mathbf{x}_p$ under the Gaussian assumption. 
 The minimization of the cost function $J$ with respect to $\mathbf{Q}$, with $\partial J / \partial \mathbf{Q}=\mathbf{0}$, produces
 \begin{equation}
 \frac{\partial J} {\partial \mathbf{Q}}= \frac{\partial} {\partial \mathbf{Q}} \Big( - \ln (\mathrm{det}(\mathbf{Q}))+\mathrm{Trace}\{\mathbf{R}_x \mathbf{Q}\}\Big) .\label{glcost2}
 \end{equation}
 In order to find this derivative, we will use the relation among the trace of a positive semidefinite matrix, the trace of its eigenvalues, $\lambda_k$, and the eigenvalue matrix, $\mathbf{\Lambda}$, in the form
 \begin{gather}
 \ln (\mathrm{det}(\mathbf{Q}))=\sum_{i=1}^{N} \ln (\lambda_k) \nonumber \\
 =\mathrm{Trace}(\ln (\mathbf{\Lambda}))=\mathrm{Trace}(\ln (\mathbf{Q})). \label{Traceeig}
 \end{gather}
 Note also that for a differentiable  matrix function $f(\mathbf{Q})$ the following holds   
 \begin{equation}
  \frac{\partial} {\partial \mathbf{Q}} \Big( \mathrm{Trace}\{f (\mathbf{Q})\}\Big)= \frac{\partial f (\mathbf{Q})} {\partial \mathbf{Q}} . \label{tracederiv}
 \end{equation}
 Having in mind the  properties in (\ref{Traceeig}) and (\ref{tracederiv}), we can write 
 \begin{equation}
 \frac{\partial J} {\partial \mathbf{Q}}=  - \mathbf{Q}^{-1}+\mathbf{R}_x .\label{glcost3}
 \end{equation}
 The best estimate of $\mathbf{Q}$ follows from  $\partial J / \partial \mathbf{Q=0}$ and has the form
\begin{equation}
\mathbf{Q}=\mathbf{R}_x^{-1}.\label{gl_solution}
\end{equation}

 \begin{Remark}
 	Therefore the solution in (\ref{gl_solution}), being equal to the precision matrix,
 can be used as the generalized Laplacian estimate in order to obtain the graph structure. 
 \end{Remark}

 \begin{Example}\label{ExPrecMatr} The weight matrix which corresponds to the inverse of the correlation matrix $\mathbf{R}_x$, for which the positive and small off-diagonal values were set to zero, is shown in Fig. \ref{topo_el_circ_LASSO_VVb} (right).
 Here, we consider the graph from {\color{red} Fig. 2 in Part I} and $P=10,000$ observations. The observations were simulated by assuming white Gaussian external sources with zero-mean and variance $\sigma^2=1$, located at a randomly chosen vertex (as described in more detail in Section \ref{CorrGraphLear}).
 
 \begin{figure}
 \centering
 \includegraphics[]{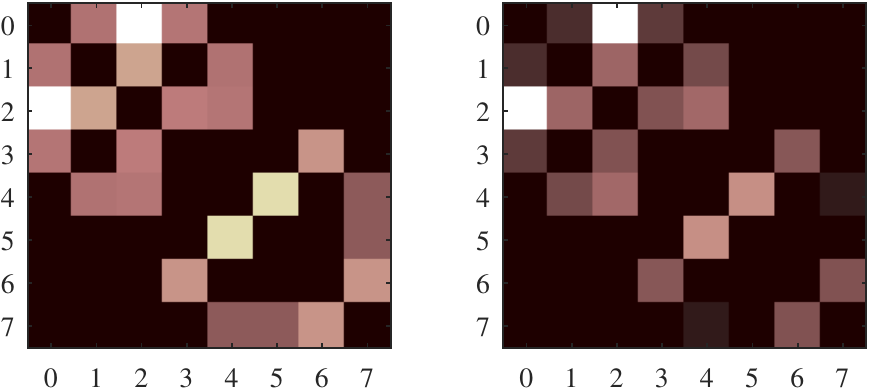}
 \\
 (a) \hspace{40mm} (b) \\
 
 \caption{Weight matrix for the graph from {\color{red} Fig. 2 in Part I}.  (a) Ground truth weight matrix. (b) Estimated weight matrix using the inverse correlation (precision) matrix. } 
 \label{topo_el_circ_LASSO_VVb}
 \end{figure}
 \end{Example}

 \begin{Remark}\label{remarksing} Notice that the correlation matrix, $\mathbf{R}_x$, may be singular.  The correlation matrix, $\mathbf{R}_x$, is always singular when the number of observations, $P$, is lower than the number  vertices (dimension of the correlation matrix, $N$) that is, $N>P$. This follows from the fact that the correlation matrix is formed as a combination of $P$ signals, $\mathbf{R}_x= \frac{1}{P}\sum_{p=1}^P\mathbf{x}_p\mathbf{x}_p^T$, which means that its dimensionality is spanned over at most $P$ independent vectors (eigenvectors), and that its rank is equal to or lower than $P$ (see Example \ref{ex_GLASSOPmalo} in  Section \ref{SecLassGlasso}). 
 \end{Remark}	
 	Also, this form will not produce a matrix satisfying the conditions for a generalized Laplacian. The inverse correlation function may also have positive off-diagonal values. Therefore, for a reliable solution, the cost function in (\ref{glcost}) should have additional constraints. Here, we will present two of such constraints.
 
 \bigskip

 \noindent\textbf{Graphical LASSO.} In  this approach the classical reconstruction formulation of a sparse signal is used as the additional constraint onto the precision matrix and the cost function from (\ref{glcost}). The sparsity constraint on the generalized Laplacian is added to achieve the solution with the smallest possible number of nonzero entries in the estimated graph weight matrix -- the smallest number of edges. The sparsity condition also allows for the problem solution with a reduced correlation matrix rank (as within the compressive sensing framework described in Part II). The cost function, with the included sparsity penalty function, $\Vert \mathbf{Q} \Vert _1$, is then defined as 
\begin{equation}
 J= -\ln (\mathrm{det}(\mathbf{Q}))+ \mathrm{Trace}\{\mathbf{R}_x \mathbf{Q}\}+\rho \Vert \mathbf{Q} \Vert _1. \label{minSparGlasso}
 \end{equation}
 This minimization problem can be solved using various methods. One of them is the graphical LASSO algorithm, an extension of the standard LASSO algorithm to graph problems (see Algorithm \ref{GLassoAlg} and Section \ref{SecLassGlasso}).

 \begin{algorithm}[htb]
 	\caption{\!\!\textbf{.} Graphical LASSO,  $\mathbf{Q}$=glasso($\mathbf{R},\rho$)}
 	\label{GLassoAlg}
 	\begin{algorithmic}[1]
 		\Input
 		\Statex
 		\begin{itemize}
 			\item Correlation matrix $\mathbf{R}$
 			\item Regularization parameter $\rho$
 		\end{itemize}
 		\Statex
 		\State $M_i  \gets 100$, $E_p  \gets 0.0001$
 		\State $[p,n] \gets \mathrm{size}(\mathbf{R})$
 		\State $C_p \gets  \mathrm{mean}(|\mathbf{R}-\mathrm{diag}(\mathrm{diag}(\mathbf{R}))|) E_p$
 		\State $\mathbf{V}_0= \mathbf{R}+\rho \mathbf{I}$
 		\State $\mathbf{V}= \mathbf{V}_0$
 		\For{$r = 1$ to $M_{i}$}
 		\For{$j = p$ to $1$ step $-1$}
 		\State $\mathbf{V}_{11} \gets \mathbf{V}$
 		\State $\mathbf{V}_{11} \gets \mathbf{V}_{11}$ with removed $j$th row
 		\State $\mathbf{V}_{11} \gets \mathbf{V}_{11}$ with removed $j$th column
 		\State $v_{22} \gets V(j,j)$
 		\State $\mathbf{r}_{12} \gets$ \, $j$th column of $\mathbf{R}$
 		\State $\mathbf{r}_{12} \gets \mathbf{r}_{12}$ \, with removed $j$th element
 		\State $\mathbf{A} \gets \sqrt{\mathbf{V}_{11}} $
 		\State $\mathbf{b} \gets (\sqrt{\mathbf{V}_{11}})  ^{-1}\mathbf{r}_{12}$
 		\medskip
 		\State $\boldsymbol{\beta}=\mathrm{lasso}(\mathbf{A},\mathbf{b},\rho) $, as in 	Algorithm \ref{LassoAlg}
 		\medskip	
 		\State $\mathbf{v}_{12} \gets \mathbf{V}_{11}\boldsymbol{\beta}$
 		\State $\mathbf{V} \gets \mathbf{V}$ with $\mathbf{v}_{12}$  inserted as the $j$th column	
 		\State $\mathbf{v}_{12} \gets \mathbf{v}^{T}_{12}$ with $v_{22}$ inserted as the $j$th element	
 		\State $\mathbf{V} \gets \mathbf{V}$ with $\mathbf{v}_{12}$ inserted as the $j$th row	 	
 		\EndFor
 		\State \textbf{if} $\mathrm{mean}(|\mathbf{V}-\mathbf{V}_0|< C_p)$ \textbf{break}, \textbf{end}
 		\State $\mathbf{V}_0= \mathbf{V}$
 		\EndFor
 		\State $\mathbf{Q}=\mathbf{V}^{-1}$
 		\Statex
 		\begin{itemize}
 			\item Estimated precision matrix $\mathbf{Q}$
 		\end{itemize}
 	\end{algorithmic}
 \end{algorithm}

 \begin{Example}
 	For the same signal as in Example \ref{ExPrecMatr}, the weight matrix obtained using the graphical LASSO,
 $$\mathbf{W}=\mathrm{glasso}(\mathbf{R}_x,0.3).$$ 
  with both positive and small values set to zero, is shown in Fig. \ref{topo_el_circ_LASSO_VV_glasso} (b) (see also Example \ref{ex_GLASSOPmalo}).
  
  \begin{figure}
  	\centering
  	\includegraphics[]{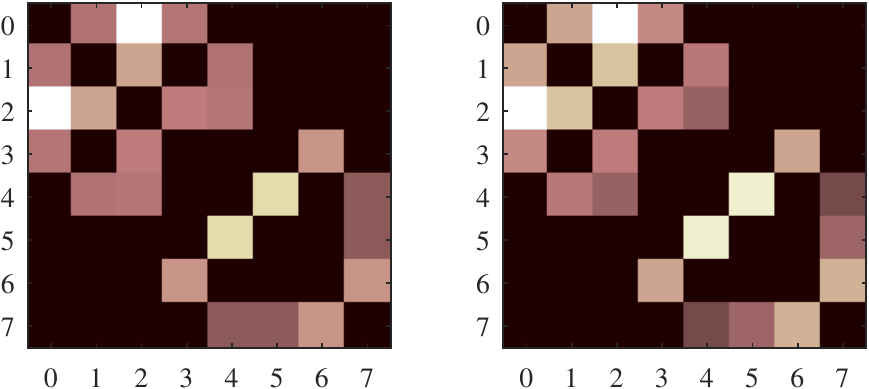}
  	\\
  	(a) \hspace{40mm} (b) \\
  	
  	\caption{Weight matrix for the graph from {\color{red} Fig. 2 in Part I}.  (a) Ground truth weight matrix. (b) Estimated weight matrix using the graphical LASSO and inverse correlation (precision) matrix. } 
  	\label{topo_el_circ_LASSO_VV_glasso}
  \end{figure}

\end{Example}

\noindent\textbf{Generalized Laplacian constrained approach.} Another possible approach employs  to the Lagrange multipliers, $\mathbf{B}$, which are added in such a way that these values do not change the diagonal elements of $\mathbf{Q}$, and ensure that all 
 $$Q_{mn}=Q_{nm}\le 0$$
 for $n\ne m$, with $B_{nm}=B_{mn} \ge 0$. The diagonal elements of matrix $\mathbf{B}$ are   $B_{nn}=0$. Finally, the condition $B_{nm}Q_{nm}=0$ for all $n$ and $m$ is used.  In this case, the minimization solution for the generalized Laplacian is obtained as
 $$\mathbf{Q}=(\mathbf{R}_x+\mathbf{B})^{-1}$$
based on the cost function
 $$
 J= -\ln (\mathrm{det}(\mathbf{Q}))+\mathrm{Trace}\{\mathbf{R}_x \mathbf{Q}\}+\mathrm{Trace}\{\mathbf{B} \mathbf{Q}\}.
 $$
 The results obtained  in this case are similar to those obtained with the graphical LASSO approach.

 \subsection{Graph Topology Learning Based on the Eigenvectors}\label{graphtopeig}
 
 Assume that the available observations of a graph signal, $x_p(n)$, are graph wide sense stationary (GWSS), that is, they can be considered as the outputs of a linear system  
 $H(\mathbf{A})$,    
 driven by white noise,  $\boldsymbol{\varepsilon}_p$,  as the input.
 In other words, the signal on a graph is formed using a linear combination of a white noise realization, $\boldsymbol{\varepsilon}_p$,  and its graph shifted versions. The output signal after $M$ such graph shifts, defined by the normalized Laplacian is given by
 \begin{gather}
 \mathbf{x}_p=
 (h_M\mathbf{L}^M+h_{M-1}\mathbf{L}^{M-1}+\cdots+h_{1}\mathbf{L}^{1}+h_0\mathbf{L}^0)\boldsymbol{\varepsilon}_p.\label{graphSigDefiH}
 \end{gather} 
 The resulting graph signal can be written in the form
 $$\mathbf{x}_p=H(\mathbf{L})\boldsymbol{\varepsilon}_p,$$
 with its correlation matrix given by (for $\sigma^2_{\varepsilon}=1$)
 \begin{gather}\mathbf{R}_x= \frac{1}{P}\sum_{p=1}^P\mathbf{x}_p\mathbf{x}_p^T=\frac{1}{P}\sum_{p=1}^PH(\mathbf{L})\boldsymbol{\varepsilon}_p\boldsymbol{\varepsilon}^T_pH^T(\mathbf{L}) \nonumber \\
 =H(\mathbf{L})\Big(\frac{1}{P}\sum_{p=1}^P\boldsymbol{\varepsilon}_p\boldsymbol{\varepsilon}^T_p\Big)H^T(\mathbf{L}) \nonumber \\
 =H(\mathbf{L})H^T(\mathbf{L})=\mathbf{U}^T|H(\mathbf{\Lambda})|^2\mathbf{U} \label{RxCorrr}
 \end{gather}
 where $\boldsymbol{\varepsilon}_p$ is a white unit variance noise and $\mathbf{U}$ is the matrix of graph Laplacian eigenvectors, $\mathbf{L}=\mathbf{U}^T\mathbf{\Lambda}\mathbf{U}.$ 
 
  From (\ref{RxCorrr}) it is now obvious that  we can learn about the graph eigenvectors  from the decomposition of the autocorrelation matrix. The same holds for the precision matrix, $\mathbf{Q}=\mathbf{R}_x^{-1}$, since the inverse matrix has the same eigenvectors as the original matrix. 
  
 For the the normalized graph Laplacian, it is straightforward  to relate the Laplacian, $\mathbf{L}_N$, based shift and the normalized weight matrix, $\mathbf{W}_N$, based shift since
 \begin{gather*}
 \mathbf{L}^p_N=(\mathbf{I}-\mathbf{W}_N)^p=
 \mathbf{I}-p\mathbf{W}_N+\cdots+(-1)^p\mathbf{W}^p_N.
 \end{gather*}
 
 Therefore from (\ref{RxCorrr}), in order to estimate the graph connectivity (estimate its Laplacian or adjacency matrix) we can use the  eigenvectors of the autocorrelation matrix. 
 
 \begin{Remark}
 Since we do not know $H(\mathbf{\Lambda})$, it will be assumed that the graph is defined by the eigenvalues, $\mathbf{\Lambda}$, that produce the smallest number of edges. This can be achieved by minimizing the number of nonzero values in $\mathbf{L}$ for the given eigenvectors. 
 \end{Remark}
 
 The minimization problem now becomes
 $$\min_{\lambda_k}\left\Vert \mathbf{L} \right\Vert _0 \text{ subject to }  \mathbf{L}= \sum_{k=0}^{N-1} \lambda_k \mathbf{u}_k    \mathbf{u}_k^T,$$
 while the convex (norm-one) form of this minimization problem is  
 $$\min_{\lambda_k}\left\Vert \mathbf{L} \right\Vert _1 \text{ subject to }  \mathbf{L}= \sum_{k=0}^{N-1} \lambda_k \mathbf{u}_k    \mathbf{u}_k^T.$$
 
 \begin{Remark}
 The convex norm-one based form can produce the same solution as the original norm-zero form if the Laplacian sparsity is low and the Laplacian satisfies some other mild conditions (in the sense discussed within Section \ref{SubSGC}). 
 \end{Remark}
 
 Since the eigenvectors are obtained from the decomposition of the correlation matrix, spectral analysis performed in this way is related to  principal components analysis (PCA), where the signal is decomposed onto the set of the  eigenvectors of correlation matrix.
 
 This approach to the graph topology learning can be summarized through the following steps:
 \begin{enumerate}
 	\item For a given set of graph signal observations, $\mathbf{x}_p$, $p=1,2,\dots,P$, calculate the correlation matrix  
  \begin{gather}\mathbf{R}_x= \frac{1}{P}\sum_{p=1}^P\mathbf{x}_p\mathbf{x}_p^T. \label{LambdaCorr}
  \end{gather}
  \item Perform the eigendecomposition of the correlation matrix, in the form
   \begin{gather}\mathbf{R}_x=\mathbf{U}^T\mathbf{\Lambda}_{R_x}\mathbf{U} \nonumber \\
   \mathbf{\Lambda}_{R_x}=\mathbf{U}\mathbf{R}_x\mathbf{U}^T  \label{EigVCorr}
   \end{gather} 
   \item Find the eigenvalues, $\lambda_k$, of the graph Laplacian, $\mathbf{L}=\mathbf{U}^T\mathbf{\Lambda}\mathbf{U}$, such that it assumes the sparsest possible form, using the minimization 
   \begin{gather}\min_{\lambda_k}\left\Vert \mathbf{L} \right\Vert _1 \text{ subject to }  \mathbf{L}= \sum_{k=0}^{N-1} \lambda_k \mathbf{u}_k    \mathbf{u}_k^T. \label{Min1Lap}
   \end{gather}
  \end{enumerate}

\noindent\textbf{Dimensionality-reduced methods.} It is often reasonable to assume that the observed graph signals are generated by exciting a low-order graph system with white noise as the input. However, the problem of estimating the polynomial coefficients from its samples at unknown (eigenvalue) positions is under-determined and cannot be directly solved. However, by adding the constraint that true eigenvalue positions should produce a sparse graph Laplacian, the solution becomes tractable within the compressive sensing framework.  

In this way, instead of the minimization over $N$ variables, $\lambda_k$, $k=0,1,\dots,$ $N-1$, we can find the Laplacian eigenvalues starting from the eigendecomposition of the correlation matrix of a signal produced by a system on a graph, that is,
\begin{equation}
	\mathbf{R}_x =\mathbf{U}|H(\mathbf{\Lambda})|^2\mathbf{U}^T=\mathbf{U}\mathbf{\Lambda}_{R_x}\mathbf{U}^T.
	\label{Reig}
\end{equation} 
Assume that the transfer function of the graph system is of a  polynomial form
\begin{equation}
	H(\mathbf{\lambda}_k)=h_0+h_1\lambda_k+h_2\lambda_k^2+\cdots+h_{M}\lambda_k^{M}
	\label{poly}
\end{equation}
with $M\ll N$.
From the correlation matrix eigendecomposition in (\ref{Reig}), we have $N$ values of $H(\lambda_k)$ obtained as square roots of the eigenvalues of the correlation matrix,  $\lambda^{(\mathbf{R}_x)}_k$.
Without loss of generality, we will assume that nondecreasing $H(\lambda_{k})$, that is $H(\lambda_{k-1}) \le H(\lambda_k)$.
The problem now boils down to the determination of the  Laplacian eigenvalues, $\lambda_k$, $k=0,1,\ldots,N-1$, having in mind that $\lambda_0=0$, $\sum_{k=0}^{N-1}\lambda_k=N$ and that there exist (unknown) coefficients $h_i$, $i=0,1,\ldots,M$ such that (\ref{poly}) is satisfied for each $k$, while the true values $\lambda_k$ produce the sparsest graph Laplacian, $\mathbf{L}$.

The estimation of the system coefficients, Laplacian eigenvalues and Laplacian itself is performed using this \textbf{polynomial fitting method} in the following way:

\begin{enumerate}
	\item Select $M+1$ indices $m_0=0<m_1<\cdots<m_M=N$ with the corresponding transfer function values $H(\lambda_{m_i})$, for $i=0,1,\ldots,M$. Assume that $(M+1)$ eigenvalues are $\bar{\lambda}_{0}=0$, $\bar{\lambda}_{m_1}=\xi_1$, $\bar{\lambda}_{m_2}=\xi_2$, \dots, $\bar{\lambda}_{m_{M-1}}=\xi_{M-1}$, $\bar{\lambda}_{m_M}=1$, where $0<\xi_1<\xi_2<\cdots<\xi_{M-1}<1$. 
	\item Then, the coefficients of an $M$-th order polynomial
	$$P(\bar{\lambda})=a_0 + a_1\bar{\lambda}+a_2\bar{\lambda}^2 + \cdots + a_M\bar{\lambda}^M
	$$
	can be found such that  $P(\hat{\lambda}_i)=H(\lambda_{m_i})$, for $i=0,1,\ldots,M$, is a Lagrange polynomial of $M$-th order defined by $(M+1)$ points.  
	\item Now the eigenvalues of $P(\hat{\lambda})$, $\bar{\lambda}_k$, for each $k$, can be calculated as a solution of
	$$P(\bar{\lambda})=H(\lambda_k), \quad 0\le \bar{\lambda} \le 1$$
	for the unknown $\bar{\lambda}$. Note that this solution is unique if the polynomial $P(\hat{\lambda})$ is an increasing function for $0\le \hat{\lambda}\le 1$. 
	\item Having in mind that $\sum_{k=0}^{N-1}\lambda_k=N$,  the eigenvalues, $\hat{\lambda}_k$, can be found by scaling the obtained values, $\bar{\lambda}_k$, for each $k$, as 
	$
	\hat{\lambda}_k=N\bar{\lambda}_k/\sum_{k=0}^{N-1} \bar{\lambda}_k.
	$
	\item For the so obtained estimates of the eigenvalues, $\hat{\lambda}_k$, the normalized graph Laplacian can be calculated as $\mathbf{L}=\mathbf{U}\mathbf{\hat{\Lambda}}\mathbf{U}^T$, where $\mathbf{\hat{\Lambda}}$ is a diagonal matrix with  $\hat{\lambda}_k$ on the diagonal.
	\item The described procedure should be repeated for various $0<\xi_1<\xi_2<\cdots<\xi_{M-1}<1$ and the final solution is obtained by minimizing the energy normalized sparsity condition, given by
	$$
	\min_{\xi_1,\xi_2,\ldots,\xi_{M-1}} \frac{\Vert \mathbf{L}\Vert_1}{\sqrt{\Vert \mathbf{L}\Vert_2}}.
	$$
\end{enumerate}

Notice that for $M=1$ we should consider only two points in Step 1, and there is no need for the minimization of variables $\xi_i$. For $M=2$, we have one minimization variable $0<\xi_1<1$. For $M=3$, the minimization is performed over only two variables, $0<\xi_1<\xi_2<1$. The dimensionality of the minimization problem is $(M-1)$ and since $M \ll N$, the dimensionality reduction when proposed method is compared to  (\ref{Min1Lap}) is evident.

The spectral indices $0=m_0,m_1,\dots,m_{M}=N$, selected in Step 1, should be equally spaced over $N$ possible indices. For $M=2$, the index $m_1$ should be close to $(N-1)/2$, while for $M=3$ the indices $m_1$ and $m_2$ should be close to $(N-1)/3$ and $2(N-1)/3$, respectively.

\begin{Example}\label{Expoly1}
	Consider a graph with $N=8$ vertices, for which the weight matrix is given in Fig. \ref{par2r}(a). An $N\times P$ matrix of the simulated signal, $\mathbf{X}_P$, was  formed by calculating the graph signal as in (\ref{graphSigDefiH}), with a given graph, its weight matrix, $\mathbf{W=L-I}$, the normalized Laplacian, $\mathbf{L}$,  system order $M$, and system coefficients, $h_0,h_1,\dots,h_M$. White Gaussian external sources, $\boldsymbol{\varepsilon}_p$, with zero-mean and variance $\sigma^2=1$ were assumed in all $P=10,000$ realizations. 

In the first experiment, the proposed method was implemented for the assumed degree $M=2$ of the polynomial $H(\lambda)$, with $h_0=0.3$, $h_1=0.2$, and $h_2=0.5$ used in the graph signal simulation, according to (\ref{graphSigDefiH}).  By forming $\mathbf{R}_x$ from $\mathbf{X}_P$  and after its eigendecomposition, the eigenvectors $\mathbf{U}$ were estimated, while the eigenvalues of the correlation matrix were used to calculate
$H(\lambda_k)=\sqrt{\lambda^{(\mathbf{R}_x)}_k}$. 

\begin{figure}
	\centering
	\includegraphics[scale=0.95]{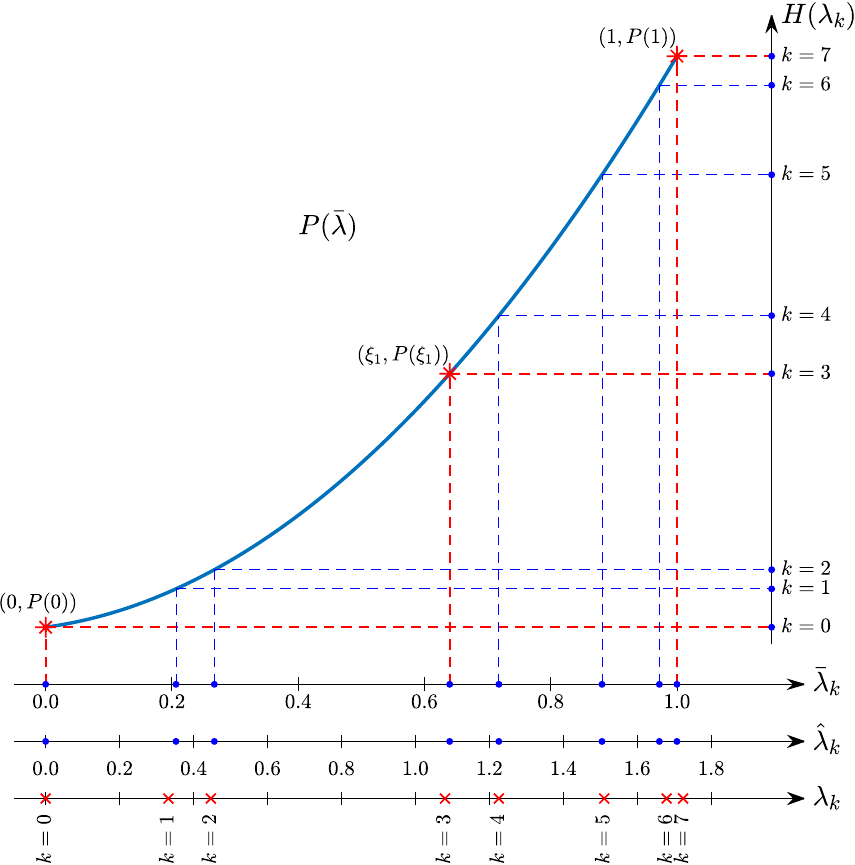}
	\caption{Illustration of eigenvalue calculation based on their second order polynomial obtained from $H(\lambda_k)=\sqrt{\lambda^{(\mathbf{R}_x)}_k}$. } 
\label{par_ilustr}
\end{figure}

Observe that the polynomial fitting method reduces to one-dimensional minimization over variable $0<\xi_1<1$, shown in Fig. \ref{par_ilustr}. After the minimum value of the sparsity measure is found, the eigenvalues are calculated with the corresponding parameter,  $\xi_1$. The Laplacian then follows from $\mathbf{L}=\mathbf{U}^T\mathbf{\Lambda}\mathbf{U}$. 

In this case, the obtained error in the weight matrix elements (absolute value of the off-diagonal elements of the Laplacian) is characterized by $MSE=-35.1$ dB, with the results presented in Fig. \ref{par2r}. The true weight matrix,  $\mathbf{W=L-I}$,  along with estimated one, is given in Fig. \ref{par2r}(a) and (b), the sparsity measure function is plotted in Fig. \ref{par2r}(c), while the true and the estimated Laplacian eigenvalues are given in Fig. \ref{par2r}(d).

\begin{figure}
	\centering

	\includegraphics[scale=0.95]{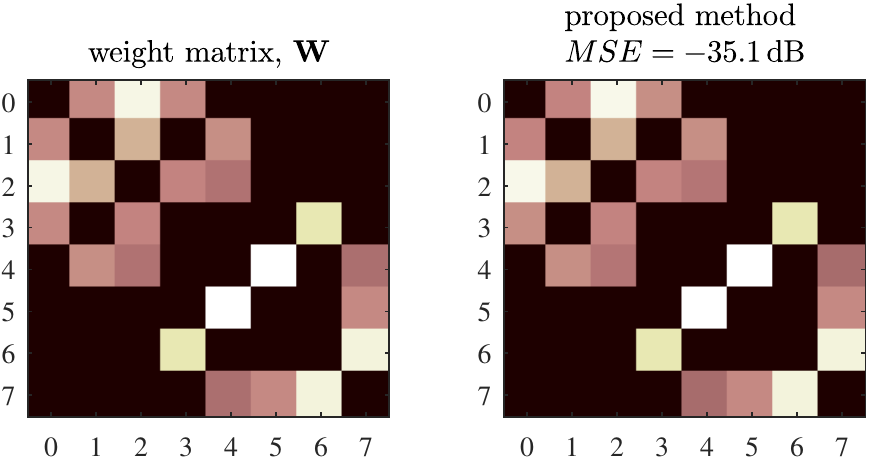}
	\\
	\hspace*{3.5mm}	(a) \hspace{38mm} (b)
	
	\vspace{2mm}

	\includegraphics[scale=0.95]{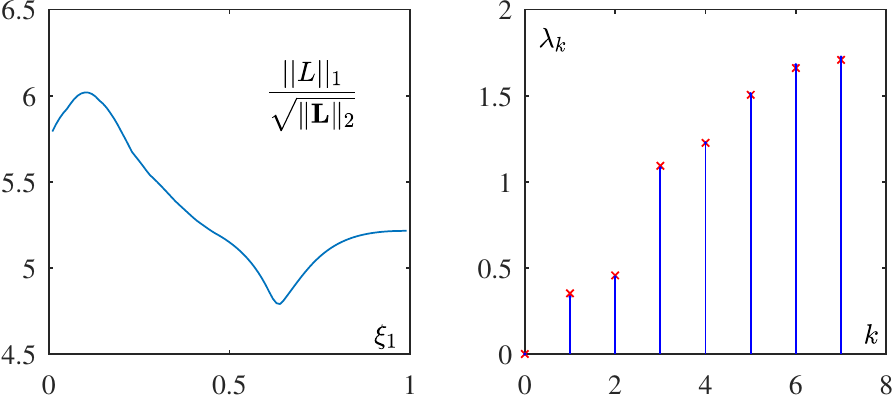}
	\\
	\hspace*{3.5mm}	(c) \hspace{38mm} (d)
	
	\caption{Estimation of the weight matrix, $\mathbf{W=L-I}$, for the graph with $N=8$ vertices.  (a) Ground truth weight matrix. (b) Estimated weight matrix using the sparsity minimization of the normalized Laplacian. (c) Sparsity measure minimization, as a function of parameter $\xi_1$. (d) The exact (blue lines) and estimated (red crosses) eigenvalues of the normalized Laplacian.} 
	\label{par2r}
\end{figure}

\begin{figure}
	\centering

	\includegraphics[scale=0.95]{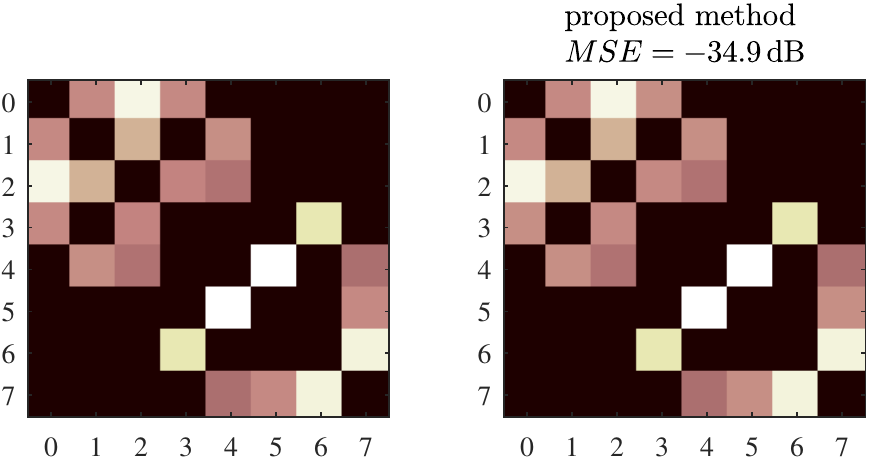}
	\\
	\hspace*{3.5mm}	(a) \hspace{38mm} (b)
	
	\vspace{2mm}

	\includegraphics[scale=0.95]{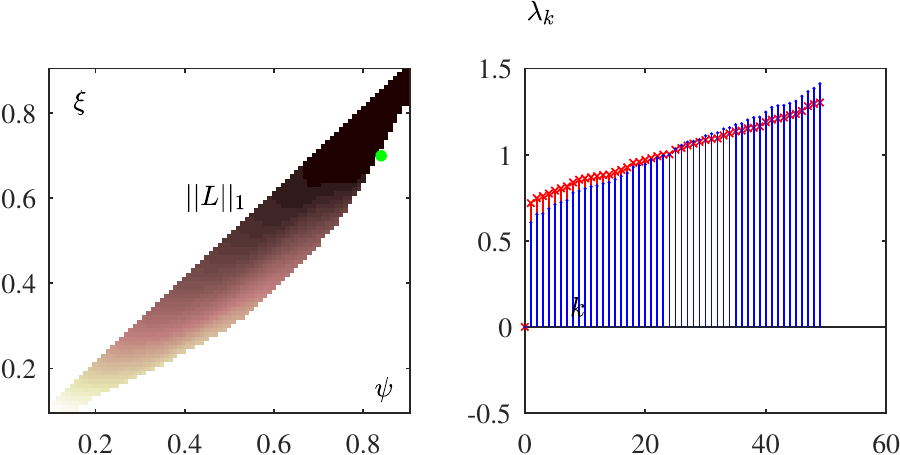}
	\\
	\hspace*{3.5mm}	(c) \hspace{38mm} (d)

	\vspace{2mm}

	\includegraphics[scale=0.95]{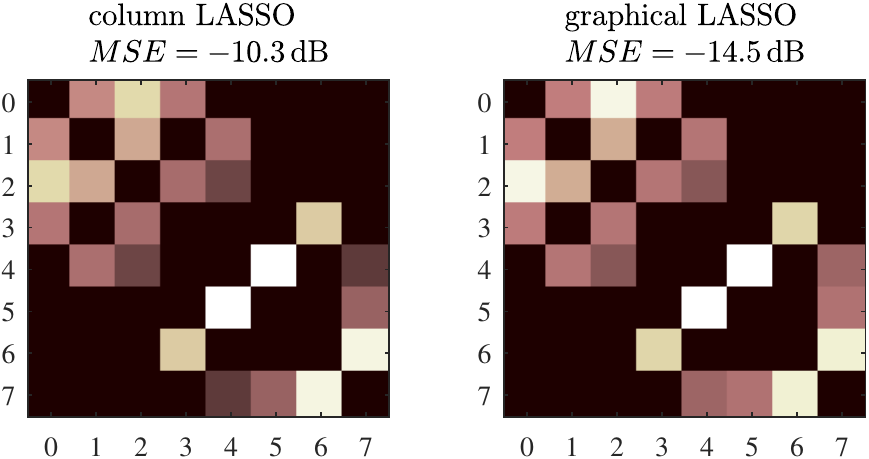}
	\\
	\hspace*{3.5mm}	(e) \hspace{38mm} (f)
	
	\caption{Estimation of the weight matrix for the graph with $N=8$ vertices.  (a) Ground truth weight matrix. (b) Estimated weight matrix using the polynomial fitting method. (c) Sparsity measure minimization, as a function of parameters $\xi_1$ and $\xi_2$. (d) The exact (blue lines) and estimated (red crosses) eigenvalues of the normalized Laplacian. (e)  The estimated weight matrix using the LASSO minimization. (f)  The estimated weight matrix using the graphical LASSO.  } 
	\label{par3r}
\end{figure}

\begin{figure}
	\centering
	\includegraphics[scale=0.95]{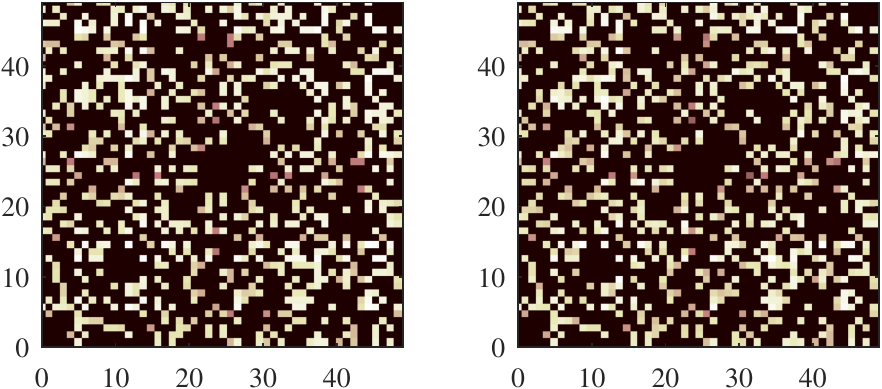}
	\\
	\hspace*{3.5mm}	(a) \hspace{38mm} (b)

	\caption{Estimation of the weight matrix for a graph with $N=50$ randomly positioned vertices.  (a) Ground truth weight matrix. (b) Estimated weight matrix using the sparsity minimization of the normalized Laplacian and the polynomial fitting method.} 
	\label{par3r_N50}
\end{figure}
\end{Example}

\begin{Example}
The experiment from Example \ref{Expoly1} was repeated for a low number of observations, $P=8N_L=256$, where $N_L=32$ is the sparsity of the Laplacian matrix according to practical hints for the number of measurements and sparsity \cite{candes2006robust}. The reconstruction using the polynomial fitting was  with $MSE=-18.0$~dB.

In the second experiment, we assumed $M=3$ and $h_0=0.4$, $h_1=0.5$, $h_2=0.4$, and $h_3=0.2$ when simulating the graph signal, $\mathbf{X}_P$. The correlation matrix was estimated using this simulated signal, along with its eigenvectors and eigenvalues. 
We now have two minimization variables $\xi_1$ and $\xi_2$, $0<\xi_1<\xi_2<1$. 
The results for the polynomial fitting method are presented in Fig. \ref{par3r}(a)-(d). The obtained estimation error was $MSE=-34.9$ dB.
The sparsity measure function (Fig. \ref{par3r}(c)) is now two-dimensional and is calculated only when unique solutions are obtained in Step 3 of the polynomial fitting method. These results were compared with those obtained using the rows of the correlation matrix, $\boldsymbol{\beta}_{n}=\mathrm{lasso}(\mathbf{Y}_n^T, \mathbf{y}_n^T,0.2)$ (Fig. \ref{par3r}(e))  and graphical LASSO, $\mathbf{Q}=\mathrm{glasso}(\mathbf{R}_x,0.3)$ (Fig. \ref{par3r}(f)), with optimized values of the parameter $\rho$. In these cases, the obtained error in the weight matrix elements was characterized by $MSE=-10.3$ dB and $MSE=-14.5$ dB, respectively. 
\end{Example} 

\begin{Example}
Finally, the polynomial fitting method was tested on a larger scale graph, with $N=50$ and $M=2$. The original and estimated weight matrices are shown in Fig.\ref{par3r_N50}.
\end{Example}

 So far, for the examples related to classical data analytics, we have used Fourier analysis and a circular directed graph. The problem formulation presented in this section can also be used to define a graph such that the spectral analysis on this graph leads to some other well known transforms.

 \begin{Example} 
 We will illustrate the method of defining a graph which corresponds to a given classical signal transform on the examples of Hadamard transform with $N=8$, and with the eigenvectors 
 \[
 \mathbf{U}=\frac{1}{\sqrt{8}} \left[  {%
 	\begin{tabular}
 	[c]{rrrrrrrr}%
 	$1$ & $1$ & $1$ & $1$ & $1$ & $1$ & $1$ & $1$\\
 	$1$ & $-1$ & $1$ & $-1$ & $1$ & $-1$ & $1$ & $-1$\\
 	$1$ & $1$ & $-1$ & $-1$ & $1$ & $1$ & $-1$ & $-1$\\
 	$1$ & $-1$ & $-1$ & $1$ & $1$ & $-1$ & $-1$ & $1$\\
 	$1$ & $1$ & $1$ & $1$ & $-1$ & $-1$ & $-1$ & $-1$\\
 	$1$ & $-1$ & $1$ & $-1$ & $-1$ & $1$ & $-1$ & $1$\\
 	$1$ & $1$ & $-1$ & $-1$ & $-1$ & $-1$ & $1$ & $1$\\
 	$1$ & $-1$ & $-1$ & $1$ & $-1$ & $1$ & $1$ & $-1$%
 	\end{tabular}
 }\right]. 
 \] 
 If the eigenvalues are found so as to minimize the number of nonzero elements in the Laplacian, we obtain the graphs for $N=8$ and $N=16$, as shown in Fig. \ref{fig:DadamardT}.  
 
 \begin{figure}
 \centering
 	\includegraphics[scale=1]{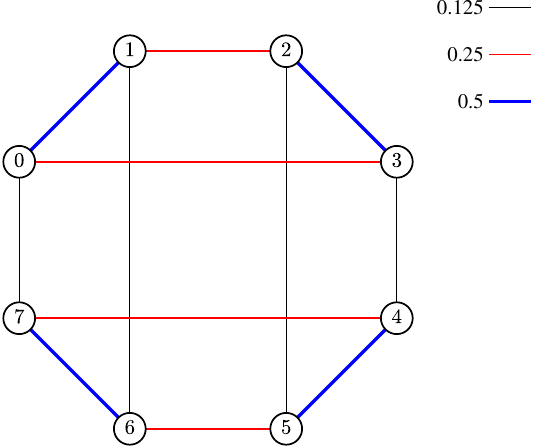} \\
 	\includegraphics[scale=1]{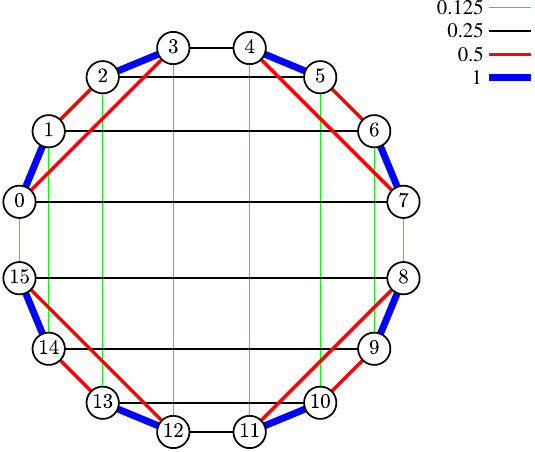}
 	
 	\caption{Graph for which the Laplacian eigenvectors are the Hadamard transform basis functions for $N=8$ and $N=16$.}
 	\label{fig:DadamardT}
 \end{figure}
\end{Example}

 \section{Physically Well Defined Graphs}
 
 The simplest scenario of  graph connectivity is when the graph associated with a problem is \textit{physically well defined}. Examples of such graphs are manifold, including electric circuits, power networks, linear heat transfer, social and computer networks, and spring-mass systems, which will be presented in this section. 
 
 \subsection{Resistive Electrical Circuits}
 
 Graph theory based methods for the analysis and transformations of electrical circuits are already part of classical courses and textbooks. It is also interesting that some general information theory problems can be interpreted and solved within the graph approach to the basic electrical circuits framework. In these cases, the underlying graph topology is well defined and is a part of the problem statement.
 
 The graph Laplacian can also be considered within the basic
 electric circuit theory. In this case, since it can be derived based on the Kirchhoff's laws, the graph Laplacian is also known as the \textit{Kirchhoff matrix} in electric circuit theory. 
 
\noindent\textbf{Graph representation of electric circuits.} Consider a resistive electric circuit, and the electric potential in the circuit vertices (nodes), denoted by $x(n)$. The vertices in an electrical circuit are connected with edges, where the weight of an edge connecting the vertices $n$ and $m$  is defined by the edge conductance, $W_{nm}$. The conductances are the reciprocal values
 to edge resistances 
 $$W_{nm}=\frac{1}{R{_{nm}}}.$$
   The current in the edge from vertex $n$ to vertex $m$ is then equal to
 $$i_{nm}=\frac{x(n)-x(m)}{R_{nm}}=W_{nm}\Big(x(n)-x(m)\Big).$$
 
 \begin{figure*}
 	\centering
 	\includegraphics[]{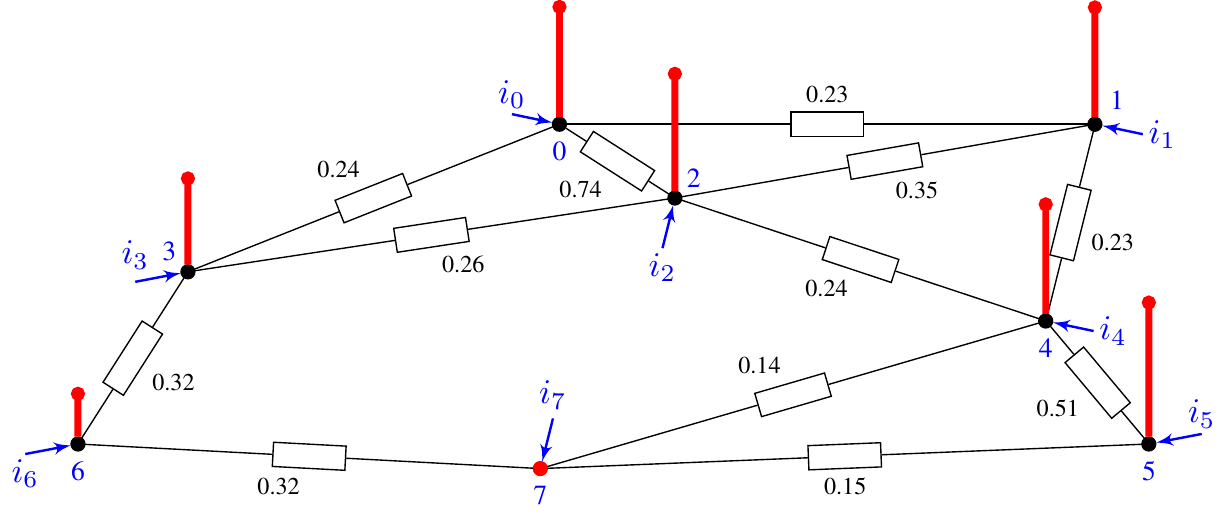}
 	\caption{Electric potential, $x(n)$, as a signal on an electric circuit graph.}
 	\label{fig:ec}
 \end{figure*}

 In addition to the edge currents, an external current generator may be attached to every vertex, and can be considered as a source of the signal change in the vertices. The external current  at a vertex $n$ is denoted  by $i_n$. 
 
 Since the sum of all currents going from a vertex $n$, $n=0,1,\dots,N-1$, must be $0$, that is
 $$
 -i_n+\sum_{m}i_{nm}=0,
 $$
 the current of the external 
 generator at a vertex $n$ must be equal to the sum of all edge currents
 going from this vertex, to give
 \begin{gather*}
 i_n=\sum_{m}W_{nm}\Big(x(n)
 -x(m)\Big)=d_nx(n)
 -\sum_{m}W_{nm}x(m), \\
 n=0,1,\dots,N-1,
 \end{gather*}
 where 
 $$d_n=\sum_{m}W_{nm}=\sum_{m=0}^{N-1}W_{nm}$$ 
 is the degree of vertex $n$. The summation over $m$ can be extended to all vertices, $m=0,1,\ldots, N-1$, since $W_{nm}=0$  if there is no edge between vertices $n$ and $m$. 
 
 The above equations can be written in a matrix form as
 $$\mathbf{i}=\mathbf{D}\mathbf{x}-\mathbf{W}\mathbf{x}$$
 or
 $$\mathbf{L}\mathbf{x}=\mathbf{i}$$
 where $\mathbf{L}=\mathbf{D}-\mathbf{W}$ is the Laplacian of a graph representing an electric circuit.

 If the Laplacian matrix is decomposed as $\mathbf{L}=\mathbf{U} \mathbf{\Lambda}  \mathbf{U}^T$ we have
 $ {\mathbf{\Lambda}}  \mathbf{U}^T
 \mathbf{x=}\mathbf{U}^T\mathbf{i}$, or $$ {\mathbf{\Lambda}}  \mathbf{X}=\mathbf{I}$$
 where $\mathbf{X}=\mathbf{U}^T\mathbf{x}$ and $\mathbf{I}=\mathbf{U}^T\mathbf{i}$ are GDFT of graph signals $\mathbf{x}$ and  $\mathbf{i}$ (see {\color{red}Part II, Section \ref{II-SecGLDFT}}). 
 
 Components of the spectral transform vector $\mathbf{X}$ are such that
 $$\lambda_k X(k)=I(k)$$
 for each $k$.
 
 A signal measured on an electrical circuit graph can be related to the above theory in several ways. For example, potentials on all vertices could be measured under some measurement noise, which calls for application of filtering on a graph. Another possible case is when the external conditions are imposed, for example external sources  are applied to some vertices. We are then interested in potential values at all vertices. This problem corresponds to graph signal reconstruction. 
 
 For nontrivial solutions, there should be an external source on at least two vertices. If we assume that a vertex with an external source is chosen as the reference vertex, then the signal or external source values at these vertices are sufficient to find signal values at all other vertices.
 
 \begin{Example}\label{ExamActive} Consider the graph and signal sensed on the graph presented in Fig. \ref{fig:ec}.
 The signal values are 
 $$\mathbf{x}=[6.71,    6.88,    7.13,    5.25,    6.67,    8.18,    2.62, 0]^T$$
 and the graph Laplacian (as a matrix operator) applied to the signal yields
 $$\mathbf{Lx}=[0, 0, 1, 0, 0, 2, 0,-3]^T.$$
 This means that in this case the vertices denoted by $0,1,3,4,6$ are \textit{not active}, and their values can be obtained as linear combinations of the signal at neighboring active vertices:
 \begin{footnotesize}	
	\begin{align}
	1.21x(0)-0.23 x(1)-0.74x(2) - 0.24 x(3) & =0 \nonumber \\
	-0.23 x(0)+0.81 x(1)-0.35x(2)-0.23 x(4) & =0 \nonumber \\
	-0.24x(0)-0.26x(2)+0.82 x(3)-0.32 x(6) & =0 \nonumber \\
	-0.23 x(1) -0.24x(2) +1.12x(4)-0.51x(5)-0.14x(7)& =0 \nonumber \\
	-0.32 x(3)+0.64 x(6) -0.32x(7)& =0
	\end{align}\label{systElec}
\end{footnotesize}

 \begin{figure*}
 	\centering
 	\includegraphics[]{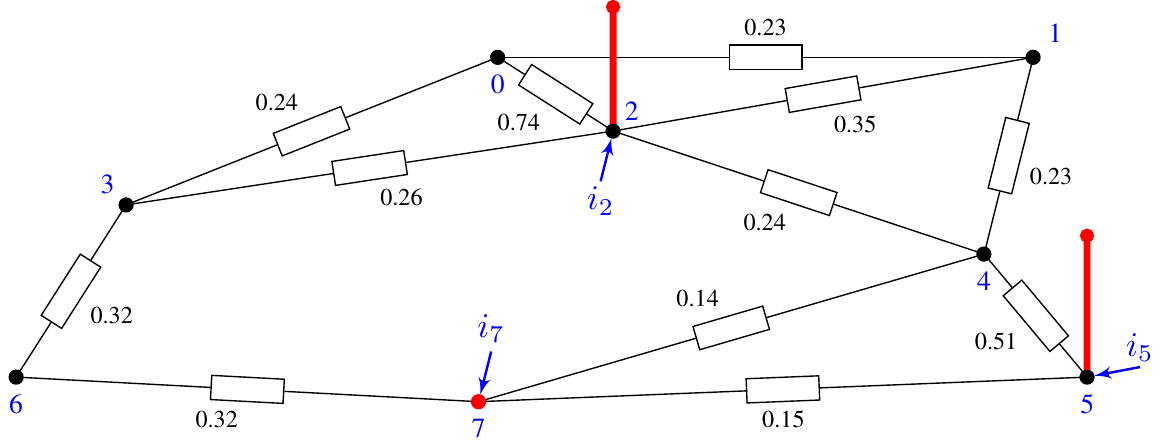}
 	\caption{Electric potential, $x(n)$, as a signal on an electric circuit graph at the three vertices with nonzero external sources. For this graph, all other values of $x(n)$ in Fig. \ref{fig:ec} can be calculated based on the signal values at vertices $n=2$, $n=5$, and $n=7$.  }
 	\label{fig:ec2}
 \end{figure*}

 After solving this system with known signal values $x(2)=7.13$, $x(5)=8.18$, and $x(7)=0$ at the active vertices, we obtain the remaining signal values
 \begin{gather*}\mathbf{x}_p=[x(0), x(1), x(3), x(4), x(6)]^T \\
 =[6.71,    6.88,    5.25,    6.67,      2.62]^T.
 \end{gather*}
  \end{Example}
 
 \subsubsection{Graph transformations} 
 A graph with one or more inactive vertices (where the elements of $\mathbf{Lx}$ are equal o zero) can be simplified by removing these vertices using the well-known transformations of edges connected in series, parallel, or start-to-mesh transforms. This process corresponds to the \textit{downsampling} of the graph  signal.
 
 Similar procedure can be used to add inactive vertices, either by inserting a vertex within an edge or by transforming meshes to stars, in what corresponds to the \textit{interpolation} of the graph signal.
 
\begin{Example} 
For the graph and the graph signal from Example \ref{ExamActive}, the  active vertices are $n=0,5,7$, as shown in Fig. \ref{fig:ec2}, while the signal values at all vertices are given in Fig. \ref{fig:ec}. Notice that the existing signal values will not change, for the given external sources, if the graph is \textit{“downsampled”}, as shown in Fig. \ref{fig:ec_red}, or if the graph signal is \textit{“interpolated”}  by adding new vertices, as shown in Fig. \ref{fig:ec_int}. 

\begin{figure*}
	\centering
	\includegraphics[]{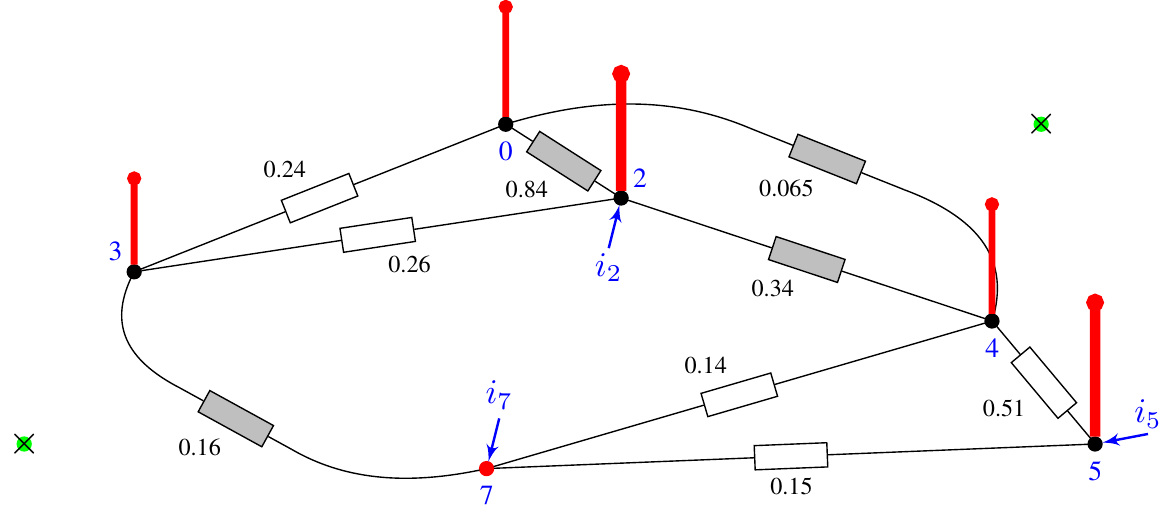}
	\caption{Graph signal, $x(n)$,  from Fig. \ref{fig:ec}, observed on a graph  with a reduced number of vertices (“downsampling”), whereby the vertices $n=6$ and $n=1$ are removed (\textit{crosses in green dots}). Observe that the signal values at the active vertices, $n=0$, $n=5$, and $n=7$, are not changed. The edge weights in gray shade are the equivalent values obtained using the standard resistor, $R_{mn}=1/W_{mn}$, transformations.}
	\label{fig:ec_red}
\end{figure*}

\begin{figure*}
	\centering
	\includegraphics[]{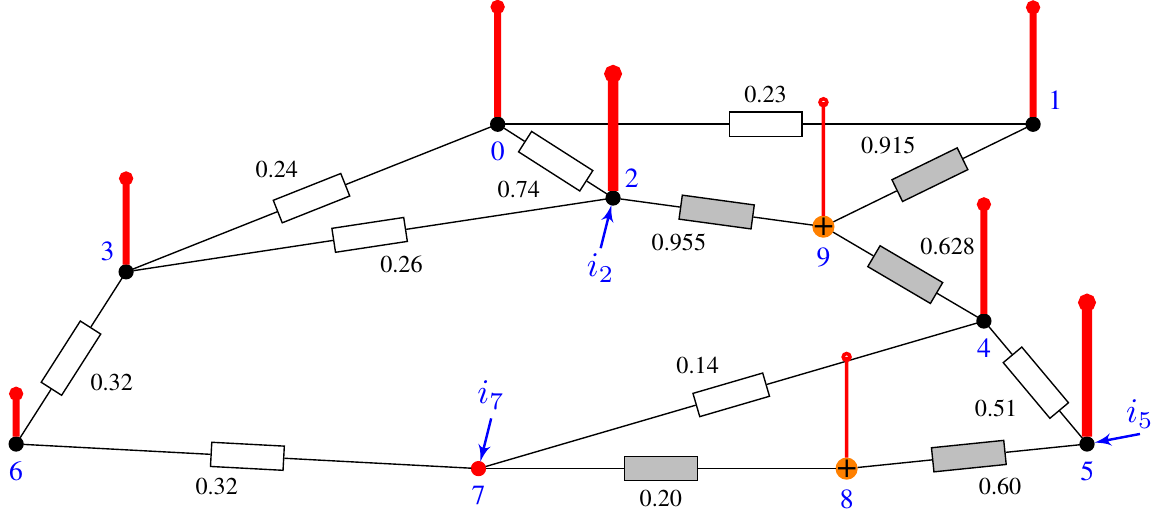}
	\caption{Graph signal, $x(n)$,  from Fig.  \ref{fig:ec} observed on a graph with an extended number of vertices (“interpolation”). Observe that the signal values at all vertices, $n=0,1,2,3,4,5,6,7$, from Fig. \ref{fig:ec2} are not changed. In the locations where the new vertices $n=8$ and $n=9$ are added, the graph signal is interpolated using $x(2)$, $x(5)$, and $x(7)$, as in (\ref{systElec}), and the corresponding edge weights are shown in gray.}
	\label{fig:ec_int}
\end{figure*}

\end{Example}	   
 
\subsubsection{Graph Data Denoising for Sparse External Sources}  

The external sources are considered sparse if their number is much smaller than the number of vertices, $N$. For this scenario, the norm-zero of the external sources vector, $\mathbf{Lx}$, is such that $\Vert \mathbf{Lx} \Vert_0 \ll N$. If the noisy observations, $\mathbf{y}$, of data on graph, $\mathbf{y}$, are available and we know that the number of external sources is small, then the cost function for denoising  can be written in the form
\begin{equation}
J=\Vert \mathbf{y}-\mathbf{x} \Vert_2^2 + \rho \Vert \mathbf{Lx} \Vert_0.
\end{equation}   
This minimization problem can be solved either by writing through the corresponding norm-one form 
\begin{equation}
J=\Vert \mathbf{y}-\mathbf{x} \Vert_2^2 + \rho \Vert \mathbf{Lx} \Vert_1
\end{equation}
or using a kind of matching  pursuit which will be presented in the next example with classical data denoising scenario, since this is not one of the standard approaches in classical data analysis.

\begin{Example}Consider the classical time domain and a piece-wise linear signal, of which noisy observations are available, as shown in Fig. \ref{Denoising_ext_sources}(a). In standard analysis, the graph representation of the domain of this signal is an undirected and unweighted path graph, where the elements of $\mathbf{Lx}$ play the role of external sources, as shown in Fig. \ref{Denoising_ext_sources}(b). We shall assume that $n=0$ is the reference vertex with $x(0)=0$. 

The data denoising problem is then solved in the following way. The initial estimate of the external sources is calculated as $\mathbf{Ly}$. Since we assumed that the external sources are sparse we will consider the positions, $k_1$, $k_2$, $k_3$, $k_4$, and  $k_5$, of $K=5$ largest absolute values of the initial estimate. 

The largest $K$ nonzero values of the external source vector, $\mathbf{Ly}$, are denoted by  $\mathbf{J}_K$, with the elements $i(k_1)$, $i(k_2)$, $\dots$, $i(k_K)$. The value of  $\mathbf{J}_K$ is found in such a way that it minimizes the difference between the estimated data, $\mathbf{L}_K^{(-1)}\mathbf{J}_K$, and the observations, $\mathbf{y}$, that is 
$$\mathrm{min}_{\mathbf{J}_K } \Vert \mathbf{y}-\mathbf{L}_K^{(-1)}\mathbf{J}_K \Vert_2^2,$$
where  $\mathbf{L}_K^{(-1)}$ is obtained from the inverse transform of the graph Laplacian (after the reference row and column, $n=0$ are omitted) by keeping only $K$ columns which correspond to the nonzero positions in the external source vector, $\mathbf{J}_K$.
The solution therefore becomes 
$$\mathbf{J}_K=\mathrm{pinv}(\mathbf{L}_K^{(-1)}) \mathbf{y}.$$
After the nonzero external sources are found, the full external source vector, $[i(1),i(2),\dots,i(N-1)]^T$, is formed using the calculated nonzero values in $\mathbf{J}_K$ and inserting zero values at the remaining positions, as shown in  Fig. \ref{Denoising_ext_sources}(c).  

Finally, the reconstructed signal is obtained from
$$
\setlength{\arraycolsep}{2pt}
\begin{bmatrix}L_{11} & L_{12} & \dots & L_{1,N-1} \\
L_{21} & L_{22} & \dots & L_{2,N-1} \\
\vdots & \vdots & \ddots & \vdots \\
 L_{N-1,2} &  L_{N-1,2} & \dots &  L_{N-1,N-1}
\end{bmatrix}\begin{bmatrix} x(1) \\ x(2) \\  \vdots \\x(N-1)\end{bmatrix}=\begin{bmatrix} i(1) \\ i(2) \\ \vdots \\i(N-1)\end{bmatrix} $$
as, $\mathbf{x}=\mathbf{L}^{-1}\mathbf{J}$, with  the result  shown in Fig. \ref{Denoising_ext_sources}(d). 	
 \begin{figure}
	\centering
	\includegraphics[]{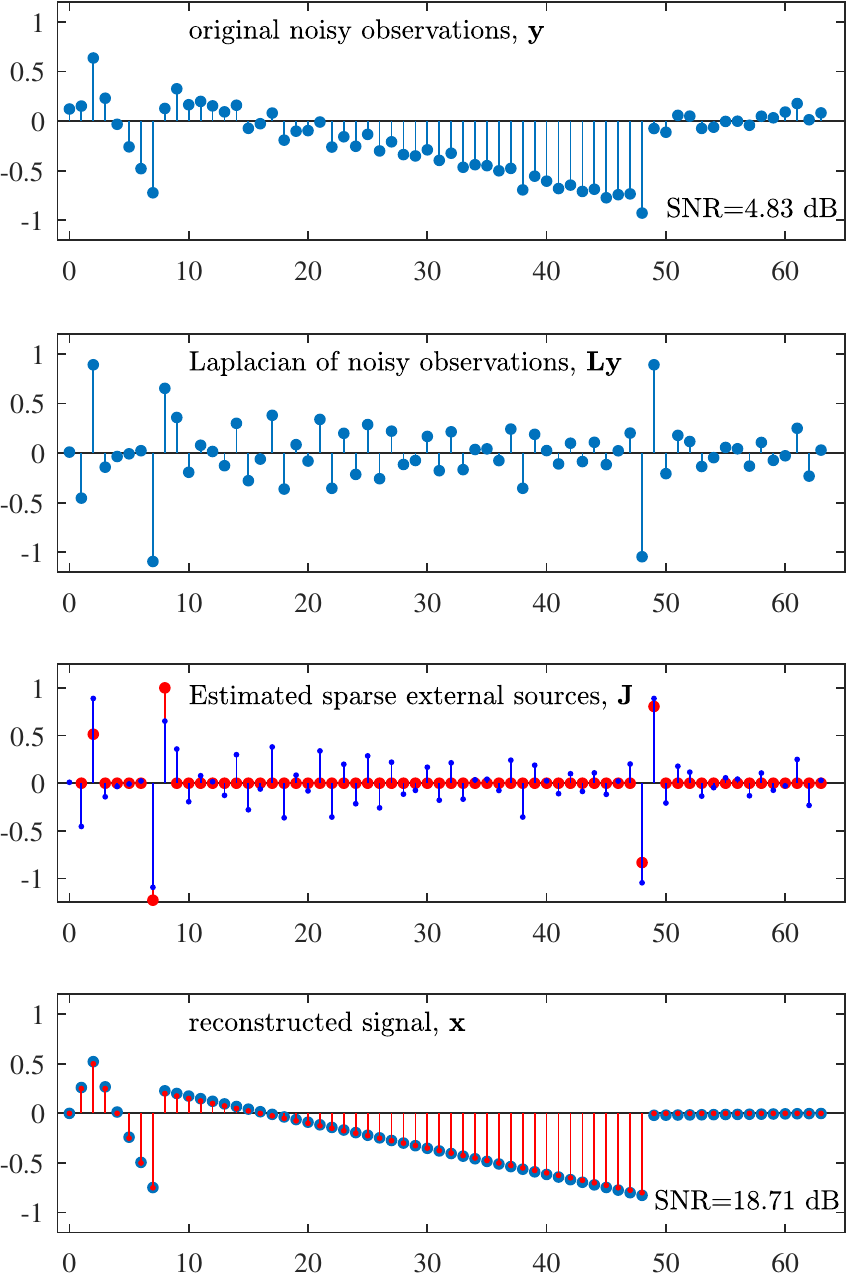}
	\caption{Original piece-wise linear noisy signal (top) and the reconstructed signal (bottom), with the Laplacian of the noisy observations and its re-estimated sparse version (middle panels).}
	\label{Denoising_ext_sources}
\end{figure}

\begin{Remark}
The crucial advantage over the standard total variation (TV) minimization approach in the compressive sensing based denoising is that  the cost function used in this example does not penalize for the linear changes of the signal, while the TV approach promotes piece-wise constant signals. 	
\end{Remark}
	 
\end{Example}	
    
 \subsection{Heat Transfer}
 
 \begin{figure*}
 	\centering
 	\includegraphics[]{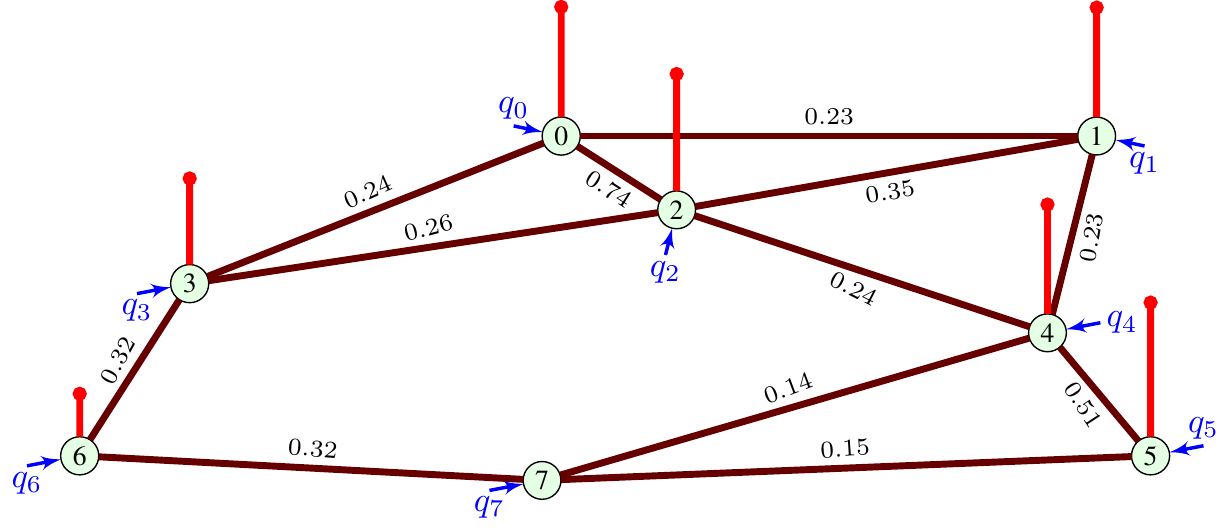}
 	
 	\caption{Temperature, $x(n)=T(n)$, as a signal on a heat transfer graph.}
 	\label{fig:heat}
 \end{figure*}

 The same model as in resistive electrical circuits can be used for a heat transfer network. In this case, the signal values are the measured temperatures, $x(n)=T(n)$, while the heat flux is defined as 
 $$q_{nm}=\Big(T(n)-T(m)\Big)C_{nm}=\Big(x(n)-x(m)\Big)W_{nm},$$
 where $C_{nm}$ are the heat transfer constants, which represent edge weights in the underlying graph,  $C_{nm}= W_{nm}$. 
  
 Then, the input heat flux in the  vertex $n$ can be written as
 $$q_n=\sum_{m}W_{nm}(x(n)
 -x(m))=d_nx(n)
 -\sum_{m=0}^{N-1}W_{nm}x(m),$$
 with 
 $$\mathbf{q}=\mathbf{L}\mathbf{x}$$
 Active vertices are those with an external heat flux, while the passive vertices are those where all heat flux coming to a vertex is forwarded to other vertices, through the edges.
 An example of a heat transfer graph is given in Fig. \ref{fig:heat}.

 \subsubsection{Spring-Mass Systems}
 
 \begin{figure}
 	\centering
 	\includegraphics[]{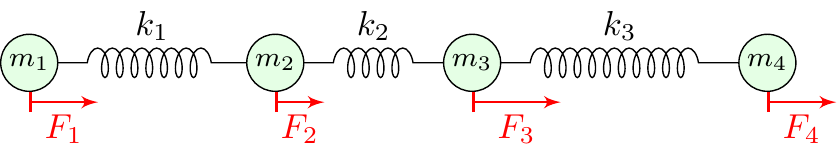}
 	\caption{Spring-mass system on a path graph.}
 	\label{fig:spring-mass}
 \end{figure}

 A spring mass system can also be modeled as a graph. Consider a system of $N=4$ masses which correspond to the path graph, as in Fig. \ref{fig:spring-mass}. Assume that all displacements and forces are in the direction of the system line. According to Hook's law, in a steady state the displacements, $x(n)$, and the forces, $F_n$,   are related as
 \begin{align*}
 k_1(x(1)-x(2)) & =F_1 \\
 k_1(x(2)-x(1))+k_2(x(2)-x(3)) & =F_2 \\
 k_2(x(3)-x(2))+k_3(x(3)-x(4)) & =F_3 \\
 k_3(x(4)-x(3)) & =F_4
 \end{align*}
 or in a matrix form
 \begin{align*}
 \begin{bmatrix}
 k_1 & -k_1 & 0 & 0\\
 -k_1 & k_1+k_2 & -k_2 \\
 0 & -k_2 & k_2+k_3 & -k_3 \\
 0 & 0 & -k_3 & k_3
 \end{bmatrix}
 \begin{bmatrix}
 x_1\\
 x_2\\
 x_3\\
 x_4\\
 \end{bmatrix}
 & =
 \begin{bmatrix}
 F_1\\
 F_2\\
 F_3\\
 F_4\\
 \end{bmatrix} \\
 \mathbf{L} \mathbf{x} & = \mathbf{F}
 \end{align*}
 These equations define a weighted graph and its corresponding graph Laplacian.
 
 Given that the graph Laplacian is singular matrix, in order to solve this system for unknown displacements (graph signal), we should introduce a reference vertex with a fixed position (zero displacement). Then, the system $\mathbf{L} \mathbf{x}  = \mathbf{F}$ can be solved.

 \subsubsection{Social Networks and Linked Pages}
 Social networks are also examples of well defined graphs, where the vertices are network members and the edges define their relationships in a social network. If two members are related, then the corresponding edge weight is $1$, and the weight matrix is equal to the adjacency matrix. An example of a small social network with the corresponding member links is shown in Fig. \ref{fig:social-network}.
 
 \begin{figure*}
 	\centering
 	\includegraphics[]{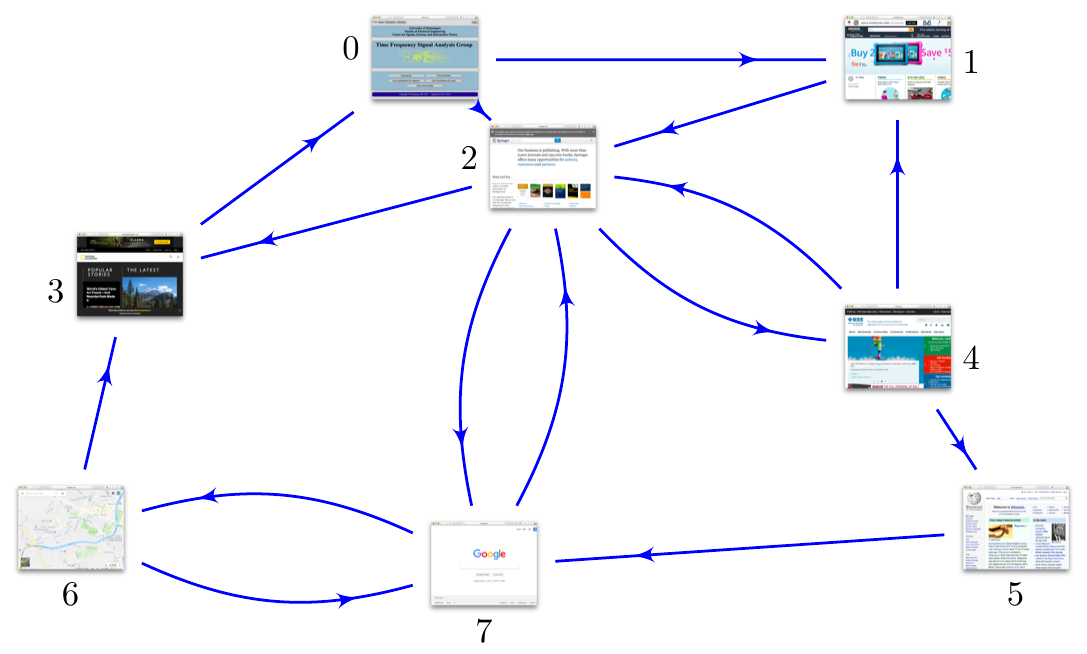}
 	\caption{Hyper-linked pages represented as a directed graph.}
 	\label{fig:linked-pages}
 \end{figure*}

 Pages with hyper-links can also be considered as a well defined directed graph; an example of links between $N=8$ pages is given in Fig. \ref{fig:linked-pages}.
 An interesting parameter for this kind of graphs is the PageRank.

 \subsection{PageRank} \label{PageRank}

The PageRank was defined by Google to rank the web pages. For a directed graph, PageRank of vertex $n$ is defined as a graph signal satisfying the relation  
 $$x(n)=\sum_m \frac{1}{d_{m}} W_{mn}x(m),$$ 
 where $ W_{mn}$ are weights of the directed edges connecting the vertex $m$ to vertex $n$ and $d_m$ is the outgoing degree of the vertex $m$. This means that
 the PageRank of each vertex is related to the PageRank of the vertices connected to it. 
 
 The PageRank is usually calculated using an iterative procedure  defined by
\begin{equation}x_{k+1}(n)=\sum_m \frac{1}{d_{m}} W_{mn}x_{k}(m), \label{pagerank0} 
 \end{equation}
 starting from an arbitrary PageRank, for example $x_0(n)=1$.  
  In the original definition by Google scaling factors  $0.15$ and $0.85$ were added, to give
\begin{equation}x_{k+1}(n)=0.15+0.85\sum_m \frac{1}{d_{m}} W_{mn}x_{k}(m). \label{pagerankgoogle}
\end{equation}
 
 \begin{Example}\label{ExampPager}
 Consider the graph from  Fig. \ref{fig:linked-pages} (the same graph as in {\color{red} Part I, Fig. \ref{I-GSPb_ex1a}(b))}. In this case the vertices represent pages on the Internet, while the directed edges designate their relations. For example, the page which corresponds to vertex $0$ cites (gives a hyper-link to) pages marked with $1$ and $2$, while it is cited (hyper-linked) by a page at vertex $3$. All other vertices are connected by the edges in the same way. Intuitively, we can expect that the rank in this network is higher for the pages that are highly cited (hyper-linked) with other also highly cited (hyper-linked) pages. To find the rank of the pages in this graph/network, we shall calculate the PageRank for all pages/vertices. The weight/adjacency matrix of this graph, $\mathbf{W}=\mathbf{A}$, is given by (see also {\color{red} Part I,  equation (\ref{I-AdjMtxFirs})})
 \begin{gather}
 \hspace{-1.8mm}   \mathbf{W}  = \!
 \begin{array}{cr}
 & \\
 {
 	\color{blue}
 	\begin{matrix}
 	\text{\footnotesize 0}\\
 	\text{\footnotesize 1}\\
 	\text{\footnotesize 2}\\
 	\text{\footnotesize 3}\\
 	\text{\footnotesize 4}\\
 	\text{\footnotesize 5}\\
 	\text{\footnotesize 6}\\
 	\text{\footnotesize 7}\\
 	\end{matrix}
 } &  \!
 \begin{bmatrix}
 \ 0\  & \ 1\  & \ 0\  & \ 0\  & \ 0\  & \ 0\  & \ 0\  & \ 0\  \\
 \ 0\  & \ 0\  & \ 1\  & \ 0\  & \ 0\  & \ 0\  & \ 0\  & \ 0\  \\
 \ 1\  & \ 0\  & \ 0\  & \ 1\  & \ 1\  & \ 0\  & \ 0\  & \ 1\  \\
 \ 1\  & \ 0\  & \ 0\  & \ 0\  & \ 0\  & \ 0\  & \ 0\  & \ 0\  \\
 \ 0\  & \ 1\  & \ 1\  & \ 0\  & \ 0\  & \ 1\  & \ 0\  & \ 0\  \\
 \ 0\  & \ 0\  & \ 0\  & \ 0\  & \ 0\  & \ 0\  & \ 0\  & \ 1\  \\
 \ 0\  & \ 0\  & \ 0\  & \ 1\  & \ 0\  & \ 0\  & \ 0\  & \ 1\  \\
 \ 0\  & \ 0\  & \ 1\  & \ 0\  & \ 0\  & \ 0\  & \ 1\  & \ 0\ 
 \end{bmatrix}
 \end{array}\!\!\!. \label{AdjMtxFirsW}
 \end{gather} 
 
  The outgoing vertex degrees are calculated as the sum of columns of the matrix $\mathbf{W}^T$, that is 
  $d_m=\sum_{n=0}^7 W_{mn}.$
 Their values are
 $$\mathbf{d}= [ 1 \ \ 1 \ \ 4 \ \ 1 \ \ 3 \ \ 1 \ \ 2 \ \ 2 ]. $$ 
 Now, the PageRank values for vertices can be obtained through an iterative procedure starting with the initial page ranks $\mathbf{x}_0= [1, 1, 1, 1, 1, 1, 1, 1].$ After a few iterations, the results for PageRank are
 $$
 \begin{bmatrix}
 \mathbf{x}_0^T \\
 \mathbf{x}_1^T  \\
 \mathbf{x}_2^T   \\
 \vdots \\
 \mathbf{x}_5^T    \\
 \vdots \\
 \mathbf{x}_{11} ^T
 \end{bmatrix}
 =
 \begin{bmatrix}
 1.00 \ \ 1.00 \ \ 1.00 \ \ 1.00 \ \ 1.00 \ \ 1.00 \ \ 1.00 \ \ 1.00 \\
 1.25   \ \  1.33   \ \  1.83   \ \  0.75   \ \  0.25  \ \   0.33   \ \  0.50   \ \  1.75 \\
 1.21   \ \  1.33  \ \ 2.29  \ \  0.71   \ \  0.46   \ \  0.08   \ \  0.87  \ \   1.04 \\
 \vdots \\
 1.29  \ \   1.68   \ \  2.10   \ \  0.80   \ \  0.52   \ \  0.17   \ \  0.46   \ \  0.99 \\
 \vdots \\
 1.33   \ \  1.53  \ \   2.14   \ \  0.80   \ \  0.55  \ \   0.18   \ \  0.48   \ \  0.99
 \end{bmatrix}.
 $$
 \end{Example}

 The matrix form of the iterations in (\ref{pagerank0}) is 
 $$\mathbf{x}_{k+1}= \mathbf{W}_N\mathbf{x}_{k},$$ 
 where $\mathbf{W}_N$ is obtained from $\mathbf{W}^T$ by dividing all elements of the $m$th column, $m=0,1,\dots,N-1$, by $d_m$. The mean-values of matrix  $\mathbf{W}_N$ columns are normalized. 
 
 \begin{Example} 
 	In Example \ref{ExampPager}, the normalized adjacency/weighing matrix is  
 $$
 \mathbf{W}_N=
 \begin{bmatrix}
 \ 0\  & \ 0 \  & \ \frac{1}{4}\  & \ 1\  & \ 0\  & \ 0\  & \ 0\  & \ 0\  \\
 \ 1\  & \ 0\  & \ 0\  & \ 0 \  & \ \frac{1}{3}\  & \ 0\  & \ 0\  & \ 0  \  \\
 \ 0\  & \ 1 \  & \ 0\  & \ 0\  & \ \frac{1}{3}\  & \ 0\  & \ 0\  & \ \frac{1}{2}\  \\
 \ 0\  & \ 0\  & \ \frac{1}{4}\  & \ 0\  & \ 0\  & \ 0\  & \ \frac{1}{2}\  & \ 0\  \\
 \ 0\  & \ 0\  & \ \frac{1}{4}\  & \ 0\  & \ 0\  & \ 0\  & \ 0\  & \ 0\  \\
 \ 0\  & \ 0\  & \ 0\  & \ 0\  & \ \frac{1}{3}\  & \ 0\  & \ 0\  & \ 0\  \\
 \ 0\  & \ 0\  & \ 0\  & \ 0\  & \ 0\  & \ 0 \  & \ 0\  & \ \frac{1}{2}\  \\
 \ 0\  & \ 0\  & \ \frac{1}{4}\  & \ 0\  & \ 0\  & \ 1\  & \ \frac{1}{2}\  & \ 0\ 
 \end{bmatrix}.
 $$
 \end{Example}

 The final, steady state, PageRank can be obtained from
 $$\mathbf{x}= \mathbf{W}_N\mathbf{x}.$$
 The final PageRank, $\mathbf{x}$, is the eigenvector of matrix $\mathbf{W}_N$ corresponding to the eigenvalue equal to $1$. 
 
 \begin{Example} The eigenvalue decomposition of the matrix $\mathbf{W}_N$ in Example \ref{ExampPager}  results in the eigenvector which corresponds to eigenvalue $\lambda_k=1$, whose elements are 
 $$\mathbf{x}^T=[1.33  \ \    1.52 \ \    2.18  \ \   0.79  \ \   0.55  \ \   0.18  \ \   0.48  \ \   0.97].$$
 The eigenvector is normalized with its mean value. It corresponds to the iterative solution obtained after 11 iterations. 
\end{Example}

 \subsection{Random Walk}\label{RandomWalk}

 Assume that the signal, $x(n)$, represents the probabilities that a random walker is present at a vertex $n$. The random walker will then transit from the vertex $n$ to one of its neighboring vertices, $m$, with probability $p_{nm}$. There are several ways to define this probability and the corresponding forms of random walk; for an extensive review see \cite{masuda2017random}. Here, we consider two random-walk definitions:
 	\begin{itemize}
 		\item vertex-centric random walk, and
 		\item edge-centric random walk.
 		\end{itemize}
 In the \textbf{vertex-centric random walk} the probability,  $p_{nm}$, that a random walker will transit from the vertex $n$ to one of its neighboring vertices, $m$, is defined by
 \begin{equation}
 p_{nm}=\frac{W_{nm}}{\sum_m W_{nm}}=\frac{1}{d_n}W_{nm}, \label{probRW}
\end{equation}
 where $W_{nm}$ are the affinities of the walker to transit from a vertex $n$ to a vertex $m$ and ${d_n}=\sum_m W_{nm}$ is the degree of a vertex $n$. The probability, $x_{p+1}(m)$, that a walker is at the vertex $m$ at the step $(p+1)$ is then equal to the sum of all probabilities that a walker was in one the vertices $n$ at the distance equal to one (neighboring vertices to the vertex $m$) multiplied by the probabilities that the walker transits from the vertex $n$ to the vertex $m$, that is 
 \begin{equation}x_{p+1}(m)=\sum_nx_{p}(n)p_{nm}= \sum_nx_{p}(n)\frac{1}{d_n}W_{nm}.\label{walker1d}
 \end{equation}
 
 The calculation of the signal $x(n)$ can now be naturally considered within the graph framework, where $W_{nm}$ are edge weights.
 
 The probabilities in the stage $(p+1)$ of the random walk transition are calculated starting from the probabilities at the previous stage as in (\ref{walker1d}), with the matrix form given by 
 $$\mathbf{x}_{p+1}=\mathbf{W}\mathbf{D}^{-1}\mathbf{x}_p$$
 or 
 $$\mathbf{D}^{-1/2}\mathbf{x}_{p+1}=\mathbf{D}^{-1/2}\mathbf{W}\mathbf{D}^{-1/2}\mathbf{D}^{-1/2}\mathbf{x}_{p},$$ 
 where the matrix $\mathbf{W}$ is a matrix of weighting coefficients and $\mathbf{D}$ is the degree matrix.
 
 In the steady state, when $\mathbf{x}_{p+1}=\mathbf{x}_{p}=\mathbf{x}$, we have
 $$ \mathbf{y} = \mathbf{D}^{-1/2}\mathbf{W}\mathbf{D}^{-1/2} \mathbf{y}$$
where $ \mathbf{y} =\mathbf{D}^{-1/2}\mathbf{x}$. The solution is the smoothest eigenvector of the normalized Laplacian, $\mathbf{L}_N=\mathbf{I}- \mathbf{D}^{-1/2}\mathbf{W}\mathbf{D}^{-1/2}$, calculated from
$$ (\mathbf{I}- \mathbf{D}^{-1/2}\mathbf{W}\mathbf{D}^{-1/2}) \mathbf{y}=\mathbf{0},$$
and given by $\mathbf{y}=[1,1,\dots,1]^T/\sqrt{N}$ or $$\mathbf{x}=\mathbf{D}^{1/2}[1,1,\dots,1]^T/\sqrt{N}.$$
Note that the vector $\mathbf{x}$ is not constant, and its elements are given by $x(n)=\sqrt{d_n/N}$.
  
  In the \textbf{edge-centric random walk} the probability,  $p_{nm}$, is defined by 
  \begin{equation}x_{p+1}(m)=\sum_nx_{p}(n)p_{nm}=\frac{1}{d_m} \sum_nx_{p}(n)W_{nm}.\label{walker1dEC}
\end{equation}
In this case, the in-flow probability $\sum_nx_{p}(n)W_{nm}$ for the vertex $m$ is equal (balanced) to the out-flow probability of this vertex, $x_{p+1}(m)d_m=\sum_nx_{p+1}(m)W_{nm}$. This model of random walk is also called fluid model and it has a simple interpretation within the electric circuits framework, since the probabilities (if considered as the electric potentials) satisfy the first Kirchoff low for the vertex $m$ serving as an electric circuit node, that is 
$$\sum_n\Big(x_{p+1}(m)-x_{p}(n)\Big)W_{nm}=0.$$ 
The matrix form of the edge-centric random walk is given by 
 $$\mathbf{x}_{p+1}=\mathbf{D}^{-1}\mathbf{W}\mathbf{x}_p$$
 or $\mathbf{D}\mathbf{x}_{p+1}=\mathbf{W}\mathbf{x}_{p}$.
 In the steady state, for $\mathbf{x}_{p+1}=\mathbf{x}_{p}=\mathbf{x}$, we have
 $$ \mathbf{D} \mathbf{x} = \mathbf{W} \mathbf{x}$$
 or
 \begin{equation}\mathbf{L}\mathbf{x}=\mathbf{0}. \label{Lxjed0}
 \end{equation}
 The solution of this equation is the smoothest (constant) eigenvector of the graph Laplacian, $\mathbf{x}=[1,1,\dots,1]^T/\sqrt{N}$. 
 
The presented graph theory framework admits for various problem formulations and solutions.

\begin{Example}
Consider the  graph from {\color{red} Fig. 2 in Part I} and the case where we desire  to find the probabilities, $x(n)$, that the walker reaches vertex $5$ before he reaches vertex $7$, starting from any vertex $n$, assuming that transition probabilities may be defined according the edge-centered random walk model. We therefore have to solve the system $\mathbf{L}\mathbf{x}=\mathbf{0}$,
 with $x(5)=1$ and $x(7)=0$. 
 
In the same way, we can solve another practically interesting problem. An information has reached a member of social network in Fig. \ref{fig:social-network} at vertex $4$, but it has not reached the member at vertex $3$. The task is to find probabilities that the information is known to a vertex $n$.

\begin{figure*}
	\centering
	\includegraphics[scale=1]{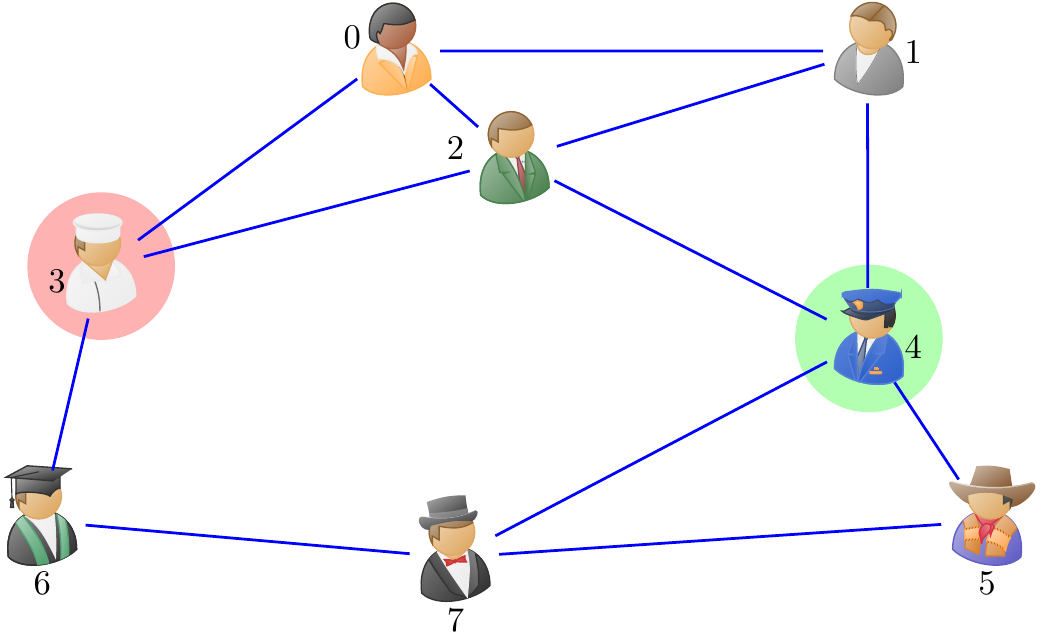}
	\caption{An example of a small social network represented as an undirected graph.}
	\label{fig:social-network}
\end{figure*}

Since the information is present at vertex $4$, then $x(4)=1$ is a certain event, and the fact that the information has not reached vertex $3$ means that $x(3)=0$. Again, according to the analysis from (\ref{probRW}) to (\ref{Lxjed0}), we have to solve the system $\mathbf{L}\mathbf{x}=\mathbf{0}$,
with $x(4)=1$ and $x(3)=0$, that is
\begin{gather}
 \begin{bmatrix}
 \ 3\  & \!\!-1\  &  \!\!-1\  &  \!\!{\color{red}-1}\  & \ {\color{green}0}\  & \ 0\  & \ 0\  & \ 0\  \\
\!\!-1\  & \ 3 \  & \!\!-1\  & \ {\color{red}0}\  & \!\!{\color{green}-1}\  & \ 0\  & \ 0\  & \ 0\  \\
\!\!-1\  &  \!\!-1\  & \ 4\  &  \!\!{\color{red}-1}\  &  \!\!{\color{green}-1}\  & \ 0\  & \ 0\  & \ 0\  \\
 {\color{red}-1}\  & \ {\color{red}0}\  & \!\! {\color{red}-1}\  & \ {\color{red}3}\  & \ {\color{red}0}\  & \ {\color{red}0}\  &  \!\! {\color{red}-1}\  & \ {\color{red}0}\  \\
 \ {\color{red}0}\  & \ \!\! {\color{red}-1}\  & \ \!\! {\color{red}-1}\  & \ {\color{red}0}\  & \ {\color{red}4}\  & \ \!\! {\color{red}-1}\  & \ {\color{red}0}\  & \ \!\!\!\! {\color{red}-1}\  \\
 \ 0\  & \ 0\  & \ 0\  & \ {\color{red}0}\  &  \!\!{\color{green}-1}\  & \ 2\  & \ 0\  & \!\!-1\  \\
 \ 0\  & \ 0\  & \ 0\  &  \!\!{\color{red}-1}\  & \ {\color{green}0}\  & \ 0\  & \ 2\  &  \!\!-1\  \\
 \ 0\  & \ 0\  & \ 0\  & \ {\color{red}0}\  &  \!\!{\color{green}-1}\  & \!\!-1\  & \!\!-1\  & \ 3\ 
 \end{bmatrix}\begin{bmatrix}x(0)\\ x(1) \\ x(2) \\ {\color{red}0} \\ {\color{green}1} \\ x(5) \\ x(6) \\ x(7)
  \end{bmatrix} = \mathbf{0},
\label{LapmmMtxFirsW}
 \end{gather}  
where columns and rows in red font are to be removed (rows for the known signal  values, $x(3)$ and $x(4)$, and  column for the zero-valued signal, $x(3)$), while the green font marks the column to be moved on the right side of equation for the known signal value, $x(4)=1$.  
The solution is obtained from
\begin{gather}
 \begin{bmatrix}
 \ 3\  & \!\!-1\  &  \!\!-1\  &    \ 0\  & \ 0\  & \ 0\  \\
\!\!-1\  & \ 3 \  & \!\!-1\  &  \ 0\  & \ 0\  & \ 0\  \\
\!\!-1\  &  \!\!-1\  & \ 4\    & \ 0\  & \ 0\  & \ 0\  \\
\ 0\  & \ 0\  & \ 0\  &  \ 2\  & \ 0\  & \!\!-1\  \\
 \ 0\  & \ 0\  & \ 0\  &  \ 0\  & \ 2\  &  \!\!-1\  \\
 \ 0\  & \ 0\  & \ 0\  & \!\!-1\  & \!\!-1\  & \ 3\ 
 \end{bmatrix}\begin{bmatrix}x(0)\\ x(1) \\ x(2)  \\ x(5) \\ x(6) \\ x(7)
 \end{bmatrix} = \begin{bmatrix} {\color{green}0} \\ {\color{green}1} \\ {\color{green}1} \\ {\color{green}1} \\{\color{green}0} \\{\color{green}1} \end{bmatrix},
 \end{gather}  
with the inserted values $x(4)=1$ and $x(3)=0$, in the following form
$$\mathbf{x}=[0.375,  \  0.625, \   0.5, \ {\color{red}0}, \ {\color{green}1},  \  0.875, \   0.375,  \  0.75]^T.$$
This means that the information is most probably available to the vertex $5$, with probability $x(5)=0.875$, while the lowest probability is that the information is available to the vertices $0$ or $6$, with probability $x(0)=x(6)=0.375$, as it can be expected from an intuitive analysis of this graph with small number of vertices.

\end{Example}

\subsection{Hitting and Commute Time} \label{Hitting}

The random walk problem is closely related to the hitting and commute time. The \textit{hitting time}, $h(m,n)$, from a vertex $m$ to any vertex $n$ is defined as the expected number of steps for a random walker to travel from the vertex $m$ to a  vertex $n$. Denote by $x^{(m)}_{p}(l)$ the hitting time from the reference vertex $m$ to the vertices $l$ which are the neighboring vertices of the considered vertex $n$. Then, the random walker will arrive from a vertex $l$ to the vertex $n$ in one step with the probability that he chooses to transit from the specific $l$ to the considered $n$. The probability that a random walker is at the neighboring vertex $l$ and transits to vertex $n$ is then
$$p_{ln}=\frac{W_{ln}}{\sum_kW_{nk}}=\frac{1}{d_n}W_{ln}.$$   The hitting time for vertex $n$ is equal to the sum of all hitting times of neighboring vertices with one step added
$$x^{(m)}_{p+1}(n)=\sum_{l}x^{(m)}_p(l)p_{ln}+1=\frac{1}{d_n}\sum_{l}x^{(m)}_p(l)W_{ln}+1.$$  The matrix form of this equation is 
$$\mathbf{x}^{(m)}_{p+1}=\mathbf{D}^{-1}\mathbf{W}\mathbf{x}^{(m)}_p+\begin{bmatrix}1 \\ 1\\ \vdots \\1 \end{bmatrix}.$$  
In the steady state, we have 
$$\mathbf{D}\mathbf{x}^{(m)}=\mathbf{W}\mathbf{x}^{(m)}+\mathbf{d},$$  
where $\mathbf{d}=\mathbf{D}[1,1,\dots,1]^T$ is a degree vector. Finally the hitting time, $h(m,n)=x^{(m)}(n)$, is a solution of the linear system of equations 
\begin{equation}\mathbf{L}_m\mathbf{x}^{(m)}=\mathbf{d} \label{randWhitTim}
\end{equation} 
with the reference  vertex $m$, where $x(m)=0$ is removed from the vector $\mathbf{x}$ to form $\mathbf{x}^{(m)}$ with elements $h(m,n)$, $n=0,1,\dots,N-1$, $n \ne m$. The equation for vertex $m$ is also removed, so that the system is of an $(N-1)$-order and the matrix $\mathbf{L}_m$ is obtained from the graph Laplacian, $\mathbf{L}$, by removing its $m$th row and $m$th column. 

\begin{Example}\label{ExamHittt4}
	We shall calculate the hitting time for all vertices, $n$, from the vertex $m=3$ for the graph from {\color{red} Fig. 2 in Part I}. For this graph, we have
\begin{equation*}
\footnotesize
\setlength{\arraycolsep}{1.8pt}
\begin{bmatrix*}[r]
1.21 & -0.23 & -0.74 & 0 & 0 & 0 & 0 \\
-0.23 & 0.81 & -0.35 & -0.23 & 0 & 0 & 0\\
-0.74 & -0.35 & 1.59 & -0.24 & 0 & 0 & 0\\
0 & -0.23 & -0.24 & 1.12 & -0.51 & 0 & -0.14\\
0 & 0 & 0 & -0.51 & 0.66 & 0 & -0.15\\
0 & 0 & 0 &  0 & 0 & 0.64 & -0.32\\
0 & 0 & 0 &  -0.14 & -0.15 & -0.32 & 0.61
\end{bmatrix*}\!\! \begin{bmatrix}h(0,3) \\ h(1,3) \\ h(2,3) \\ h(4,3)\\ h(5,3) \\ h(6,3) \\ h(7,3) \end{bmatrix}\!\!\!=\!\!\!\begin{bmatrix}1.21 \\ 0.81 \\ 1.59 \\ 1.12\\ 0.66 \\ 0.64 \\ 0.61 \end{bmatrix}
\end{equation*}	
and this matrix is obtained from the graph Laplacian by removing the row and column corresponding to $m=3$. The hitting times from the vertex $M=3$ are then obtained as
\begin{equation*}
\footnotesize
\setlength{\arraycolsep}{2.8pt}
 \begin{bmatrix}h(0,3) \\ h(1,3) \\ h(2,3) \\ h(4,3)\\ h(5,3) \\ h(6,3) \\ h(7,3) \end{bmatrix}=\begin{bmatrix}9.0155 \\ 11.3003 \\ 9.5942 \\ 12.6594\\ 13.1427 \\ 6.1930
  \\ 10.3860 \end{bmatrix}.
\end{equation*}

\end{Example}	

The \textit{commute time}, $CT(m,n)$  between vertices $m$ and $n$ is defined as the expected time for the random walker to reach vertex $n$ starting from vertex $m$, and then to return (see {\color{red}Part 1, Section \ref {I-Commteig}}), to give
$$CT(m,n)=h(m,n)+h(n,m).$$ 
 
\begin{Example}We consider the task of finding the commute time between the vertices $m=0$ and $n=N-1=7$ for the graph from  {\color{red} Fig. 2 in Part I}, Fig. \ref{fig-ec2CT}. If we desire to use the full Laplacian matrix and the electric circuit framework for the hitting time, then we should include the $m$th equation with $h(m,m)=x(m)=0$. Since the sum of all external sources (on the right side of the equation (\ref{randWhitTim})) must be zero, this means that for the vertex $m=0$ the right side terms should be $d_0-D$, and the full Laplacian form of (\ref{randWhitTim}) for the vertex $m=0$ becomes
	$$\mathbf{L}\begin{bmatrix} 0 \\ h(0,1) \\ \vdots \\ h(0,6) \\  h(0,7) \end{bmatrix}=\begin{bmatrix} d_0-D \\ d_1  \\ \vdots \\d_{6} \\ d_{7} \end{bmatrix},$$   
where $D=\sum_{i=0}^{N-1} d_i$, and $d_i=\sum_nW_{in}$ are the degrees of vertices, $i$. 	

	The same relation can be written for $m=7$ (or any other vertex $m$), to yield
	$$\mathbf{L}\begin{bmatrix} h(7,0) \\ h(7,1) \\ \vdots \\ h(7,6) \\  0 \end{bmatrix}=\begin{bmatrix} d_0 \\ d_1  \\ \vdots \\d_{6} \\ d_{7} -D \end{bmatrix}.$$  
The difference between the two previous systems of equation is
	$$\mathbf{L}\begin{bmatrix} -h(7,0) \\ h(0,1)-h(7,1) \\ \vdots \\h(0,6)-h(7,6) \\  h(0,7) \end{bmatrix}=\begin{bmatrix} -D \\ 0  \\ \vdots \\0 \\  D \end{bmatrix},$$  
This system can be interpreted within the electric circuit framework as the electric circuit  with an external source at $m=0$ whose current is $i(0)=-D$. This external source is closed at $m=7$  with the current $i(7)=D$, while there are no sources at any other vertex. The difference of voltages  in this electric circuit at $m=0$ and $m=7$ is equal to the difference of the seventh element, $h(0,7)$, and the first element, $-h(0,7)$, to yield
\begin{gather*}x_{0,7}=h(0,7) -\big(-h(7,0) \Big)
=h(0,7)+h(7,0)\\
=CT(7,0)=R^{(7,0)}_{\mathrm{eff}}i(0)
\end{gather*}
where $R^{(7,0)}_{\mathrm{eff}}$ is the effective electric resistance between $m=0$ and $m=N-1=7$, as illustrated in Fig. \ref{fig-ec2CT}. 

Finally, the previous relation holds for any two vertices, $m$ and $n$,  that is
$$CT(m,n)=DR^{(m,n)}_{\mathrm{eff}}$$
where $D=\sum_{i=0}^{N-1} d_i$.

\begin{figure*}
	\centering
	\includegraphics[scale=1]{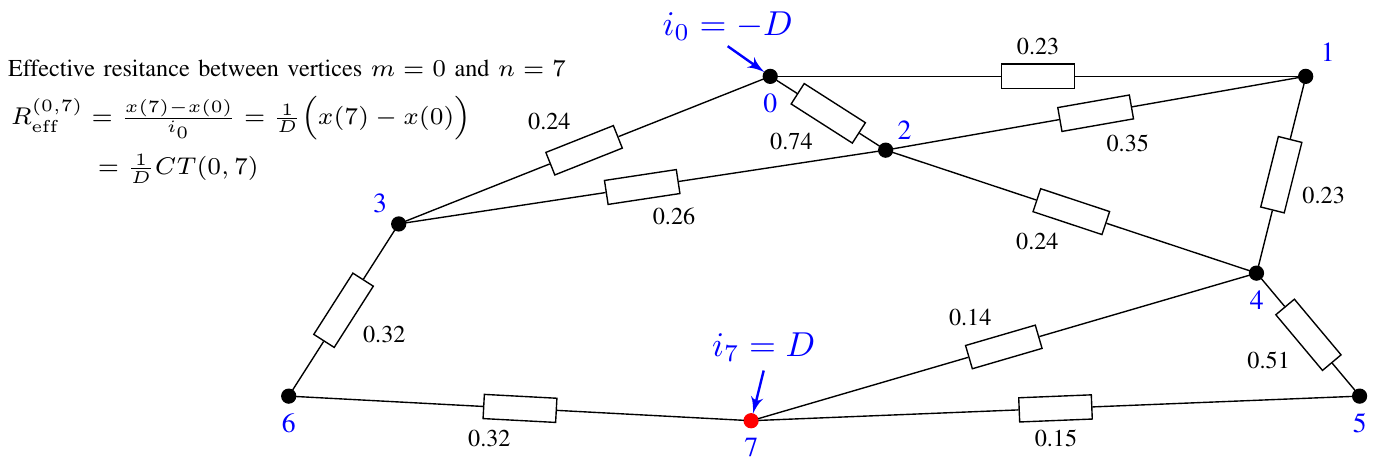}
	\caption{Electric circuit interpretation of the commute time,  $CT(m,n)=DR^{(m,n)}_{\mathrm{eff}}$.}
	\label{fig-ec2CT}
\end{figure*}

	\end{Example} 
 
 \begin{Example}The commute time between vertices $m=0$ and $n=7$ for the graph from {\color{red} Fig. 2 in Part I}, Fig. \ref{fig-ec2CT}, can be obtained by calculating the hitting times $h(0,7)$ and $h(7,0)$, as in Example \ref{ExamHittt4}. The result is
 $$CT(7,0)=h(0,7)+h(7,0)= 10.7436+19.6524=30.3960.$$
 
 The same result can be obtained by finding the effective resistance between vertices $m=0$ and $n=7$ in the electric circuit from Fig. \ref{fig-ec2CT} using the elementary calculations for the effective resistance, $R_{\mathrm{eff}}^{(7,0)}$, given by 	
$$R_{\mathrm{eff}}^{(7,0)}=4.0745.$$
With $D=\sum_{i=0}^{7}d_i=7.46$, the commute time, $CT(7,0)=DR^{(7,0)}_{\mathrm{eff}}=30.3960$, follows. 	 
 	
 		\end{Example} 
 
 \subsection{Gaussian Random Signal}\label{GrafRanDSIG}
 
 Consider a random graph signal, $x(n)$, and assume that each sample is Gaussian distributed with mean $\mu_n$ and standard deviation $\sigma_n$. Assuming that the signal values are correlated, the pdf of the signal $\mathbf{x}$ is given by
 \begin{equation}P(\mathbf{x}) =\frac{1}{\sqrt{(2\pi)^N}} \textrm{det}(\boldsymbol{\Sigma}^{-1}_x) \exp{(-\frac{1}{2}(\mathbf{x}-\boldsymbol{\mu})\boldsymbol{\Sigma}^{-1}_x(\mathbf{x}-\boldsymbol{\mu}))}.\label{MultiPdft}
 \end{equation}		
 The inverse of the autocovariance matrix is the precision matrix $\mathbf{Q}=\boldsymbol{\Sigma}^{-1}_x$.  Note that the name precision comes from the one-dimensional case  where the precision is inversely proportional to the variance, that is $Q=1/\sigma^2$.
 
 The maximum likelihood estimate of $\mathbf{x}$ is then obtained from (\ref{MultiPdft}) by minimizing 
 $$E_x=\frac{1}{2}(\mathbf{x}-\boldsymbol{\mu})\boldsymbol{\Sigma}^{-1}_x(\mathbf{x}-\boldsymbol{\mu})$$
 and the solution is 
\begin{equation}\boldsymbol{\Sigma}^{-1}_x(\mathbf{x}-\boldsymbol{\mu})=0.\label{minenergrand}
\end{equation}
 For a zero-mean random signal, $\boldsymbol{\mu}=\mathbf{0}$ and $\boldsymbol{\Sigma}^{-1}_x\mathbf{x=0}$, and the solution in (\ref{minenergrand}) corresponds to minimizing the energy of change  (maximal smoothness) in the graph.
 
 The generalized Laplacian corresponding to the precision matrix is defined by
 $$ \boldsymbol{\Sigma}^{-1}_x = \mathbf{Q}= \mathbf{L} + \mathbf{P}$$
 where $\mathbf{P}$ is a diagonal matrix such that the sum of columns of the Laplacian is zero. 
  
  Now, the edge weights can be extracted from the Laplacian matrix. Since the Laplacian is defined using the observed graph signal values, this is a point where the presented analysis meets the discussion from the previous section (see also Example \ref{ExamGaussMD1} and Example \ref{ExamGaussMD2}). The electric circuit form of the minimization condition is obtained from 
 $$(\mathbf{L} + \mathbf{P})(\mathbf{x}-\boldsymbol{\mu})=\mathbf{0}$$
 or
 $$\mathbf{L} \mathbf{x}=-\mathbf{Px}+(\mathbf{L+P})\boldsymbol{\mu}.$$
In terms of the external current generators we can define the problem as  
$$\mathbf{L} \mathbf{x}=\mathbf{i}_x+\mathbf{i}_g,$$
where $\mathbf{i}_x=-\mathbf{Px}$ are voltage-driven current generators and $\mathbf{i}_g=(\mathbf{L+P})\boldsymbol{\mu}=\mathbf{Q}\boldsymbol{\mu}$ are constant external current generators.   Therefore, the steady-state solution  can be interpreted and solved in the same way as the described electric circuit is solved. For example, if the observed state is $x(7)=1$ and $\mu(n)=0$, we can solve the system for other values of $x(n)$ for a given matrix $ \boldsymbol{\Sigma}^{-1}_x = \mathbf{Q}= \mathbf{L} + \mathbf{P}$. 
 
\section{Graph Learning from Data and External Sources}
In the previous section, learning of graph topology from data on the graph has been considered  using the correlation and precision matrices. The basic additional assumption which has been used in the estimation is that the graph signal is smooth. If we can measure the graph signal and external sources in the vertices, then it is possible to learn   graph topology in an exact way. 

Consider the $p$th observation of the data on a graph, $[x_p(0), x_p(1),\dots,x_p(N-1)]^T$, and the corresponding external sources, $[i_p(0), i_p(1),\dots,i_p(N-1)]^T$. Without loss of generality assume that the $(N-1)$th vertex is a reference, where $x_p(N-1)=0$ and $i_p(N-1)=-\sum_{n=1}^{N-2}i_p(n)$. These elements will be removed from the data and equations and only the data on remaining  vertices will be considered, and denoted as $\mathbf{x}_p=[x_p(0), x_p(1),\dots,x_p(N-2)]^T$  and $\mathbf{i}_p=[i_p(0), i_p(1),\dots,i_p(N-2)]^T$. The equation for these reduced sets of data is then   
$$\begin{bmatrix}\boldsymbol{l}_0 \\
\boldsymbol{l}_1\\
\vdots \\
\boldsymbol{l}_{N-2}
\end{bmatrix}\mathbf{x}_p=\mathbf{i}_p,$$
where 
$$\boldsymbol{l}_i=[L_{i0}, \  L_{i1}, \dots, L_{N-2,0}]$$
are the rows of the graph Laplacian matrix, $\mathbf{L}$, with the elements from $n=0$ to $n=N-2$. The last element and the last row in the graph Laplacian, which correspond to the reference vertex, $n=N-1$, are omitted.

If $P$ sets of observations are available, then we can write a system in the form 
$$\begin{bmatrix}\boldsymbol{l}_0 \\
\boldsymbol{l}_1\\
\vdots \\
\boldsymbol{l}_{N-2}
\end{bmatrix}[\mathbf{x}_1, \mathbf{x}_2,\dots,\mathbf{x}_P]=[\mathbf{i}_1,\mathbf{i}_2,\dots,\mathbf{i}_P].$$
or
$$\begin{bmatrix}\boldsymbol{l}_0 \\
\boldsymbol{l}_1\\
\vdots \\
\boldsymbol{l}_{N-2}
\end{bmatrix}\mathbf{X}_{N-1,P}=\mathbf{J}_{N-1,P} 
$$
The matrices $\mathbf{X}_{N-1,P}$ and $\mathbf{J}_{N-1,P}$ represent respectively the signal on graph and external sources matrices of dimensionality $(N-1)\times P$. 

Now, we can consider two cases:
\begin{itemize}
	\item When there are $P \ge N-1$ independent observations then the exact form of the graph Laplacian (its first $(N-1)$ rows and columns) follows from 
$$\begin{bmatrix}\boldsymbol{l}_0 \\
	\boldsymbol{l}_1\\
	\vdots  \\
	\boldsymbol{l}_{N-2}
\end{bmatrix}=\mathbf{J}_{N-1,P} \  \mathrm{pinv}\{\mathbf{X}_{N-1,P}\}
.$$	
The last column and the last row of the graph Laplacian, $\mathbf{L}$, are formed so that the sum over every column or row is zero. 

\item A more complex case arises when $P < N-1$. Then, there is a sufficient number of observations to recover the graph Laplacian. However, if we assume that the graph Laplacian is sparse, with a small number of nonzero elements (edges), the solution is possible within the compressive sensing framework. In order to adapt the system for the standard LASSO algorithm, we shall rewrite it in the form
$$\mathbf{X}^T_{N-1,P}\begin{bmatrix}
\boldsymbol{l}_0 \\
\boldsymbol{l}_1\\
\vdots  \\
\boldsymbol{l}_{N-2}
\end{bmatrix}^T=\mathbf{J}^T_{N-1,P}. 
$$
Now, we can perform LASSO minimization for each column, $\boldsymbol{l}^T_k$, and the corresponding column of the matrix $\mathbf{J}^T_{N-1,P}$, denoted by $\boldsymbol{i}_{k}$, in the form 
$$\boldsymbol{l}_{k}=\mathrm{lasso}( \mathbf{X}^T_{N-1,P},\boldsymbol{i}_{k},\rho).
$$
Another approach would be to transform the matrices with graph Laplacian rows, $\boldsymbol{i}_{k}$, and external sources matrix, $\mathbf{J}^T_{N-1,P}$, into column vectors to have 
$$
\setlength{\arraycolsep}{1.6pt}
\begin{bmatrix}
\mathbf{X}^T_{N-1,P} & \mathbf{0} & \dots & \mathbf{0}\\
\mathbf{0}  & \mathbf{X}^T_{N-1,P} &  \dots & \mathbf{0} 
\\
\vdots & \vdots & \ddots & \vdots  
 \\
\mathbf{0} & \mathbf{0} & \dots & \mathbf{X}^T_{N-1,P}
\end{bmatrix}\begin{bmatrix}\boldsymbol{l}^T_0 \\
\boldsymbol{l}^T_1\\
\vdots  \\
\boldsymbol{l}^T_{N-2}
\end{bmatrix}=\begin{bmatrix}\boldsymbol{i}_0 \\
\boldsymbol{i}_1\\
\vdots  \\
\boldsymbol{i}_{N-2}
\end{bmatrix}.
$$ 
Using the notation
$$\Big(\mathbf{I}_{N-1,N-1} \otimes \mathbf{X}^T_{N-1,P}\Big) \ \boldsymbol{l}_{\mathrm{vec}} =\boldsymbol{i}_{\mathrm{vec}}$$
for the matrix and the vectors in the above equation (where $\mathbf{I}_{N-1,N-1}$ is the identity matrix), this system can be solved using
 $$\boldsymbol{l}_{\mathrm{vec}}=\mathrm{lasso}\Big(\mathbf{I}_{N-1,N-1} \otimes \mathbf{X}^T_{N-1,P},\boldsymbol{i}_{\mathrm{vec}},\rho\Big).
 $$
 
\end{itemize}

\begin{Example}\label{EstLaph28}
	Consider a graph with $N=50$ vertices, with small number of edges. Such a sparse graph Laplacian, $\mathbf{L}$, is shown in Fig. \ref{Learning_ext_sources}(a). 
	
	The graph Laplacian was estimated using a large number, $P=60$, of observations of the graph signal and external sources. Both
 the norm-two and the LASSO estimates of the graph Laplacian were accurate, as shown in  Fig. \ref{Learning_ext_sources}(b) and (c).  Next, the number of observations was  reduced to $P=40<N=50$. In this case, the sparsity of the graph Laplacian is crucial for the solution. The LASSO algorithm, which has included the sparsity constraint, can produce a good result in this case, as can be seen from  Fig. \ref{Learning_ext_sources}(e). The norm-two was calculated using the pseudo-inverse of the data matrix, $\mathbf{X}^T_{N-1,P}$, and cannot be used as the graph Laplacian estimate, as seen in Fig. \ref{Learning_ext_sources}(d).   
	\begin{figure}
		\centering
		
		\includegraphics[scale=0.95]{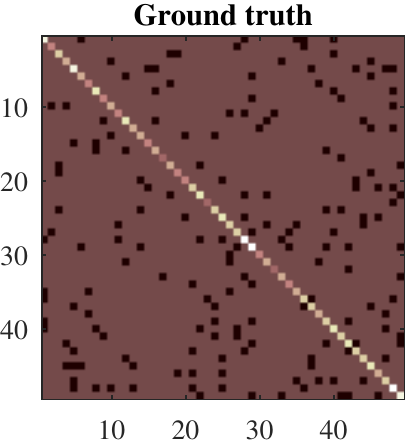}
		\\
		(a)  \\
		\vspace{5mm}
			\includegraphics[scale=0.95]{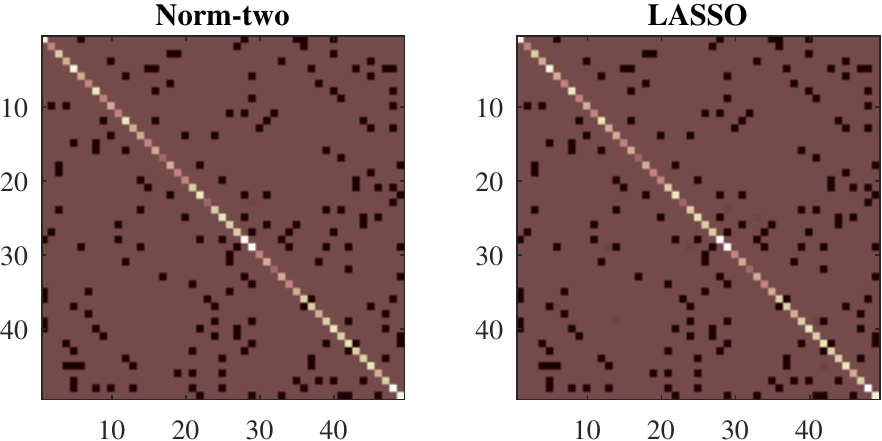}
		\\
	\hspace*{3.5mm}	(b) \hspace{38mm} (c) \\
		\vspace{5mm}
		\includegraphics[scale=0.95]{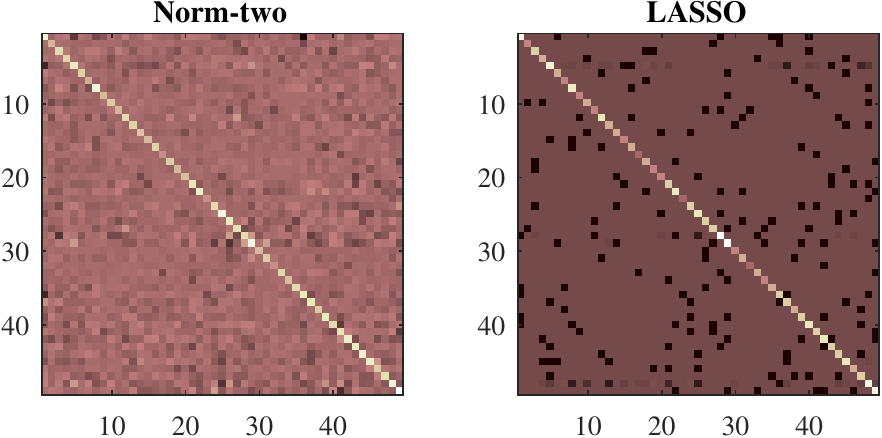}
		\\
		\hspace*{3.5mm}	(d) \hspace{38mm} (e)\\
	
		\caption{Estimation of the graph Laplacian, $\mathbf{L}$, for a graph with $N=50$ randomly positioned vertices and a small number of edges.  (a) Ground truth graph Laplacian, $\mathbf{L}$. (b) Estimated graph Laplacian using the norm-two for a large number of observations, $P=60 > N=50$. (c) Estimated graph Laplacian using the LASSO, for a large number of observations, $P=60 > N=50$. (d) Estimated graph Laplacian using the norm-two, for a small number of observations, $P=40 < N=50$. (c) Estimated graph Laplacian using the LASSO, for a small number of observations, $P=40 < N=50$.} 
		\label{Learning_ext_sources}
	\end{figure}
\end{Example}
Finally, we shall mention that in Example \ref{EstLaph28} we have not used the conditions that the graph Laplacian is a symmetric matrix and that the elements of the weight matrix, $W_{mn}$, from which the graph Laplacian elements are formed, are nonnegative. These conditions can be used within linear programming formulations to improve the estimation.   


\section{Random Signal Simulation on Graphs}

The presentation of a graph and graph signal within the circuit theory framework can be used to simulate random signals on graphs. While several approaches are possible, we will here present some of the most frequently used ones.

1) Assume that the graph is initiated by \textbf{external sources} that are random variables. In that case, the $p$th observation of a random signal on this graph is simulated as a solution of the system of equations 
  $$\mathbf{L}\mathbf{x}_p=\boldsymbol{\varepsilon}_p,$$
with $\mathbf{i}_p=\boldsymbol{\varepsilon}_p$. Note that one of the external sources (randomly chosen for each observation $p$) should compensate for all other sources, to ensure $\sum_{n=0}^{N-1}{\varepsilon}_p(n)=0$.

Since the graph Laplacian is singular, the graph signal value (the electric potential in the electric circuit case) at a vertex, for example, $n=0$, should be considered as \textit{a reference} and its value assumed, $x(0)=0$. This should be the case whenever the inversion of the graph Laplacian is required.

2) The graph is  \textbf{initiated at only one of its vertices} (and the reference vertex) with a random  external zero-mean white source. The position of these vertices is randomly selected for each $p$. Then, the random signal observation on a graph is obtained as a solution to 
 $$\mathbf{L}\mathbf{x}_p=\mathbf{i}_p$$
 where $i_p(n)=\varepsilon_p \delta(n-n_i)-\varepsilon_p \delta(n-n_j)$ and $n_i$ and $n_j$ are two randomly selected vertices in each observation.  

3) A minimal information needed to calculate a random graph signal is to know its values at two randomly positioned vertices. Assuming that $x_p(n)=\varepsilon_p \delta(n-n_i)+\epsilon_p \delta(n-n_j)$ and $n_i$ and $n_j$ are two randomly selected vertices at each observation, we may solve the system for all other signal samples, based on 
$$\mathbf{L}\mathbf{x}_p=\mathbf{0}.$$
With the two assumed values, $x_p(n)$, at $n=n_i$ and $n=n_j$,  we can solve this system for all other signal values. In the case of external sources the values should be compensated, as mentioned earlier. In this case, there is no need for compensation, which means that $\varepsilon_p$ and $\epsilon_p$ could be independent random variables. 

4) The signal on a graph is formed using a linear combination of white noise $\boldsymbol{\varepsilon}_p$  and its graph shifted versions. The output signal after $M$ such graph shifts, defined by the normalized Laplacian, is given by
 \begin{gather}
 \mathbf{x}_p=
 (h_M\mathbf{L}^M+h_{M-1}\mathbf{L}^{M-1}+\cdots+h_{1}\mathbf{L}^{1}+h_0\mathbf{L}^0)\boldsymbol{\varepsilon}_p.\label{graphSigDefiHa}
 \end{gather} 
 The resulting graph signal can be written in the form
 $$\mathbf{x}_p=H(\mathbf{L})\boldsymbol{\varepsilon}_p.$$

5) Analysis based on the adjacency matrix and graph shifts. Assume that an undirected graph with the adjacency matrix $\mathbf{A}$, is initiated at $N_a$ randomly chosen vertices $n_1,n_2,\dots,n_{N_a}$, $\eta=N_a/N$, with spikes $\delta(n-n_i)$, $i=1,2,\dots,N_a$. After shifting these spikes  $K$ times we obtain 
$$\mathbf{x}=\mathbf{A}^K \sum_{i=1}^{N_a}\delta(n-n_i).$$
 The parameters $K$ and $N_a$ define the resulting signal smoothness. An example of one realization of such a signal is presented in Part II, Fig. \ref{II-fig:sig-gr1} for $\eta=1/8$, $K=1$ (upper subplots) and $\eta=2/8$, $K=1$ (lower subplots) using the spikes $a_i\delta(n-n_i)$, where $a_i$ are the spike amplitudes.
 
 6) Signals are commonly simulated as sums of the harmonic basis functions, as in classical Fourier analysis. This kind of simulation may be used in graph signal processing, too. Such a signal on a graph can be written as
 $$\mathbf{x}=\sum_{i=1}^Ka_{k_i}\mathbf{u}_{k_i}$$
 where $\mathbf{u}_{k}$ are the eigenvectors of the  Laplacian or adjacency matrix eigenvectors, and $a_k$ are random constants. This kind of graph signal simulation, with or without an additive noise, has been often used in this chapter.

  \section{From Newton Minimization to Graphical LASSO, via LASSO}\label{SecLassGlasso}

Currently the most important approaches to the learning of graph topology from the available data are based on the regression analysis method of the least absolute shrinkage and selection operator (LASSO). Its extension to graphs is called the graphical LASSO (GLASSO). These methods will be derived and explained in this section, starting from the simple, one-dimensional Newton minimization method. 

\subsection{Newton Method}

We will first briefly review the Newton iterative algorithm for finding the minimum of a convex function. Consider a function $f(x)$ and assume that it is differentiable. Denote the position of the minimum of $f(x)$ by $x^*$. The first derivative of $f(x)$ at the minimum point position 
$$x^*=x+\Delta x$$
 can be expanded into a Taylor series around
an arbitrary position $x$, using the linear model (which is exact if $f'''(x)=0$ for all $x$), as 
\begin{equation}
f'(x^*)=f'(x)+f''(x)\Delta x. \label{NLF}
\end{equation}
Since $f'(x^*)=0$, by definition, with $\Delta x=x^*-x$, the relation in (\ref{NLF}) can be rewritten as
$$
 x^{*}-x=-\frac{f'(x)}{f''(x)}.
$$
This formula is used to define an iterative procedure (called the \textbf{Newton's iterative method}) for finding the position of the minimum of function $f(x)$, $x^*$, starting from an $x=x_0$ as  
$$
x_{k+1}=x_k-\alpha f'(x_{k}).
$$
The parameter $\alpha $ is commonly used instead of $1/f''(x)$ to control the iteration step, and its value should be 
$$0<\alpha\le \max(|1/f''(x)|),$$ 
for the considered interval of $x$. This is the form of the well-known \textbf{steepest descend method} for convex function minimization. 

Notice that the value $x^*=x-\alpha f'(x)$ would also be obtained as a result of the minimization of a cost function defined by the quadratic form
\begin{gather*}
x^*=\arg \min _{z} G(z)\\
=\arg \min _{z} (f(x)+f'(x)(z-x)+\frac{1}{2\alpha}(z-x)^2),
\end{gather*}
namely, from the zero-value of the derivative of this cost function  
\begin{gather*}\frac{d}{dz}\Big(f(x)+f'(x)(z-x)+\frac{1}{2\alpha}(z-x)^2\Big)=0\end{gather*}
 we would arrive at
 \begin{gather*}z=x-\alpha f'(x)=x^*.\end{gather*} 

Next, assume that we wish to minimize the cost function
$$
J(x)=\frac{1}{2\alpha}(x-y)^2+\rho |x|,
$$
where $\rho$ is a parameter. This cost function corresponds to the minimization of the squared difference between $x$ and $y$, that is $(x-y)^2$, with an addition sparsity constraint on $x$, given by $|x|$. From 
\begin{gather*}
\frac{dJ(x)}{dx}=\frac{1}{\alpha}(x-y)+\rho \mathrm{sign}(x)=0
\end{gather*} 
we obtain 
$$   
x+\rho \alpha \mathrm{sign}(x)=y.
$$
Soft-thresholding, denoted as $\mathrm{soft}
(y,\alpha \rho)$, 
 may be used as a 
solution to this equation, and it is defined by
\begin{gather}
x=\mathrm{soft}(y,\alpha \rho)=\left\{
\begin{array}
[c]{ccc}%
y+ \alpha \rho, & \mathrm{ for } & y<-\alpha\rho \\
0, & \mathrm{for} & \left\vert y\right\vert \leq\alpha \rho\\
y-\alpha  \rho,& \mathrm{for} & y>\alpha  \rho.
\end{array}
\right.  \label{soft1D}
\end{gather}
This form could be considered as the LASSO method for one-dimensional variables. Now, we can proceed with deriving the LASSO method for $N$-dimensional variables. 

\subsection{LASSO}
	
For the LASSO minimization of $N$-dimensional variables we will consider the cost function
\begin{gather*}J(\mathbf{X})= \Vert \mathbf{y}-\mathbf{A}\mathbf{X}
 \Vert _2 ^2 + \rho \Vert \mathbf{X}
 \Vert _1 \\
 =
  \Vert \mathbf{y} \Vert _2^2 -2\mathbf{}\mathbf{X}^{T}\mathbf{A}^T\mathbf{y}
 +\mathbf{X}^T\mathbf{A}^T\mathbf{A}\mathbf{X} + \rho \Vert \mathbf{X}
 \Vert _1,
 \end{gather*}
 where $\mathbf{y}$ is an $M\times 1$ column vector,  $\mathbf{X}$ is an $N\times 1$ column vector, and $\mathbf{A}$ is an $M\times N$ matrix \cite{stankovic2015digital}. 
 
The minimization of this cost function with respect to the $N$-dimensional variable $\mathbf{X}$ will produce a value which minimizes $\Vert \mathbf{y}-\mathbf{A}\mathbf{X}
  \Vert _2 ^2$, meaning that $\mathbf{A}\mathbf{X}$ is \textit{as close to} $\mathbf{y}$ as possible, while at the same time \textit{promoting the sparsity} of $\mathbf{X}$, by including the term $\Vert \mathbf{X}
  \Vert _1$ in the minimization. The balance between these two requirements is defined by the parameter $\rho$.     
 
 Consider first the differentiable part of the cost function $J(\mathbf{X})$ denoted by  
 \begin{gather*}
 J_{D}(\mathbf{X})=\Vert \mathbf{y}-\mathbf{A}\mathbf{X}
 \Vert _2 ^2 =(\mathbf{y}-\mathbf{A}\mathbf{X})(\mathbf{y}-\mathbf{A}\mathbf{X})^T.
 \end{gather*}
   Its derivatives are 
 $$
 \frac{\partial J_{D}(\mathbf{X})}{\partial \mathbf{X}^T}=-2\mathbf{A}^T \mathbf{y}
+ 2\mathbf{X}^T\mathbf{A}^T\mathbf{A} 
 $$
 and
 $$
 \frac{\partial^2 J_{D}(\mathbf{X})}{(\partial \mathbf{X}^T)^2}=2\mathbf{A}^T\mathbf{A}.
$$
The linear model for the first derivative of $J_{D}(\mathbf{X})$ around its minimum, which corresponds to (\ref{NLF}), is  
 $$ 
\frac{\partial J_{D}(\mathbf{X}^*)}{\partial
\mathbf{X}^T}=\frac{\partial J_{D}(\mathbf{X})}{\partial
\mathbf{X}^T}+ (\Delta \mathbf{X}) \frac{\partial^2
J_{D}(\mathbf{X})}{(\partial \mathbf{X}^T)^2}.
$$
By replacing the inverse of the second order derivative by a constant diagonal matrix $\alpha \mathbf{I}$ we have
$$
\Delta \mathbf{X}=\mathbf{X}^*-\mathbf{X}=-\alpha\frac{\partial J_{D}(\mathbf{X})}{\partial
\mathbf{X}^T},
 $$
or
\begin{equation}
\mathbf{X}^*=\mathbf{X}-\alpha\frac{\partial
J_{D}(\mathbf{X})}{\partial
\mathbf{X}^T}, \label{DiffItSol}
\end{equation}
with 
\begin{gather*}
0<\alpha< \frac{1}{\max \Vert 2\mathbf{A}^T\mathbf{A} \Vert} = \frac{1}{2\lambda_{\max}},
\end{gather*}
 where $\lambda_{\max}$ is the maximum eigenvalue of matrix $\mathbf{A}^T\mathbf{A}$.

In order to find $\mathbf{Z}=\mathbf{X}^*$ that minimizes the complete cost function $J(\mathbf{X})$, we can minimize the squared difference  $$\mathbf{Z}-(\mathbf{X}-\alpha\mathbf{I}\frac{\partial
	J_{D}(\mathbf{X})}{\partial
	\mathbf{X}^T})$$
	 and the norm-one of $\mathbf{Z}$, by forming the cost function $G(\mathbf{Z})$ as
\begin{gather*}
G(\mathbf{Z})
=\frac{1}{2\alpha }\Vert \mathbf{Z}-\mathbf{(X}-\alpha\mathbf{I}\frac{\partial
J_{D}(\mathbf{X})}{\partial
\mathbf{X}^T} )\Vert ^2_2 + \rho \Vert \mathbf{Z} \Vert _1.
\end{gather*}
The minimization of $G(\mathbf{Z})$ will produce  $\mathbf{Z}$ which is as close as possible to the desired solution in (\ref{DiffItSol}), while minimizing its norm-one at the same time. The balance parameter is $\rho$. 

If we use the notation 
$$\mathbf{Y}=\mathbf{(X}-\alpha\mathbf{I}\frac{\partial
 	J_{D}(\mathbf{X})}{\partial
 	\mathbf{X}^T} ),$$
 the solution of  
 $$
 \mathbf{X}^*=\arg \min _{\mathbf{Z}} G(\mathbf{Z})=\arg \min _{\mathbf{Z}}\frac{1}{2\alpha }\Vert \mathbf{Z}-\mathbf{Y}\Vert ^2_2 + \rho \Vert \mathbf{Z} \Vert _1
 $$
is obtained from
$$\frac{1}{\alpha }(\mathbf{X}^*-\mathbf{Y})+\rho \mathrm{sign}(\mathbf{X}^*)=\mathbf{0}.
$$
Using the soft function as in (\ref{soft1D}) we can further write 
$$ \mathbf{X}^* = \mathrm{soft} (\mathbf{Y}, \alpha \rho).
$$
Next, we will replace the value of $\mathbf{Y}$ by  
 \begin{gather*}
 \mathbf{Y}=\mathbf{(X}-\alpha\mathbf{I}\frac{\partial
 	J_{D}(\mathbf{X})}{\partial
 	\mathbf{X}^T} )=\mathbf{X}-\alpha\mathbf{I}(-2\mathbf{A}^T \mathbf{y}
	+ 2\mathbf{X}^T\mathbf{A}^T\mathbf{A}) \\ 
	=2 \alpha \mathbf{A}^T  \mathbf{y}+(\mathbf{I}-2\alpha \mathbf{A}^T \mathbf{A} )\mathbf{X}.
\end{gather*}
The iterative formula for the solution of the defined minimization problem is obtained by replacing $\mathbf{X}^*=\mathbf{X}_{k+1}$ and $\mathbf{X}=\mathbf{X}_{k}$ to yield
\begin{equation}
\mathbf{X}_{k+1}=\mathrm{soft}(2 \alpha \mathbf{A}^T(\mathbf{y}-
\mathbf{A} \mathbf{X}_k)
+\mathbf{X}_k,\alpha \rho). \label{Lasoodef}
\end{equation}
This formula can be easily written for each element of $\mathbf{X}_{k}$ and implemented as in  Algorithm \ref{LassoAlg}. This is the essence of the LASSO (Least Absolute Shrinkage and Selection Operator) iterative algorithm. As the initial estimate, $\mathbf{X}_0=\mathbf{A}^T\mathbf{y}$, is commonly used.

\begin{Example}
	Consider a sparse signal, $X(k)$, with $N=60$ elements. In general to calculate the signal elements we need at least $M=60$ measurements (linear combinations of signal elements). A signal can be reconstructed with reduced $M<N$ measurements if it is sparse, with $K \ll N$ nonzero elements at unknown positions.
	
	Assume that the original sparse signal
	of the total length $N=60$ is $X(k)=0$ for all $k$ except for $X(5)=1$, $X(12)=0.5$, $X(31)=0.9$, and $X(45)=-0.75$, in the transform domain, and that it is measured with a
	matrix $\mathbf{A}$ with only $M=40<N$ measurements stored in vector $\mathbf{y}$.
	
	 The measurement matrix $\mathbf{A}$ is formed as a Gaussian random matrix of the
	size $40\times60$, with elements
	$N(0,\sigma^{2})$, where $\sigma^{2}=1/40$ is used.  
	
	All
	$60$ signal values were reconstructed using these $40$ measurements
	$\mathbf{y}$ and the matrix $\mathbf{A}$, in $1000$ iterations. In the initial
	iteration $\mathbf{X}_{0}=\mathbf{A}^T\mathbf{y}$ was used. Then for each next iteration $k$ the new
	values of $\mathbf{X}$ were calculated using (\ref{Lasoodef}) and Algorithm \ref{LassoAlg}, given the data
	$\mathbf{y}$ and matrix $\mathbf{A}$. 
	
	The results for
	$\rho=0.1$ and $\rho=0.001$ are shown in Fig. \ref{LASSO}. For very
	small $\rho=0.001$, the result is not sparse, since the constraint is too
	weak.
		\begin{figure}
		\begin{center}
			\includegraphics[]{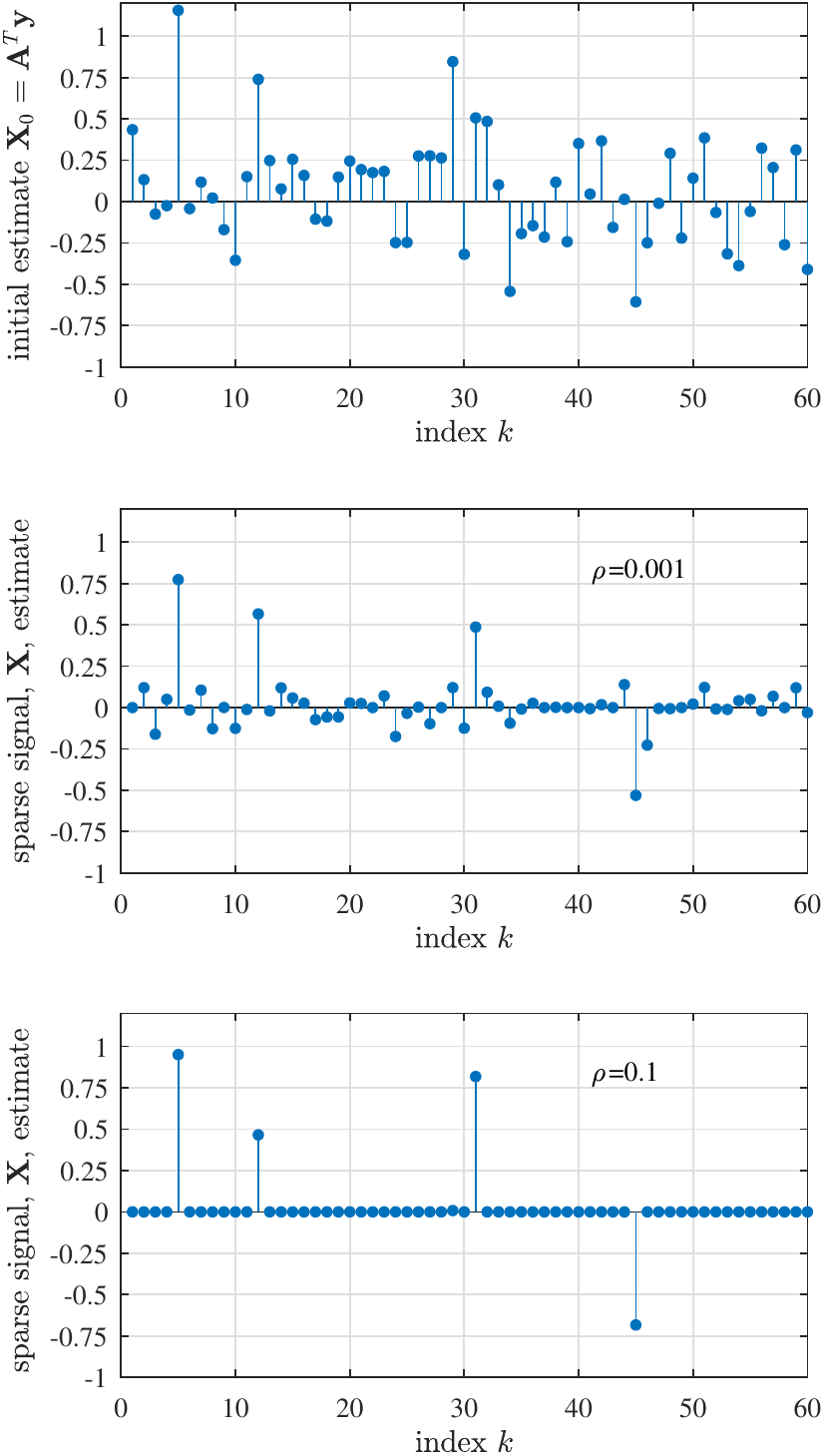}%
			\caption{A sparse signal with $N=60$ and $K=4$, which is reconstructed using a reduced
				set of $M=40$ observations and the LASSO iterative algorithm. The results for  the matched filter (initial estimate), $\mathbf{X}_0=\mathbf{A}^T\mathbf{y}$, and LASSO iterative algorithm with  
				$\rho=0.01$ and $\rho=0.0001$ are shown.}%
			\label{LASSO}%
		\end{center}
	\end{figure}

\end{Example}

\subsection{Graphical LASSO}

In graph model learning, the cost function of the form
$$
J(\mathbf{Q})=-\ln (\det \mathbf{Q}) + \mathrm{Trace}( \mathbf{Q}\mathbf{R}_x)+ \rho \Vert \mathbf{Q} \Vert _1
$$
may be used. Here, $\mathbf{Q}$ is the $N\times N$ generalized Laplacian  matrix, while $\mathbf{R}_x$ is the available  $N\times N$ correlation  matrix. Physical meaning of these terms is explained in Section \ref{secGLASSO}.

The derivative of the cost function with respect to the elements of $\mathbf{Q}$ can be written as
\begin{equation}
-\mathbf{Q}^{-1}+\mathbf{R}_x+\rho \mathrm{sign}(\mathbf{Q})=\mathbf{0} \label{dereq}
\end{equation}
at $\partial J(\mathbf{Q})/\partial \mathbf{Q}=\mathbf{0}$. 

Upon introducing the notation 
\begin{gather*}
\mathbf{V}=\mathbf{Q}^{-1}
\end{gather*}
 or   
$$
\mathbf{V}\mathbf{Q}=\mathbf{I}
$$
we can write
\begin{gather}
\mathbf{V}=\begin{bmatrix}
\mathbf{V}_{11} & \mathbf{v}_{12} \\
\mathbf{v}^T_{12} & v_{22} 
\end{bmatrix} \hspace{3mm} \mathbf{Q}=\begin{bmatrix}
\mathbf{Q}_{11} & \mathbf{q}_{12} \\
\mathbf{q}^T_{12} & q_{22} 
\end{bmatrix}
\end{gather}
and
\begin{gather}
\begin{bmatrix}
\mathbf{V}_{11} & \mathbf{v}_{12} \\
\mathbf{v}^T_{12} & v_{22} 
\end{bmatrix}
\begin{bmatrix}
\mathbf{Q}_{11} & \mathbf{q}_{12} \\
\mathbf{q}^T_{12} & q_{22} 
\end{bmatrix}=
\begin{bmatrix}
\mathbf{I}& \mathbf{0} \\
\mathbf{0}^T & 1 
\end{bmatrix}, \label{QQWW}
\end{gather}
where $\mathbf{Q}_{11}$ and $\mathbf{V}_{11}$ are  $(N-1)\times (N-1)$ matrices,   $\mathbf{v}_{12}$ and $ \mathbf{q}_{12}$ are $(N-1)\times 1$ column vectors, and $v_{22}$ and $q_{22}$ are scalars.
 
After multiplying the first row of blocks in $\mathbf{V}$ with the last column of blocks in 
$\mathbf{Q}$,
we have 
$$\mathbf{V}_{11}\mathbf{q}_{12}+ \mathbf{v}_{12} q_{22}=\mathbf{0}$$
which gives
\begin{equation}\mathbf{v}_{12}=-\mathbf{V}_{11}\mathbf{q}_{12}/q_{22}= \mathbf{V}_{11}\boldsymbol{\beta}, \label{w12rel}
\end{equation}
where
\begin{equation}
\boldsymbol{\beta}=-\mathbf{q}_{12}/q_{22} \label{q12rel}
\end{equation}
is normalized with $q_{22}>0$.

Now, from the derivative equation (\ref{dereq}) we may write
$$
-\begin{bmatrix}
\mathbf{V}_{11} & \mathbf{v}_{12} \\
\mathbf{v}^T_{12} & v_{22} 
\end{bmatrix}+
\begin{bmatrix}
\mathbf{R}_{11} & \mathbf{r}_{12} \\
\mathbf{r}^T_{12} & r_{22} 
\end{bmatrix}
+\rho \mathrm{sign}({\begin{bmatrix}
	\mathbf{Q}_{11} & \mathbf{q}_{12} \\
	\mathbf{q}^T_{12} & q_{22} 
	\end{bmatrix}})=\mathbf{0}.
$$
For the upper right block we have
$$
-\mathbf{v}_{12}+\mathbf{r}_{12}+\rho \mathrm{sign}(\mathbf{q}_{12})=\mathbf{0},
$$
while after replacing $\mathbf{v}_{12}= \mathbf{V}_{11}\boldsymbol{\beta}$  and $
\mathbf{q}_{12}=-\boldsymbol{\beta}/q_{22}$ from (\ref{w12rel}) and (\ref{q12rel}) we arrive at
\begin{equation}
-\mathbf{V}_{11}\boldsymbol{\beta}+\mathbf{r}_{12}-\rho \mathrm{sign}(\boldsymbol{\beta})=\mathbf{0}. \label{ggllls}
\end{equation}
The solution to this equation for $\boldsymbol{\beta}$ has been already defined within the LASSO framework,
\begin{equation}
\beta_i V_{11}(i)=\mathrm{soft} \Big({r}_{12}(i)-\sum_{k\ne i} V_{11}(k,i)\beta_k, \rho \Big). \label{softGLASSO}
\end{equation}
In order to apply the LASSO as in (\ref{Lasoodef}),  we can interpret the minimization of difference 
$$\mathbf{A}^T(\mathbf{y}-
\mathbf{A} \mathbf{X})=\mathbf{A}^T\mathbf{y}-\mathbf{A}^T\mathbf{A} \mathbf{X}$$
 in (\ref{Lasoodef})  as the goal to find the least-squares regression estimate of $\mathbf{A}^T\mathbf{y}$ by
$\mathbf{A}^T\mathbf{A} \mathbf{X}$. Now, we can adjust (\ref{ggllls}) to assume a similar form
\begin{equation}
-\mathbf{V}^{1/2}_{11}\mathbf{V}^{1/2}_{11}\boldsymbol{\beta}+\mathbf{V}^{1/2}_{11}\mathbf{V}^{-1/2}_{11}\mathbf{r}_{12}-\rho \mathrm{sign}(\boldsymbol{\beta})=\mathbf{0}. \label{gglllsA}
\end{equation}
In this case, the matrix $\mathbf{V}^{1/2}$ plays the role of $\mathbf{A}$ in (\ref{Lasoodef}) and  $\mathbf{V}^{-1/2}_{11}\mathbf{r}_{12}$ plays the role of $\mathbf{y}$. Therefore, the standard LASSO should be calculated using \begin{equation}\boldsymbol{\beta}=\mathrm{lasso}(\mathbf{V}^{1/2}_{11},\mathbf{V}^{-1/2}_{11}\mathbf{r}_{12},\rho) \label{lassoPoziv}
\end{equation}
 as in Algorithm \ref{GLassoAlg}.

Now, we may summarize the graphical LASSO (GLASSO) iterative algorithm as:

\begin{itemize}

\item 
In the initial step, use
$$
\mathbf{V}=\mathbf{R}_{x}+\rho \mathbf{I}.
$$

\item For each coordinate $j=1,2,\dots,N$, the matrix equation of the form (\ref{QQWW}) is written. For each $j$, the reduced matrix $\mathbf{V}_{11}$ is formed by omitting the $j$th row and the $j$th column. Then, the matrix $\mathbf{R}_{x}$ is rearranged accordingly. 

\item 
Equation  (\ref{softGLASSO})
is solved using (\ref{lassoPoziv}).

\item
The matrix $\mathbf{V}$ is updated for each $j$ by inserting the $j$th column 
$$\mathbf{v}_{12}= \mathbf{V}_{11}\boldsymbol{\beta}.$$
and inserting at the $j$th row $\mathbf{v}^T_{12}$ with the element $v_{22}$ at the $j$ position.
\item
After all $j$ indices are used in the calculation, the final estimate of the generalized Laplacian is  $\mathbf{Q}=\mathbf{V}^{-1}.$
\end{itemize}

This calculation procedure is also presented in Algorithm \ref{GLassoAlg}.

\begin{Remark}
Notice that the value of matrix $\mathbf{Q}=\mathbf{V}^{-1}$ is updated for each $j$ and in the last iteration, using the column vector 
$$\mathbf{q}_{12}= -\boldsymbol{\beta}q_{22}$$
where $q_{22}$ can be calculated from  $\mathbf{v}^T_{12}\mathbf{q}_{12}+v_{22}q_{22}=1$ or $-\mathbf{v}^T_{12} \boldsymbol{\beta}q_{22}+v_{22}q_{22}=1$, finally producing the value $$q_{22}=\frac{1}{v_{22}-\mathbf{v}^T_{12}\boldsymbol{\beta}}.$$ 
and 
$$\mathbf{q}_{12}= \frac{ \boldsymbol{\beta}}{\mathbf{v}^T_{12}\boldsymbol{\beta}-v_{22}}$$
which are used to update the $j$th column and row of the matrix $\mathbf{Q}$ in the same was as the update of matrix $\mathbf{V}$. 

This algorithm can be used for iterative matrix inversion with $\rho=0$.
\end{Remark}

\begin{Example}\label{ex_GLASSOPmalo}
	Consider a graph with $N=50$ vertices, with a small number of edges, and for which the weight matrix, $\mathbf{W}$, is sparse.
	The ground truth weight matrix, $\mathbf{W}$,  is shown in Fig. \ref{par3r_N50+part3}(a). This matrix is estimated first from a large number, $P=1000$, of observations of a signal on this graph. Both the precision matrix, $\mathbf{R}^{-1}$, and the graphical LASSO, given in  Fig. \ref{par3r_N50+part3}(b) and (c), produce good estimations of the weight matrix, $\mathbf{W}$.  Next, the number of observations was significantly reduced to $P=40<N=50$, a case when the correlation matrix, $\mathbf{R}$, is singular and of rank lower or equal to $P=40$. In this case, the sparsity of the weight matrix is crucial for solution. Here, only the the graphical LASSO, which includes the sparsity constraint, was able to produce good result, as shown in  Fig. \ref{par3r_N50+part3}(e), while the precision matrix can be calculated only as a pseudo-inverse, and cannont be used as the weight matrix estimate, as can be seen from \ref{par3r_N50+part3}(d).

\begin{figure}
	\centering
	
	\includegraphics[scale=0.95]{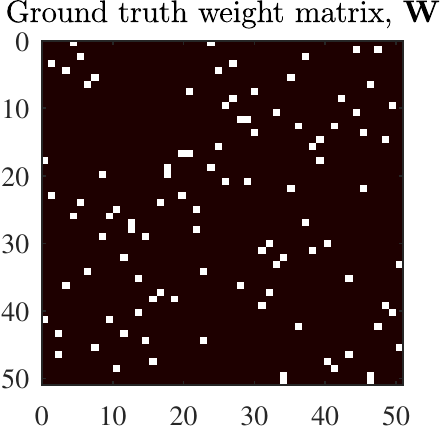}
		\\
	(a)
	
	\vspace{5mm}
	\includegraphics[scale=0.95]{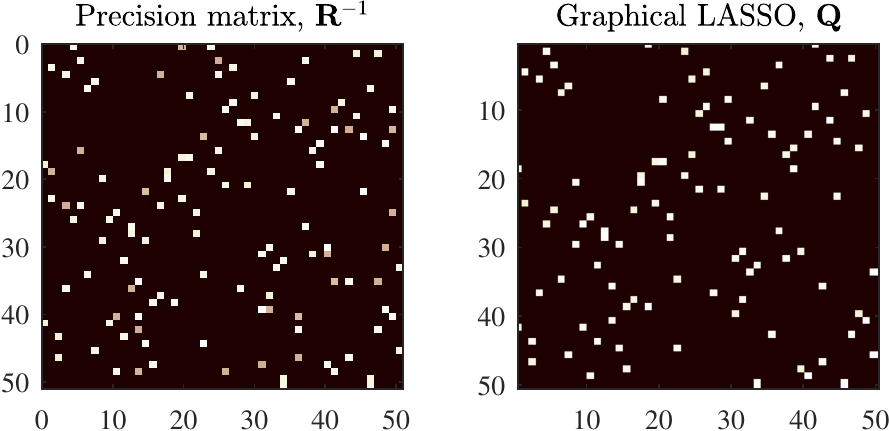}
	\\
	\hspace*{3.5mm}	(b) \hspace{38mm} (c)\\
	
	\vspace{5mm}

		\includegraphics[scale=0.95]{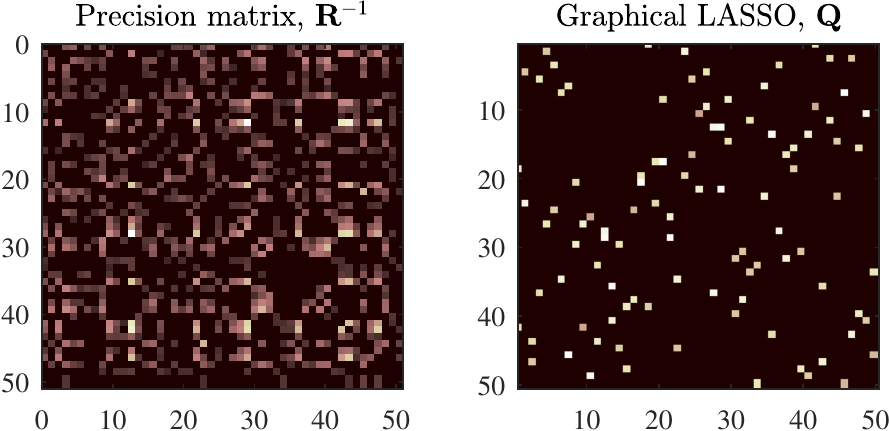}
			\\
		\hspace*{3.5mm}	(d) \hspace{38mm} (e)
	
	\caption{Estimation of the weight matrix, $\mathbf{W}$, for a graph with $N=50$ randomly positioned vertices.  (a) Ground truth weight matrix, $\mathbf{W}$. (b) Precision matrix, for a large number of observations, $P=1000 \gg N=50$. (c) Estimated weight matrix using the graphical LASSO, for a large number of observations, $P=1000 \gg N=50$. (d) Precision matrix,  for a small number of observations, $P=40 < N=50$ (the correlation matrix, $\mathbf{R}$, is singular and with a rank lower or equal to $P$, so that pseudo-inversion is used). (e) Estimated weight matrix using the graphical LASSO, for a small number of observations, $P=40 < N=50$.} 
	\label{par3r_N50+part3}
\end{figure}
\end{Example}

\section{Summary of Graph Learning from Data Using Probabilistic Generative Models}
Graph data analytics with known or given topologies is feasible for applications that involve physically meaningful structures, such as citation networks, transport networks and observable social networks. In those applications,  various vertex or spectral domain techniques, as mentioned in Part 2 of this monograph, have been successfully implemented and developed to filter, analyse or visualise graph signals. However, in many situations where the graph topology cannot be directly observed or even when the data is partially observed, the inference of graph structure is a key first step. This is because different graph structures can lead to totally different results, as discussed earlier in this part of the monograph.

Unfortunately, given the observed graph data, graph learning is an ill-posed problem. In other words, different types of graphs can generate the same data we have observed and the same data can result in different graphs depending on the graph learning method used. Thus, to infer graph topology we need to employ some priors, for example, to match statistics via imposing sparsity or smoothness conditions on the graph. Previous sections in this part of the monograph have introduced various techniques such as the graphical LASSO and smoothness constrained graph learning, mostly from the perspective of linear algebra  \cite{dong2019learning, mateos2019connecting, giannakis2018topology}. 

However, it is more natural to connect and summarise those techniques under the umbrella of probabilistic generative models. A straightforward approach would be on the basis of some fundamental statistical models, such as the covariance or precision matrices of the Gaussian distribution (due to their positive definiteness property), the Gaussian Markov random field with local independence prior, or a factor analysis model with smoothness assumption. We also envisage further progress of generative models to be based on the concept of diffusion processes on graphs in the graph signal processing, whereby the signal generating process can be regarded as the graph signal that has been diffused by some graph kernels (the polynomial kernels) from a white Gaussian distribution.

Generally speaking, graph learning can be treated as an inverse problem to a graph data generation process, that is, $\mathbf{x} = f_{\mathcal{G}}(\mathbf{z})$, where $\mathbf{x}$ denotes the observed data, while the data are considered to be the output of an unknown transform (denoted by $f_{\mathcal{G}}$) of some initial state, $\mathbf{z}$, on the graph  $\mathcal{G}$. The existing literature of learning a graph can be thought of as an attempt to infer the generative process, $f_{\mathcal{G}}$, by matching the data statistics, $\mathbf{x}$, with different priors on $\mathbf{z}$. It needs to be pointed out that in this section we discuss the problem of learning graphs with fully observed graph data because this is a fundamental role in many advanced techniques, such as graph learning with partially observed data \cite{wai2019community,grotas2019power} and dynamic graph learning \cite{kaplan2008structural,chen2011vector,ioannidis2019semi}.

\subsection{Basic Gaussian models}\label{Sec_BasicGaussian}
The simplest way of constructing a graph would be to associate edge weights with the covariance of graph data; this is reasonable under the Gaussian assumption, since the first two moments fully capture the whole statistics of the data. Indeed, given the covariance matrix, its non-zero elements naturally provide consistent estimation of the connectivity within a graph. This method is explained within the introductory part of Section \ref{CorrGraphLear}. 

Given a set of $P$ independent and identically distributed (i.i.d.) observed data vectors, $\mathbf{x}_1$, $\mathbf{x}_2$, $\ldots$, $\mathbf{x}_P$, the empirical covariance is calculated as 
\begin{equation}
{\Sigma}_x(m,n)=\frac{1}{P}\sum_{p=1}^P \Big(x_p(m) -\mu (m)\Big) \Big(x_p(n)-\mu (n)\Big), \label{2CorrMatrixPPP}
\end{equation}
or
\begin{equation}\label{EmpiCov}
\mathbf{\Sigma} = \frac{1}{P}\sum_{p=1}^P(\mathbf{x}_p-\mathbf{\overline{x}})(\mathbf{x}_p-\mathbf{\overline{x}})^T,
\end{equation}
where $\mathbf{\overline{x}}$ is the mean value of the observed samples. Alternatively, a normalised version of $\mathbf{ \Sigma}$ can also be employed in order to produce the edge weights:
\begin{equation}
\sigma_x(m,n) = \frac{{\Sigma}_x(m,n)}{\sqrt{{\Sigma}_x(m,m){\Sigma}_x(n,n)}}.
\end{equation}
Then, for this empirical statistics, we can employ a threshold, $\tau$, to designate the non-zero connections of the adjacency weight via, (similar to (\ref{GaussWE})),
\begin{equation}
{W}_{m,n}=\left\{
\begin{aligned}
{\sigma}_x(m,n), &  ~~~~|{\sigma}_x(m,n)| \geq \tau \\
0, &  ~~~~|{\sigma}_x(m,n)| < \tau \\
\end{aligned}
\right.
\end{equation}
Furthermore, a more sophisticated approach would be to use hypothesis testing via setting a false alarm rate, whereby 
\begin{equation}
\mathcal{H}_0:~{\sigma}_x(m,n) = 0 ~~~~\mathrm{versus}~~~~\mathcal{H}_1:~{\sigma}_x(m,n) \neq 0.
\end{equation}
In these scenarios, the empirical covariance is a common choice of the test statistics. Although the density of ${\sigma}_x(m,n)$ may have closed-form representations, it typically needs numerical integration when calculating the p-values; however, transformations of ${\sigma}_x(m,n)$ can relieve this issue to obtain closed-form densities. For example, under the Gaussian distribution and $\mathcal{H}_0$,
the weighting 
$$s(m,n) = \frac{\sigma_x(m,n)\sqrt{k-2}}{\sqrt{1-\sigma_x^2(m,n)}}$$
would satisfy a student t-distribution of $(k-2)$ degrees of freedom, and 
$$s(m,n) = \mathrm{tanh}^{-1}(\sigma_x(m,n))$$
would then result in a Gaussian distribution with zero mean and $\nicefrac{1}{(k-3)}$ variance (see Chapter 7.3.1 \cite{kolaczyk2014statistical}). In those transformed test statistics, the significance can be easily adjusted to meet the false alarm rate. However, the limitation of this model is that by employing individual tests, the number of implementations in inferring the graph grows up to $\mathcal{O}(N^2)$. This results in high computational complexity in relatively large graphs; on the other hand, this leads to increasingly false judgements even with a constant false alarm rate. 

A further possible misleading of the correlation models is due to the fact that the \textit{correlation does not mean the causation}. In other words, the $m$-th and $n$-th vertices can show a strong correlation when they are all highly influenced by a middle vertex, however, they are not the causation of one another, as illustrated in Example \ref{SimCooEX}.

\subsection{Gaussian graphical model}\label{Sec_PartialCorrModel}
To address the issues with the correlation and causation, and to be able to construct a graph that reflects a direct relationship among vertices, one classical method would be to use the partial correlation, whereby the correlation of two vertices is calculated by eliminating associations of other contributing vertices. Under the assumption that vertices satisfy some mild distributions such as elliptical distributions, the partial correlation coincides with the conditional correlation \cite{baba2004partial}, and further equals to the conditional independence under the Gaussian assumption on vertices, so that the partial correlation can be explicitly related to the precision matrix. The so established relationship is crucial in understanding other techniques such as the graphical LASSO, graph regression and other generative models. 

\subsubsection{Partial correlation model}
In order to simplify the notation we will consider vertices $n=0$ and $m=1$. The set of all other vertices, except for the $m$-th and the $n$-th vertex, are denoted by $\mathcal{V}{\backslash \{m,n\}}=\{2,3,\dots,N-1\}$. Define the data vectors at each vertex by $\mathbf{y}_n$, as in (\ref{verttoR}). 
Let the values $\hat{\mathbf{y}}_0$ and $\hat{\mathbf{y}}_1$  the best linear approximations to the signal samples $\mathbf{y}_0$  and $ \mathbf{y}_1$ based on the data at other vertices,  $\mathbf{y}_2,\mathbf{y}_3,\dots,\mathbf{y}_{N-1}$. The new data values are then defined as 
\begin{gather*}
\mathbf{z}_0=\mathbf{y}_0-\hat{\mathbf{y}}_0\\
\mathbf{z}_1=\mathbf{y}_1-\hat{\mathbf{y}}_1
\end{gather*}

Now, the (empirical) partial correlation between vertices $m=0$ and $n=1$ can be defined as,
\begin{equation}
\sigma_z(0,1)= \frac{{\Sigma}_z(0,1)}{\sqrt{{\Sigma}_z(0,0)}\sqrt{{\Sigma}_z(1,1)}}
\end{equation}
In a similar way, all other partial correlations, $\sigma_z(m,n)$, between pairs of vertices $m$ and $n$ are calculated. Then, one way of hypothesis testing can be conducted as follows,
\begin{equation}
\mathcal{H}_0:~\sigma_z(m,n) = 0 ~~~~\mathrm{versus}~~~~\mathcal{H}_1:~\sigma_z(m,n) \neq 0,
\end{equation}
where $\sigma_z(m,n)$ can be employed as the test statistics. Moreover, other choices such as the Fisher's transform  $s(m,n) = \mathrm{tanh}^{-1}(\sigma_z(m,n))$ also obtain an asymptotically Gaussian null distribution (Chapter 7.3.2 \cite{kolaczyk2014statistical}).  

\subsubsection{Gaussian Markov random field}
A further assumption for the partial correlation model may be that it is under the Gaussian distribution, which in many cases is a common setting as this facilitates closed-form solutions and ease of analysis. For example, under the Gaussian distribution, the partial correlation coincides with the conditional correlation \cite{baba2004partial}, or equivalently, conditional independence; this in turn forms the pairwise Markov property of random fields, which constitutes a Gaussian Markov random field.

We shall denote the $m$-th and the $n$-th elements of signal samples as $\mathbf{y}_A$ and all other elements except for the $m$-th and the $n$-th elements as $\mathbf{y}_B$. The covariance of $\mathbf{y}_A$ is then represented as $\mathbf{\Sigma}_{AA}$, which is of the size $2\times2$. Then, in a in a block-wise manner, (\ref{EmpiCov}) turns to
\begin{equation}\label{Eq_BlockCov}
\mathbf{\Sigma} = \begin{bmatrix}
\mathbf{\Sigma}_{AA} & \mathbf{\Sigma}_{AB} \\
\mathbf{\Sigma}_{BA} & \mathbf{\Sigma}_{BB} \\
\end{bmatrix}.
\end{equation}
The covariance of the corresponding $\mathbf{y}_A$ conditioned on $\mathbf{y}_B$ is then easily obtained as
\begin{equation}\label{Eq_Partialcov}
\mathbf{\Sigma}_{A|B} = \mathbf{ \Sigma}_{AA} - \mathbf{ \Sigma}_{AB}\mathbf{ \Sigma}_{BB}^{-1}\mathbf{ \Sigma}_{BA},
\end{equation}
which is also called the Schur complement. On the other hand, to rewrite the expression in \eqref{Eq_BlockCov} with regard to the precision matrix, $\mathbf{Q} = \mathbf{\Sigma}^{-1}$, we can use the following block-wise matrix property,

\begin{gather}
\mathbf{Q} = \mathbf{\Sigma}^{-1} = \begin{bmatrix}
\mathbf{\Sigma}_{A|B}^{-1} & -\mathbf{\Sigma}_{A|B}^{-1}\mathbf{\Sigma}_{AB}\mathbf{\Sigma}_{BB}^{-1} \\
 -\mathbf{\Sigma}_{BB}^{-1}\mathbf{\Sigma}_{AB}\mathbf{\Sigma}_{A|B}^{-1} & \mathbf{\Sigma}_{BB}^{-1}\mathbf{\Sigma}_{AB}\mathbf{\Sigma}_{A|B}^{-1}\mathbf{\Sigma}_{AB}\mathbf{\Sigma}_{BB}^{-1} \\
\end{bmatrix} \nonumber \\ =\begin{bmatrix}
\mathbf{Q}_{AA} & \mathbf{Q}_{AB} \\
\mathbf{Q}_{BA} & \mathbf{Q}_{BB} \\
\end{bmatrix}.\label{Eq_invCov}
\end{gather}

From (\ref{Eq_invCov}), observe that $\mathbf{\Sigma}_{A|B}=\mathbf{Q}_{AA}^{-1}$ if the inverse of $\mathbf{Q}_{AA}$ exists. In other words, to obtain the partial correlation in \eqref{Eq_Partialcov}, it is more convenient to use the precision matrix than the covariance matrix. Thus, one feasible way to associate the edge weights is via 
\begin{equation}
W_{m,n} = -\frac{{Q}(m,n)}{\sqrt{{Q}(m,m){Q}(n,n)}},
\end{equation}
where $\mathbf{Q} = \mathbf{\Sigma}^{-1}$ is the empirical precision matrix. Then, the association of edge weights can be used to infer non-zero elements of $W_{m,n}$, which is also known as the covariance selection problem \cite{dempster1972covariance}. One feasible method is to recursively update the graph is by testing the hypotheses in the form 
\begin{equation}
\mathcal{H}_0:~W_{m,n} = 0 ~~~~\mathrm{versus}~~~~\mathcal{H}_1:~W_{m,n} \neq 0,
\end{equation}
where the $W_{m,n}$ is used as the test statistic. For large-scale graphs, however, this model also shows limitations that are similar to those of correlation models in Section \ref{Sec_BasicGaussian}. Although this model can relieve the vagueness between correlation and causation, it has one more additional limitation, in that it requires the number of samples to be larger than the dimension of covariance to ensure a proper inverse of covariance; this does not necessarily hold, especially for large-scale graphs, as stated in Remark \ref{remarksing}. The graphical LASSO and linear regression methods may be used to solve this issue. 

\subsubsection{Graphical LASSO and regression}
A common way of overcoming the problem of rank deficiency is to add a regularisation term when estimating the precision matrix. 

\noindent\textbf{Graphical LASSO.} Given the set of independent and identically distributed samples, $\mathbf{x}_1, \mathbf{x}_2, \ldots, \mathbf{x}_P$, the log-likelihood of a Gaussian distribution with zero mean and precision matrix $\mathbf{Q}$ is represented as in (\ref{LLfun})
\begin{gather}\label{Eq_LL}
J = \sum_{p=1}^P \big(-\frac{1}{2}\mathbf{x}_p^T\mathbf{Q}\mathbf{x}_p -\frac{P}{2}\ln(2\pi) + \frac{1}{2}\ln|\mathbf{Q}|\big) \\ 
\propto P\ln|\mathbf{Q}| - \sum_{p=1}^P \big( \mathbf{x}_p^T\mathbf{Q}\mathbf{x}_p\big),
\end{gather}
where $|\mathbf{Q}|=\textrm{det}(\mathbf{Q})$. By maximising this log-likelihood, the attained optimum is $$\mathbf{Q}^{-1} = \frac{1}{P}\sum_{p=1}^P\mathbf{x}_p\mathbf{x}_p^T,$$ 
as in (\ref{glcost3})-(\ref{gl_solution}).

However, when $P$ is smaller than the dimension of $\mathbf{x}_p$, the term $\sum_{p=1}^P(\mathbf{x}_p\mathbf{x}_p^T)$ is not full rank, thus leading to the singularity of $\mathbf{Q}$. One way of avoiding this issue is to use the ${l}_1$ norm to promote sparsity in \eqref{Eq_LL}, in a similar form to (\ref{minSparGlasso}), to yield
\begin{equation}\label{Eq_ReguLL}
\overline{J} = P\ln|\mathbf{Q}| - \sum_{p=1}^P \big( \mathbf{x}_p^T\mathbf{Q}\mathbf{x}_p\big) - \rho ||\mathbf{Q}||_1,
\end{equation}
which is known as a graphical LASSO problem. As shown in \cite{yuan2007model}, the correct graph can be inferred with probability approaching one, when choosing $\rho$ that satisfies $\rho\cdot P\rightarrow \infty$ and $\rho\cdot \sqrt{P}\rightarrow 0$, for $P\rightarrow \infty$. 
\begin{Remark}
	Other than the $l_1$ norm, other regularisations can also be employed in \eqref{Eq_ReguLL}. For example, solving \eqref{Eq_ReguLL} could result in negative values, which of course are of no meaning for associating the edge weights. Thus, constraining edge weights to be non-negative is also a common regularisation type in graphical learning (see Section \ref{energConL}). For more detail, we refer to \cite{friedman2008sparse, banerjee2008model,  yuan2006model}.
\end{Remark}

\noindent\textbf{Graph regression.} Another perspective of learning the Gaussian graphical model (described in Section \ref{SpartImpos} and Example \ref{ExLassoSim}) is via a regression of data observed at each vertex, $\mathbf{y}_m$, given the data observations at other vertices, $\mathbf{y}_n$, $n\in \{0,1,2,\dots,m-1,m+1,\dots,N-1\}=\mathcal{V} \backslash \{m\}$. The aim of the regression here is to learn a graph that can generate the optimal mean square error given the observed samples. More specifically, the values $\beta_{nm}$, $n=\mathcal{V} \backslash \{m\}$, that minimize
\begin{equation}
J_m=\Vert \mathbf{y}_m - \sum_{n=1,n\ne m}^{N-1} \beta_{nm} \mathbf{y}_n \Vert_2^2 \label{Eq_Regression}
\end{equation}
follow from 
$$\Big(\mathbf{y}_m - \sum_{n=1,n\ne m}^{N-1} \beta_{nm} \mathbf{y}_n \Big)\mathbf{y}_k^T =\mathbf{0}$$
or 
$\sum_{n=1,n\ne m}^{N-1} \beta_{nm} \Sigma_x(n,k)=\Sigma_x(m,k)$, for $k,n\in\mathcal{V} \backslash \{m\}$. A matrix solution of this equation is 
$$\bm{\beta}_m=\mathbf{\Sigma}^{-1}_{{m}{m}}\mathbf{\Sigma}_{1{m}},$$
where $\mathbf{\Sigma}_{1{m}}$ is a vector with $(N-1)$ elements $\Sigma_x(m,k)$, $k=\mathcal{V} \backslash \{m\}$, and $\mathbf{\Sigma}_{{m}{m}}$ is an $(N-1)\times (N-1)$ matrix with elements $\Sigma_x(n,k)$, $k,n=\mathcal{V} \backslash \{m\}$. 
On the other hand, under the Gaussian assumption, the conditional mean of $\mathbf{y}_m$ on $\mathbf{y}_n$ is given by
$$\mathrm{E}_{p(\mathbf{y}_m|\mathbf{y}_n)}\{\mathbf{y}_m\}=(\mathbf{\Sigma}^{-1}_{{m}{m}}\mathbf{\Sigma}_{1{m}})^T\mathbf{X}_{P,m},$$
where 
$$\mathbf{X}_{P,m}=
\begin{bmatrix}
\mathbf{y}_0 \\  \mathbf{y}_1 \\    \vdots   \\ \mathbf{y}_{m-1} \\\mathbf{y}_{m+1} \\ \vdots \\  \mathbf{y}_{N-1}    
\end{bmatrix} ,
$$
with
\begin{equation}
\mathbf{y}_n=\begin{bmatrix}
\ x_1(n), \  \ x_2(n), \   \ \dots,   \ x_P(n)    
\end{bmatrix}. \label{verttoR2}
\end{equation}

Therefore, given the data observed on a graph, $\mathbf{x}_1, \mathbf{x}_2, \ldots, \mathbf{x}_P$, to infer $\mathbf{Q}$, we can regress $x_m$ for each vertex, $m$, on the basis of \eqref{Eq_Regression} as follows,
\begin{equation}
x_m = \bm{\beta}_m^T \mathbf{X}_{P,m} + \bm{\epsilon}_{m},
\end{equation}
where $\bm{\epsilon}_{m}$ is independent Gaussian noise.

Therefore, the problem of learning $\mathbf{Q}$ turns into the regression problem on $\bm{\beta}_m$, for each vertex, and non-zero elements in $\bm{\beta}_m$ also indicate the corresponding non-zero elements in $\mathbf{Q}$, namely, the edges in the graph. 

The main advantage of the regression-style methods is that the regressions for each vertex can be computed in parallel, which provides computational ease when learning large graphs.  However, additional attention should be paid to the symmetry of the learnt regression coefficients when dealing with an undirected graph, for example as in (\ref{betadiregm}), more detail can be found in \cite{meinshausen2006high}. The condition of coefficient sparsity could also be included, which leads to the LASSO formulation and solution of this problem, as in Section \ref{SpartImpos}.

\subsection{Factor analysis model}
In Sections \ref{Sec_BasicGaussian} and \ref{Sec_PartialCorrModel}, the Gaussian distribution is assumed and on the basis of this distribution, most methods have been proposed to learn the graph edges in a recursive manner, i.e., by learning an edge per iteration. On the other hand, such methods can be regarded as a generative process via a basic Gaussian distribution, whereby the covariance or the precision matrix is nontrivially associated with the graph edges. It is thus natural to adopt more general and sophisticated models in graph learning. One important model in probabilistic generative models is the factor analysis model, which forms the basis of many important tools, such as the probabilistic principal component analysis. 
Therefore, the observed data on a graph, $\mathbf{x}$, is assumed to be generated via a factor model that can be represented as
\begin{equation}\label{Eq_Factoranalysis}
\mathbf{x}= \mathbf{U}\mathbf{v} + \mathbf{\epsilon},
\end{equation}
where $\mathbf{U}$ is a unitary matrix of the graph Laplacian eigenvectors, and $\mathbf{v}$ is a vector of latent variables (or factor loadings) which is Gaussian distributed with zero mean and a diagonal precision matrix corresponding to the graph Laplacian eigenvalues $\bm\Lambda$, that is, $$\mathbf{v}\sim\mathcal{N}(\mathbf{0}, \bm\Lambda^{-1}),$$ 
where $\Lambda^{-1}$ is the Moore-Penrose pseudoinverse of $\Lambda$, while $\mathbf{\epsilon}\sim\mathcal{N}(\mathbf{0}, \alpha^2 \mathbf{I})$ is also Gaussian distributed but independent of latent variables $\mathbf{{v}}$.

On the basis of this factor model, it is easy to obtain the distribution of the observations, $\mathbf{x}$, as $$\mathbf{x}\sim\mathcal{N}(\mathbf{0}, \mathbf{U}\mathbf{\Lambda}^{-1}\mathbf{U}^T+\alpha^2\mathbf{I}).$$ The term $(\mathbf{U}\mathbf{\Lambda}^{-1}\mathbf{U}^T)^{-1} = \mathbf{U}^T\mathbf{\Lambda}\mathbf{U}$ uniquely defines a Laplacian matrix, $\mathbf{L}$, of a graph. This allows us to infer the graph structure by learning $\mathbf{L} = \mathbf{U}^T\mathbf{\Lambda}\mathbf{U}$ from the factor model via maximising the posterior distribution of $\mathbf{x}$, given by 
$$P(\mathbf{v}|\mathbf{x}) \propto P(\mathbf{x}|\mathbf{v})P(\mathbf{v}) \propto e^{-\frac{(\mathbf{x}-\mathbf{U}\mathbf{v})^T(\mathbf{x}-\mathbf{U}\mathbf{v})}{\alpha^2}}e^{-\mathbf{v}^T\mathbf{\Lambda}\mathbf{v}}.$$
Its log-likelihood form is formulated as \cite{dong2016learning}
\begin{equation}\label{Eq_factormin}
\min_{\mathbf{\Lambda}, \mathbf{U}, \mathbf{v}} ||\mathbf{x} - \mathbf{U}\mathbf{v}||^2 + \rho \cdot \mathbf{v}^T\mathbf{\Lambda}\mathbf{v},
\end{equation}
where $\rho$ is a hyperparameter that balances between the mean square error $||\mathbf{x} - \mathbf{U}\mathbf{v}||^2$ and the positive definite constraint $\mathbf{v}^T\mathbf{\Lambda}\mathbf{v}$. Expression \eqref{Eq_factormin} can be further rewritten  using the notation $\mathbf{y}=\mathbf{U}\mathbf{v}$, as
\begin{equation}\label{Eq_factormin2}
\min_{\mathbf{L}, \mathbf{y}} ||\mathbf{x} - \mathbf{y}||^2 + \rho\cdot \mathbf{y}^T\mathbf{L}\mathbf{y}.
\end{equation}
By inspection of \eqref{Eq_factormin2} we see that $\mathbf{y}^T\mathbf{L}\mathbf{y}$ indicates the smoothness of the signal $\mathbf{y}$ on the graph; this means that \eqref{Eq_factormin2} minimises the error between the observed samples and the generated signals, whilst imposing the smoothness on the generated signals, as discussed in Section \ref{SubSGC}. Other regularisations can also be imposed onto this model, such as that trace($\mathbf{L}$) is equal to the dimension of the graph, in order to avoid a trivial all zero optimum and non-positive values in the non-diagonal elements of $\mathbf{L}$, and to learn a feasible graph \cite{dong2016learning}. Finally, \eqref{Eq_factormin2} can be optimised in an alternative manner, as discussed in Section \ref{SubSGC} and Algorithm \ref{ReconMetAlg}, namely, by alternatively optimising one of the two parameters ($\mathbf{L}$ and $\mathbf{y}$) while fixing the other one.

Further improvements following the factor model of learning a smooth graph include the use of a more flexible smoothness prior when optimising $\mathbf{L}$ in alternative optimisation, as various constrains on the $\mathbf{L}$ can lead to a complicated optimisation implementation \cite{kalofolias2016learn}. This is achieved by rewriting the smoothness prior, $\mathbf{y}^T\mathbf{L}\mathbf{y}$ in \eqref{Eq_factormin2}, as $\mathbf{y}^T\mathbf{L}\mathbf{y} = \frac{1}{2} \sum_{m,n}\mathbf{A}_{mn}({y}(m)-{y}(n))^2$ so that the constrains can be explicitly imposed on the adjacency matrix $\mathbf{A}$, instead of on the Laplacian $\mathbf{L}$. It is also possible to learn the graph by selecting the edges from atoms in a dictionary (called the incidence matrix) \cite{chepuri2017learning}. Although this strategy can explicitly control the sparsity of the graph, it cannot optimise the edge weights \cite{mateos2019connecting}.
\begin{Example}
	Fig. \ref{Fig_graphconnection2} and Fig. \ref{Fig_graphsmooth} show that different graph connections can have different smoothness features, given the same observed samples, $\mathbf{x}$, in Fig. \ref{Fig_graphconnection2} (a). As indicated in Figure \ref{Fig_graphsmooth}, the observed sample retains the lowest frequency components for the graph in Fig. \ref{Fig_graphconnection2} (b) and the highest frequency components for the graph in  Fig. \ref{Fig_graphconnection2} (c). This also results in the smaller smoothness, $\mathbf{x}^T\mathbf{L}\mathbf{x}$, for the graph in Fig. \ref{Fig_graphconnection2} (b). This exemplifies that, given the observed graph samples, the smoothness prior is convenient for learning a graph.
	
	\begin{figure*}
	\centering
	\includegraphics[scale=0.75]{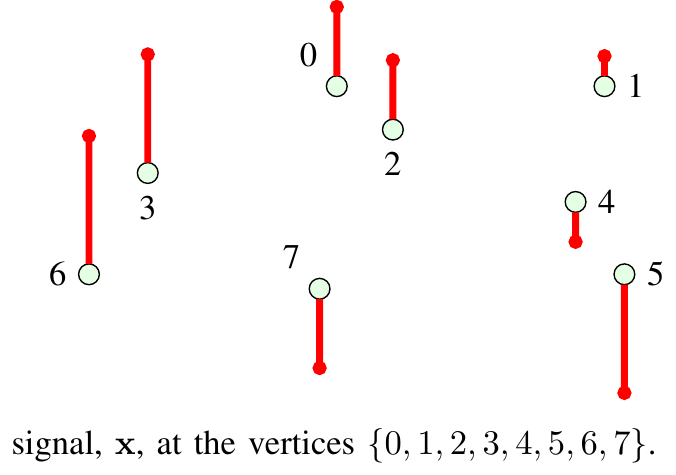} \includegraphics[scale=0.75]{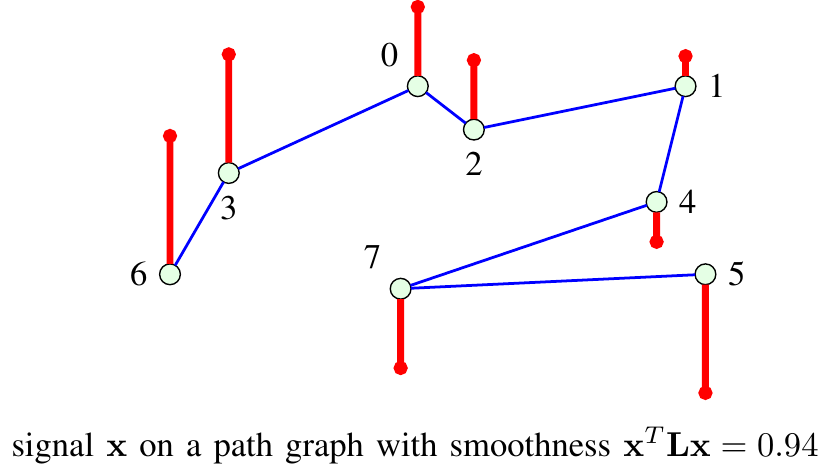} \includegraphics[scale=0.75]{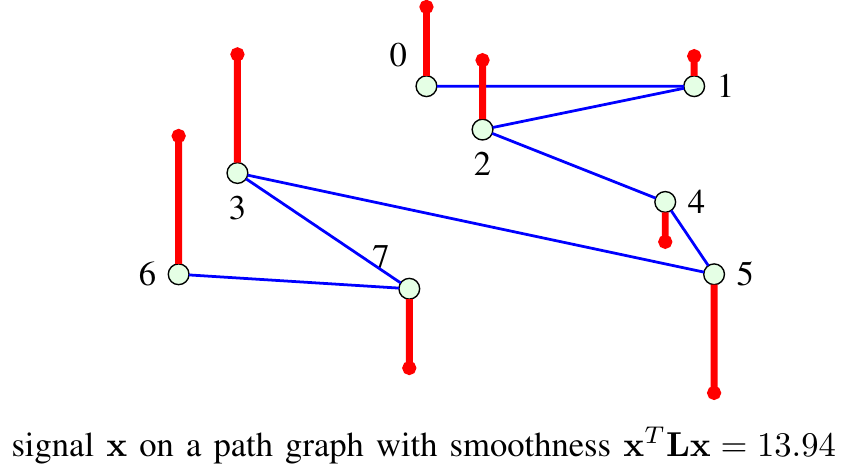} \\  (a) \hspace{50mm} (b) \hspace{50mm} (c)
	\caption{Smoothness and graph learning. (a) The observed graph signal  $\mathbf{x}=[0.7, 0.2, 0.6, 1.1 -0.3, -1.1, 1.3, -0.7]^T$, with (b)-(c) two types of possible path graph connections resulting in different smoothness values, $\mathbf{x}^T\mathbf{L}\mathbf{x}$.}
	\label{Fig_graphconnection2}
\end{figure*}
	
	\begin{figure}
		\centering
		\includegraphics{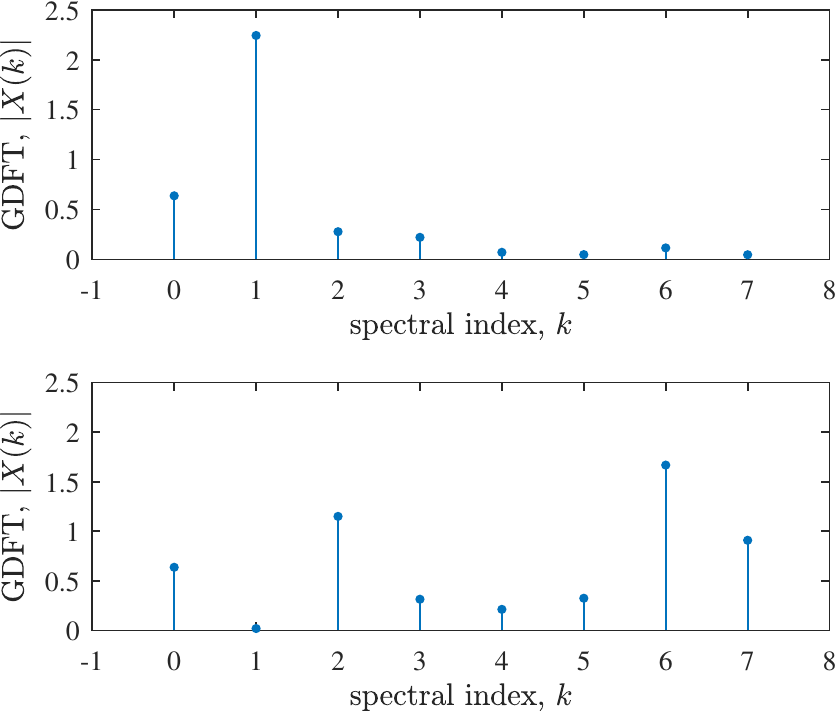}
		\caption{The graph signal spectrum values corresponding to the two types of graph connections in Figure \ref{Fig_graphconnection2}. The top panel corresponds to Fig. \ref{Fig_graphconnection2}-(b) and the bottom panel to  Fig. \ref{Fig_graphconnection2}-(c). The energy is calculated via $\mathbf{x}^T\mathbf{L}\mathbf{x}$, where small values indicate a smooth graph.}
		\label{Fig_graphsmooth}
	\end{figure}
\end{Example}
\subsection{Diffusion models}
It is important to notice that the smoothness that arises from the factor model is imposed in a global manner, which is effective in learning the main structure of a graph. However, the global smoothness can also overestimate the details within a graph. To resolve this issue, we can further assume that the observed graph signals are generated via a more complex and powerful model, such as the diffusion model. As shall be discussed in detail in Section \ref{Subsec_SpectralPoly}, the polynomial filter is a typical choice of treating the diffusion from a graph signal processing perspective. The benefits related to learning a graph are mainly three-fold: 
\begin{itemize}
	\item Analytical and computational ease during learning;
	\item This ensures the (weak) stationarity in the generation system \cite{mateos2019connecting};
	\item Ability to control the local smoothness in the model.
\end{itemize}
The diffusion model is given by (\ref{graphSigDefiH})
\begin{equation}\label{Eq_DeepFactorAnalysis}
\mathbf{x} = \sum_{m=0}^{M} h_m \mathbf{S}^m\mathbf{v} + \mathbf{\epsilon},
\end{equation}
where $\mathbf{v}$ is white Gaussian noise $\mathbf{v}\sim\mathcal{N}(\mathbf{0},\mathbf{I})$, while similar to the factor model in \eqref{Eq_Factoranalysis}, $\mathbf{\epsilon}\sim \mathcal{N}(\mathbf{0}, \alpha^2\mathbf{I})$. From \eqref{Eq_DeepFactorAnalysis}, recall that $\mathbf{S}$ is the (symmetric) shift operator which can be chosen as e.g., the adjacency matrix, or the Laplacian matrix, to name but a few. Here, we will continue to use 
$$\mathbf{S}=\mathbf{L}$$
as in Section \ref{graphtopeig}. Furthermore, \eqref{Eq_DeepFactorAnalysis} can be compactly written in the form of (\ref{graphSigDefiH})
\begin{equation}\label{Eq_DiffusionModel}
\mathbf{x} = \sum_{m=0}^M h_m \mathbf{L}^m\mathbf{v},
\end{equation}
where $\mathbf{L}^0 = \mathbf{I}$ and $h_0=\alpha^2$ retain the same statistics as those in \eqref{Eq_DeepFactorAnalysis}.

On the basis of \eqref{Eq_DiffusionModel}, the covariance of $\mathbf{x}$ can be calculated as 
\begin{gather}
\mathbf{\Sigma} = \mathrm{E}\{\mathbf{x}\mathbf{x}^T\} = (\sum_{m=0}^M h_m \mathbf{L}^m)\mathrm{E}\{\mathbf{v}\mathbf{v}^T\}(\sum_{m=0}^M h_m \mathbf{L}^m)^T \nonumber \\  \label{Eq_DiffusionCov}= \sum_{m=0}^M h_m \mathbf{L}^m(\sum_{m=0}^M h_m \mathbf{L}^m)^T 
= \mathbf{U}^T\big(\sum_{m=0}^M h_m\mathbf{\Lambda}^m\big)^2\mathbf{U},
\end{gather}
where we have used the eigendecomposition $\mathbf{L} = \mathbf{U}^T\mathbf{\Lambda}\mathbf{U}$. 

\noindent\textbf{Eigenvector estimation.} From \eqref{Eq_DiffusionCov}, we can see that the eigenvectors of $\mathbf{L}$ are the same as those of the covariance of $\mathbf{x}$. This means, in a straightforward way, that we can infer the eigenvectors of the $\mathbf{L}$ by the empirical covariance from the observed data, $\mathbf{x}_1, \mathbf{x}_2, \ldots, \mathbf{x}_P$.

\noindent\textbf{Eigenvalue estimation.} After obtaining the eigenvectors, the remaining task is to estimate the eigenvalues of $\mathbf{L}$. Without any additional constraints, it is obvious that arbitrary values can be chosen as the eigenvalues of $\mathbf{L}$ because we can always find a corresponding set of $h_0, h_1, \ldots, h_M$ that satisfies \eqref{Eq_DiffusionCov}. Thus, to achieve a unique solution, we need to employ some prior on the function $f(\cdot)$ \cite{segarra2017network}, which is modelled as 
\begin{equation}
\min_{\mathbf{L}, \mathbf{\Lambda}} f(\mathbf{L}), \mathrm{~subject~to~} \mathbf{L} = \mathbf{U}^T\mathbf{\Lambda}\mathbf{U}.
\end{equation}
For example, when $f(\mathbf{L}) = ||\mathbf{L}||_0$, the objective function minimises the number of edges, whereas $f(\mathbf{L}) = ||\mathbf{L}||_2$ minimises the energy of graph edges. The number of edges can also be minimized using convex relation of $f(\mathbf{L}) = ||\mathbf{L}||_0$ in the form $f(\mathbf{L}) = ||\mathbf{L}||_1$, as explained in Part 2 of this monograph and Section \ref{graphtopeig}.

Equation \eqref{Eq_DiffusionModel} assumes that the diffusion process starts from the same initial status, of white Gaussian noise. An enhanced diffusion model has been proposed in \cite{thanou2017learning} by assuming that the signals are generated from multiple heat diffusion processes
\begin{equation}
\mathbf{x} = \sum_{m=0}^M e^{-h_m\mathbf{L}}\mathbf{v}_m.
\end{equation}
Here, $\mathbf{v}_m$ represents the initial state that can also be optimised, and $h_m$ controls the diffusion time (depth). This means that with a small $h_m$, the $k$-th column of $e^{-\lambda_m\mathbf{L}}$ is localised at the $k$-th vertex. This model can be solved via a dictionary-learning solver by regarding $[e^{-h_0\mathbf{L}}, e^{-h_1\mathbf{L}}, \ldots, e^{-h_M\mathbf{L}}]$ as the dictionary $\mathbf{D}$ and $[\mathbf{v}_0, \mathbf{v}_1, \ldots, \mathbf{v}_M]$ as coefficients $\mathbf{V}$. The objective function can now be formulated as 
\begin{gather*}
\min_{\mathbf{L},\mathbf{X}, h_m} ||\mathbf{\mathbf{X}}-\mathbf{D}\mathbf{V}||_F^2 + \mathrm{reg}(\mathbf{V}) + \mathrm{reg}(\mathbf{L}), \\ \mathrm{subject ~ to~} \{h_m\}_{m=0}^M\leq0,
\end{gather*}
where $\mathrm{reg}(\cdot)$ denotes a certain regularisation; for more detail, we refer to \cite{thanou2017learning}.

\section{Graph Neural Networks}

An emerging field that is closely related to graphs is that of graph neural networks, with the aim of benefiting from the universal approximation property exhibited by multiple stacked layers of neurons. This area has witnessed many breakthroughs in recent years, facilitated by growing computational powers and the increasing amount of available data. The beginning of graph neural networks (GNNs) can be traced back to vanilla network structures \cite{gori2005new, scarselli2008graph, micheli2009neural} one decade ago, while recent developments have been centred around convolutional graph networks (GCNs). The GCNs benefit from their intrinsic graph structure, which allows for complex and implicit connections and information aggregation when processing (or filtering) each vertex. This is particularly desirable in deep neural network (DNN) techniques, where the involvement of graphs provides a balance between the ``black-box'' (but powerful) DNNs and the purely mathematical tools such as manifold optimisation and manifold learning. Benefiting from prior information embedded into the graph structure, GCNs can not only handle irregular data but also help convert the ``black-box'' nature of NNs into a ``grey-box'' model, two major issues with current DNN operation.

The recent literature on GCNs \cite{zhou2018graph, wu2019comprehensive} typically considers the learning aspect, while highlighting two key properties of CNNs: i) stationarity (via shift invariance of convolution operations) and ii) compositionality (via downsampling of pooling operations). Taking a sightly different viewpoint, we start from the graph itself and proceed to illuminate that certain types of graphs correspond to major trends in GCNs. We also outline the advantage of treating GCNs in this way, such as the possibility to open avenues for novel types of GCNs.

This section first introduces some basic elements of graph data analysis that will be used to understand GCNs. Then, we embark upon the ability of graphs to provide intrinsic structures when aggregating information, to describe recurrent GNNs as a kind of diffusion processes of task-oriented models. We further employ the concept of system on a graph to understand  spectral GCNs, while spatial GCNs are shown to admit interpretation as a relaxation of spectral GCNs on the localisation in graphs. 

\subsection{Basic graph elements related to GCNs}\label{Sec_Basics}
The following properties of graphs are helpful in understanding the GCNs (for more detail we refer to  Section 2.1 of  Part 1):
\begin{itemize}
	\item \textbf{Property 1}: When $\mathbf{A}$ is binary, i.e., representing the connection of vertices (adjacency matrix), the number of walks of a length $k$, between two vertices $m$ and $n$, is equivalent to the value of the corresponding element $a_{mn}$ of the $k$-th power of $\mathbf{A}$, that is, of $\mathbf{A}^k$. The number of walks between the vertices $m$ and $n$, that are of length not higher than $k$, is given by the corresponding element of $\mathbf{B}_k$, where $\mathbf{B}_{k} = \mathbf{A}+\mathbf{A}^2+\cdots+\mathbf{A}^k$. Matrix $\mathbf{B}_{k}$ gives the $k$-neighborhood of a vertex, which is a set of vertices that are reachable from this vertex in walks within $k$ steps.
	\item \textbf{Property 2}: For any signal on graph, $\mathbf{x}$, the quadratic form of the Laplacian, $\mathbf{x}^T\mathbf{L}\mathbf{x}$, has the form, 
	\begin{equation}
	\mathbf{x}^T\mathbf{L}\mathbf{x} = \frac{1}{2} \sum_{m=0}^{N-1}\sum_{n=1}^{N-1}A_{mn}(x(m)-x(n))^2.
	\end{equation}
	This indicates that: 1) The Laplacian matrix, $\mathbf{L}=\mathbf{D}-\mathbf{A}$, is positive semi-definite because $A_{mn}(x(m)-x(n))^2\geq 0$; 2) The smoothness of graph signal, $\mathbf{x}$, can be quantified via $\mathbf{x}^T\mathbf{L}\mathbf{x}$, which ensures that the quadratic form  $\mathbf{x}^T\mathbf{L}\mathbf{x}$ is equivalent to the Dirichlet energy of $\mathbf{x}$, which has been widely used in probabilistic graph models.
\end{itemize}

The smoothness of graph signal, $\mathbf{x}$, implies that the signal value would not change much from one vertex to another within the neighbourhood of vertex $n$ (assessed by $(x(m)-x(n))^2$). However, signal values are allowed to change significantly when the two vertices are not connected (indicated by zeros values of $A_{mn}$). Therefore, the minimisation on $\mathbf{x}^T\mathbf{L}\mathbf{x}$ finds the smoothest signal $\mathbf{x}$ on the graph. 

Note that the absolute minimum of smoothness is achieved for the constant signal over all vertices, being equal to the eigenvector corresponding to the smallest eigenvalue, $\lambda_0=0$, of the graph Laplacian, $\mathbf{L}$ (owing to the Rayleigh quotient). More importantly, this yields $\mathbf{1}^T\mathbf{L}\mathbf{1}=\mathbf{1}^T(\mathbf{D}-\mathbf{A})\mathbf{1} = 0$, which means that the smallest eigenvalue is $0$ and the corresponding normalised eigenvector is $\mathbf{x}=\mathbf{u}_0=\mathbf{1}/\sqrt{N}$, where $\mathbf{1}$ denotes an $N$-dimensional vector whose elements equal to $1$.

\subsubsection{Connection to the Laplacian operator in function analysis}
A way of understanding the role of the Laplacian matrix in measuring the smoothness is via its counterpart -- the Laplacian operator in functional analysis. The Laplacian operator over a function $f(\vec{r})$ in the  Euclidean space is defined as 
$$ \mathrm{div}(\mathrm{grad}(f(\vec{r})))=\nabla(\nabla f(\vec{r}))=\Delta f(\vec{r}),$$
 where $\mathrm{grad}(\cdot)$ is the gradient operation and $\mathrm{div}(\cdot)$ is the divergence operation. 
For example, in Cartesian coordinates of two dimensions, $\vec{r}=(x,y)$, we have
\begin{equation}
\Delta f(x,y) = \frac{\partial^2 f(x,y)}{\partial x^2} + \frac{\partial^2 f(x,y)}{\partial y^2}.
\end{equation}

Similarly, we can also define the Laplacian operation on the graph, whereby the different and difficult part is the differential operation. Namely, while as in the Euclidean space, the differential operation is defined as $\nabla f(x) = f(x+1) - f(x)$, which calculates the difference between $f(x+1)$ and $f(x)$, the differential operation on a graph is defined for each edge, that is 
$$\nabla f_{mn} = f(m) - f(n).$$ 
This means that, in general, the differential on a graph allows for a different number of directions at each point (vertex), while for the path graph, $\nabla f_{mn}$  naturally simplifies into the standard differential in the Euclidean space. 

\begin{figure}[!htb]
	\begin{center}
		\includegraphics{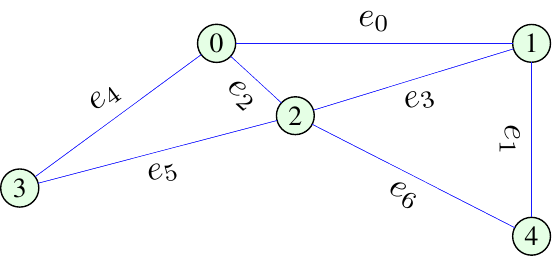}
	\end{center}
	\caption{An illustrative graph, which is a simplified version of Figure 1-(a) of Part 1.}\label{Fig_Graph}
\end{figure}
\begin{Example}
	To demystify the Laplacian operator, consider a graph in Figure \ref{Fig_Graph}, which is a simplified version of Figure 1-(a) of Part 1. Its adjacency matrix and the corresponding graph Laplacian matrix are given by
	\begin{equation}\label{Eq_Grapg_laplacian}
	\mathbf{A} = \begin{bmatrix}
	0 & 1 & 1 & 1 & 0\\
	1 & 0 & 1 & 0 & 1\\
	1 & 1 & 0 & 1 & 1\\
	1 & 0 & 1 & 0 & 0\\
	0 & 1 & 1 & 0 & 0\\
	\end{bmatrix},~~ \mathbf{L} =
	\begin{bmatrix}
	\begin{array}{rrrrr}
	3 & -1 & -1 & -1 & 0\\
	-1 & 3 & -1 & 0 & -1\\
	-1 & -1 & 4 & -1 & -1\\
	-1 & 0 & -1 & 2 & 0\\
	0 & -1 & -1 & 0 & 2\\
	\end{array}
	\end{bmatrix}.
	\end{equation}
	To calculate the gradient of a signal $\mathbf{f}$ on this graph,
	\begin{equation}
	\mathbf{f} = \begin{bmatrix}
	f(0) \\
	f(1) \\
	f(2) \\
	f(3) \\
	f(4) \\
	\end{bmatrix},
	\end{equation}
	which shall represent the differential on each edge, we introduce the so called incidence matrix, $\mathbf{K}$, given by 
	\begin{equation}
	\mathbf{K} = \bbordermatrix{~ & e_0 & e_1 & e_2 & e_3 & e_4 & e_5 & e_6 \cr
		0 &1&0&1&0&1&0&0\cr
		1 &-1&1&0&1&0&0&0\cr
		2 &0&0&-1&-1&0&1&1\cr
		3 &0&0&0&0&-1&-1&0\cr
		4 &0&-1&0&0&0&0&-1\cr},
	\end{equation}
	The gradient on the graph now becomes
	\begin{equation}
	\mathrm{grad}(\mathbf{f}) = \mathbf{K}^T\mathbf{f} = \bbordermatrix{~ & \nabla f \cr
		e_0 & f(0) -f(1) \cr
		e_1 & f(1) -f(4) \cr
		e_2 & f(0) -f(2) \cr
		e_3 & f(1) -f(2) \cr
		e_4 & f(0) -f(3) \cr
		e_5 & f(2) -f(3) \cr
		e_6 & f(2) -f(4) \cr}.
	\end{equation}
	Due to the adjoint property of the divergence operator with regard to inner products, the graph Laplacian for this graph becomes
	\begin{gather*}
	\Delta \mathbf{f} = \mathrm{div}(\mathrm{grad}(\mathbf{f})) = \mathbf{K}(\mathbf{K}^T\mathbf{f}) = (\mathbf{K}\mathbf{K}^T)\mathbf{f} \\ = \begin{bmatrix}
	\begin{array}{rrrrr}
	3 & -1 & -1 & -1 & 0\\
	-1 & 3 & -1 & 0 & -1\\
	-1 & -1 & 4 & -1 & -1\\
	-1 & 0 & -1 & 2 & 0\\
	0 & -1 & -1 & 0 & 2\\
	\end{array}
	\end{bmatrix}  \begin{bmatrix}
	f({v}_0) \\
	f({v}_1) \\
	f({v}_2) \\
	f({v}_3) \\
	f({v}_4) \\
	\end{bmatrix}.
	\end{gather*}
	It is now obvious that $\mathbf{K}\mathbf{K}^T$ is equivalent to the graph Laplacian matrix $\mathbf{L}$ in \eqref{Eq_Grapg_laplacian}. 
\end{Example}
This exemplifies that a graph actually defines local coordinates with a prior or learnt linkage information, and thus in some sense it can be considered as a discrete approximation to a manifold. 

\subsection{Recurrent GNNs as a diffusion process}\label{Sec_RGCN}
Consider a physical diffusion process, and in particular the Newton's law of cooling, which states that the energy (or heat) loss rate is proportional to the temperature difference between the body (node) and its surrounding environment. The diffusion process can be understood as an iterative process that converges toward the state of minimum energy, given by $\mathbf{x}^T\mathbf{L}\mathbf{x}$, from any initial condition. Since the gradient of energy is 
$$ \mathrm{grad}(\mathbf{x}^T\mathbf{L}\mathbf{x})=\frac{\partial(\mathbf{x}^T\mathbf{L}\mathbf{x})}{\partial \mathbf{x}^T}=2\mathbf{L}\mathbf{x},$$
the iterative discrete-time solution for the diffusion process, at an instant $t+1$, is given by
\begin{equation}\label{Eq_Diffusion}
\mathbf{x}_{t+1} - \mathbf{x}_t = -\alpha \mathbf{L}\mathbf{x}_t,
\end{equation} 
or
$$\mathbf{x}_{t+1} = \mathbf{x}_t -\alpha \mathbf{L}\mathbf{x}_t,$$
where $\alpha$ is a constant. This solution to the diffusion process can also be formulated as
\begin{equation}\label{Eq_Newton_law}
\nabla x(n) \approx -\alpha\sum_{m \in \mathcal{V}_n}(x(n)-x(m)),
\end{equation}
where  $\mathcal{V}_n$ is the set of vertices within the neighborhood-one of the vertex $n$, while $\sum_{m \in \mathcal{V}_n}(x(n)-x(m))$ denotes an aggregate temperature difference between the vertex $n$ and its surrounding vertices. 

\begin{Remark}
Equation \eqref{Eq_Diffusion} models the change in temperature along time, starting from an initial state $\mathbf{x}_0$. In the following, we will show that this provides an ideal means for designing recurrent GNNs. 

The quadratic term, $\mathbf{x}^T\mathbf{L}\mathbf{x}$, is frequently used in data analytics on graphs, for example for estimating smoothness. The gradient of $\mathbf{x}^T\mathbf{L}\mathbf{x}$ is $\nicefrac{\partial(\mathbf{x}^T\mathbf{L}\mathbf{x})}{\partial \mathbf{x}} = 2\mathbf{L}\mathbf{x}$, so that the diffusion process in \eqref{Eq_Diffusion} will find the exact minimum of this quadratic form. As mentioned in Section \ref{Sec_Basics}, the minimum of $\mathbf{x}^T\mathbf{L}\mathbf{x}$ is a constant eigenvector with all elements equal to $1$, which indicates that such a diffusion process, when left without any external sources, will eventually settle to the same temperatures for all vertices. 
\end{Remark}

\subsection{Label propagation as a diffusion process}\label{Subsec_Diffusion}
The stable state (equilibrium) of a diffusion process cannot give us any useful information because in this case the data at all the vertices have the same value (i.e., the lowest entropy on the graph). In physics, we can alter the stable state by adding some constant external sources, which ensures that the final temperatures are not all the same but exhibit some fluctuations governed by their inherent relationships. This is also the basic idea behind many graph signal processing approaches, especially in semi-supervised learning tasks, such as the \textit{label propagation} given in Algorithm \ref{Alg1}.
\begin{algorithm}
	\caption{\!\!. \ Label Propagation}\label{Alg1}
	\begin{algorithmic}[1]
		\footnotesize
		\Procedure{Initialisation}{}
		\State Initialise a graph by treating each data sample separately, as a single vertex; 
		\State Connect all vertices in the graph, whereby edge weights are defined by some similarity measure;
		\State Assign the labels from the labeled samples to the corresponding vertices; 
		\State Randomly assign values to the unlabeled vertices.
		\While{Not converged:}
		\State Propagate from the labeled to the unlabeled vertices: $\mathbf{x}\leftarrow \mathbf{L}\mathbf{x}$. 
		\Comment{Diffusion process}
		\State Re-assign the original labels to the labeled vertices, $\mathbf{x}_L$. \Comment{Keep external resources}
		\EndWhile
		\State \textbf{return} $\mathbf{v}$
		\EndProcedure
	\end{algorithmic}
\end{algorithm}

The final state of this modified diffusion process can be easily shown to be \cite{zhu2005semi},
\begin{equation}
\mathbf{x}_U = (\mathbf{I} - \mathbf{L}_{UU})^{-1}\mathbf{L}_{UL}\mathbf{x}_L,
\end{equation}
where
\begin{equation}
\mathbf{L} = \begin{bmatrix}
\mathbf{L}_{LL} & \mathbf{L}_{LU} \\
\mathbf{L}_{UL} & \mathbf{L}_{UU} \\
\end{bmatrix}, 
\end{equation}
and the subscripts $U$ and $L$ designate respectively the unlabelled and labelled parts. Note that for a graph shift, instead of $\mathbf{L}$ we may also use $\mathbf{A}$. 

The final stable state will now no longer have the same signal values for all vertices (at least $\mathbf{x}_U \neq \mathbf{x}_L$). This is due to the ``external constant" sources of the labelled samples (Line 7 in Algorithm \ref{Alg1}), which ensures that the diffusion process results in stable states with signals which are different for each vertex; it also gives the predicted labels for unlabelled signal samples (or vertices) in the inner structures of the graph. 
\begin{Example}
	To provide a simple illustration of label propagation in digit recognition, we used 3 sets of handwritten digits, 1, 5 and 9, each with ten images from the MNIST database \cite{lecun1998gradient}. We adopted the structural similarity (SSIM) metric \cite{wang2004image} to measure the similarity between images and constructed a graph accordingly, shown in Figure \ref{Fig_labelprop}-(a). In this example, we chose only two labels for each digit type to act as the external sources in the diffusion process. In \eqref{Eq_Diffusion}, it needs to be pointed out that without the external sources, the final state would settle to a constant vector; this does not provide any informative predictions. 
	
	The total of six given labels are annotated in Figure \ref{Fig_labelprop}-(c) and the predicted labels are shown in Figure \ref{Fig_labelprop}-(d). 
	By comparing with the ground truth shown in Figure \ref{Fig_labelprop}-(b), we can see that  label propagation achieved adequate prediction, given the correctly constructed graph. The certainty of prediction is designated by the node color in Figure \ref{Fig_labelprop}-(d), with the provided labels (ground truth) in the red color, and the nodes on the intersections of two types of digits in green colors, indicating the large uncertainty of predictions in these vertices. On the other hand, when regarding label propagation as a diffusion process, the temperature can be interpreted as the level of certainty  whereby the external sources (the six given labels) have the highest temperature (designated via the red color) and the heat diffusion performs a certainty propagation. Vertices surrounding the external sources, as a consequence, would retain relatively high temperature (we are much more sure about the predictions on these nodes).

	\begin{figure*}
		\centering
		\includegraphics[width=0.45\textwidth]{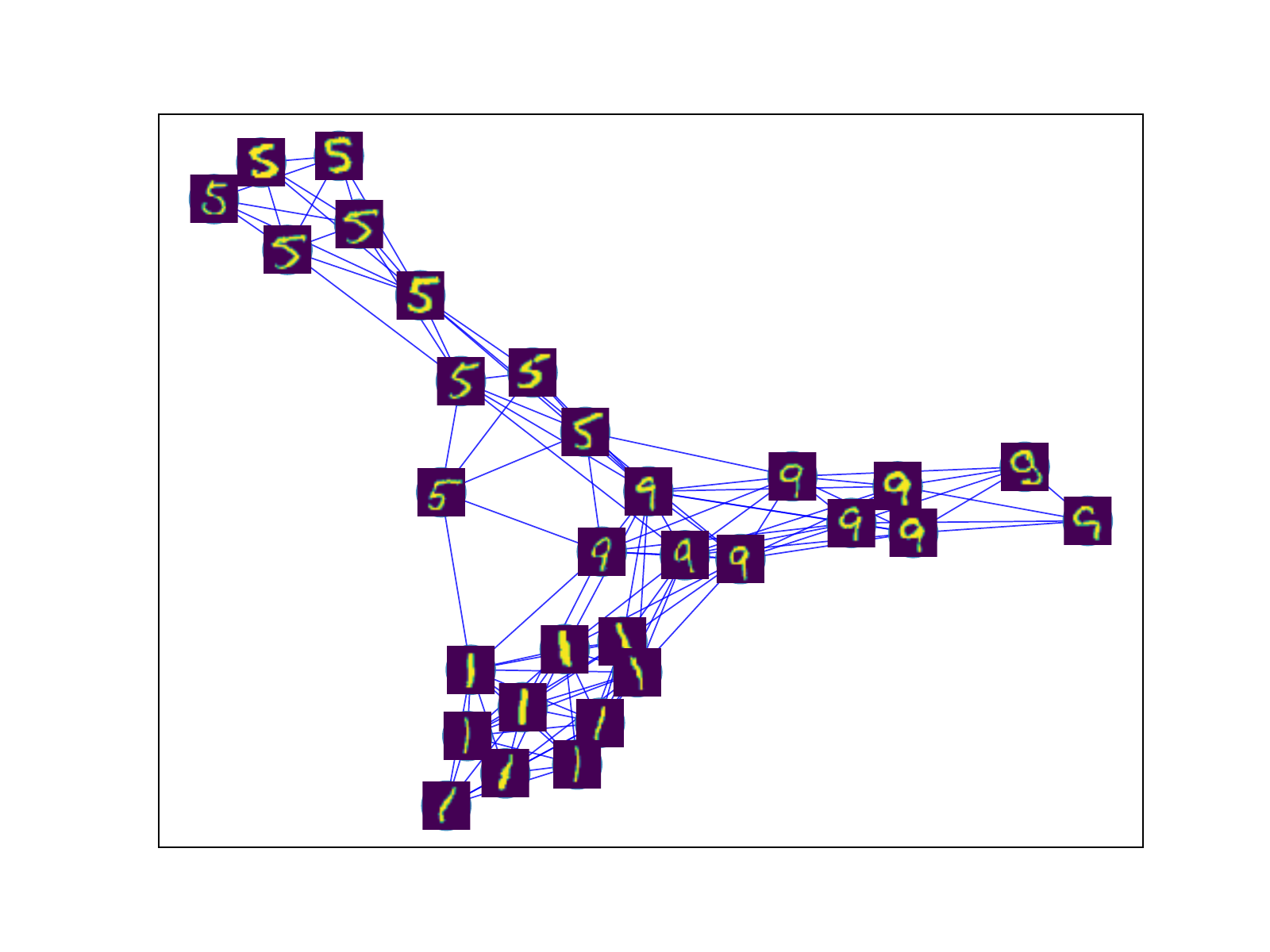}
		\includegraphics[width=0.4\textwidth]{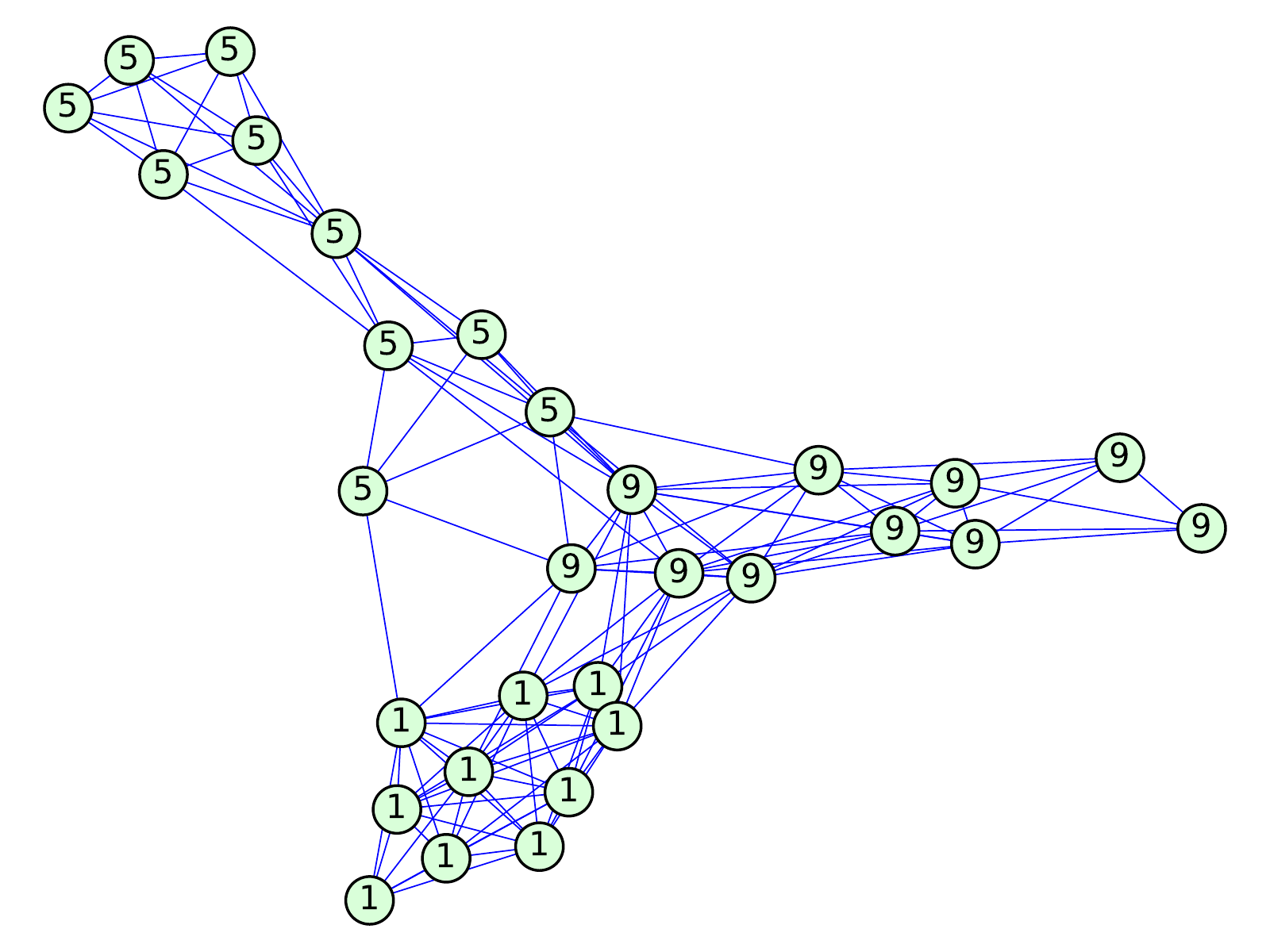}
		\\\hspace{30mm} (a) \hspace{70mm} (b) \hspace{30mm} 
		\vspace{5mm}
		
		\includegraphics[width=0.45\textwidth]{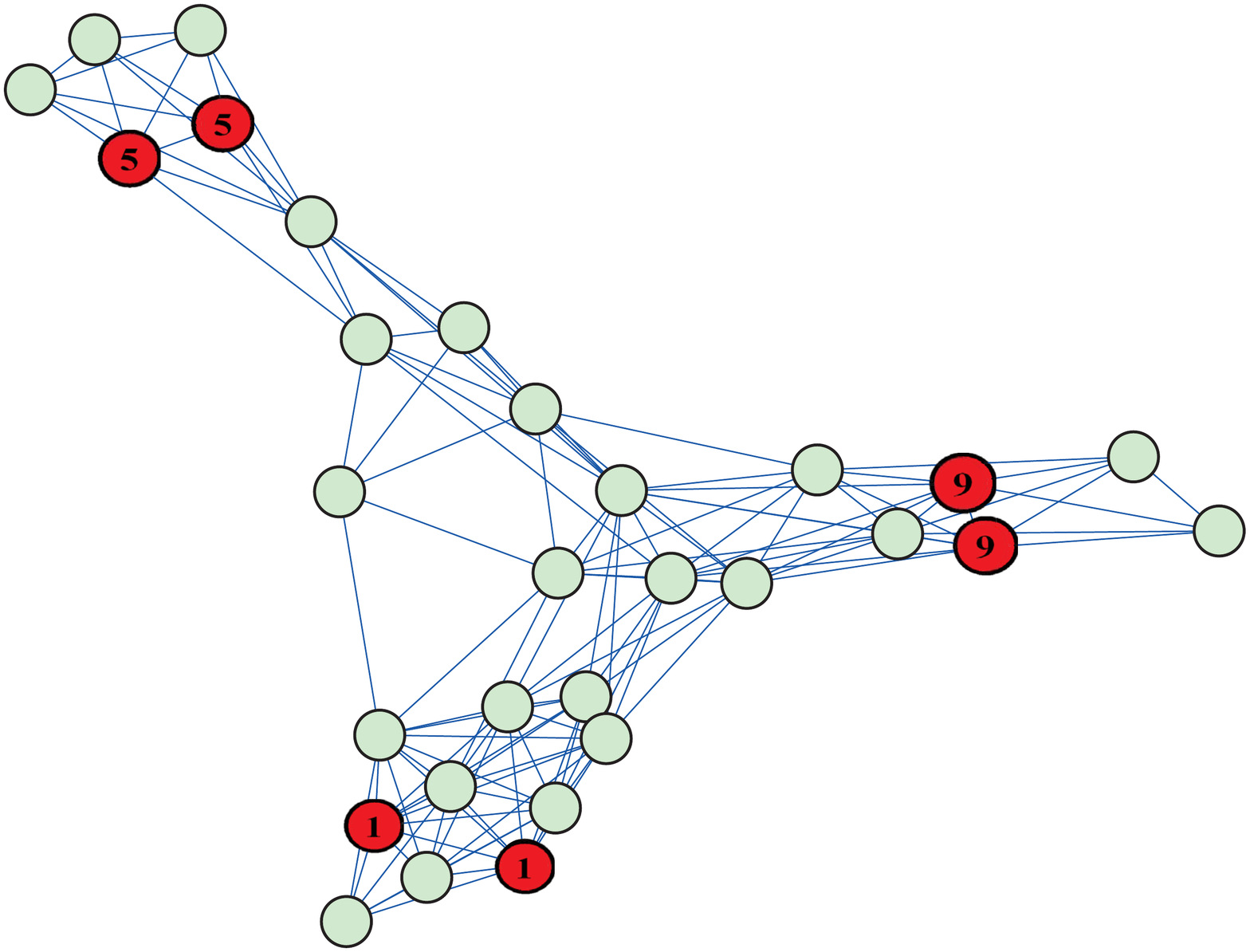}
		\includegraphics[width=0.4\textwidth]{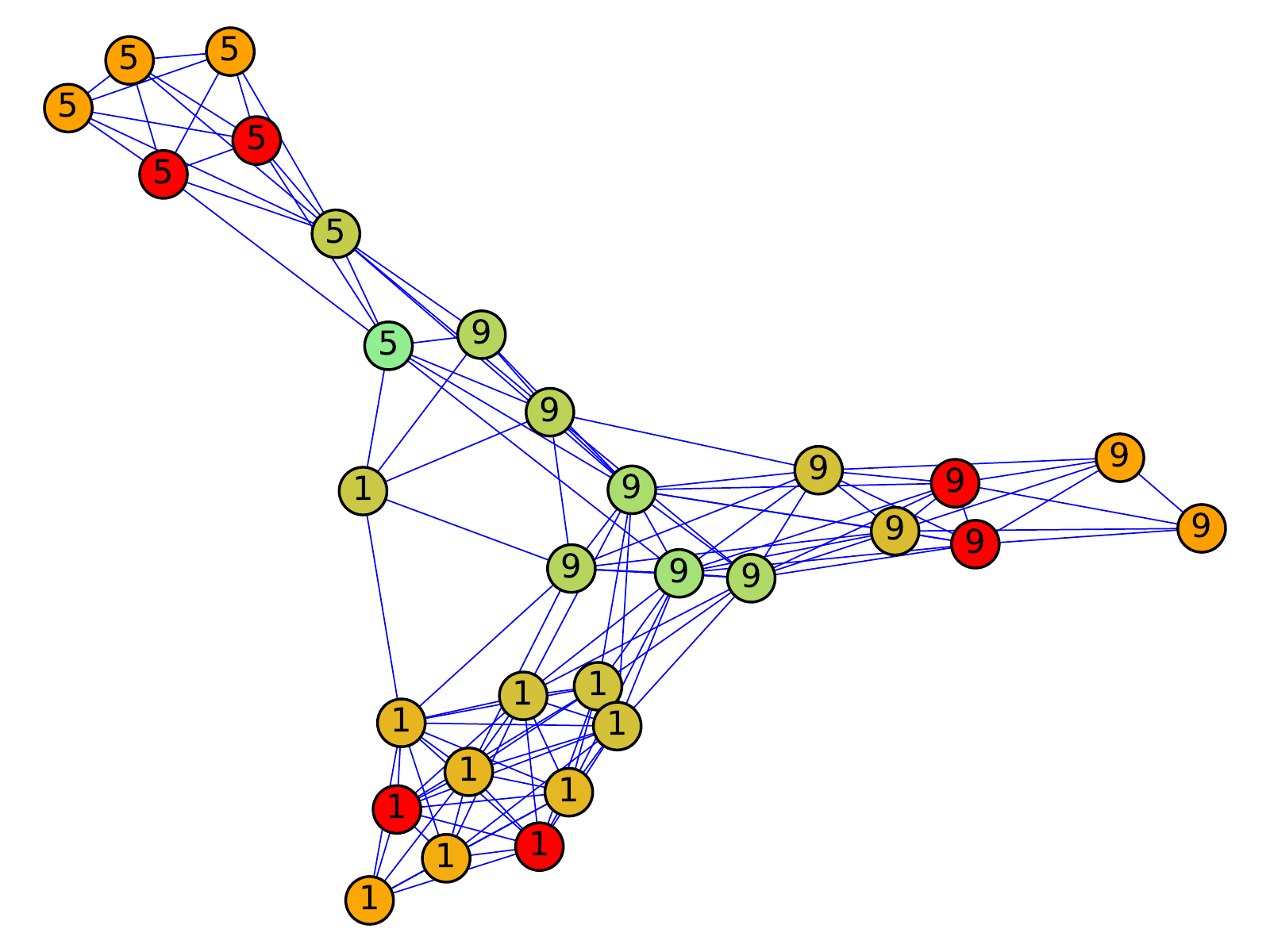}
		\includegraphics[width=0.05\textwidth]{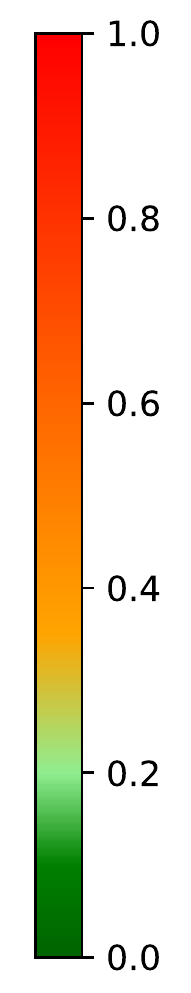}
		\\\hspace{30mm} (c) \hspace{70mm} (d) \hspace{30mm} 
			\caption{Label propagation via two labelled images out of ten available images per digit from the MNIST dataset. Three sets of digits (1, 5 and 9) are chosen and each set contains ten images. a) The constructed graph via the SSIM metric, where two images (nodes) are connected when their SSIM is larger than a threshold (set to 0.35). b) The ground truth labels for the 30 images considered. c) Only two labels are provided for each set of images, as indicated by the red color. d) Predicted labels from the given six (i.e., $2\times3$) labels via label propagation over the graph Laplacian matrix $\mathbf{L}$. The color bar designates the certainty of predictions, namely, the red color denotes an almost sure prediction with probability approaching 1 and green color poor prediction; }
		\label{Fig_labelprop}
	\end{figure*}
\end{Example}

\subsection{GNNs of a recurrent style}\label{Subsec_RecurrGCN}
Now that we have shown that different diffusion models can be utilised to aggregate information across graph vertices, we may employ diffusion to design neural networks on graphs, as NNs also rely ipon information aggregation. One such frequently used recurrent GNN was proposed by Scarselli \textit{et al.} \cite{scarselli2008graph}, which aggregates information as,
\begin{gather}\label{Eq_GNNdiffusion}
{x}_{t+1}(n) = \sum_{m\in \mathcal{V}_n} f(x_t(n), q(n), x_t(m), q(m)), \\ o(n) = \rho(x(n), l(n))
\end{gather}
where $v_t(n)$ is the signal value at the $n$-th vertex at a time instant $t$, $\mathcal{V}_n$ denotes the neighbourhood-one of the vertex $n$, $q(n)$ is a pre-defined feature of $x(n)$, $q(m)$ represents the pre-defined features at the neighbour vertices, and $o(n)$ is the output at the $n$-th vertex. The operators $f(\cdot)$ and $\rho(\cdot)$ can be chosen to be neural networks so that they can be learnt via back-propagation; in other words, the diffusion style model can be learnt from data samples. In a particular case when $q(n)$ and $q(m)$ are omitted, and  
$$f(x(n), q(n), x(m), q(m))=(x(n)-x(m)),$$
\eqref{Eq_GNNdiffusion} turns to the original diffusion process given in \eqref{Eq_Newton_law}.

The aggregation function in \eqref{Eq_GNNdiffusion} motivates much recent work on graph neural networks and spatial GCNs, however, this variant of recurrent GNNs needs to undergo the diffusion process until convergence, for every iteration of back-propagation. Moreover, the $f(\cdot)$ in \eqref{Eq_GNNdiffusion} needs to be carefully designed to be a contraction mapping to ensure convergence \cite{mandic2007machine}. More recent efforts to improve this model include the gated recurrent GNN \cite{li2015gated} that employs a gated unit as $f(\cdot)$ to ensure convergence within a fixed number of steps, while stochastic steady-state recurrent GNNs \cite{dai2018learning} perform update in \eqref{Eq_GNNdiffusion} in a stochastic manner. 

Another interesting work is the diffusion convolution neural network (DCNN) \cite{atwood2016diffusion}, which \textit{incorporates both spatial convolutions and temporal diffusions}, and can be formulated as 
\begin{equation}\label{Eq_diffGCN}
\mathbf{h}^{l} = \rho(\mathbf{w}^{l}\odot\mathbf{L}^l\mathbf{x}),
\end{equation}
where $\mathbf{h}^l$ is the hidden state of the $k$-th layer, $\mathbf{w}^{l}$ are convolution kernels that are to be learnt, and $\mathbf{L}^l$ is the $l$-th power of a certain probability transition matrix (in this case graph Laplacian $\mathbf{L}$) which is similar to Line 7 in Algorithm \ref{Alg1}; recall that $\odot$ denotes the element-wise product and $\rho$ the activation function. It should be pointed out that the model in \eqref{Eq_diffGCN} implies that $\mathbf{h}^{l}$ does not depend on the previous layer (state) $\mathbf{h}^{l-1}$, and that the dimensions of each layer need to be the same; this limits the number of degrees of freedom in the design. The overall output of this GCN is a composition of all layers $\{\mathbf{h}^l\}_{l=1}^L$, so that \eqref{Eq_diffGCN} can be understood as a set of diffusion processes of different depths (by regarding $l$ as time instant $t$). 

Another way of understanding \eqref{Eq_diffGCN} is that each diffusion step, $\mathbf{L}^l\mathbf{x}$, aggregates to a certain degree the heat (or general features and labels). This is a kind of message passing and aggregation that equips the network with the ability to extract statically salient features, which belong to spatial GCNs introduced below.

\begin{Remark}
	Almost all the available literature on recurrent GNNs aims to find an efficient and stable diffusion way to propagate and aggregate the labels or information at each vertex, thus achieving reliable and robust predictions at the final stable stage of the GNNs. 
\end{Remark}

\subsection{Spatial GCNs via localisation of graphs}\label{Sec_SpaGCN}
It is important to note that while CNNs have been an enabling technology for modern machine learning applications, they also suffer from the limitations inherited from their assumption of a regular time/space grid, such as images and videos. The effort to extend CNNs to GCNs that are able to operate on data acquired on irregular domains therefore needs to accommodate both the convolution (to learn local stationary features) and the pooling (to compose multi-scale patterns) operators. Our main focus is on ways, to accommodate the data on irregular domains, while the generalisation of pooling is naturally related to the downsampling on the graph (see Part 2 and  \cite{bacciu2019non,sakiyama2019eigendecomposition,zhang2019star,tanaka2019generalized,ioannidis2019graphsac}). The key difficulty in defining the convolution on a graph is the absence of a rigorous translation (shift) operator. To this end, the basic idea behind spatial GCNs is the information aggregation principle, which is very similar (sometimes even intertwined with) to the diffusion GCNs in Section \ref{Subsec_RecurrGCN}. 
Instead of waiting for a stable state (along the time instants) of recurrent GCNs, spatial GCNs directly aggregate information by the stacked layers, which is also called message passing.  The initial work in this area was by Alessio \cite{micheli2009neural}, the so called neural network for graphs (NN4G). A more general model is the \textit{message passing neural networks} (MPNNs) \cite{gilmer2017neural}, which is given by
\begin{equation}\label{Eq_MPNN}
{x}^{l+1}(n) = \rho\big({x}^l(n), \sum_{m\in\mathcal{V}_n}f({x}^l(n),{x}^l(m),e_{nm})\big),
\end{equation}
where ${x}^l(n)$ represents the data value at the $n$-th vertex of the $l$-th layer, $e_{nm}$ denotes the edge between the $n$-th and the $m$-th vertex, while $f(\cdot)$ is the message passing function and $\rho(\cdot)$ denotes the activation (or vertex updating) function. The model in \eqref{Eq_MPNN} caters for many GCNs, such as those in \cite{micheli2009neural} and \cite{kipf2016semi} which all have different forms of functions $f(\cdot)$ and $\rho(\cdot)$. This model also involves the basic steps for processing graph signals in the spatial domain, i.e., by aggregating the previous messages and passing to the next layer. Davide \textit{et al.} further extended this idea to a probabilistic framework \cite{bacciu2018contextual}, which enables a probabilistic explanation on each state of each layer.

Furthermore, instead of looking for all neighbours of the central vertex in \eqref{Eq_diffGCN}, the GraphSAGE approach proposes to sample several neighbours around every vertex \cite{hamilton2017inductive}, as follows
\begin{equation}
{x}^{l+1}(n) = \rho\big(\mathbf{W}^l\cdot\mathrm{concat}\{{x}^l(n), f\{{x}^l(m), {m\in{\mathcal{\widetilde V}_n}}\}\}\big),
\end{equation}
where $\mathrm{concat}\{\cdot,\cdot\}$ denotes the concatenation and $f\{\cdot\}$ the aggregation function, $\mathbf{W}^l$ is the matrix of learnable parameters, and $\mathcal{\widetilde V}_n$ denotes a randomly chosen neighbour of the $n$-th vertex. This strategy allows for a mini-batch operation on graphs, which is extremely useful for large graphs.

A further trend is to learn the weights while choosing the neighbouring vertices; this includes the graph attention network (GAT) \cite{velivckovic2017graph}, and the mixture model network (MoNet) \cite{monti2017geometric}. Within GATs, an attention weight, $\alpha_{n,m}$, is added to the parameters in \eqref{Eq_MPNN}, which allows to assign different importance levels to vertices, even within the same neighbourhood. The attention weight can be further learnt from an additional convolution sub-network, as proposed in \cite{zhang2018gaan}. On the other hand, the MoNet defines the weights of neighbouring edges as a consequence of local coordinates, which has an intrinsic link with the manifolds. More specifically, it defines the importance of the edge connecting the $n$-th and the $m$-th vertex as a probability, $p$, over some local coordinates, $\mathbf{u}(m,n)$, which reflects the difference (or distance) between the $n$-th and the $m$-th vertex. Then, the $n$-th vertex can be aggregated via a specially defined convolution, given by
\begin{equation}\label{Eq_MoNet}
(\mathbf{x} * \mathbf{g})(n) = \sum_{j=1}^J g_j \sum_{m\in\mathcal{V}_n}p\big(\mathbf{u}(m,n)\big)x(m),
\end{equation}
where $g_j$ is the $j$-th index (element) of the convolution kernel, $\mathbf{g}$. In \cite{monti2017geometric}, the probability, $p(\mathbf{u}(m,n))$, was chosen as a Gaussian mixture model, which has $J$ clusters to cater for the size of convolution kernel. It has also been shown that the framework of \eqref{Eq_MoNet} accounts for various geometric deep neural networks, through a choice of different local coordinates and weight functions. 

\subsection{Spectral GCNs via graph Fourier transform}\label{Sec_SpecGCN}
As shown in Section \ref{Sec_SpaGCN}, message passing via the convolution operation plays a crucial role in spatial GCNs. Here, we focus on the methods that operate in a transfer domain and benefit from the mathematically well-defined convolution in the graph spectral domain to yield a class of spectral GCNs. 

\subsubsection{Graph Fourier transform}
Due to the positive semi-definiteness of $\mathbf{L}$, there are $N$ (the number of vertices) real-valued eigenvalues ($\lambda_0=0\!\leq\! \lambda_1 \!<\! \lambda_2 \!<\! \cdots \!<\! \lambda_{N-1}$), which correspond to $N$ distinct orthogonal eigenvectors ($[\mathbf{u}_0,\mathbf{u}_1,\ldots,\mathbf{u}_{N-1}]$). As mentioned in Section \ref{Sec_Basics}, the quadratic form $\mathbf{x}^T\mathbf{L}\mathbf{x}$ measures the smoothness of the data $\mathbf{x}$ on the graph. Further, when $\mathbf{x}$ equals one of the eigenvectors, $\mathbf{u}_j$, the term $\mathbf{x}^T\mathbf{L}\mathbf{x}$ then measures the smoothness of the eigenvectors $\mathbf{u}_j^T\mathbf{L}\mathbf{u}_j = \lambda_j$. The matrix of eigenvectors, $\mathbf{U = }[\mathbf{u}_0,\mathbf{u}_1,\ldots,\mathbf{u}_{N}]$, represents an orthogonal transform basis, which is similar to the principal component analysis (PCA), while what is more physically important and beneficial in practice is that the graph Laplacian bases indicate the smoothness of eigenvectors. 
\begin{Remark}
	Through multiplication of the data, $\mathbf{x}$, by the eigenmatrix, $\mathbf{U}\mathbf{x}$, the original data $\mathbf{x}$ are decomposed into different constituent components, which vary from the most smooth to the most non-smooth. This is exactly the principle of the Fourier transform, which transforms a signal to different frequency components (bases). In this case, $\lambda_j$ has the physical meaning of (squared)  frequency, as shown in Section 3.5.2 of Part 2. In particular, when the graph structure is the path graph, the original Fourier transform is obtained. 
\end{Remark}

Based on the graph Fourier transform, we can now define the graph convolution operator which states that the convolution in the spatial (vertex) domain is equal to the multiplication in the spectral domain. This bypasses the requirement for translation (or shift operator) to define convolution in the vertex domain, whilst maintaining the concept of ``convolution'' over graph signals. In this way, the graph convolution is given by
\begin{equation}\label{Eq_convolution}
\mathbf{U}^T(\mathbf{x} * \mathbf{g}) = (\mathbf{U}^T\mathbf{x})\odot(\mathbf{U}^T\mathbf{g}),
\end{equation}
where $\mathbf{x}$ and $\mathbf{g}$ are two vectors whose elements are the data values at vertices $n\in \mathcal{V}$. Recall that $\mathbf{U}$ in \eqref{Eq_convolution} is the Fourier basis composed by the eigenvectors of $\mathbf{L}$ and $\odot$ denotes the Hadamard (element-wise) product.

\subsubsection{Graph spectral filtering as multiple diffusion processes}\label{Subsec_SpectralPoly}
Upon inspection of the diffusion process of the cooling law in Section \ref{Subsec_Diffusion}, we can see that it actually aggregates the values of the connected vertices to process the current vertex. Consider now a polynomial filter of the diffusion process, given by
\begin{equation}
\mathbf{x} \leftarrow \mathbf{B}_k\mathbf{x} = (\mathbf{A} + \mathbf{A}^2 + \cdots + \mathbf{A}^k)\mathbf{x},
\end{equation}
where $k$ neighbouring vertex data values are aggregated to produce the current vertex data sample according to the Property 1 of Section \ref{Sec_Basics}. It can be proved that the $k$-neighbouring property also holds when $\mathbf{B}_k$ is given by the powers of the Laplacian, $\mathbf{L}^k$ (Lemma 5.4, \cite{hammond2011wavelets}), as we are still using the $k$-neighbour information when aggregating, that is
\begin{equation}\label{Eq_LaplacianPoly}
\mathbf{x} \leftarrow (\mathbf{L} + \mathbf{L}^2 + \cdots + \mathbf{L}^k)\mathbf{x}.
\end{equation}

Upon rewriting \eqref{Eq_LaplacianPoly} in the graph spectral domain, we have
\begin{equation}\label{Eq_Spatiopoly}
\mathbf{x}\leftarrow\mathbf{U}(\mathbf{\Lambda}+\mathbf{\Lambda}^2+\cdots+\mathbf{\Lambda}^k)\mathbf{U}^T\mathbf{x},
\end{equation}
or equivalently
\begin{equation}\label{Eq_Spectralpoly}
\mathbf{X} \leftarrow (\mathbf{\Lambda}+\mathbf{\Lambda}^2+\cdots+\mathbf{\Lambda}^k)\mathbf{X},
\end{equation}
where $\mathbf{X}$ is the spectral representation of $\mathbf{x}$, through $\mathbf{X} = \mathbf{U}^T\mathbf{x}$, and $\mathbf{\Lambda}$ a diagonal matrix of which the elements are the ordered eigenvalues of $\mathbf{L}$. By combining \eqref{Eq_convolution} and \eqref{Eq_Spectralpoly}, the convolution operation on the graph can be achieved as
\begin{equation}\label{Eq_Spectralpolyk}
\mathbf{g} = \mathrm{poly}(\mathbf{\Lambda}) = \mathbf{\Lambda}+\mathbf{\Lambda}^2+\cdots+\mathbf{\Lambda}^k.
\end{equation}
We should point out that although there are many choices for the convolutional filter, $\mathbf{g}$, we typically choose the polynomial kernel as $\mathrm{poly}(\mathbf{\Lambda}) = \mathbf{\Lambda}+\mathbf{\Lambda}^2+\cdots+\mathbf{\Lambda}^k$, which ensures the localisation in the vertex domain within $k$-neighbours.

\subsubsection{Graph spectral filtering via neural networks}
It is natural to employ neural networks to replace the function $\mathbf{g}$ in \eqref{Eq_Spectralpolyk}, per layer. In this way we also take advantages of the spatial convolution operations and the universal approximation property of neural networks. This forms the basis of various spectral GCN methods. 

The first spectral GCN was proposed by Bruna \textit{et al.} \cite{bruna2013spectral}, based on a simple spectral model given by
\begin{equation}\label{Eq_BrunaGCN}
\mathbf{x}^{l+1}_j = \rho(\mathbf{U}\sum_{i=1}^{c_{l}}\mathbf{\Theta}^l_{i,j}\mathbf{U}^T\mathbf{x}^l_i)~~~(j = 1,2,\ldots,c_{l+1}),
\end{equation}
where $l$ represents the index of each layer, $c_l$ is the number of filters (channels) of the $l$-th layer, $\mathbf{\Theta}_{i,j}^l$ is a diagonal matrix which contains the set of learnt parameters of the $l$-th layer, and $\rho(\cdot)$ is the activation function of neurons. In \eqref{Eq_BrunaGCN}, the summation ensures the aggregation of features filtered by different convolutional kernels, $\mathbf{\Theta}_{i,j}^l$, which is similar to a linear combination across kernels in CNNs. Although it achieves graph convolution through NNs, this work has two main limitations: i) the localisation at the vertex domain cannot be ensured by $\mathbf{\Theta}_{i,j}^l$, although it is crucial in convolutional neural networks to extract local stationary features; ii) computational complexity brought by the $\mathcal{O}(N^2)$  multiplications of $\mathbf{U}$ and $\mathbf{U}^T$, and the eigen-decomposition of $\mathbf{L}$ to obtain $\mathbf{U}$, at the first time, may be prohibitive for large graphs. 

A possible way of mitigating these issues is to employ a polynomial form similar to that of \eqref{Eq_Spectralpoly}, as mentioned in Section \ref{Subsec_SpectralPoly}. This both relieves the first issue of the localisation, and helps to control a balance between the localisation in the vertex domain and the localisation in the spectral domain (see Part 2), as the \textit{uncertainty principle} of Fourier transform states that the localisation cannot be realised simultaneously in the time and frequency domains. More specifically, to further improve the localisation in the spatial domain in order to extract local patterns, we promote smoothness in the spectral domain through filtering by $\mathrm{poly}(\mathbf{\Lambda})$, whereby the term $\mathrm{poly}(\mathbf{\Lambda})$ is designed with a set of learnable parameters $\mathbf{\Theta}=\{\theta_i\}_{i=1}^k$, in the form
\begin{equation}
\mathrm{poly}_{\mathbf{\Theta}}(\mathbf{\Lambda})={\theta}_1\mathbf{\Lambda}+{\theta}_2\mathbf{\Lambda}^2+\cdots+{\theta}_k\mathbf{\Lambda}^k.
\end{equation}
Thus, the update rule of \eqref{Eq_Spatiopoly} can now be rewritten as
\begin{equation}\label{Eq_Spectralpoly2}
\mathbf{x}\leftarrow \mathrm{poly}_{\mathbf{\Theta}}(\mathbf{L})\mathbf{x} = \mathbf{U}\mathrm{poly}_{\mathbf{\Theta}}(\mathbf{\Lambda})\mathbf{U}^T\mathbf{x}.
\end{equation}

\begin{figure*}[t]
	\begin{center}
		\includegraphics[width=1\textwidth]{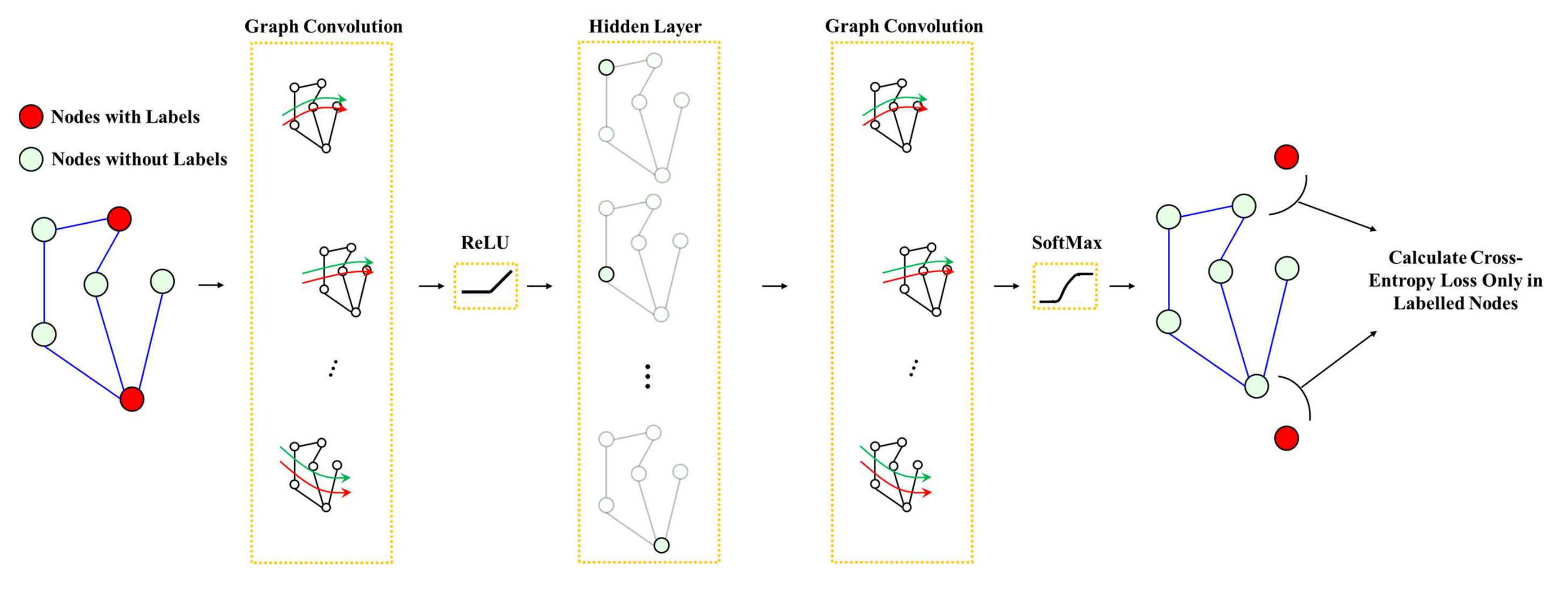}		
	\end{center}
	\caption{The structure of the GCN proposed in \cite{kipf2016semi} for semi-supervised learning.}\label{Fig_gcn}
\end{figure*}
\begin{figure}[t]
	\begin{center}	
		\includegraphics[width=0.3\textwidth]{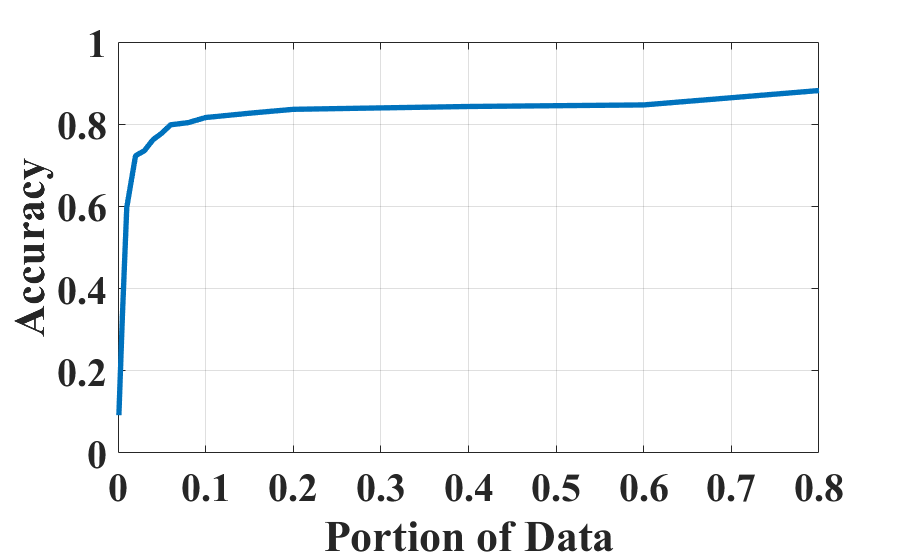}
	\end{center}
	\caption{Portions of data used for training versus the test accuracy on Cora dataset \cite{motl2015ctu} for a simple implementation of one typical GCN \cite{kipf2016semi} for semi-supervised learning.  In this example, we used one hidden layer with 256 neurons. The dropout rate was set to 0.5 and learning rate to 0.01.}\label{Fig_gcn_results}
\end{figure}

Notice that in \eqref{Eq_Spectralpoly2}, the multiplication with $\mathbf{U}$ is unnecessary at every layer, but the powers of $\mathbf{L}$ are needed and are computational demanding. 
On the basis of \eqref{Eq_Spectralpoly2}, Defferrard \textit{et al.} \cite{defferrard2016convolutional} further proposed the Chebyshev graph neural network, which employs the Chebyshev polynomial to ease the computation of $\mathrm{poly}_{\mathbf{\Theta}}(\mathbf{\Lambda})$, in the form
\begin{equation}
\mathrm{poly}_{\mathbf{\Theta}}(\mathbf{\Lambda}) = \sum_{i=1}^k {\theta}_i T_i(\mathbf{\widetilde\Lambda}),
\end{equation}
where $\mathbf{\widetilde\Lambda} = \nicefrac{2\mathbf{\Lambda}}{\lambda}_{\max} - \mathbf{I}_{N}$,  while $T_i(\mathbf{\widetilde\Lambda})$ is the Chebyshev polynomial that has an easy-to-compute recurrent form $T_i(\mathbf{\widetilde\Lambda}) = 2\mathbf{\widetilde\Lambda}T_{i-1}(\mathbf{\widetilde\Lambda}) - T_{i-2}(\mathbf{\widetilde\Lambda})$ ($T_0(\mathbf{\widetilde\Lambda})=1$, and $T_1(\mathbf{\widetilde\Lambda})=\mathbf{\widetilde\Lambda}$). With this Chebyshev polynomial, we are able to elegantly avoid the computation of the powers of $\mathbf{L}$, through
\begin{equation}\label{Eq_Chebnet}
\mathbf{x}\leftarrow \mathrm{poly}_{\mathbf{\Theta}}(\mathbf{L})\mathbf{x} = \sum_{i=1}^k {\theta}_i T_i(\mathbf{\widetilde L}) \mathbf{x},
\end{equation}
where $\mathbf{\widetilde L} = \nicefrac{2\mathbf{L}}{\lambda}_{\max} - \mathbf{I}_{N}$. This framework significantly reduces the computational complexity from $\mathcal{O}(N^2)$ to $\mathcal{O}(kN)$, and has been widely used in various graph learning tasks. Recent work \cite{kipf2016semi} further simplifies \eqref{Eq_Chebnet} by only employing the first-order Chebyshev polynomial ($k=1$), which achieves superior performances in semi-supervised learning. The authors claimed that it is unnecessary to employ a $k$-order format because the first-order Chebyshev polynomial is sufficient to mitigate overfitting, while the localisation of $k$-neighbours can be achieved by stacking layers of neural networks.

Despite mathematical beauty and physical intuition, spectral GCNs have been mainly limited to fixed network structures during both training and testing. More specifically, when employing spectral GCNs, the graph connections should be ascertained in advance because even a slight change in a graph connection would lead to a totally different eigenbasis. This, in turn, means that the whole graph needs to be loaded before training, which implies that GCNs cannot be trained in a mini-batch manner, as the trained model is domain dependent. 

\begin{Example}	
	To illustrate an implementation of one typical spectral GCN \cite{kipf2016semi} in semi-supervised learning, we employed the Cora dataset \cite{motl2015ctu} that contains 2708 machine learning related publications with 7 classes (case based, genetic algorithms, neural networks, probabilistic methods, reinforcement learning, rule learning and theory). Each publication has a feature vector that indicates whether an article includes any selected unique key-words. Furthermore, the graph is constructed via its citation relationships. 
	
	For the GCN method, we employed a Pytorch implementation of the work in \cite{kipf2016semi} which is available at https://github.com/tkipf/pygcn. The basic structure of the GCN network is illustrated in Figure \ref{Fig_gcn}. In this example, the number of hidden units was set to 256. We used different portions of data for training and plotted the test accuracy in classifying those publications into the 7 classes in Figure \ref{Fig_gcn_results}. Observe that with only $10\%$ of the available samples, a simple GCN with one hidden layer can achieve $>80\%$ classification accuracy. It is possible to further improve the test accuracy by extending the number of hidden units or increasing network depth. This simple example, however, highlights the powerful learning ability of GCNs on structural data.
\end{Example}

\section{Tensor Representation of Lattice-Structured Graphs}

In this section, we show that tensors (multidimensional data arrays) are a special class of graph signals, whereby the graph vertices reside on a high-dimensional regular lattice structure. In this way, the associated adjacency matrix exhibits a desirable structured form, referred to as \textit{Kronecker summable}, which effectively reduces the number of parameters required to model the entire graph connectivity structure.


\subsection{Tensorization of graph signals in high-dimensional spaces}

A tensor of \textit{order} $M$ is an $M$-way data array, denoted by $\boldsymbol{\mathcal{X}} \in \mathbb{R}^{I_{1} \times \cdots \times I_{M}}$. For example, a vector $\mathbf{x} \in \mathbb{R}^I$ is an order-$1$ tensor, a matrix $\mathbf{X} \in \mathbb{R}^{I_{1} \times I_{2}}$ is an order-$2$ tensor, while a 3-way array $\boldsymbol{\mathcal{X}} \in \mathbb{R}^{I_1 \times I_2 \times I_3}$ is an order-$3$ tensor. The $m$-th dimension of an order-$M$ tensor, $\boldsymbol{\mathcal{X}} \in \mathbb{R}^{I_{1} \times \cdots \times I_{M}}$, is referred to as the $m$-th \textit{mode} which is of size $I_{m}$ entries. 

To establish a relationship between graph signals and tensors, we begin by considering an $N$-vertex graph, denoted by $\mathcal{G}=\{ \mathcal{V}, \mathcal{E} \}$. With each vertex on the graph we can associate a variable (signal), denoted by $x(n) \in \mathbb{R}$, which maps a vertex number, $n \in \mathcal{V}$, to a real, that is, $x: \mathcal{V} \mapsto \mathbb{R}$. In other words, each vertex represents a scalar-valued field in a single-dimensional coordinate system. When considering all $N$ vertices in $\mathcal{V}$, we can form the vector ${\bf x} \in \mathbb{R}^{N}$ which defines the mapping ${\bf x}: \mathcal{V} \mapsto \mathbb{R}^{N}$.

On the other hand, if a graph resides in an $M$-dimensional space, then each vertex, $n \in \mathcal{V}$, has a one-to-one correspondence with a unique coordinate vector in this space, denoted by $(i_{1},...,i_{M}) \in \mathbb{N}^{M}$, where $i_{m} \in \mathbb{N}$ is the coordinate associated with the $m$-th axis. In other words, there exists a unique mapping $n \mapsto (i_{1},...,i_{M})$. In this way, the graph vertex signal can be viewed as a field in an $M$-dimensional coordinate system, that is, each vertex can be defined equivalently as $x(n) \equiv x(i_{1},...,i_{M}) \in \mathbb{R}$, that is, it induces the mapping $x: \mathbb{N}^{M} \mapsto \mathbb{R}$.

\begin{figure}[]
	\centering
	\includegraphics[width=0.4\textwidth, trim={0cm 0 0 0},clip]{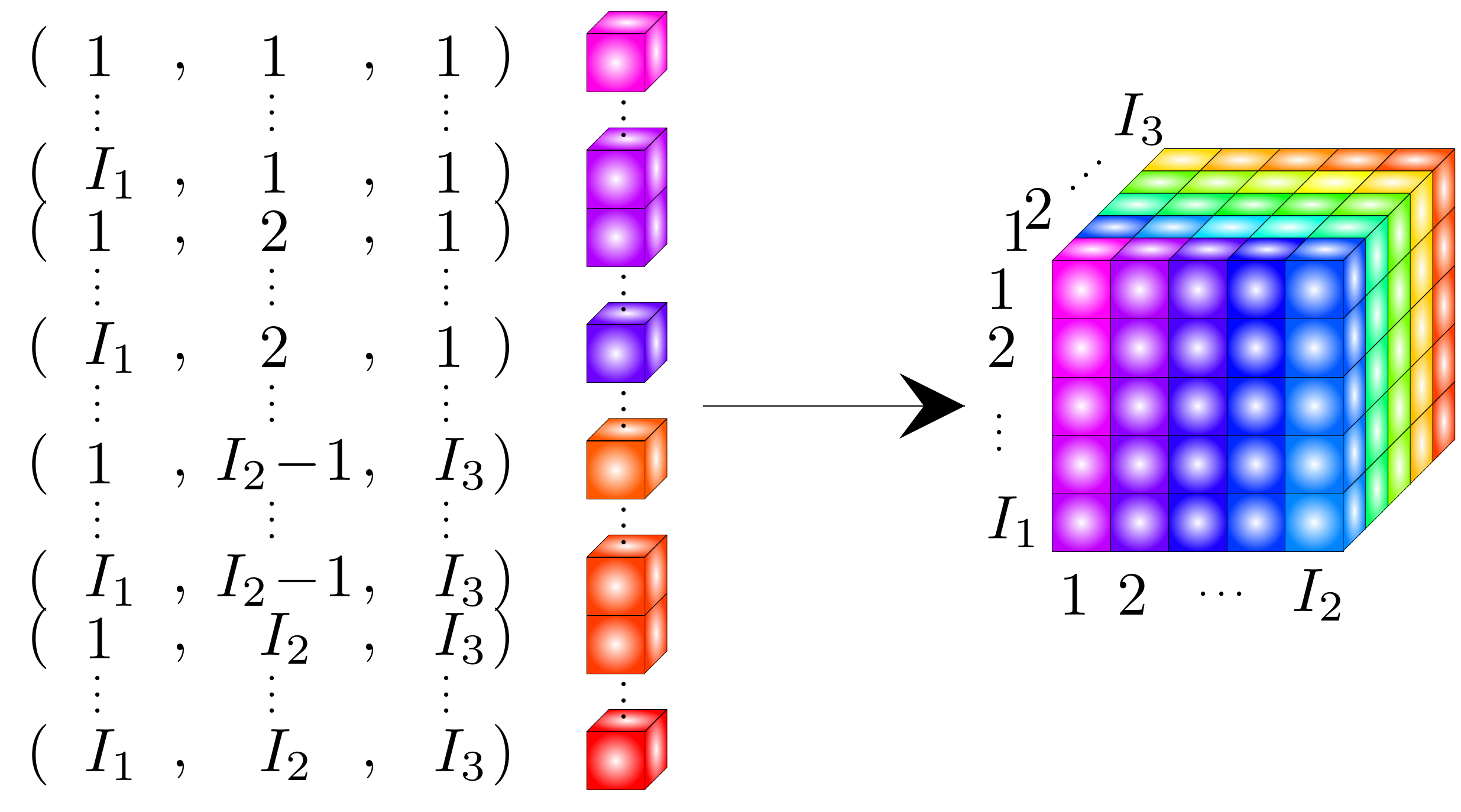}
	\caption{\label{fig:multi_index} Tensorization of discrete samples from a field $x: \mathbb{N}^{3} \mapsto \mathbb{R}$.}
\end{figure}

When discrete points in the field, $x: \mathbb{N}^{M} \mapsto \mathbb{R}$, are sampled using a regular lattice of dimensions $I_{1} \times \cdots \times I_{M}$, thereby sampling a total of  $$\prod_{m=1}^{M} I_{m} \equiv N$$ discrete points, the collection of samples naturally form the tensor ${\boldsymbol{\mathcal{X}}} \in \mathbb{R}^{I_{1} \times \cdots \times I_{M}}$, with the $(i_{1},...,i_{M})$-th entry defined as
\begin{gather}
[ {\boldsymbol{\mathcal{X}}} ]_{i_{1}...i_{M}} = x(i_{1},...,i_{M}), \quad i_{m} \in \mathbb{N},  \\ m=1,2,\dots,M. \nonumber
\end{gather}
Fig. \ref{fig:multi_index} illustrates a collection of discrete points from a field in a $3$-dimensional coordinate system, which together form an order-$3$ tensor. This procedure is referred to as \textit{tensorization}.

\begin{Remark}
	Real-world examples of a field in $M$-dimensional coordinates include:
	\begin{itemize}
		\item Netflix ratings in the \textit{user} $\times$ \textit{movie} space ($M=2$);
		\item Temperature measurements in the \textit{longitude} $\times$ \textit{latitude} $\times$ \textit{altitude} space ($M=3$);
		\item Video pixels in the \textit{time} $\times$ \textit{column} $\times$ \textit{row} $\times$ \textit{RGB} space ($M=4$);
		\item EEG signals in the \textit{time} $\times$ \textit{frequency} $\times$ \textit{channel} $\times$ \textit{subject} $\times$ \textit{trial} space ($M=5$).
	\end{itemize}
\end{Remark}

\subsection{Tensor decomposition}

\begin{figure}[]
	\centering
	\includegraphics[width=0.35\textwidth, trim={0cm 0 0 0},clip]{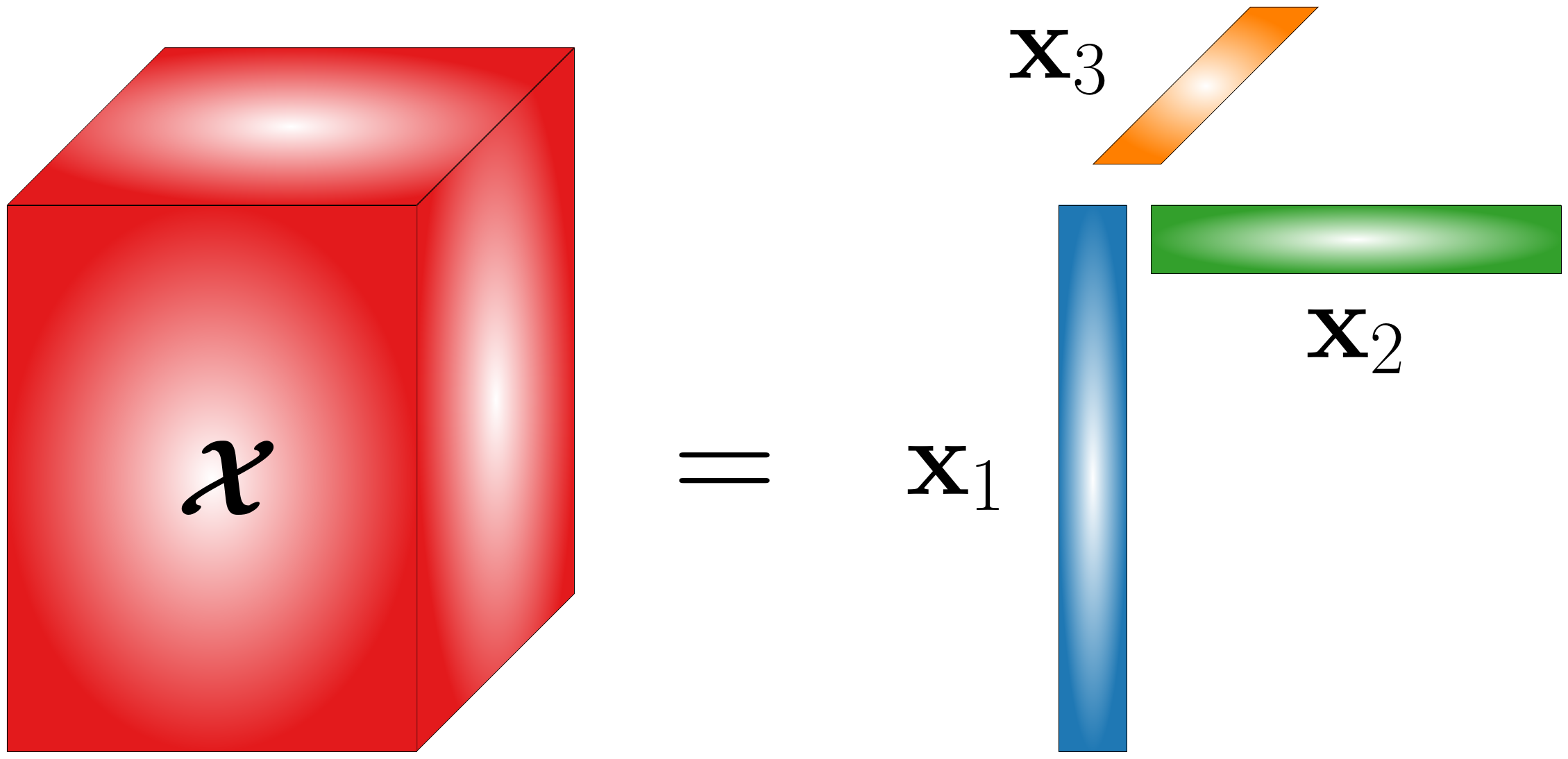}
	\caption{\label{fig:CPD_rank1} Rank-$1$ CPD of an order-$3$ tensor.}
\end{figure}

If the underlying field, $x: \mathbb{N}^{M} \mapsto \mathbb{R}$, is defined  as a \textit{multilinear map} of the form
\begin{align}
x: \underbrace{\mathbb{N} \times \cdots \times \mathbb{N}}_{\text{$M$ times}} \mapsto \mathbb{R}
\end{align}
then it is said to be \textit{linearly separable}, and therefore admits the following decomposition
\begin{align}
x(i_{1},...,i_{M}) = \prod_{m=1}^{M} x_{m}(i_{m}) \label{eq:linear_separability}
\end{align}
In other words, the value of $x(i_{1},...,i_{M})$ is given by the product of $M$ independent single-dimensional functions, $x_{m} : \mathbb{N} \mapsto \mathbb{R}$, each of which is associated with the $m$-th coordinate axis of the underlying $M$-dimensional coordinate system. In this way, a tensor, ${\boldsymbol{\mathcal{X}}} \in \mathbb{R}^{I_{1} \times \cdots \times I_{M}}$, which is sampled from a linearly separable field of the kind in (\ref{eq:linear_separability}) admits the following rank-$1$ canonical polyadic decomposition (CPD)
\begin{align}
{\boldsymbol{\mathcal{X}}} = {\bf x}_{1} \circ \cdots \circ {\bf x}_{M} \label{eq:CPD}
\end{align}
with the symbol $\circ$ denoting the outer product operator, and ${\bf x}_{m} \in \mathbb{R}^{I_{m}}$ being a parameter vector associated with the $m$-th coordinate axis. This property is referred to as the \textit{Kronecker separability} condition, which is fundamental to most tensor decompositions and algorithms. With regard to the linear separability property in (\ref{eq:linear_separability}), the $i$-th entry of ${\bf x}_{m}$ is given by $[{\bf x}_{m}]_{i} = x_{m}(i)$. Fig. \ref{fig:CPD_rank1} shows the rank-$1$ CPD of an order-$3$ tensor.

\noindent Kronecker separable tensors admit a vector representation (vectorisation), denoted by ${\bf x} = \text{vec}({\boldsymbol{\mathcal{X}}}) \in \mathbb{R}^{N}$, which can be expressed as follows
\begin{align}
{\bf x} = {\bf x}_{M} \otimes \cdots \otimes {\bf x}_{1}
\end{align}
and is a direct consequence of (\ref{eq:CPD}), where the symbol $\otimes$ denotes the Kronecker product operator.

\begin{figure}[]
	\centering
	\includegraphics[width=0.4\textwidth, trim={0cm 0 0 0},clip]{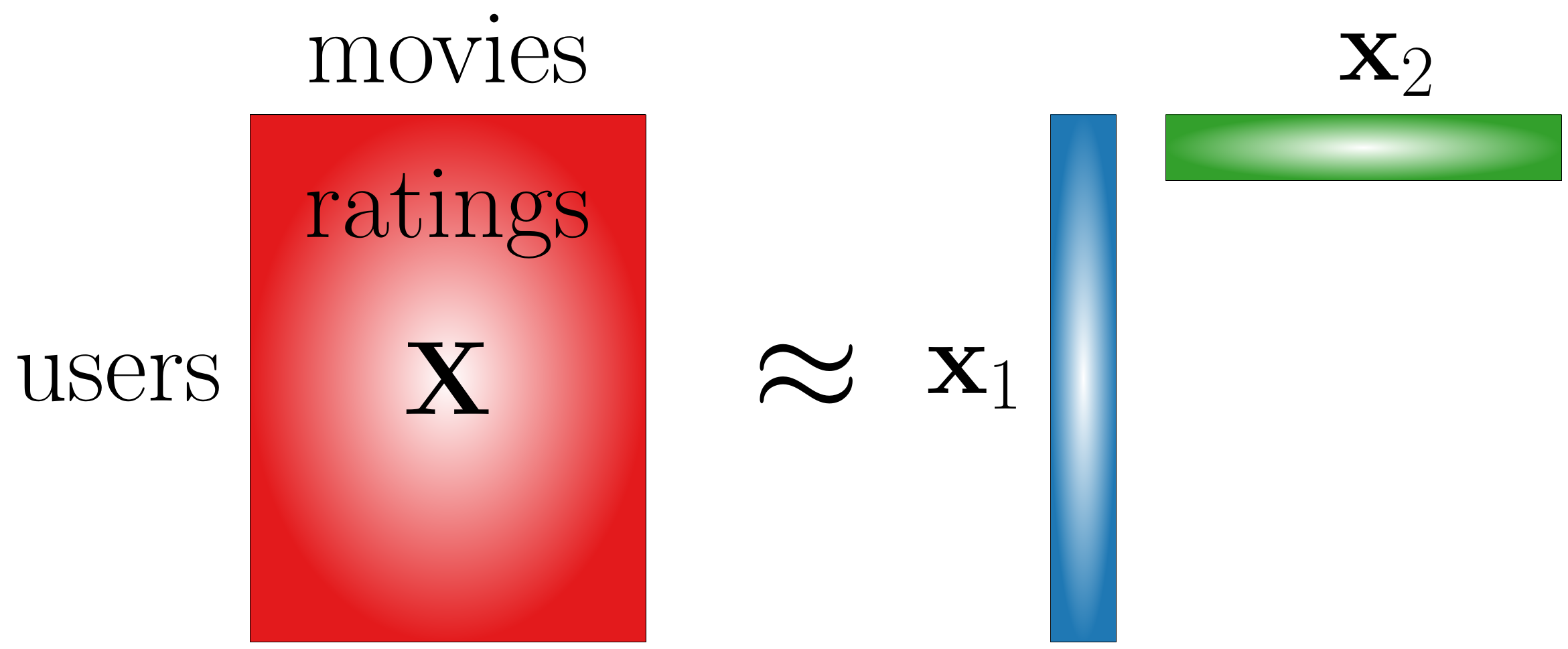}
	\caption{\label{fig:Netflix} Rank-$1$ CPD of the Netflix ratings data matrix.}
\end{figure}

\begin{Example}
	Consider the data matrix, $\mathbf{X} \in \mathbb{R}^{I \times J}$, which contains the Netflix ratings assigned by $I$ users to $J$ movies, whereby the $(i,j)$-th entry designates the rating assigned by the $i$-th user to the $j$-th movie, $x(i,j) \in \mathbb{R}$. The graph representation of this dataset consists of $(IJ)$ vertices residing in a two-dimensional space (\textit{user} $\times$ \textit{movie}). Owing to the lattice-like structure of the graph, we can employ its inherent order-$2$ tensor representation, whereby the data can be approximated using the following rank-$1$ CPD
	\begin{align}
	\mathbf{X} \approx \mathbf{x}_{1} \circ \mathbf{x}_{2} \equiv \mathbf{x}_{1} \mathbf{x}_{2}^{T}
	\end{align}
	with $\mathbf{x}_{1} \in \mathbb{R}^{I}$ being the factor associated with the \textit{user} axis, and $\mathbf{x}_{2} \in \mathbb{R}^{J}$ the factor associated with the \textit{movie} axis. Note that for order-$2$ tensors, the CPD is equivalent to the singular value decomposition (SVD). Fig. \ref{fig:Netflix} illustrates the tensor decomposition of the Netflix ratings data matrix.
	
	The factorization of $\mathbf{X}$ assumes that the rating assigned by the $i$-th user to the $j$-th movie can be approximated as
	\begin{align}
	x(i,j) \approx x_{1}(i)x_{2}(j)
	\end{align}
	where $x_{1}(i) \equiv [\mathbf{x}_{1}]_{i}$ and $x_{2}(j) \equiv [\mathbf{x}_{2}]_{j}$. In other words, the rating, $x(i,j)$, can be approximated by a rating assigned by the $i$-th user to all movies, $x_{1}(i)$, multiplied by a rating assigned to the $j$-th movie by all users, $x_{2}(j)$.
	
	The so achieved parameter reduction becomes evident, since we have reduced a fully connected $(IJ)$ parameter model to an $(I+J)$ parameter model. This parameter reduction is most pronounced for higher-order tensors, e.g. an order-$N$ tensor model with $\prod_{n=1}^{N}I_{n}$ parameters (exponential) reduces to a $\sum_{n=1}^{N}I_{n}$ parameter (linear) model.
\end{Example}

\subsection{Connectivity of a tensor}

We next show that the tensor structure inherent to ${\boldsymbol{\mathcal{X}}} \in \mathbb{R}^{I_{1} \times \cdots \times I_{M}}$ can be modelled naturally as a graph. This is achieved by exploiting the well-known property of lattice-structured graphs which can be decomposed into constituent single-dimensional path graphs.

The Cartesian product of $M$ disjoint $I_{m}$-vertex path graphs, $\mathcal{G}_{m} = (\mathcal{V}_{m}, \mathcal{E}_{m})$ for $m=1,...,M$, yields a graph with an $M$-dimensional regular lattice structure, denoted by $\mathcal{G} = \mathcal{G}_{M} \, \square \,\, \cdots \, \square \,\,  \mathcal{G}_{1} =  (\mathcal{V}, \mathcal{B})$, with the symbol $\square$ denoting the graph Cartesian product. In this way, the resulting vertex set takes the form $\mathcal{V} = \mathcal{V}_{M} \times \cdots \times \mathcal{V}_{1}$, and the resulting graph contains a total of $\prod_{m=1}^{M} I_{m} \equiv N$ vertices. 

If the adjacency matrix of the $m$-th path graph, $\mathcal{G}_{m}$, is denoted by $\mathbf{A}_{m} \in \mathbb{R}^{I_{m} \times I_{m}}$, then the adjacency matrix of the resulting $M$-dimensional regular lattice graph, $\mathcal{G}$, is given by
\begin{align}
\mathbf{A} = \left( \mathbf{A}_{M} \oplus \cdots \oplus \mathbf{A}_{1} \right) \in \mathbb{R}^{N \times N} \label{eq:kron_summability_A}
\end{align}
where the symbol $\oplus$ denotes the Kronecker sum operator. Such an adjacency matrix is said to be \textit{Kronecker summable}. 

\begin{Remark}
	The adjacency matrix, $\mathbf{A}$, when interpreted through the underlying tensor, describes the connectivity between the entries of tensor's vectorization, $\mathbf{x} \in \mathbb{R}^{N}$, while $\mathbf{A}_{m} \in \mathbb{R}^{I_{m} \times I_{m}}$ describes the connectivity between entries along the $m$-th mode. Under this model, the entries of the tensor are only connected to neighbouring entries which reside in the same \textit{fibre}.
\end{Remark}

\begin{figure}[]
	\centering
	\includegraphics[width=1\linewidth,  trim={2cm 23cm 7cm 1.8cm}, clip]{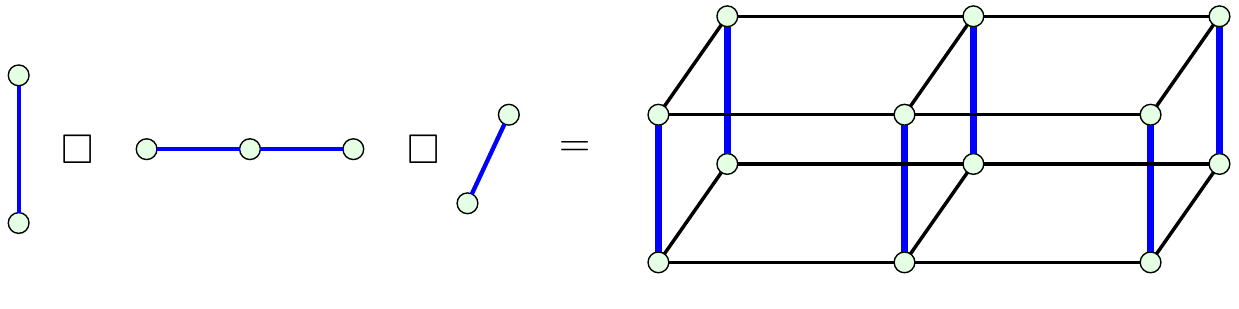}
	\caption{Cartesian product of $3$ disjoint path graphs.}
	\label{fig:line_graph}
\end{figure}

For illustration purposes, Fig. \ref{fig:line_graph} shows the Cartesian product of $3$ disjoint path graphs, which results in a graph with a $3$-dimensional lattice structure. The resulting graph would naturally represent the connectivity between the entries of an order-$3$ tensor, ${\boldsymbol{\mathcal{X}}} \in \mathbb{R}^{2 \times 3 \times 2}$.

\begin{Example} \label{ex:tensor_decomposition}
	Consider a field on a two-dimensional coordinate system, denoted by $x: \mathbb{N}^{2} \mapsto \mathbb{R}$, illustrated in Fig. \ref{fig:2D_field}.
	
	\begin{figure}[]
		\centering
		\includegraphics[width=0.6\textwidth, trim={6cm 7cm 0 7cm}, clip]{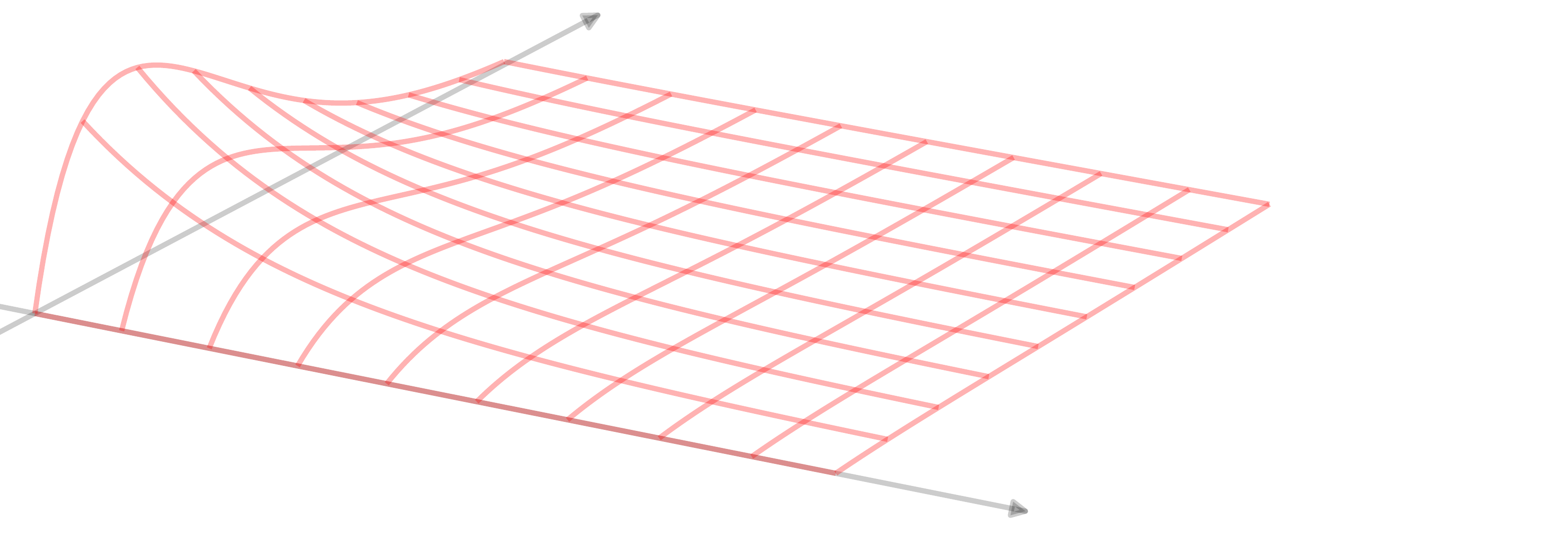}
		\caption{\label{fig:2D_field}A field, $x: \mathbb{N}^{2} \mapsto \mathbb{R}$.}
	\end{figure}
	
	Next, consider the order-$2$ tensor, ${\bf X} \in \mathbb{R}^{I_{1} \times I_{2}}$, with entries sampled from the field, $x: \mathbb{N}^{2} \mapsto \mathbb{R}$, using a $2$-dimensional regular lattice as illustrated in Fig. \ref{fig:2D_field_grid}.
	
	\begin{figure}[]
		\includegraphics[width=0.6\textwidth, trim={6cm 7cm 0 7cm}, clip]{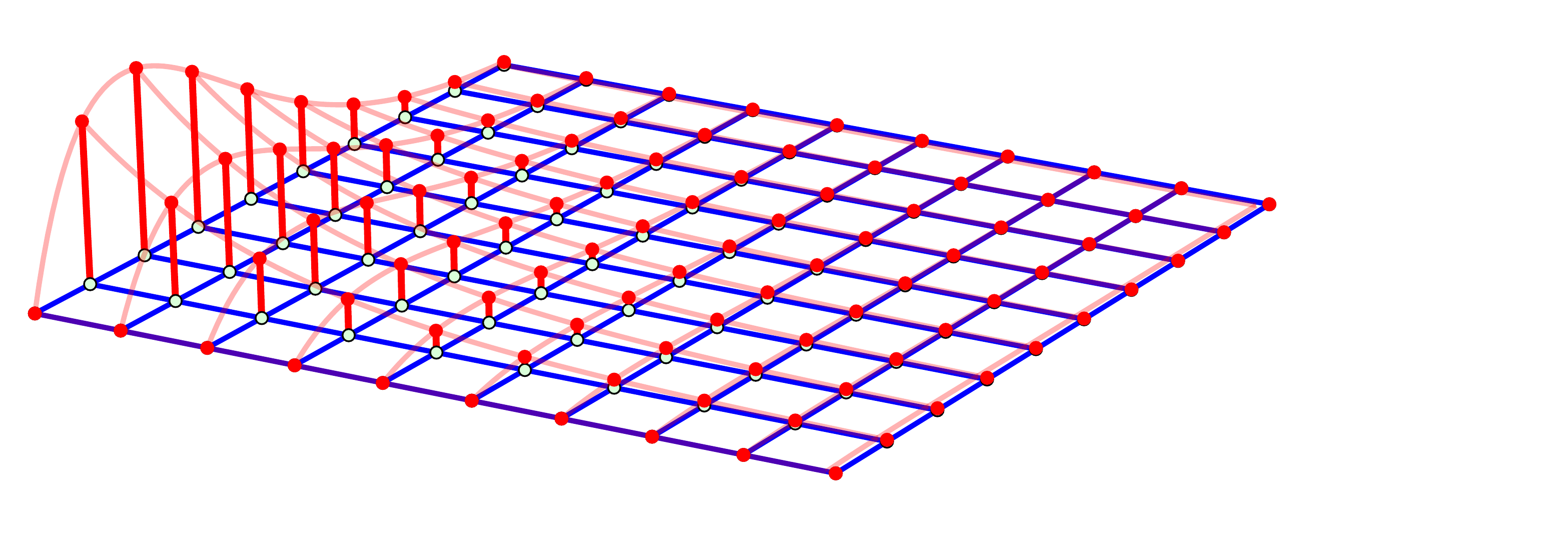}
		\caption{\label{fig:2D_field_grid}Order-$2$ tensor, ${\bf X} \in \mathbb{R}^{I_{1} \times I_{2}}$, sampled from $x: \mathbb{N}^{2} \mapsto \mathbb{R}$.}
	\end{figure}
	
	If the scalar field is linearly separable, that is, $x(t_{1},t_{2}) = x_{1}(t_{1})x_{2}(t_{2})$, then the sampled tensor, ${\bf X}$, is Kronecker separable, and can therefore be expressed as
	\begin{align}
	{\bf X} = {\bf x}_{1} \circ {\bf x}_{2} \Longleftrightarrow \text{vec}(\mathbf{X}) = {\bf x}_{2} \otimes {\bf x}_{1}
	\end{align}
	with $x_{1} \in \mathbb{R}^{I_{1}}$ and $x_{2} \in \mathbb{R}^{I_{2}}$ being data on path graphs sampled respectively from the single-dimensional fields, $x_{1}: \mathbb{N} \mapsto \mathbb{R}$ and $x_{2}: \mathbb{N} \mapsto \mathbb{R}$, as illustrated in Fig. \ref{fig:2D_field_x}--\ref{fig:2D_field_y}.
	
	\begin{figure}[]
		\centering
		\begin{minipage}[b]{0.5\textwidth}
			\centering
			\includegraphics[width=0.6\textwidth, trim={13cm 10cm 13cm 9cm}, clip]{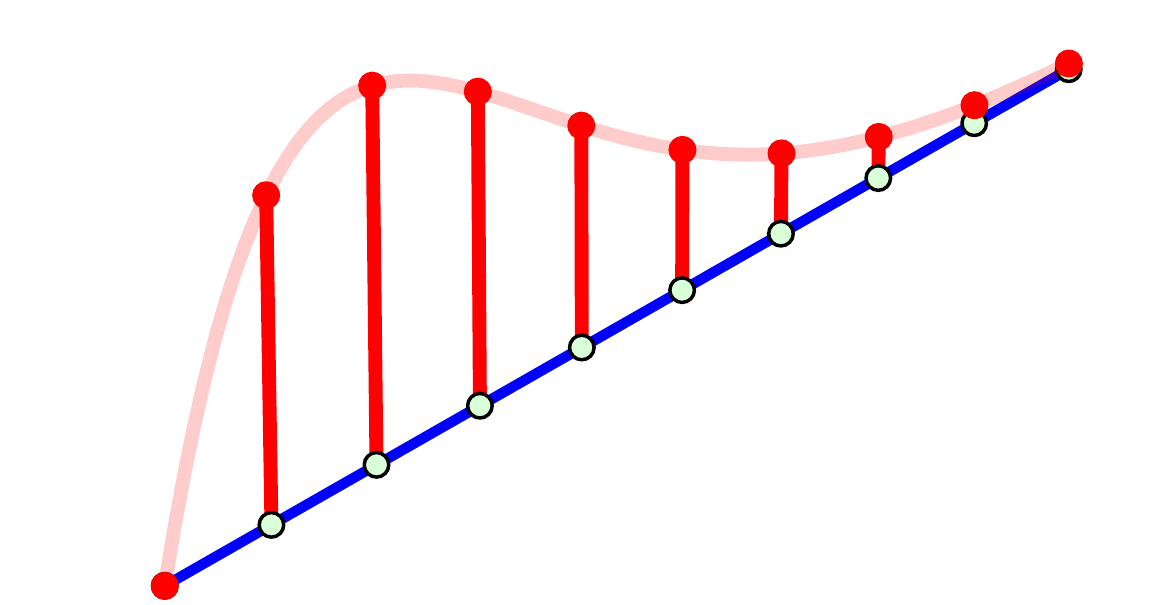}
			\caption{\label{fig:2D_field_x}Path graph signal, ${\bf x}_{1} \in \mathbb{R}^{I_{1}}$, sampled from $x_{1}: \mathbb{N} \mapsto \mathbb{R}$.}
		\end{minipage}
		\vspace{0.2cm}
		
		\begin{minipage}[b]{0.5\textwidth}
			\centering
			\includegraphics[width=0.6\textwidth, trim={11cm 10.5cm 10cm 4.5cm}, clip]{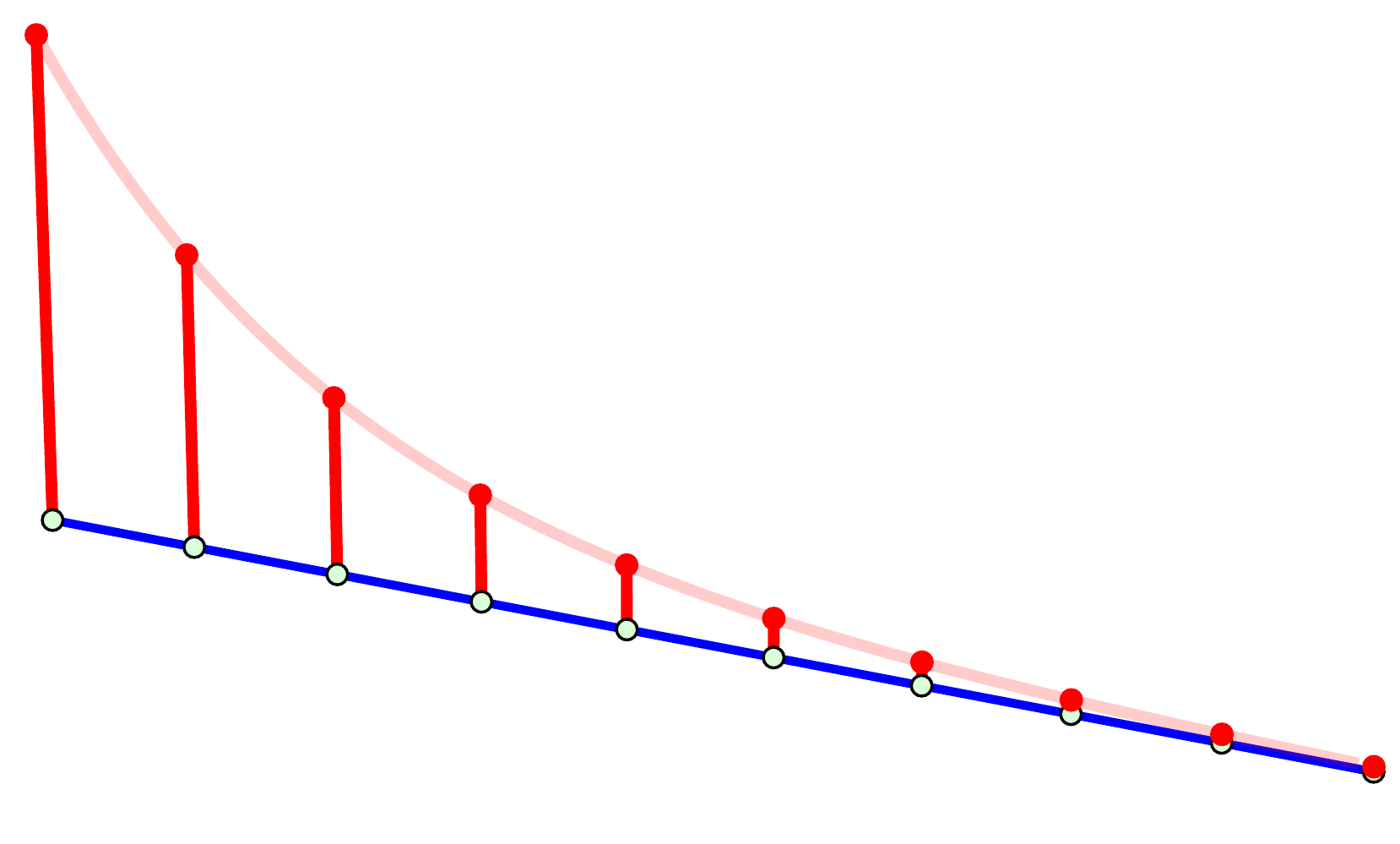}
			\caption{\label{fig:2D_field_y}Path graph signal, ${\bf x}_{2} \in \mathbb{R}^{I_{2}}$, sampled from $x_{2}: \mathbb{N} \mapsto \mathbb{R}$.}
		\end{minipage}
	\end{figure}
\end{Example}

\subsection{DFT of a tensor}

Since tensors are a special class of graphs with a Kronecker summable adjacency matrix (see (\ref{eq:kron_summability_A})), then the DFT of a tensor can be naturally obtained from the graph DFT (GDFT). The GDFT of a graph with lattice structure can be performed by evaluating the eigenvalue decomposition of the adjacency matrix $\mathbf{A}$, given by
\begin{align}
\mathbf{A} = \mathbf{U}\mathbf{\Lambda}\mathbf{U}^{-1}
\end{align}
where $\mathbf{U} \in \mathbb{R}^{N \times N}$ and $\mathbf{\Lambda} \in \mathbb{R}^{N \times N}$ respectively denote the matrix of eigenvectors and eigenvalues of $\mathbf{A}$.

Owing to the Kronecker sum structure of $\mathbf{A}$ in (\ref{eq:kron_summability_A}), the eigenvector and eigenvalue matrices of GDFT exhibit the following structure
\begin{align}
\mathbf{U} & = \left( \mathbf{U}_{M} \otimes \cdots \otimes \mathbf{U}_{1} \right) \\
\mathbf{\Lambda} & = \left( \mathbf{\Lambda}_{M} \oplus \cdots \oplus \mathbf{\Lambda}_{1} \right)
\end{align}
where $\mathbf{U}_{m} \in \mathbb{R}^{I_{m} \times I_{m}}$ and $\mathbf{\Lambda} \in \mathbb{R}^{I_{m} \times I_{m}}$ respectively denote the matrix of eigenvectors and eigenvalues of the $m$-th path graph adjacency matrix, $\mathbf{A}_{m}$, obtained through
\begin{align}
\mathbf{A}_{m} = \mathbf{U}_{m} \mathbf{\Lambda}_{m}\mathbf{U}_{m}^{-1}
\end{align}
Therefore, the eigenvectors of $\mathbf{A}$ are said to be \textit{Kronecker separable}, while the eigenvalues are \textit{Kronecker summable}.

\subsection{Unstructured graphs}

Consider an $N$-vertex graph, $\mathcal{G}$, with vertex signals sampled from the field, $x: \mathbb{R}^{M} \mapsto \mathbb{R}$, using a regular lattice, which together form the order-$M$ tensor, ${\boldsymbol{\mathcal{X}}} \in \mathbb{R}^{I_{1} \times \cdots \times I_{M}}$, with $\prod_{m=1}^{M} I_{m} \equiv N$.

Similarly, consider a $K$-vertex graph, $\tilde{\mathcal{G}}$, with vertex signals also sampled from the same field, $x: \mathbb{R}^{M} \mapsto \mathbb{R}$, but using instead an unstructured sampling scheme. In this way, the unstructured graph can be defined as a subset of a lattice-structured graph, i.e. $\tilde{\mathcal{G}} \subset \mathcal{G}$.

The vertex signals of $\tilde{\mathcal{G}}$, denoted by the vector $\tilde{\bf x} \in \mathbb{R}^{K}$, can therefore be defined as
\begin{align}
\tilde{\bf x} = {\bf \Pi} \, \text{vec}({\boldsymbol{\mathcal{X}}})
\end{align}
where ${\bf \Pi} \in \mathbb{R}^{K \times N}$ is a \textit{sampling matrix}, with entries defined as
\begin{align}
[{\bf \Pi}]_{kn} = \begin{cases}
1, & \text{if $\tilde{x}(k) \equiv x(n)$},\\
0, & \text{otherwise}
\end{cases}
\end{align}
with $\tilde{x}(k) \in \mathbb{R}$ and $x(n) \in \mathbb{R}$ denoting respectively the $k$-th vertex of $\tilde{\mathcal{G}}$ and the $n$-th vertex of $\mathcal{G}$. 

Although the lattice-structured graph, $\mathcal{G}$, exhibits a Kronecker separable signal vector and a Kronecker summable adjacency matrix, the associated unstructured graph, $\tilde{\mathcal{G}}$, does not exhibit such properties because, in general, ${\bf \Pi}$ is not separable. This can be seen from the relationship between the adjacency matrices of $\tilde{\mathcal{G}}$ and $\mathcal{G}$, which is given by
\begin{align}
\tilde{\mathbf{A}} = {\bf \Pi}\mathbf{A}{\bf \Pi}^{T} = {\bf \Pi}\left( \mathbf{A}_{M} \oplus \cdots  \oplus \mathbf{A}_{1} \right){\bf \Pi}^{T}
\end{align}
Notice that the last term above cannot be decomposed further if $\mathbf{\Pi}$ is not separable. A direct consequence of this result is that the GDFT bases of $\tilde{\mathcal{G}}$ (eigenvalue decomposition of $\tilde{\mathbf{A}}$) do not exhibit the Kronecker summability either.

\begin{Example}
	Referring back to Example \ref{ex:tensor_decomposition}, the graph signal resulting from an irregular sampling of the field $x: \mathbb{R}^{2} \mapsto \mathbb{R}$ is not Kronecker separable as it cannot be represented as a Cartesian product of two path graphs (as in Fig. \ref{fig:2D_field_grid}--\ref{fig:2D_field_y}), as illustrated in Fig. \ref{fig:2D_field_unstructured}.
	\begin{figure}[H]
		\centering
		\includegraphics[width=0.6\textwidth, trim={5cm 7cm 0 7cm}, clip]{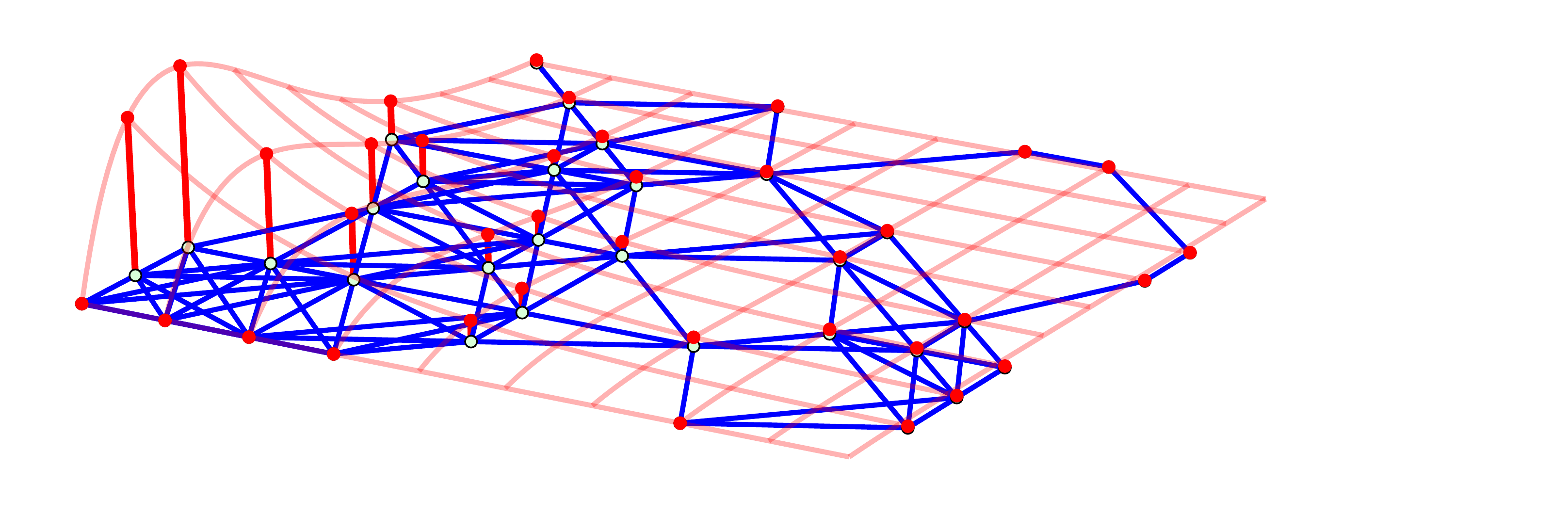}
		\caption{\label{fig:2D_field_unstructured}Unstructured graph, $\tilde{\bf x} \in \mathbb{R}^{K}$, sampled from $x: \mathbb{N}^{2} \mapsto \mathbb{R}$.}
	\end{figure}
\end{Example}

\subsection{Tensor representation of multi-relational graphs}

The rapidly growing prominence of multi-relational network data in areas as diverse as social network modeling, the semantic web, bioinformatics and artificial intelligence, has brought to light the increasing importance of Data Analytics on domains where the entities are interconnected by multiple relations. To put this into context of graphs, while traditional graph models only account for a single relation type, designated by the adjacency matrix, $\mathbf{A} \in \mathbb{R}^{N \times N}$, a multi-relational $N$-vertex graph may exhibit a large number, say $M$, of distinct relation types between vertices. In this case, a multi-relational graph would be defined by $M$ adjacency matrices, $\mathbf{A}_{m} \in \mathbb{R}^{N \times N}$ for $m=1,...,M$; one for each relation type. 

\begin{figure}[]
	\centering
	\includegraphics[width=0.25\textwidth]{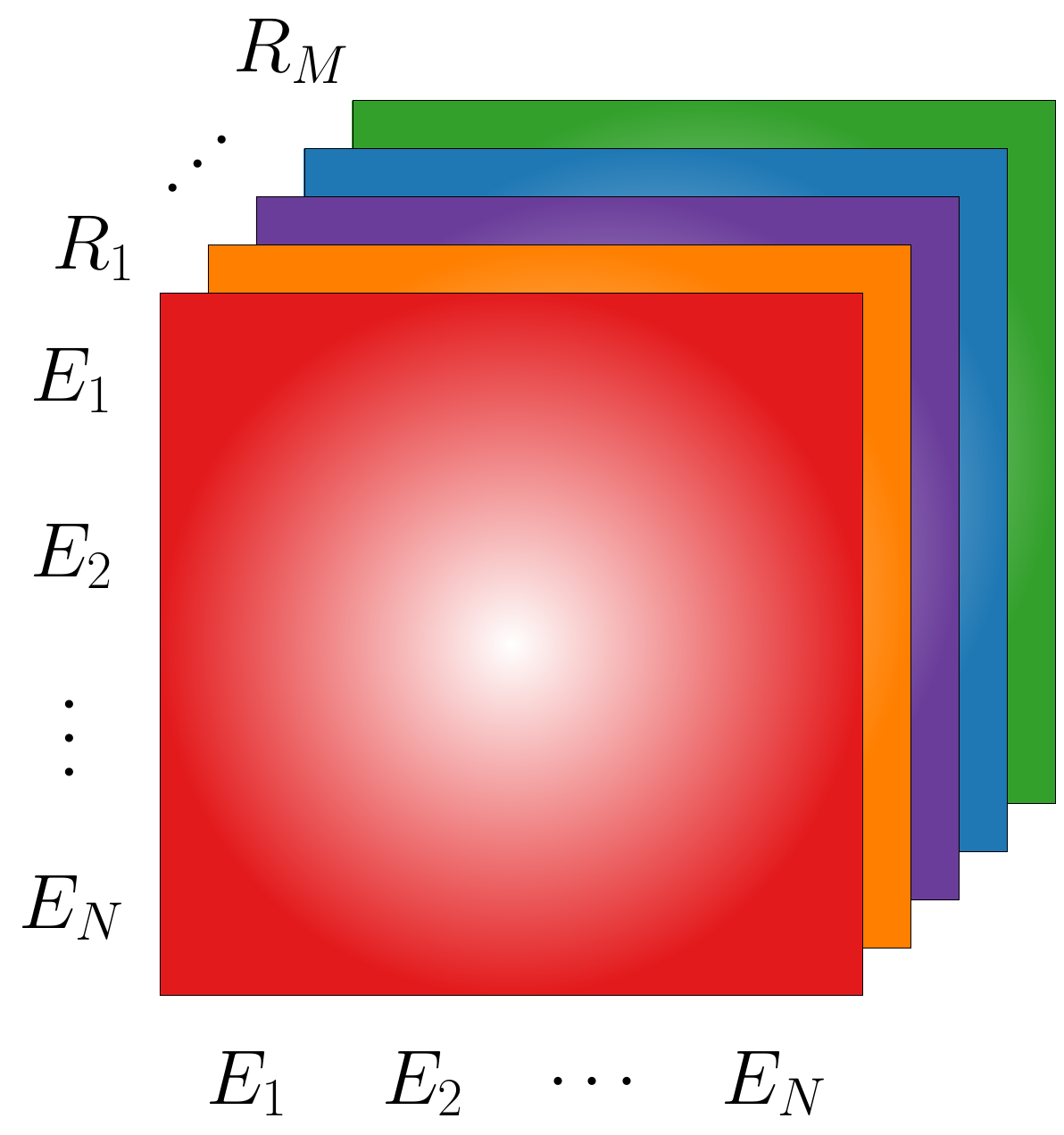}
	\caption{\label{fig:MultiRelational_Tensor}Construction of a multi-relational adjacency tensor, ${\boldsymbol{\mathcal{A}}} \in \mathbb{R}^{N \times N \times M}$, where $E_{n}$ denotes the $n$-th entity and $R_{m}$ the $m$-th relation type.}
\end{figure}

While it is possible to model this situation through a short and wide $N \times MN$ dimensional matrix, this would both involve numerical difficulties and obscure physical relevance. To this end, to model such a multi-relational graph in a parsimonious and compact manner, we may construct a three-way tensor, ${\boldsymbol{\mathcal{A}}} \in \mathbb{R}^{N \times N \times M}$, whereby its $m$-th frontal slice is given by $\mathbf{A}_{m}$. In this way, the first two modes define the entity domain, while the third mode represents the relation domain, as illustrated in Fig. \ref{fig:MultiRelational_Tensor}. The tensor entry $[{\boldsymbol{\mathcal{A}}}]_{ijk} = 1$ therefore designates the existence of a relation between the $i$-th and $j$-th entities within the $k$-th relation type; otherwise, for non-existing and unknown relations, the entry is set to zero.

\begin{figure}[]
	\centering
	\includegraphics[width=0.45\textwidth]{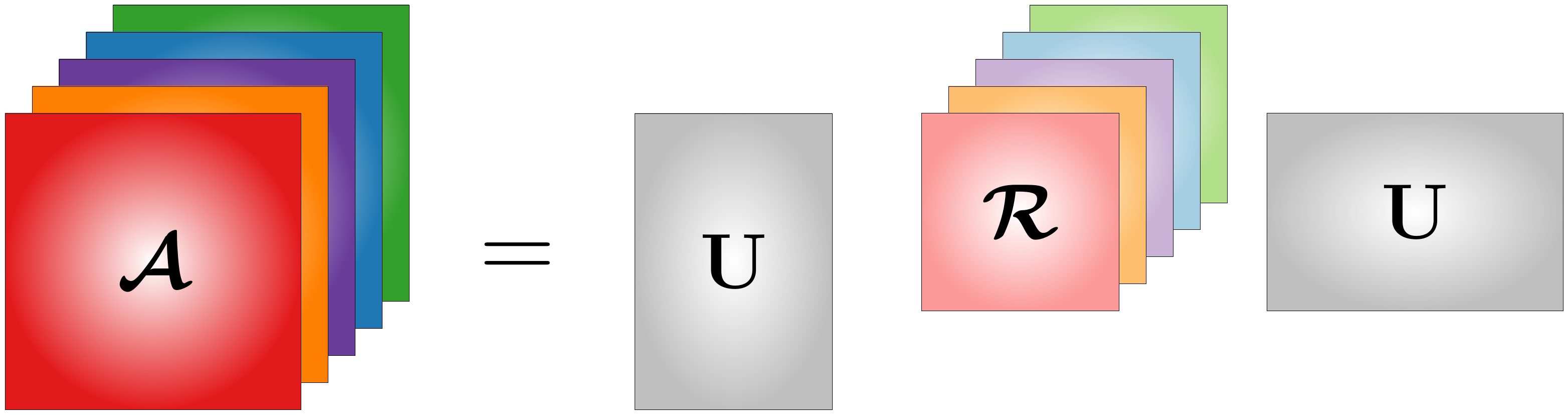}
	\caption{\label{fig:RESCAL}Factorization of a multi-relational adjacency tensor, ${\boldsymbol{\mathcal{A}}} \in \mathbb{R}^{N \times N \times M}$ as in (\ref{eq:RESCAL}).}
\end{figure}

The work in \cite{Lin2008,Lin2009,Tan2009,Nickel2011,Papalexakis2013,Gauvin2014,Verma2017,Verma2017_2,Katsimpras2019} employs such tensor model to learn an inherent structure from multi-relational data. The following rank-$L$ factorization was employed, known as the RESCAL decomposition \cite{Nickel2011}, whereby each frontal slice of ${\boldsymbol{\mathcal{A}}}$ is factorized as
\begin{align}
\mathbf{A}_{m} = \mathbf{U}\mathbf{R}_{m}\mathbf{U}^{T}, \quad m=1,...,M \label{eq:RESCAL}
\end{align}
where $\mathbf{U} \in \mathbb{R}^{N \times L}$ is a factor matrix which maps the $N$-dimensional entity space to an $L$-dimensional latent component space, and $\mathbf{R}_{m} \in \mathbb{R}^{L \times L}$ models the interactions of latent components within the $m$-th relation type. Alternatively, this can be expressed in terms of the factorization of the tensor ${\boldsymbol{\mathcal{A}}}$, that is
\begin{align}
{\boldsymbol{\mathcal{A}}} = {\boldsymbol{\mathcal{R}}} \times_{1} \mathbf{U} \times_{2} \mathbf{U}
\end{align}
where the symbol $\times_{n}$ denotes the mode-$n$ product, and ${\boldsymbol{\mathcal{R}}} \in \mathbb{R}^{L \times L \times M}$ is the latent core tensor with $\mathbf{R}_{m}$ being its $m$-th frontal slice, as illustrated in Fig. \ref{fig:RESCAL}. Such a factorization allows for link-based clustering, whereby the entities $E_{1},...,E_{N}$ are clustered according to the information in $\mathbf{U}$ only. In doing so, the similarity between entities is computed based on their similarity across multiple relations.

\begin{Example}
	Social networks play an important role in the functionality of an organisation and it is therefore of considerable interest to analyse the properties of such networks. The adoption of social networking services within organisations can largely facilitate the interaction and collaboration between employees. For example, a social network could reveal information about the characteristics of an employee which could then be used to improve efficiency and influence team structuring.
	
	A social network can be modelled as a graph, whereby each vertex represents an individual (employee) and each edge designates the existence of a social relationships between two individuals. While a conventional graph can model social networks involving one type of relationship, multi-relational graphs allow for the modelling of different types of relationships. Fig. \ref{fig:MultiRelational_Graph} illustrates a multi-relational social network involving three employees (vertices) who communicate via email (blue edge), Linkedin (green edge) and Skype (orange edge). Observe that social relationships may be directed, e.g. employee A sends emails (blue edge) to employee B but not vice versa. If the adjacency matrix associated with the $m$-th relationship type is defined as $\mathbf{A}_{m} \in \mathbb{R}^{3 \times 3}$ for $m=1,2,3,4$, then the adjacency tensor, ${\boldsymbol{\mathcal{A}}} \in \mathbb{R}^{3 \times 3 \times 4}$, be constructed to model the entire social network. Once the latent components matrix, $\mathbf{U} \in \mathbb{R}^{3 \times L}$, is inferred from ${\boldsymbol{\mathcal{A}}}$ using the factorization in (\ref{eq:RESCAL}), it is possible to apply feature-based clustering to obtain the inherent community structure in the multi-relational network. The output of this step would be a set of $K$ disjoint communities (sub-graphs), $\{\mathcal{V}_{1},...,\mathcal{V}_{K}\}$.

	\begin{figure}[]
		\centering
		\begin{subfigure}[t]{0.5\textwidth}
			\centering
			\includegraphics[width=0.6\textwidth]{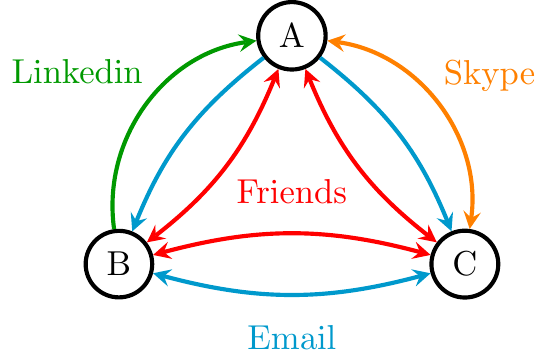}
			\caption{}
		\end{subfigure}
		
		\vspace{0.3cm}
		
		\begin{subfigure}[t]{0.5\textwidth}
			\centering
			\includegraphics[width=1\textwidth]{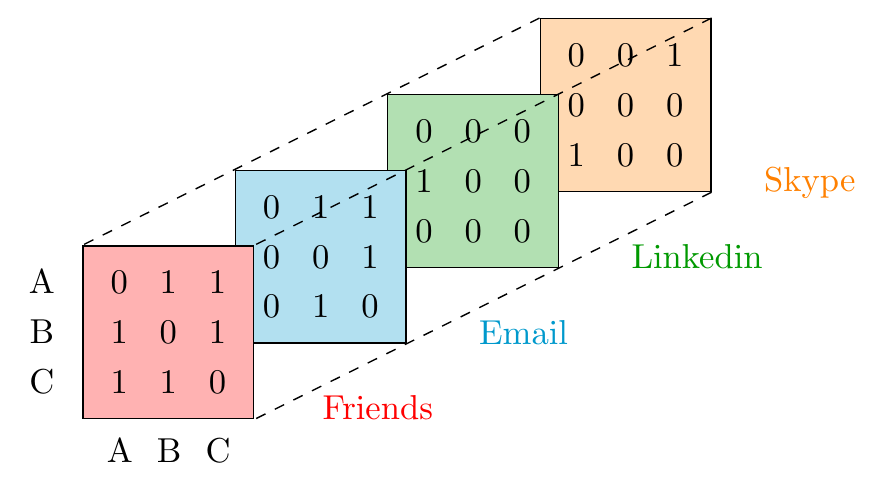}
			\caption{}
		\end{subfigure}
		
		\caption{\label{fig:MultiRelational_Graph} Social network modelled as a multi-relational graph. (a) Graph representation of the social network. (b) Adjacency tensor, ${\boldsymbol{\mathcal{A}}} \in \mathbb{R}^{3 \times 3 \times 4}$, associated with the social network in (a).} 
	\end{figure}
	
\end{Example}

\section{
	Metro Traffic Modeling through Graphs}


With the rapid development of many economies, an increasing proportion of the world's population is moving to cities, and as such urban traffic congestion is becoming a serious issue. For example, underground traffic networks routinely undergo general maintenance, frequently exhibit signal failures and train derailments, and may even occasionally experience emergency measures because of various accidents. These events ultimately require the closure of at least one station which may severely impact the traffic service across the entire network. The economic costs of these transport delays to central London business is estimated to be $\pounds 1.2$ billion per year. Hence, appropriate and physically meaningful tools to understand, quantify, and plan for the resilience of these traffic networks to disruptions are much needed.

In this section, we demonstrate how graph theory can be used to identify those stations in the London underground network which have the greatest influence on the functionality of the traffic, and proceed, in an innovative way, to assess the impact of a station closure on service levels across the city. Such underground network vulnerability analysis offers the opportunity to analyse, optimize and enhance the connectivity of the London underground network in a mathematically tractable and physically meaningful manner.

\subsection{Traffic centrality as a graph-theoretic measure}

\begin{figure}[]
	\centering
	\includegraphics[width=0.45\textwidth, trim={0.3cm 0.4cm 0.3cm 0.4cm}, clip]{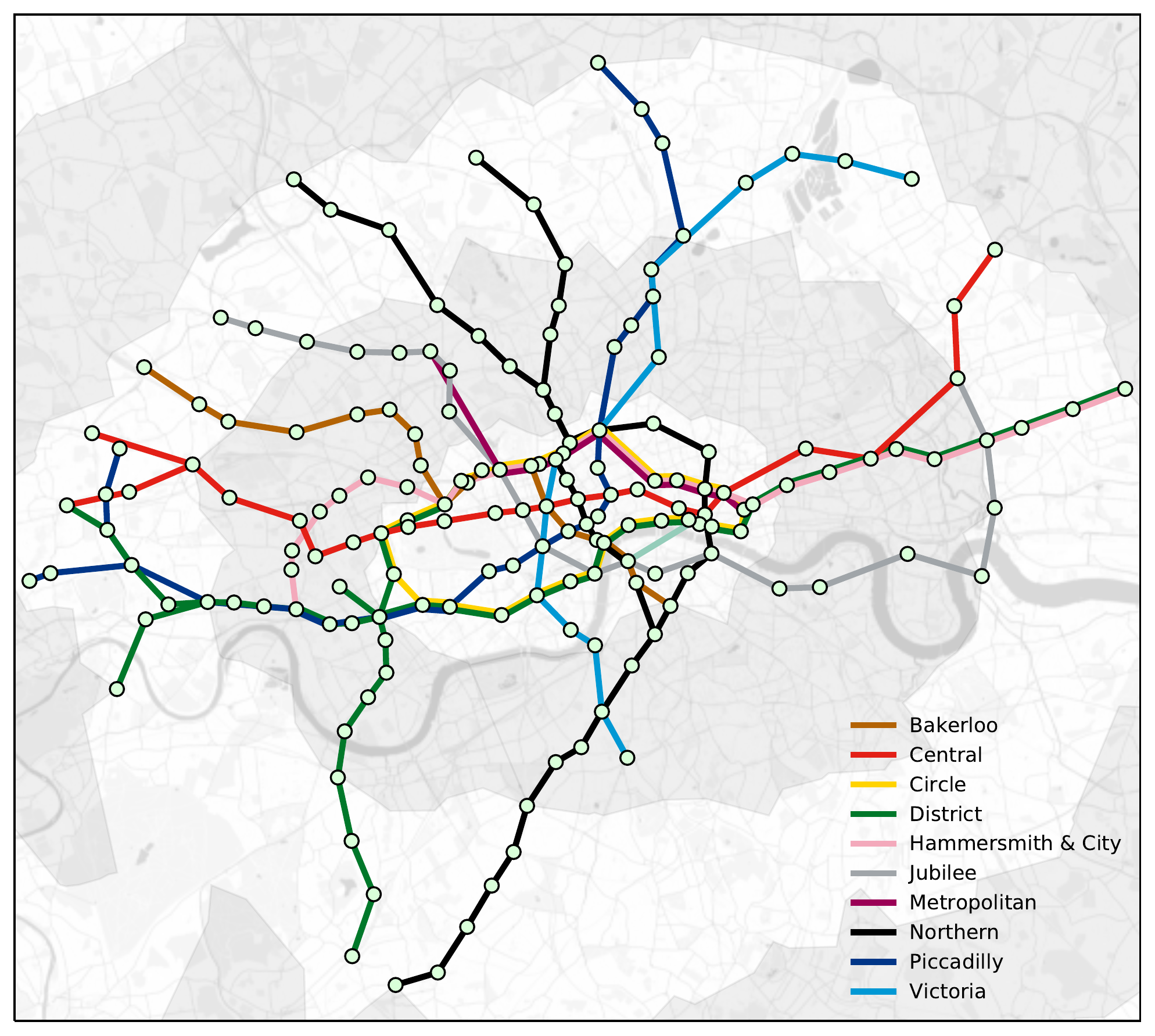}
	\caption{\label{fig:underground_graph}Graph model of the London underground network in Zones 1--3.}
\end{figure}

The underground network can be modelled as an undirected $N$-vertex graph, denoted by $\mathcal{G} = \{ \mathcal{V}, \mathcal{E} \}$, with $\mathcal{V}$ being the set of $N$ vertices (stations) and $\mathcal{E}$ the set of edges (underground lines) connecting the vertices (stations). The connectivity of the network is designated by the (undirected) adjacency matrix, $\mathbf{A} \in \mathbb{R}^{N \times N}$. Fig. \ref{fig:underground_graph} illustrates the proposed graph model of the London underground network, with each vertex representing a station, and each edge designating the underground line connecting two adjacent stations. Notice that standard data analytics domains are ill-equipped to deal with this class of problems.

We employ the following metrics to characterize the topology of the network and model its vulnerability:
\begin{itemize}
	\item  \textit{Betweenness centrality}, which reflects the extent to which a given vertex lies in between pairs or groups of other vertices of the graph, and is given by
	\begin{align}
	B_{n} = \sum_{k,m \in \mathcal{V}} \frac{\sigma(k,m|n)}{\sigma(k,m)}
	\end{align}
	where $\sigma(k,m)$ denotes the number of shortest paths between vertices $k$ and $m$, and $\sigma(k,m|n)$ the number of those paths passing through vertex $n$ \cite{Freeman1977}. In terms of the actual metro traffic, this can also be interpreted as the extent to which a vertex is an \textit{intermediate in the communication over the network}. Fig. \ref{fig:betweenness} shows that, as expected, the stations at the centre of the city exhibit the largest betweenness centrality, and their disconnection would therefore severely impact the communication over the underground network.
	
	\begin{figure}[]
		\centering
		\includegraphics[width=0.45\textwidth, trim={0.3cm 0.4cm 0.3cm 0.4cm}, clip]{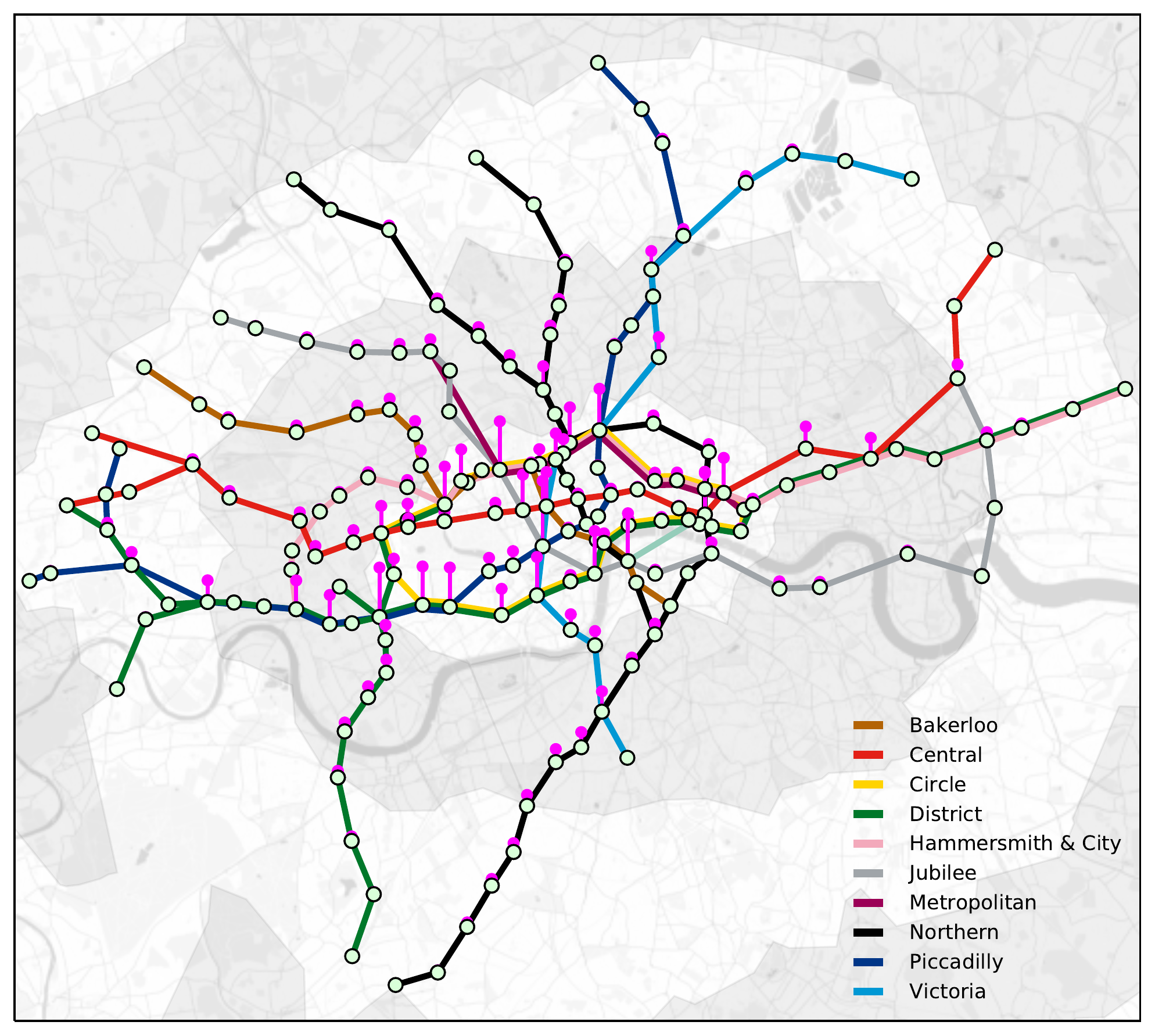}
		\caption{\label{fig:betweenness}Betweenness centrality, designated by magenta-coloured bars, of the London underground network in Zones 1--3. The largest betweenness centrality is observed for the following stations: Green Park, Earl's Court, Baker Street, Waterloo and Westminster.}
	\end{figure}
	
	\item \textit{Closeness vitality}, which represents the change in the sum of distances between all vertex pairs after excluding the $n$-th vertex \cite{Brandes2005}. Fig. \ref{fig:vitality} shows that the stations located in the more remote areas of Zones 2--3 exhibit the largest closeness vitality measure. This is because their removal from the network would disconnect the stations located at the boundaries from the rest of the network.
	
	\begin{figure}[]
		\centering
		\includegraphics[width=0.45\textwidth, trim={0.3cm 0.4cm 0.3cm 0.4cm}, clip]{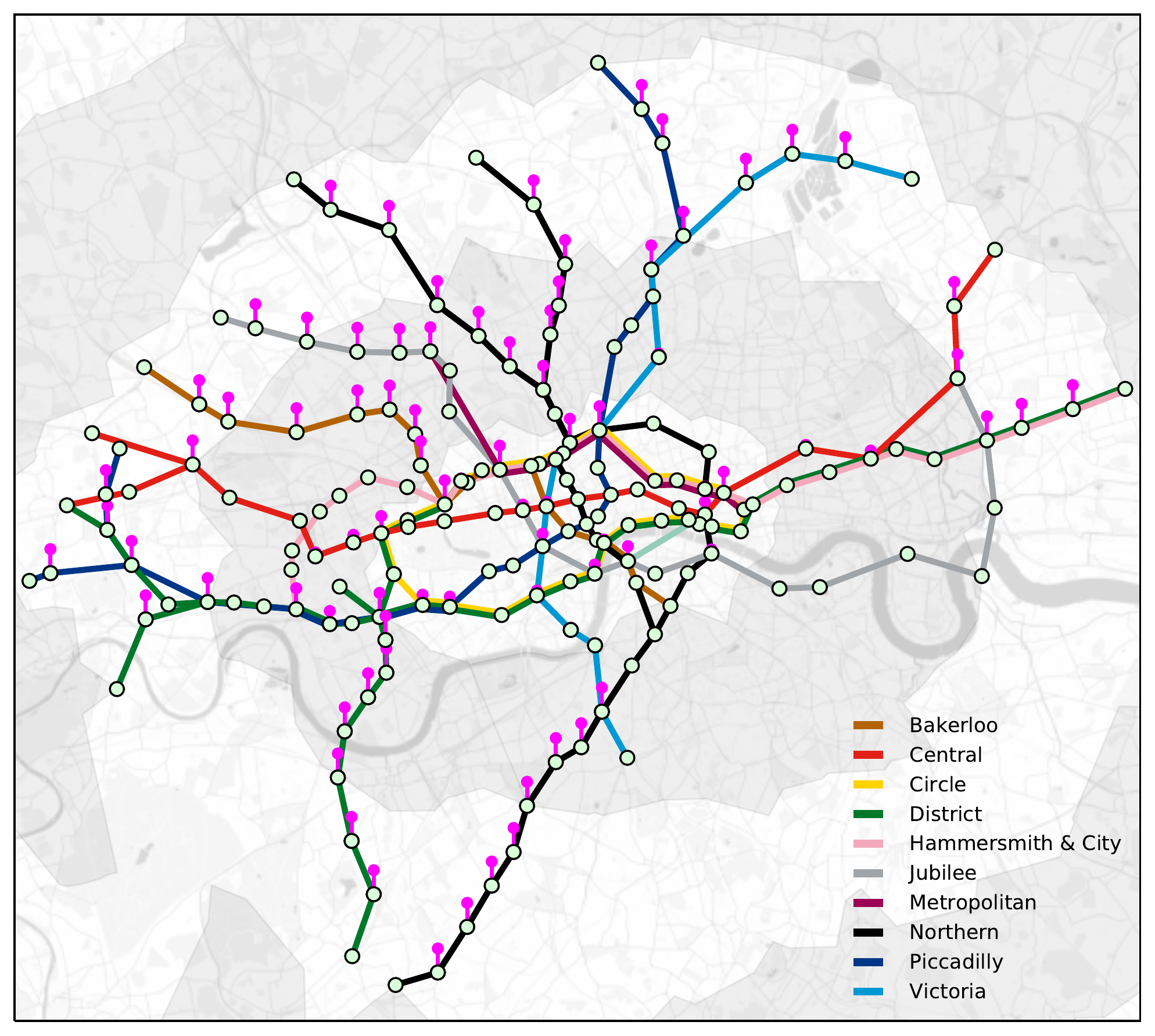}
		\caption{\label{fig:vitality}Closeness vitality, designated in magenta bars, of the London underground network in Zones 1--3.}
	\end{figure}
	
\end{itemize}

\subsection{Modeling commuter population from net passenger flow}

In this section, we employ graph theory to analyse the net passenger flow at all stations of the London underground network. In particular, we demonstrate that it is possible to infer the resident population surrounding each station based on the net passenger flow during the morning rush hour alone.

\begin{figure}[]
	\centering
	\includegraphics[width=0.4\textwidth, trim={0 3cm 0 4cm}, clip]{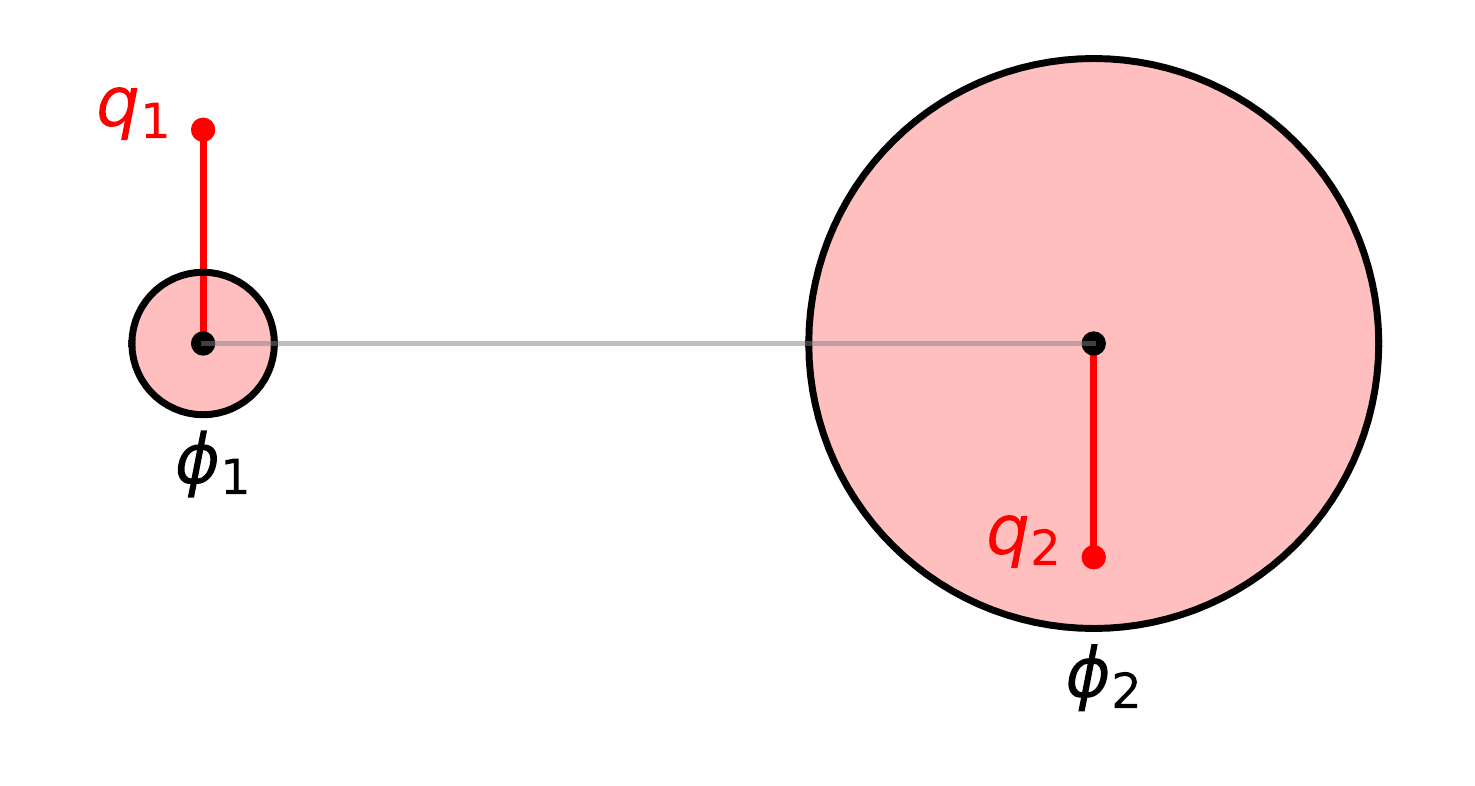}
	\caption{\label{fig:ficks_law}Towards a graph representation of the London underground network. A simplified path graph with two stations surrounded by the respective populations, $\phi(1)$ and $\phi(2)$, exhibits the corresponding net fluxes, $q(1)$ and $q(2)$. Intuitively, stations surrounded by large populations experience net in-flows of passengers, whereas stations surrounded by low populations experience net out-flows of passengers. }
\end{figure}

To derive the corresponding graph model, we employed the \textit{Fick law of diffusion} which relates the diffusive flux to the concentration of a given vector field, under the assumption of a steady state. This model asserts that the flux flows from regions of high concentration (population) to regions of low concentration (population), with a magnitude that is proportional to the concentration gradient. Mathematically, the Fick law is given by
\begin{align}
\mathbf{q} = - k \nabla \boldsymbol{\phi}
\end{align}
where 
\begin{itemize}
	\item $\mathbf{q}$ is the flux which measures the amount of substance per unit area per unit time (mol m$^{-2}$ s$^{-1}$);
	\item  $k$ is the coefficient of diffusivity, with its dimension equal to area per unit time (m$^{2}$ s$^{-1}$);
	\item $\boldsymbol{\phi}$ represents the concentration (mol m$^{-3}$).
\end{itemize}
In this way, we can model the passenger flows in the London underground network as a diffusion process, whereby during the morning rush hour the population mainly flows from concentrated residential areas to sparsely populated business districts. Therefore, the variables in our model are:
\begin{itemize}
	\item $\mathbf{q} \in \mathbb{R}^{N}$ is the net passenger flow vector, with the $i$-th entry representing the net passenger flow at the $i$-th station during the morning rush hour, that is
	\begin{align}
	q(i) = \text{(passengers exiting station $i$)} \nonumber \\ - \text{(passengers entering station $i$)} 	\label{eq:london_underground_flux}
	\end{align}
	with its dimension equal to ``passengers per station per unit time'';
	\item  $k=1$ is the coefficient of diffusivity, with its dimension equal to ``stations per unit time'';
	\item $\boldsymbol{\phi}  \in \mathbb{R}^{N}$ represents the resident population in the area surrounding the station.
\end{itemize}
This model therefore suggests that, in the morning, the net passenger flow at the $i$-th station, $q(i)$, is proportional to the population difference between the areas surrounding a station $i$ and the adjacent stations $j$, that is
\begin{align}
q(i) & = - k \sum_{j} A_{ij}(\phi(i) - \phi(j)) \notag\\
& = - k ( \phi(i) \sum_{j} A_{ij} - \sum_{j} A_{ij}\phi(j) ) \notag\\
&  =  - k ( \phi(i) D_{ii} - \sum_{j} A_{ij} \phi(j) ) \notag\\
& = - k \sum_{j}\left( \delta_{ij} D_{ii} - A_{ij} \right) \phi(j) 
 = - k \sum_{j} L_{ij} \phi(j)
\end{align}
When considering $N$ stations together, we obtain the model in the matrix form
\begin{align}
\mathbf{q} = -k\mathbf{L}\boldsymbol{\phi} \label{eq:graph_Fick_law}
\end{align}
where $\mathbf{L} = (\mathbf{D}-\mathbf{A}) \in \mathbb{R}^{N \times N}$ is the Laplacian matrix of the graph model. For clarity, Fig. \ref{fig:ficks_law} illustrates a signal within this diffusion model on a $2$-vertex path graph obeying the Fick law. 

\begin{figure}[]
	\centering
	\includegraphics[width=0.45\textwidth, trim={0.3cm 0.4cm 0.3cm 0.4cm}, clip]{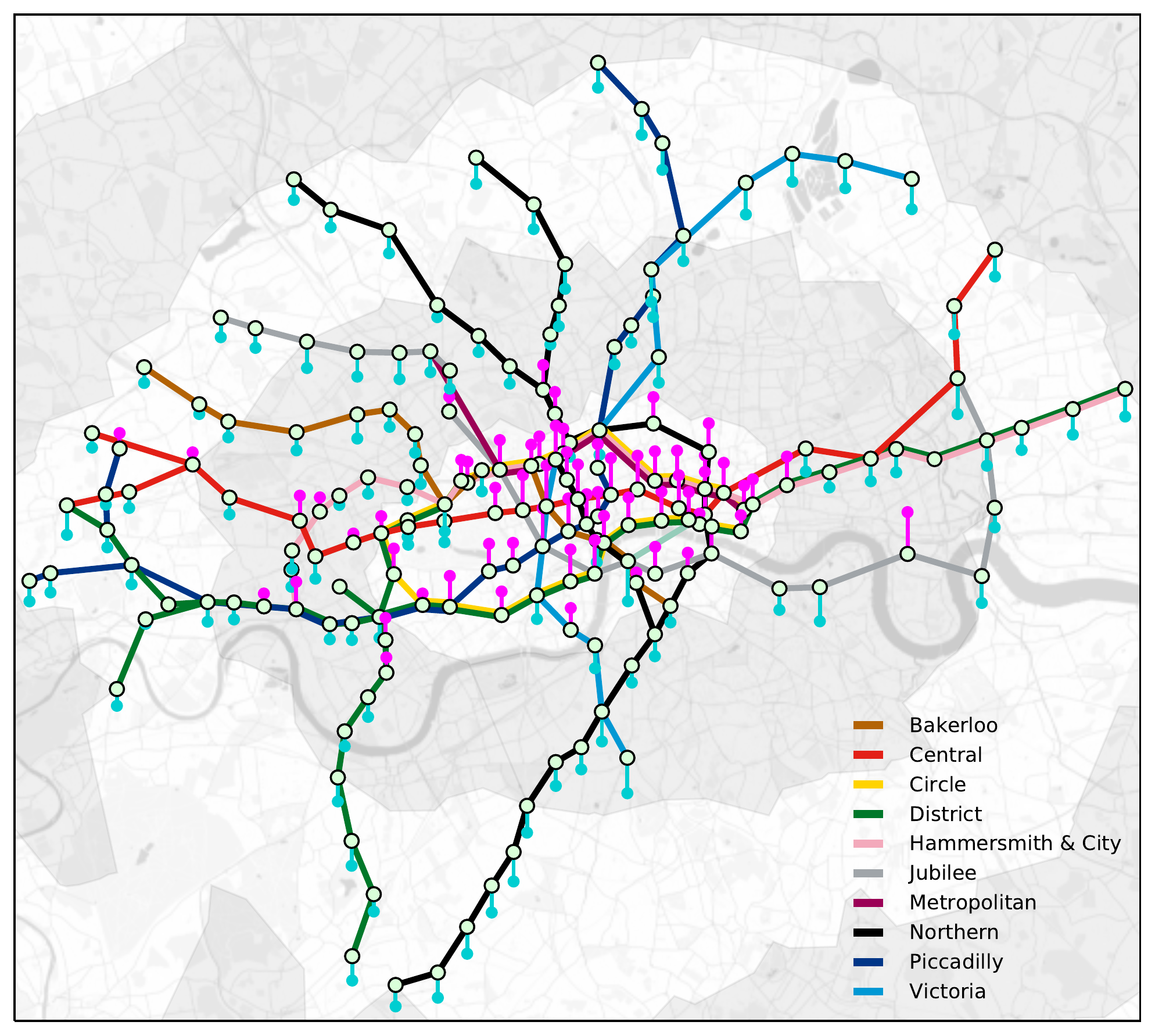}
	\caption{\label{fig:tube_net_flow}Net passenger outflow during the morning rush hour within Zones 1--3 of the London underground network. The magenta bars designate a net outflow of passengers while the cyan bars designate a net inflow of passengers. Stations located within business districts exhibit the greatest net outflow of passengers, while stations located in residential areas toward the boundaries of Zones 2--3 exhibit the largest net inflow of passengers.}
\end{figure}

The data for the average daily net flow of passengers during the morning rush hour at each station in $2016$ was obtained from Transport for London (TFL) \cite{TFL}, and is illustrated as a signal on the underground graph model in Fig. \ref{fig:tube_net_flow}. For illustration purposes, Table \ref{table:netflows} shows the daily average net flow of passengers per zone. As expected, Zone 1 is the only zone to exhibit a net outflow of passengers, while Zones 2--10 show a net inflow of passengers. In particular, Zone 3 exhibits the largest inflow. In an ideal scenario, the total net outflow across Zones 1--10 should sum up to $0$, however, the residual net outflow is attributed to passengers entering the underground network through other transport services not considered in our model, i.e. rail services. 

Moreover, Table \ref{table:netflows_top5} shows the average net flow of passengers for the top $5$ stations with the greater net inflow and outflow. The stations which the greatest net outflow of passengers are located within financial (Bank, Canary Wharf, Green Park) and commercial (Oxford Circus, Holborn) districts. In contrast, the greatest net inflow of passengers is attributed to the contribution from the railway stations located in residential areas.

\begin{figure}[]
	\centering
	\includegraphics[width=0.45\textwidth, trim={0.3cm 0.4cm 0.3cm 0.4cm}, clip]{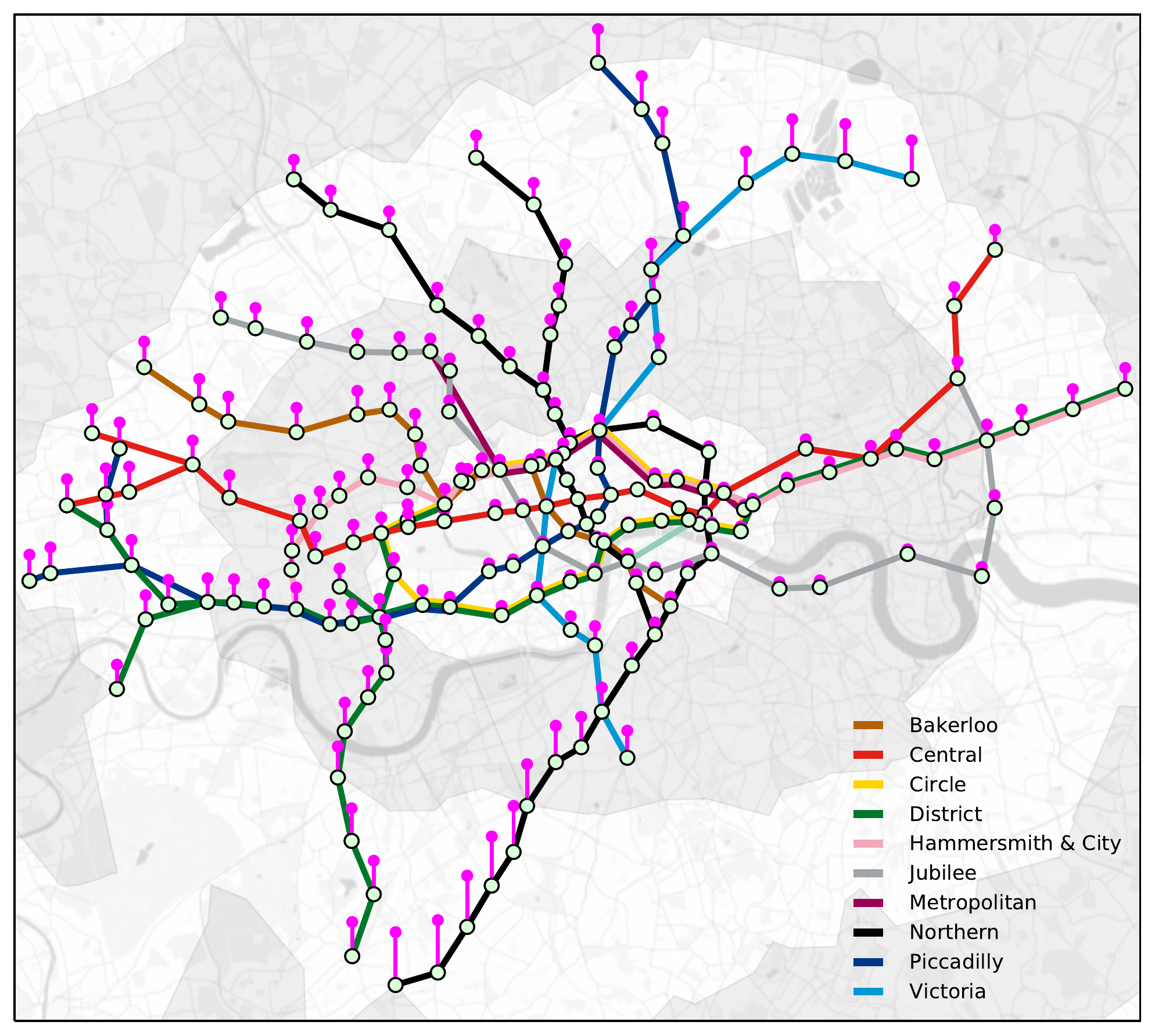}
	\caption{\label{fig:population}Population distribution implied by our graph model in (\ref{eq:population_estimate}), calculated from the net passenger outflow during the morning rush hour within Zones 1--3. As expected, business districts exhibit the lowest population density, while residential areas (Zones 2--3) exhibit the highest commuter population density.}
\end{figure}

To obtain an estimate of the resident population surrounding each station, we can simply rearrange (\ref{eq:graph_Fick_law}) to obtain
\begin{align}
\hat{\boldsymbol{\phi}} = -\frac{1}{k} \mathbf{L}^{+}\mathbf{q} \label{eq:population_estimate}
\end{align}
where the symbol $(\cdot)^{+}$ denotes the matrix pseudo-inverse operator. However, notice that the population vector can only be estimated up to a constant, hence the vector $\hat{\boldsymbol{\phi}}$ actually quantifies the \textit{relative} population between stations, whereby the station with the lowest estimated surrounding population takes the value of $0$. The so estimated resident population, based on the morning net passenger flow, is displayed in Fig. \ref{fig:population} as a signal on a graph. Observe that the estimates are reasonable since most of the resident population in London is concentrated toward the more remote areas of Zones 2--3, while business districts at the centre of Zone 1 are sparsely populated in the evening.


\begin{table}[h!]
	\setlength{\tabcolsep}{7pt}
	\renewcommand{\arraystretch}{1.2}
	\begin{center}
		\caption{\label{table:netflows}Daily average passenger flows during the morning rush hour per Zone.}
		\begin{tabular}{c|rrr}
			\hline
			\textbf{Zone} & \textbf{Entries} & \textbf{Exits} & \textbf{Net Outflow} \\
			\hline
			$1$    & $455,704$ & $844,123$ & $388,419$  \\
			$2$    & $343,145$ & $264,732$ & $-78,413$  \\
			$3$    & $275,965$ & $104,414$ & $-171,551$ \\ \hline
			$4$--$10$    & $206,408$ & $72,152$  & $-134,256$  \\\hline
			\textbf{Total} & $1,281,222$ & $1,285,421$ & $4,199$ \\
			\hline
		\end{tabular}
	\end{center}
\end{table}

\begin{table}[h!]
	\setlength{\tabcolsep}{7pt}
	\renewcommand{\arraystretch}{1.2}
	\begin{center}
		\caption{\label{table:netflows_top5}Stations with most net passenger outflow and inflow during the morning rush hour.}
		\begin{tabular}{l|r r r}
			\hline
			\textbf{Station} & \textbf{Entries} & \textbf{Exits} & \textbf{Net Outflow} \\
			\hline
			Bank          & $17,577$   & $69,972$  & $52,395$   \\
			Canary Wharf  & $8,850$    & $56,256$  & $47,406$   \\
			Oxford Circus & $3,005$    & $44,891$  & $41,886$   \\
			Green Park    & $2,370$    & $30,620$  & $28,250$   \\
			Holborn       & $1,599$    & $25,294$  & $23,695$   \\ \hline
			Finsbury Park & $20,773$   & $8,070$   & $- 12,703$ \\
			Canada Water  & $31,815$   & $14,862$  & $- 16,953$ \\
			Brixton       & $24,750$   & $4,369$   & $- 20,381$ \\
			Stratford     & $43,473$   & $22,360$  & $- 21,113$ \\
			Waterloo      & $61,129$   & $22,861$  & $- 38,268$ \\\hline
		\end{tabular}
	\end{center}
\end{table}

\section{
	Portfolio Cuts}

Investment returns naturally reside on irregular domains, however, standard multivariate portfolio optimization methods are agnostic to data structure. To this end, we investigate ways for domain knowledge to be meaningfully incorporated into the analysis, by means of \textit{portfolio cuts}. Such a graph-theoretic portfolio partitioning technique would allow the investor to devise robust and tractable asset allocation schemes, by virtue of a rigorous graph framework for considering smaller, computationally feasible, and economically meaningful clusters of assets, based on graph cuts. In turn, this makes it possible to fully utilize the covariance matrix of asset returns for constructing the portfolio, even without the requirement for its inversion. 

\textit{Modern portfolio theory} suggests an optimal strategy for minimising the investment risk, which is based on the second-order moments of asset returns \cite{Markowitz1952}. The solution to this optimization task is referred to as the \textit{minimum-variance} (MV) portfolio. Consider the vector, $\mathbf{r}(t) \in \mathbb{R}^{N}$, which contains the returns of $N$ assets at a time $t$, the $i$-th entry of which is given by
\begin{equation}
r_{t}(i) = \frac{p_{t}(i) - p_{t-1}(i)}{p_{t-1}(i)}
\end{equation}
where $p_{t}(i)$ denotes the value of the $i$-th asset at a time $t$. The MV portfolio asserts that the optimal vector of asset holdings, $\mathbf{w} \in \mathbb{R}^{N}$, is obtained through the following optimization problem
\begin{equation}
\min_{\mathbf{w}} \;\;  \mathbf{w}^{T}\mathbf{\Sigma}\mathbf{w}, \quad \text{subject to} \;\; \mathbf{w}^{T}\mathbf{1} = 1
\end{equation}
where $\mathbf{\Sigma} = \text{cov}\{\mathbf{r}\} \in \mathbb{R}^{N \times N}$ is the covariance matrix of returns, $\mathbf{1}=[1,...,1]^{T}$, and the constraint, $\mathbf{w}^{T}\mathbf{1} = 1$, enforces full investment of the capital. The optimal portfolio holdings (using the method of Lagrange multipliers) then become
\begin{equation}
\mathbf{w} = \frac{\mathbf{\Sigma}^{-1}\mathbf{1}}{\mathbf{1}^{T}\mathbf{\Sigma}^{-1}\mathbf{1}} \label{eq:minimum_variance}
\end{equation}
It is important to highlight that the matrix inversion of $\mathbf{\Sigma}$ required in (\ref{eq:minimum_variance}) may lead to significant errors for ill-conditioned matrices. These instability concerns have received substantial attention in recent years \cite{Kolm2014}, and alternative procedures have been proposed to promote robustness by either incorporating additional portfolio constraints \cite{Clarke2002}, introducing Bayesian priors \cite{Black1992} or improving the numerical stability of covariance matrix inversion \cite{Ledoit2003}. A more recent approach has been to model assets using \textit{market graphs} \cite{Boginski2003}, that is, based on graph-theoretic techniques. Intuitively, a universe of assets can be naturally modelled as a network of vertices on a graph, whereby an edge between two vertices (assets) designates both the existence of a link and the degree of similarity between assets \cite{Simon1962}. 

\begin{Remark}
	A graph-theoretic perspective offers an interpretable explanation for the underperformance of MVO techniques in practice. Namely, since the covariance matrix $\mathbf{\Sigma}$ is dense, standard multivariate models implicitly assume full connectivity of the graph, and are therefore not adequate to account for the structure inherent to real-world markets \cite{LopezdePrado2014,LopezdePrado2014_2,LopezdePrado2016}. \textit{Moreover, it can be shown that the optimal holdings under the MVO framework are inversely proportional to the vertex centrality, thereby over-investing in assets with low centrality} \cite{Peralta2016,Li2019}.
\end{Remark}

Intuitively, it would be highly desirable to remove unnecessary edges in order to more appropriately model the underlying structure between assets (graph vertices); this can be achieved through \textit{vertex clustering} of the market graph \cite{Boginski2003}. Various portfolio diversification frameworks employ this technique to allocate capital within and across clusters of assets at multiple hierarchical levels. For instance, the \textit{hierarchical risk parity} scheme \cite{LopezdePrado2016} employs an inverse-variance weighting allocation which is based on the number of assets within each asset cluster. Similarly, the \textit{hierarchical clustering based asset allocation} in \cite{Raffinot2017} finds a diversified weighting by distributing capital equally among each of the cluster hierarchies. 

Despite mathematical elegance and physical intuition, direct vertex clustering is an NP hard problem. Consequently, existing graph-theoretic portfolio constructions employ combinatorial optimization formulations \cite{Boginski2003,Boginski2005,Boginski2006,Gunawardena2012,Boginski2014,Kalyagin2014}, which too become computationally intractable for large graph systems. To alleviate this issue, we employ the \textit{minimum cut} vertex clustering method to the graph of portfolio assets, to introduce the concept of \textit{portfolio cut}. In this way, smaller graph partitions (cuts) can be evaluated quasi-optimally, using algebraic methods, and in an efficient and rigorous manner.

\subsection{Structure of market graph}

A universe of $N$ assets can be represented as a set of vertices on a \textit{market graph} \cite{Boginski2003}, whereby the edge weight, $W_{mn}$, between vertices $m$ and $n$ is defined as the absolute correlation coefficient, $|\rho_{mn}|$, of their respective returns of assets $m$ and $n$, that is
\begin{equation}
W_{mn} = \frac{|\sigma_{mn}|}{\sqrt{ \sigma_{mm}\sigma_{nn} }} = |\rho_{mn}|
\end{equation}
where $\sigma_{mn}=\text{cov}\{r_{t}(m),r_{t}(n)\}$ is the covariance of returns between the assets $m$ and $n$. In this way, we have $W_{mn}=0$ if the assets $m$ and $n$ are statistically independent (not connected), and $W_{mn}>0$ if they are statistically dependent (connected on a graph). Note that the resulting weight matrix is symmetric, $\mathbf{W}^{T}=\mathbf{W}$.

\subsection{Minimum cut based vertex clustering}

Vertex clustering aims to group together vertices from the asset universe $\mathcal{V}$ into multiple disjoint \textit{clusters}, $\mathcal{V}_i$. For a market graph, assets which are grouped into a cluster, $\mathcal{V}_i$, are expected to exhibit a larger degree of mutual within-cluster statistical dependency than with the assets in other clusters, $\mathcal{V}_j$, $j\ne i$. The most popular classical graph cut methods are based on finding the minimum set of edges whose removal would disconnect a graph in some ``optimal'' sense; this is referred to as \textit{minimum cut} based clustering \cite{Schaeffer2007}. 

Consider an $N$-vertex market graph, $\mathcal{G} = \{\mathcal{V},\mathcal{E}\}$, which is grouped into $K=2$ disjoint subsets of vertices, $\mathcal{V}_{1} \subset \mathcal{V}$ and $\mathcal{V}_{2} \subset \mathcal{V}$, with  $\mathcal{V}_{1} \cup \mathcal{V}_{2}=\mathcal{V}$ and $\mathcal{V}_{1} \cap \mathcal{V}_{2}=\emptyset$. A cut of this graph, for the given clusters, $\mathcal{V}_{1}$ and $\mathcal{V}_{2}$, is equal to a sum of all weights that correspond to the edges which connect the vertices between the subsets, $\mathcal{V}_{1}$ and $\mathcal{V}_{2}$, that is 
\begin{equation}
Cut(\mathcal{V}_{1},\mathcal{V}_{2})=\sum_{m \in \mathcal{V}_{1}} \sum_{n \in \mathcal{V}_{2} } W_{mn}  \label{eq:ideal_minimum_cut}
\end{equation} 
A cut which exhibits the minimum value of the sum of weights between the disjoint subsets, $\mathcal{V}_{1}$ and $\mathcal{V}_{2}$, considering all possible divisions of the set of vertices, $\mathcal{V}$, is referred to as \textit{the minimum cut}.

Finding the minimum cut in (\ref{eq:ideal_minimum_cut}) is an NP-hard problem, whereby the number of combinations to split an even number of vertices, $N$, into \textit{any} two possible disjoint subsets is given by $C = 2^{(N-1)}-1$.

\begin{Remark}
	Table \ref{table:cuts_combinations} depicts the computational burden associated with this brute force graph cut approach.
	
	\begin{table}[h!]
		\setlength{\tabcolsep}{10pt}
		\renewcommand{\arraystretch}{1.2}
		\begin{center}
			\caption{\label{table:cuts_combinations} The number of combinations, $C$, to split the vertices of an $N$-vertex market graph into two subsets.}
			\begin{tabular}{c | c | c}
				\hline
				Market & $N$ & $C$ \\ \hline
				S\&P 500 & $500$ & $1.64 \times 10^{150}$ \\
				Nikkei 225 & 225 & $2.70 \times 10^{67}$\\
				FTSE 100 & 100 & $6.34 \times 10^{29}$ \\
				EUROSTOXX 50 & 50 & $5.63 \times 10^{14}$ \\\hline			
			\end{tabular}
		\end{center}
	\end{table}
	
\end{Remark}

Within graph cuts, a number of optimization approaches may be employed to enforce some desired properties on graph clusters:

\smallskip

\noindent (i) \textit{Normalized minimum cut}. The value of $Cut(\mathcal{V}_{1},\mathcal{V}_{2})$ is regularised by an additional term to enforce the subsets, $\mathcal{V}_{1}$ and $\mathcal{V}_{2}$, to be \textit{simultaneously as large as possible}. The normalized cut formulation is given by \cite{Hagen1992} \vspace{-0.1cm}
\begin{equation}
CutN(\mathcal{V}_{1},\mathcal{V}_{2})=\Big(\frac{1}{N_{1}}+\frac{1}{N_{2}} \Big)\sum_{m \in \mathcal{V}_{1}} \sum_{n \in \mathcal{V}_{2} } W_{mn} \label{CutN}
\end{equation}
where $N_{1}$ and $N_{2}$ are the respective numbers of vertices in the sets  $\mathcal{V}_{1}$ and $\mathcal{V}_{2}$. Since $N_{1}+N_{2}=N$, the term $\frac{1}{N_{1}}+\frac{1}{N_{2}}$ reaches its minimum for $N_{1}=N_{2}=\frac{N}{2}$.  

\noindent (ii) \textit{Volume normalized minimum cut}. Since the vertex weights are involved when designing the size of subsets $\mathcal{V}_{1}$ and $\mathcal{V}_{2}$, then by defining \textit{the volumes} of these sets as $V_{1}=\sum_{n \in \mathcal{V}_{1}}D_{nn}$ and $V_{2}=\sum_{n \in \mathcal{V}_{2}}D_{nn}$, we arrive at \cite{Shi2000} \vspace{-0.2cm}
\begin{equation} 
CutV(\mathcal{V}_{1},\mathcal{V}_{2})=\Big(\frac{1}{V_{1}}+\frac{1}{V_{2}} \Big)\sum_{m \in \mathcal{V}_{1}} \sum_{n \in \mathcal{V}_{2} } W_{mn} \label{CutV}
\end{equation}
Since $V_{1}+V_{2}=V$, the term $\frac{1}{V_{1}}+\frac{1}{V_{2}}$ reaches its minimum for $V_{1}=V_{2}=\frac{V}{2}$. Notice that vertices with a higher degree, $D_{nn}$, are considered as structurally more important than those with lower degrees. In turn, for market graphs, assets with a higher average statistical dependence to other assets are considered as more \textit{central}.


\begin{Remark}
	It is important to note that clustering results based on the two above graph cut forms are different. While the method (i) favours the clustering into subsets with (almost) equal number of vertices, the method (ii) favours subsets with (almost) equal volumes, that is, subsets with vertices exhibiting (almost) equal average statistical dependence to the other vertices.
\end{Remark}

%

\subsection{Spectral bisection based minimum cut}

To overcome the computational burden of finding the normalized minimum cut, we may opt for an approximative spectral solution which clusters vertices using the eigenvectors of the graph Laplacian, $\mathbf{L}$. The algorithm employs the second (Fiedler \cite{Fiedler1973}) eigenvector of the graph Laplacian,  $\mathbf{u}_{2} \in \mathbb{R}^{N}$,  to yield a \textit{quasi-optimal} vertex clustering on a graph. Despite its simplicity, the algorithm is typically accurate and gives a good approximation to the normalized cut \cite{Ng2002,Spielman2007}. 

To relate the problem of the minimum cut in (\ref{CutN}) and (\ref{CutV}) to that of eigenanalysis of graph Laplacian, we employ an \textit{indicator vector}, denoted by $\mathbf{x} \in \mathbb{R}^{N}$ \cite{Stankovic2019_1}, for which the elements take sub-graph-wise constant values within each disjoint subset (cluster) of vertices, with these constants taking different values for different clusters of vertices. In other words, the elements of $\mathbf{x}$ uniquely reflect the assumed cut of the graph into disjoint subsets $\mathcal{V}_{1},\mathcal{V}_{2} \subset \mathcal{V}$. 

For a general graph, we consider two possible solutions for the indicator vector, $\mathbf{x}$, that satisfy the subset-wise constant form:

\smallskip

\noindent (i) \textit{Normalized minimum cut}. It can be shown that if the indicator vector is defined as \cite{Stankovic2019_1}
\begin{equation}
x(n)=\begin{cases} 
\frac{1}{N_{1}}, & \text{for } n \in \mathcal{V}_{1},  \\
-\frac{1}{N_{2}}, & \text{for } n \in \mathcal{V}_{2},
\end{cases} 
\end{equation}
then the normalized cut, $CutN(\mathcal{V}_{1}, \mathcal{V}_{2})$ in (\ref{CutN}), is equal to the Rayleigh quotient of $\mathbf{L}$ and $\mathbf{x}$, that is
\begin{equation}
CutN(\mathcal{V}_{1}, \mathcal{V}_{2})=\frac{\mathbf{x}^{T}\mathbf{L}\mathbf{x}}{\mathbf{x}^{T}\mathbf{x}} \label{LqfCut}
\end{equation}
Therefore, the indicator vector, $\mathbf{x}$, which minimizes the normalized cut also minimizes (\ref{LqfCut}). This minimization problem, for the unit-norm form of the indicator vector,  can also be written as
\begin{equation}
\min_{\mathbf{x}} \;\; \mathbf{x}^{T}\mathbf{L}\mathbf{x}, \quad \text{subject to} \;\; \mathbf{x}^{T}\mathbf{x}=1 \label{idicativeMIN}
\end{equation}
which can be solved through the eigenanalysis of $\mathbf{L}$, that is
\begin{equation}
\mathbf{L}\mathbf{x}= \lambda_{k} \mathbf{x}
\end{equation} 
After neglecting the trivial solution $\mathbf{x}=\mathbf{u}_{0}$, ($k=0$), since it produces a constant eigenvector, we next arrive at $\mathbf{x}=\mathbf{u}_{1}$, ($k=1$).

\noindent (ii) \textit{Volume normalized minimum cut}. Similarly, by defining $\mathbf{x}$ as
\begin{equation}
x(n)=\begin{cases} 
\frac{1}{V_{1}}, & \text{for } n \in \mathcal{V}_{1},  \\
-\frac{1}{V_{2}}, & \text{for } n \in \mathcal{V}_{2},
\end{cases} 
\end{equation}
the volume normalized cut, $CutV(\mathcal{V}_{1}, \mathcal{V}_{2})$ in (\ref{CutV}), takes the form of a generalised Rayleigh quotient of $\mathbf{L}$, given by \cite{Stankovic2019_1,stankovic2019graph}
\begin{equation}
CutV(\mathcal{V}_{1}, \mathcal{V}_{2})=\frac{\mathbf{x}^{T}\mathbf{L}\mathbf{x}}{\mathbf{x}^{T}\mathbf{D}\mathbf{x}} \label{LqfCutV}
\end{equation} 
The minimization of (\ref{LqfCutV}) can be formulated as
\begin{equation}
\min_{\mathbf{x}} \;\; \mathbf{x}^{T}\mathbf{L}\mathbf{x}, \quad \text{subject to} \;\; \mathbf{x}^{T}\mathbf{D}\mathbf{x}=1 \label{cutD}
\end{equation}
which reduces to a generalized eigenvalue problem of $\mathbf{L}$, given by
\begin{equation}
\mathbf{L}\mathbf{x} =\lambda_{k} \mathbf{D} \mathbf{x}
\end{equation}
Therefore, the solution to (\ref{cutD}) becomes the generalized eigenvector of the graph Laplacian which corresponds to its lowest non-zero eigenvalue, that is, $\mathbf{x}=\mathbf{u}_{1}$, ($k=1$).

For the spectral solutions above, the membership of a vertex, $n$, to either the subset $\mathcal{V}_{1}$ or $\mathcal{V}_{2}$ is uniquely defined by the \textit{sign} of the indicator vector, $\mathbf{x}=\mathbf{u}_{1}$, that is
\begin{equation}
\mathrm{sign}(x(n))=\begin{cases} 
1, & \text{for } n \in \mathcal{V}_{1},  \\
-1, & \text{for } n \in \mathcal{V}_{2}.
\end{cases} 
\end{equation}
Notice that a scaling of $\mathbf{x}$ by any constant would not influence the solution for clustering into the subsets $\mathcal{V}_{1}$ or $\mathcal{V}_{2}$.

\subsection{Repeated portfolio cuts}

Although the above analysis has focused on the case with $K=2$ disjoint sub-graphs, it can be straightforwardly generalized to $K \geq 2$ disjoint sub-graphs through the method of \textit{repeated bisection}. 

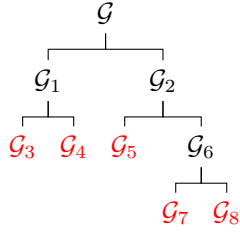
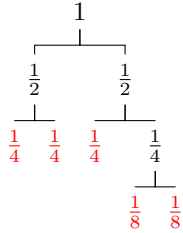
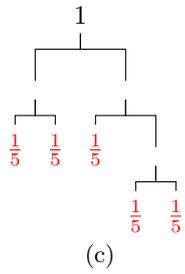
\begin{figure}[]
	\vspace{-0.3cm}
	\centering
	\begin{subfigure}[t]{0.15\textwidth}
		\centering
		\begin{tikzpicture}
		\tikzset{level distance=25pt}
		\tikzset{edge from parent/.style=
			{draw,
				edge from parent path={(\tikzparentnode.south)
					-- +(0,-6pt)
					-| (\tikzchildnode)}}}
		\Tree [.$\mathcal{G}$ 
		[.$\mathcal{G}_{1}$ 
		[.\color{red}$\mathcal{G}_{3}$ ] 
		[.\color{red}$\mathcal{G}_{4}$ ] ]
		[.$\mathcal{G}_{2}$
		[.\color{red}$\mathcal{G}_{5}$ ] 
		[.$\mathcal{G}_{6}$ 
		[.\color{red}$\mathcal{G}_{7}$ ] 
		[.\color{red}$\mathcal{G}_{8}$ ] 
		] 
		] 
		]
		\end{tikzpicture}
		\vspace{-0.6cm}
		\caption{\label{fig:dendrogram}}
	\end{subfigure}

		\vspace{0.5cm}
		
	\begin{subfigure}[t]{0.15\textwidth}
		\centering
		\begin{tikzpicture}
		\tikzset{level distance=25pt}
		\tikzset{edge from parent/.style=
			{draw,
				edge from parent path={(\tikzparentnode.south)
					-- +(0,-6pt)
					-| (\tikzchildnode)}}}
		\Tree [.$1$ 
		[.$\frac{1}{2}$ 
		[.\color{red}$\frac{1}{4}$  ] 
		[.\color{red}$\frac{1}{4}$  ] ]
		[.$\frac{1}{2}$ 
		[.\color{red}$\frac{1}{4}$  ] 
		[.$\frac{1}{4}$  
		[.\color{red}$\frac{1}{8}$  ] 
		[.\color{red}$\frac{1}{8}$  ] 
		] 
		] 
		]
		\end{tikzpicture}
		\vspace{-0.2cm}
		\caption{\label{fig:asset_allocation_i}}
	\end{subfigure}
	
	\vspace{0.5cm}
	
	\begin{subfigure}[t]{0.15\textwidth}
		\centering
		\begin{tikzpicture}
		\tikzset{level distance=25pt}
		\tikzset{edge from parent/.style=
			{draw,
				edge from parent path={(\tikzparentnode.south)
					-- +(0,-6pt)
					-| (\tikzchildnode)}}}
		\Tree [.$1$ 
		[.{}
		[.\color{red}$\frac{1}{5}$  ] 
		[.\color{red}$\frac{1}{5}$  ] ]
		[.{} 
		[.\color{red}$\frac{1}{5}$  ] 
		[.{} 
		[.\color{red}$\frac{1}{5}$  ] 
		[.\color{red}$\frac{1}{5}$  ] 
		] 
		] 
		]
		\end{tikzpicture}
		\vspace{-0.2cm}
		\caption{\label{fig:asset_allocation_ii}}
	\end{subfigure}
	\vspace{-0.2cm}
	\caption{Graph cut based asset allocation strategies. (a) Hierarchical graph structure resulting from $K=4$ portfolio cuts. (b) A graph tree based on the $\frac{1}{2^{K_{i}}}$ scheme. (c) A graph tree based on the $\frac{1}{K+1}$ scheme.} 
\end{figure}

A single application of the portfolio cut on the market graph, $\mathcal{G}$, produces two disjoint sub-graphs, $\mathcal{G}_{1}$ and $\mathcal{G}_{2}$, as illustrated in Fig. \ref{fig:dendrogram}. Notice that in this way we construct a hierarchical binary tree structure, whereby the union of the \textit{leaves} of the network is equal to the original market graph, $\mathcal{G}$. We can then perform a subsequent portfolio cut operation on one or both of the leaves based on some suitable criterion (e.g. the leaf with the greatest number of vertices or volume). Therefore, $(K+1)$ disjoint sub-graphs (leaves) can be obtained by performing the portfolio cut procedure $K$ times. 


\begin{Example}
	Fig. \ref{fig:dendrogram} illustrates the hierarchical structure resulting from $K=4$ portfolio cuts of a market graph, $\mathcal{G}$. The leaves of the resulting binary tree are denoted by $\{\mathcal{G}_{3},\mathcal{G}_{4},\mathcal{G}_{5},\mathcal{G}_{7},\mathcal{G}_{8}\}$ (in red), whereby the number of disjoint sub-graphs is equal to $(K+1)=5$. Notice that the union of the leaves amounts to the original graph, i.e.  $\mathcal{G}_{3} \cup \mathcal{G}_{4} \cup \mathcal{G}_{5} \cup \mathcal{G}_{7} \cup \mathcal{G}_{8} = \mathcal{G}$.
\end{Example}

\subsection{Graph asset allocation schemes}

\label{sec:asset_allocation}

We next elaborate upon some intuitive asset allocation strategies, inspired by the work in \cite{LopezdePrado2016,Raffinot2017}, which naturally builds upon the portfolio cut. The aim is to determine a diversified weighting scheme by distributing capital among the disjoint clusters (leaves) so that highly correlated assets within a given cluster receive the same total allocation, thereby being treated as a single investment entity.

By denoting the portion of the total capital allocated to a cluster $\mathcal{G}_{i}$ by $w_{i}$, we consider two simple asset allocation schemes:

\smallskip

\noindent \textbf{(AS1)} $w_{i} = \frac{1}{2^{K_{i}}}$, where $K_{i}$ is the number of portfolio cuts required to obtain a sub-graph $\mathcal{G}_{i}$;

\smallskip

\noindent \textbf{(AS2)} $w_{i} = \frac{1}{K+1}$, where $(K+1)$ is the number of disjoint sub-graphs.

\smallskip

\begin{Remark}
	An equally-weighted asset allocation strategy may now be employed within each cluster, i.e. every asset within the $i$-th cluster, $\mathcal{G}_{i}$, will receive a weighting equal to $\frac{w_{i}}{N_{i}}$.
\end{Remark}

\begin{Remark}
	The weighting scheme in AS1 above is closely related to the strategy proposed in \cite{Raffinot2017}, while the scheme in AS2 is inspired by the generic equal-weighted (EW) allocation scheme \cite{DeMiguel2009}. These schemes are convenient in that they require no assumptions regarding the across-cluster statistical dependence. In addition, unlike the EW scheme, they implicitly consider the inherent market risks (asset correlation) by virtue of the portfolio cut formulation, which is based on the eigenanalysis of the market graph Laplacian, $\mathbf{L}$.
\end{Remark}

\begin{Example}
	Fig. \ref{fig:asset_allocation_i} and \ref{fig:asset_allocation_ii} demonstrate respectively the asset allocation schemes in AS1 and AS2 for $K=4$ portfolio cuts, based on the market graph partitioning in Fig. \ref{fig:dendrogram}. Notice that the weights associated to the disjoint sub-graphs (leaves in red) sum up to unity.
\end{Example}

\subsection{Numerical Example}

The performance of the portfolio cuts and the associated graph-theoretic asset allocation schemes was investigated using historical price data comprising of the $100$ most liquid stocks in the S\&P 500 index, based on average trading volume, in the period 2014-01-01 to 2018-01-01. The data was split into: (i) the \textit{in-sample} dataset (2014-01-01 to 2015-12-31) which was used to estimate the asset correlation matrix and to compute the portfolio cuts; and (ii) the \textit{out-sample} dataset (2016-01-01 to 2018-01-01), used to objectively quantify the profitability of the asset allocation strategies.

\begin{figure}[]
	\centering
	\begin{subfigure}[t]{0.2\textwidth}
		\centering
		\includegraphics[width=1\textwidth, trim={0 0 0 0}, clip]{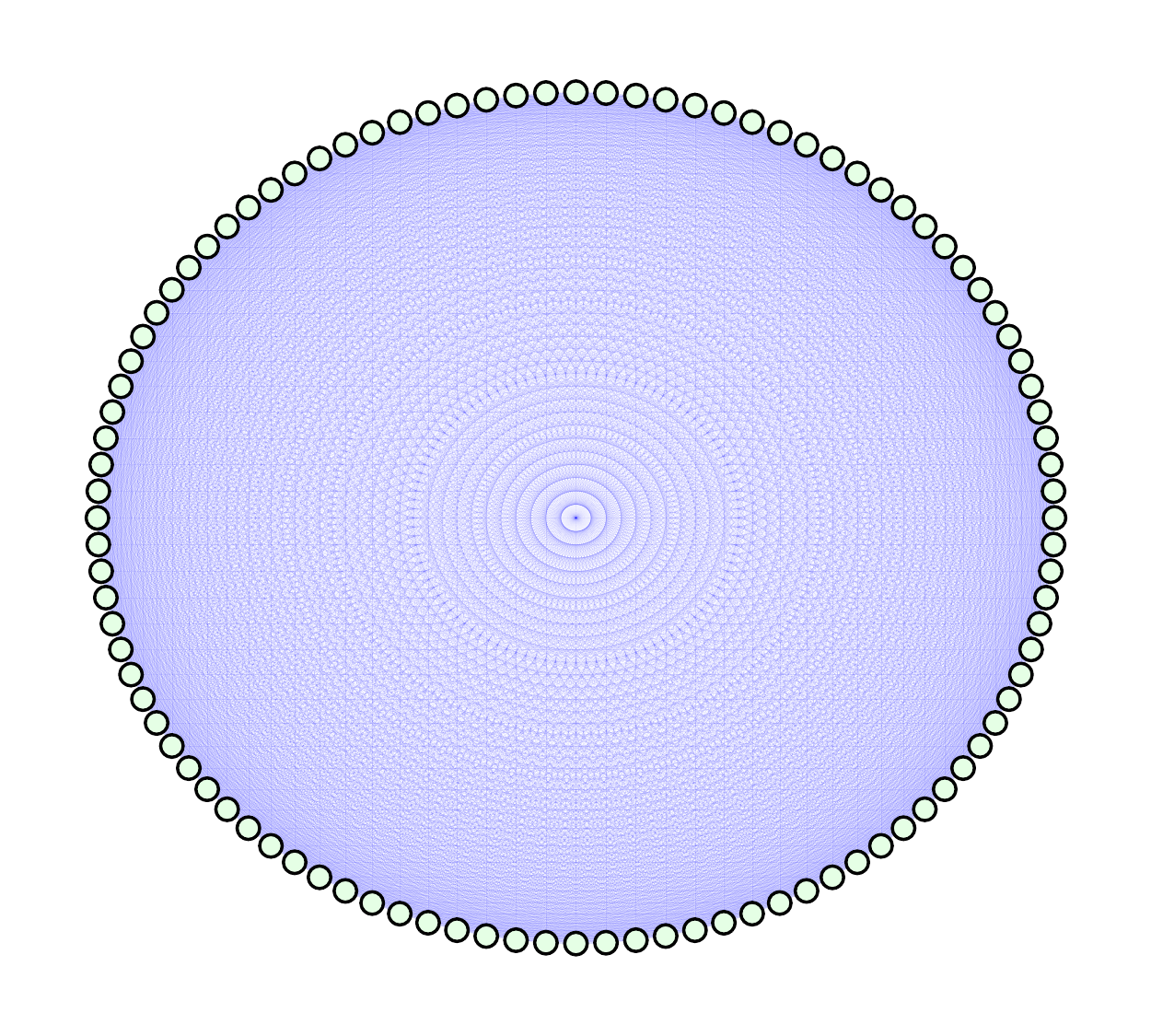} 
		\vspace{-0.7cm}
		\caption{\label{fig:fully_connected_graph}}
	\end{subfigure}
	\begin{subfigure}[t]{0.2\textwidth}
		\centering
		\includegraphics[width=1\textwidth, trim={0 0 0 0}, clip]{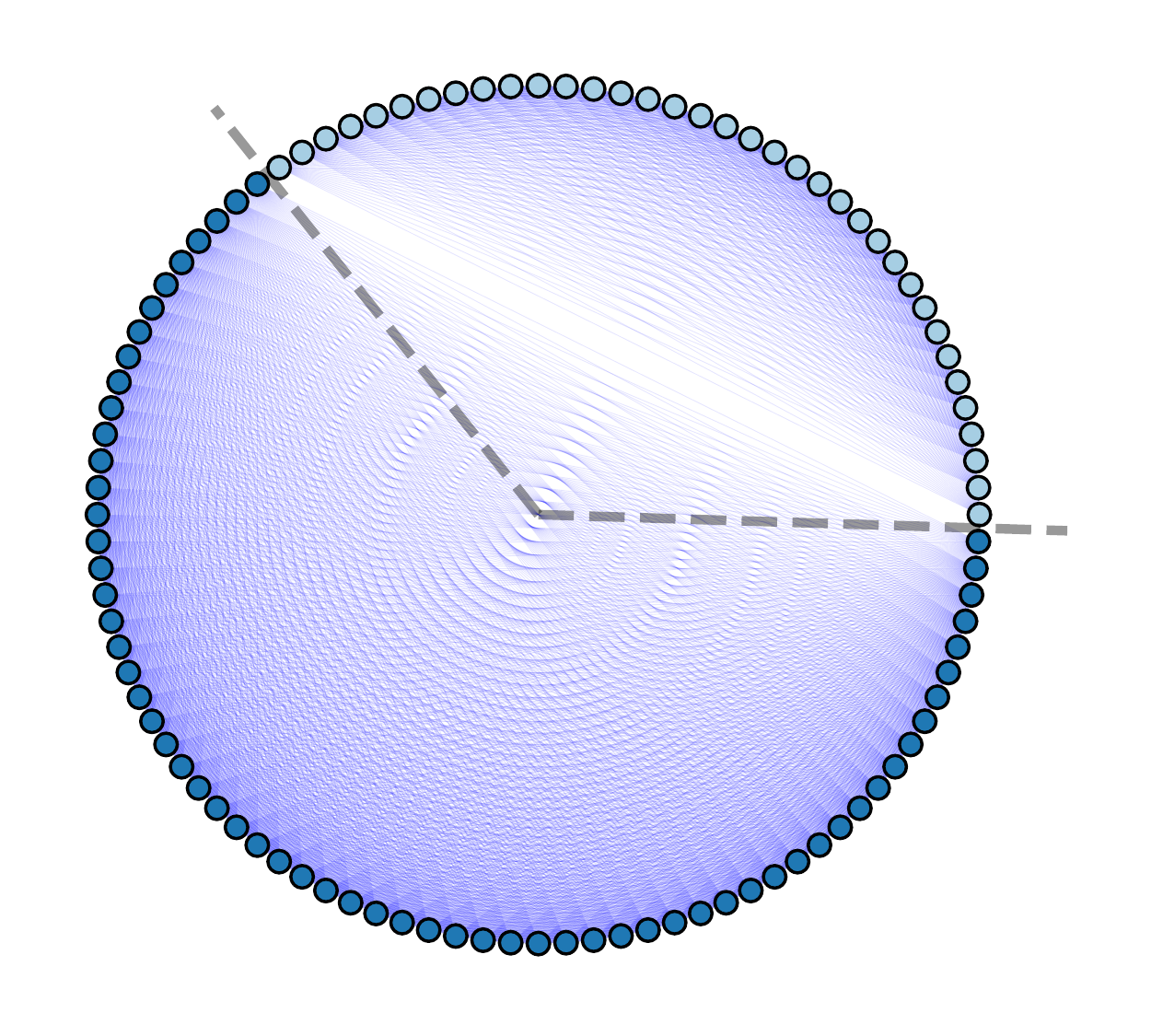} 
		\vspace{-0.7cm}
		\caption{}
	\end{subfigure}
	
	\begin{subfigure}[t]{0.2\textwidth}
		\centering
		\includegraphics[width=1\textwidth, trim={0 0 1cm 0cm}, clip]{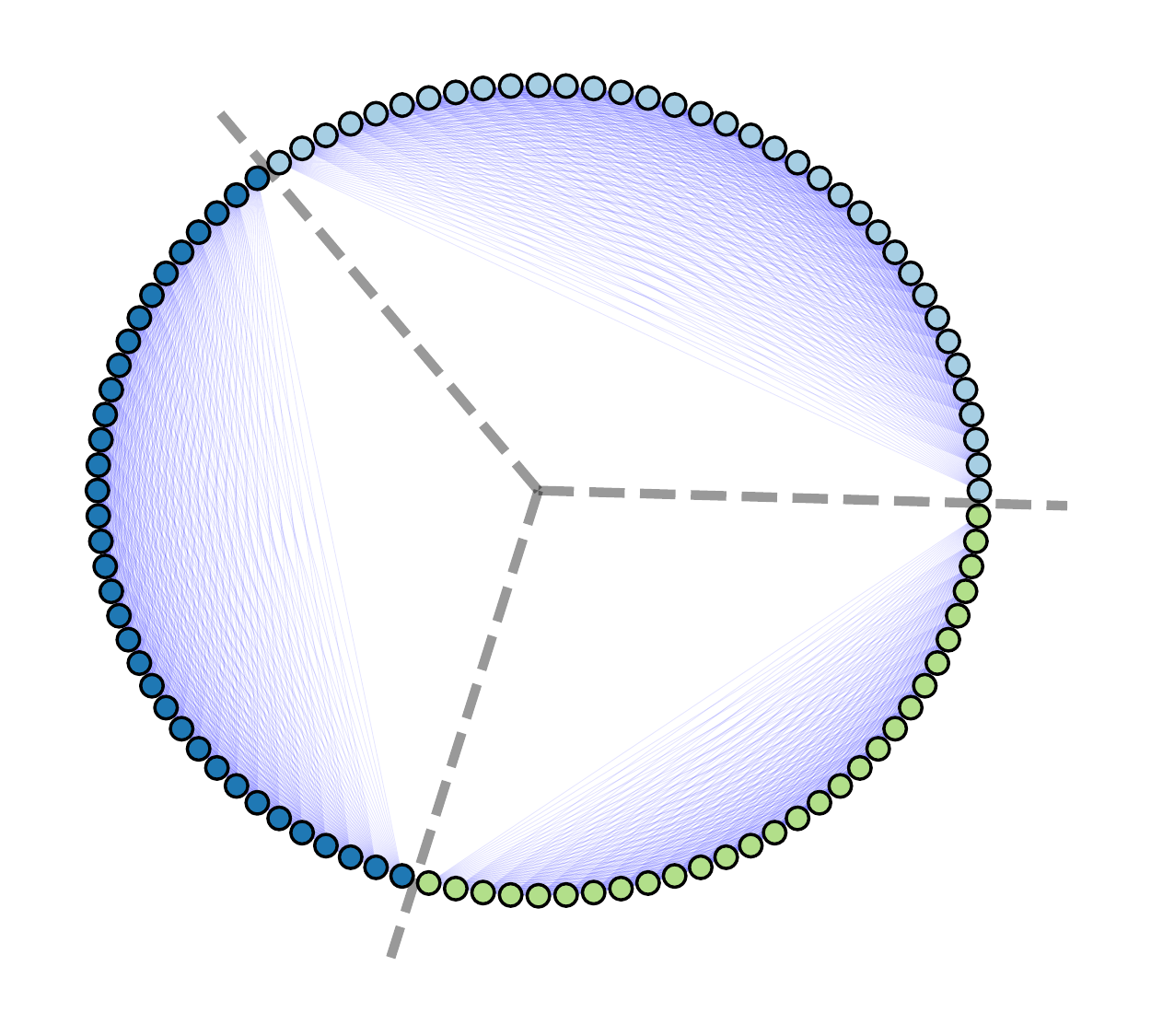} 
		\vspace{-0.7cm}
		\caption{}
	\end{subfigure}
	\begin{subfigure}[t]{0.2\textwidth}
		\centering
		\includegraphics[width=1.05\textwidth, trim={0 0 0 0cm}, clip]{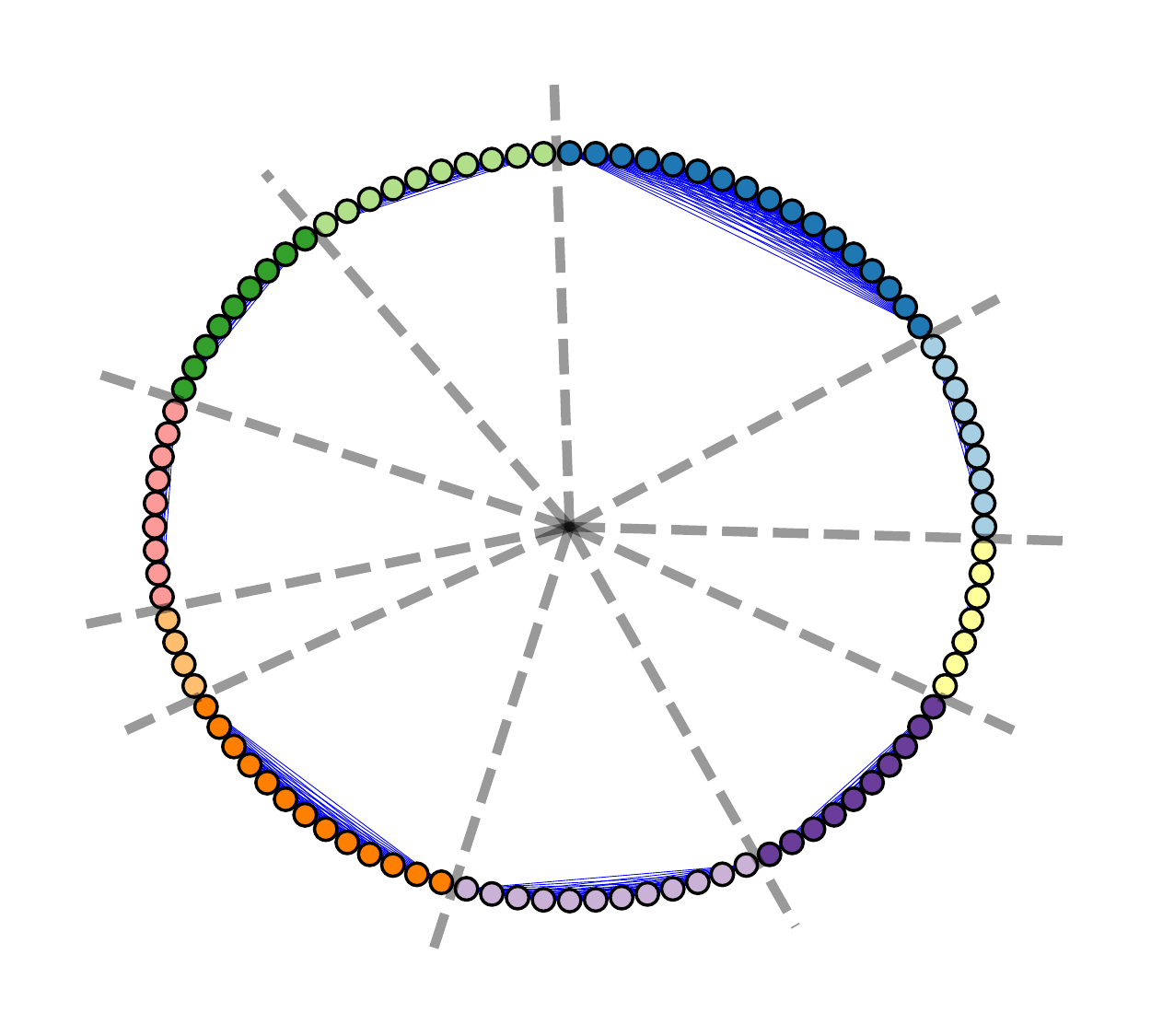} 
		\vspace{-0.7cm}
		\caption{\label{fig:connectivity_K10}}
	\end{subfigure}
	\vspace{-0.2cm}
	\caption{\label{fig:connectivity} Visualisation of the $100$-vertex market graph connectivity for the $100$ most liquid stocks in S\&P 500 index, and its partitions into disjoint sub-graphs (separated by dashed grey lines). The edges (blue lines) were calculated based on the correlation between assets. (a) Fully connected market graph with $5050$ edges. (b) Partitioned graph after $K=1$ portfolio cuts (CutV), with $2746$ edges. (c) Partitioned graph after $K=2$ portfolio cuts (CutV), with $1731$ edges. (d) Partitioned graph after $K=10$ portfolio cuts (CutV), with $575$ edges. Notice that the number of edges required to model the market graph is significantly reduced with each subsequent portfolio cut, since $\sum_{i=1}^{K+1} \! \frac{1}{2}(N_{i}^{2} \! + \! N_{i}) < \frac{1}{2}(N^{2} \! + \! N)$, $\forall K>0$.} 
\end{figure}

Fig. \ref{fig:connectivity} displays the $K$-th iterations of the normalised portfolio cut in (\ref{LqfCut}), for $K=1,2,10$, applied to the original $100$-vertex market graph obtain from the in-sample data set.

\begin{figure}[]
	\centering
	\begin{subfigure}[t]{0.5\textwidth}
		\centering
		\includegraphics[width=1\textwidth, trim={0 0 0 0}, clip]{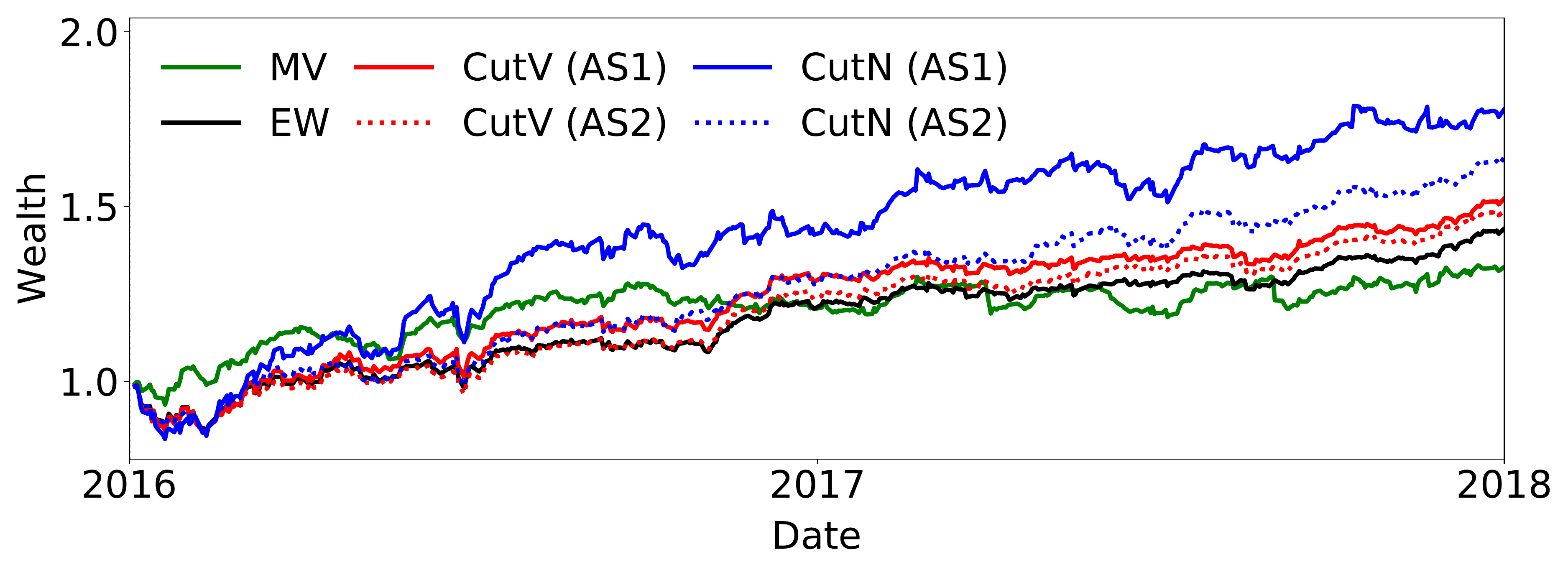} 
		\vspace{-0.6cm}
		\caption{Evolution of wealth for both the traditional (EW and MV) and graph-theoretic asset allocation strategies, based on ($K=10$) portfolio cuts.} 
		\vspace{0.2cm}
	\end{subfigure}
	\begin{subfigure}[t]{0.5\textwidth}
		\centering
		\scriptsize
		\renewcommand{\arraystretch}{1}
		\setlength{\tabcolsep}{3pt}
		\begin{tabular}[H!]{ c || c || c | c | c | c | c | c }
			\hline
			Cut Method & Allocation &  $K\!=\!1$ &  $K\!=\!2$ &  $K\!=\!3$ &  $K\!=\!4$ &  $K\!=\!5$ &  $K\!=\!10$ \\
			\hline
			CutV & AS1 &  $1.82$ &  $1.80$  &  $1.80$  &  $1.93$  &  $1.96$ &  $\boldsymbol{1.98}$  \\
			CutV & AS2 &  $1.82$ &  $1.81$ &  $1.94$ &  $2.03$ &  $1.95$ &  $\boldsymbol{2.05}$ \\
			CutN & AS1 &  $1.93$ &  $2.01$ &  $2.08$ &  $2.23$ &  $2.22$ &  $\boldsymbol{2.25}$ \\
			CutN & AS2 &  $1.93$ &  $2.04$ &  $2.17$ &  $\boldsymbol{2.65}$ &  $2.51$ &  $2.48$ \\
			\hline
		\end{tabular}
		\caption{Sharpe ratios attained for a varying number of portfolio cuts $K$.}
		\vspace{-0.2cm}
	\end{subfigure}
	\caption{\label{fig:performance} Out-sample performance of the asset allocation strategies. Notice that the Sharpe ratio typically improves with each subsequent portfolio cut. The traditional portfolio strategies, EW and MV, attained the respective Sharpe ratios of $\text{SR}_{\text{EW}}=1.85$ and $\text{SR}_{\text{MV}}=1.6$.} 
	\label{fig:1}
\end{figure}

Next, for the out-sample dataset, graph representations of the portfolio, for the number of cuts $K$ varying in the range $[1,10]$, were employed to assess the performance of the asset allocation schemes described in Section \ref{sec:asset_allocation}. The standard equally-weighted (EW) and minimum-variance (MV) portfolios were also simulated for comparison purposes, with the results displayed in Fig. \ref{fig:performance}. 

Conforming with the findings in \cite{LopezdePrado2016,Raffinot2017}, the proposed graph asset allocations schemes consistently delivered lower out-sample variance than the standard EW and MV portfolios, thereby attaining a higher \textit{Sharpe ratio}, i.e. the ratio of the mean to the standard deviation of portfolio returns. This verifies that the removal of possibly spurious statistical dependencies in the ``raw'' format, through the portfolio cuts, allows for robust and flexible portfolio constructions.

Such an approach enables the creation of graph-theoretic capital allocation schemes, based on measures of connectivity which are inherent to the portfolio cut formulation. In addition, the proposed portfolio construction employs full information contained in the asset covariance matrix, and without requiring its inversion, even in the critical cases of limited data length or singular covariance matrices.

\section{Conclusion}

In many modern applications, graph topology is not known a priori and hence its determination is part of the problem definition, rather than  serving as prior knowledge in problem solution. The focus of this part is therefore on simultaneous estimation of data on a graph and the underlying graph topology. To this end, without loss of generality, we assume that the vertices are given, while the edges and their associate weights are part of the  solution to the problem considered.  Three possible scenarios for the estimation of graph edges have been considered, as follows. In various sensing network setups (temperature, pressure, transportation) the locations of the sensing positions (vertices) are known and the vertex distances convey physical meaning about data dependence and  thus may be employed within a model for weight determination.  Besides, the covariance and precision matrices are most commonly used to measure data similarity and are thus a natural choice of metric for learning graph topology from data. Finally, examples of graphs where the relations among the sensing positions are physically well defined, such as in electric circuits, power networks, linear heat transfer, social and computer networks, spring-mass systems, have been given. The problem of simulation of a graph signal has been addressed and a detailed derivation and explanation of the LASSO and graphical LASSO has been given. The inherent connection between graphs and deep neural networks has been further addressed and enormous potential of the combination  of universal function approximation of neural networks and the elegance of graph models has been demonstrated on an example of semi-supervised learning. Next, the application of graphs in BigData scenarios has been illuminated through their link with tensors, in particular tensor factorizations. In particular, multidimensional graphs are extremely  important in practice but are inadequately modeled through the corresponding imbalanced adjacency matrices (for example, short and wide). On the other hand, we show that a multilinear algebra, whereby multidimensional graphs are modeled through the corresponding adjacency tensor, is particularly well suited to discover intrinsic relations  in multidimensional data. Finally, innovative examples ranging from portfolio cuts in finance to the modeling of vulnerability of stations in underground metro traffic support the approach. 

\section*{Bibliography}
\bibliographystyle{ieeetr}

\bibliography{graph-signal-processing}

\begin{thebibliography}{100}

\bibitem{friedman2008sparse}
J.~Friedman, T.~Hastie, and R.~Tibshirani, ``Sparse inverse covariance
  estimation with the graphical {LASSO},'' {\em Biostatistics}, vol.~9, no.~3,
  pp.~432--441, 2008.

\bibitem{meinshausen2006high}
N.~Meinshausen, P.~B{\"u}hlmann, {\em et~al.}, ``High-dimensional graphs and
  variable selection with the {LASSO},'' {\em The Annals of Statistics},
  vol.~34, no.~3, pp.~1436--1462, 2006.

\bibitem{pavez2016generalized}
E.~Pavez and A.~Ortega, ``Generalized {L}aplacian precision matrix estimation
  for graph signal processing,'' in {\em Proc. IEEE International Conference on
  Acoustics, Speech and Signal Processing (ICASSP), 2016}, pp.~6350--6354,
  IEEE, 2016.

\bibitem{pourahmadi2011covariance}
M.~Pourahmadi, ``Covariance estimation: The {GLM} and regularization
  perspectives,'' {\em Statistical Science}, pp.~369--387, 2011.

\bibitem{epskamp2018tutorial}
S.~Epskamp and E.~I. Fried, ``A tutorial on regularized partial correlation
  networks.,'' {\em Psychological Methods}, 2018.

\bibitem{das2017interpretation}
A.~Das, A.~L. Sampson, C.~Lainscsek, L.~Muller, W.~Lin, J.~C. Doyle, S.~S.
  Cash, E.~Halgren, and T.~J. Sejnowski, ``Interpretation of the precision
  matrix and its application in estimating sparse brain connectivity during
  sleep spindles from human electrocorticography recordings,'' {\em Neural
  Computation}, vol.~29, no.~3, pp.~603--642, 2017.

\bibitem{dong2016learning}
X.~Dong, D.~Thanou, P.~Frossard, and P.~Vandergheynst, ``Learning {L}aplacian
  matrix in smooth graph signal representations,'' {\em IEEE Transactions on
  Signal Processing}, vol.~64, no.~23, pp.~6160--6173, 2016.

\bibitem{Dong}
X.~Dong, D.~Thanou, P.~Frossard, and P.~Vandergheynst, ``Learning graphs from
  signal observations under smoothness prior.'' {June,} 2015 [online].
  {Available:} http://arXiv.org/abs/1406.7842.

\bibitem{stankovic2017LLLvertex}
L.~Stankovi{\'c}, E.~Sejdi{\'c}, and M.~Dakovi{\'c}, ``Vertex-frequency energy
  distributions,'' {\em IEEE Signal Processing Letters}, vol.~25, no.~3,
  pp.~358--362, 2017.

\bibitem{stankovic2018reduced}
L.~Stankovi{\' c}, E.~Sejdi{\' c}, and M.~Dakovi{\' c}, ``Reduced interference
  vertex-frequency distributions,'' {\em IEEE Signal Processing Letters},
  vol.~25, no.~9, pp.~1393--1397, 2018.

\bibitem{stankovic2019intuitive}
L.~Stankovic, D.~Mandic, M.~Dakovic, and I.~Kisil, ``An intuitive derivation of
  the coherence index relation in compressive sensing,'' {\em IEEE Signal
  Processing Magazine, arXiv preprint arXiv:1903.11136}, 2019.

\bibitem{rabiei2019estimating}
H.~Rabiei, F.~Richard, O.~Coulon, and J.~Lef{\`e}vre, ``Estimating the
  complexity of the cerebral cortex folding with a local shape spectral
  analysis,'' in {\em Vertex-Frequency Analysis of Graph Signals},
  pp.~437--458, Springer, 2019.

\bibitem{hamon2019transformation}
R.~Hamon, P.~Borgnat, P.~Flandrin, and C.~Robardet, ``Transformation from
  graphs to signals and back,'' in {\em Vertex-Frequency Analysis of Graph
  Signals}, pp.~111--139, Springer, 2019.

\bibitem{cioacua2019graph}
T.~Cioac{\u{a}}, B.~Dumitrescu, and M.-S. Stupariu, ``Graph-based wavelet
  multiresolution modeling of multivariate terrain data,'' in {\em
  Vertex-Frequency Analysis of Graph Signals}, pp.~479--507, Springer, 2019.

\bibitem{slawski2015estimation}
M.~Slawski and M.~Hein, ``Estimation of positive definite m-matrices and
  structure learning for attractive {G}aussian {M}arkov random fields,'' {\em
  Linear Algebra and its Applications}, vol.~473, pp.~145--179, 2015.

\bibitem{ubaru2017fast}
S.~Ubaru, J.~Chen, and Y.~Saad, ``Fast estimation of tr(f(a)) via stochastic
  {L}anczos quadrature,'' {\em SIAM Journal on Matrix Analysis and
  Applications}, vol.~38, no.~4, pp.~1075--1099, 2017.

\bibitem{caetano2009learning}
T.~S. Caetano, J.~J. McAuley, L.~Cheng, Q.~V. Le, and A.~J. Smola, ``Learning
  graph matching,'' {\em IEEE Transactions on Pattern Analysis and Machine
  Intelligence}, vol.~31, no.~6, pp.~1048--1058, 2009.

\bibitem{Thanou2014}
D.~Thanou, D.~I. Shuman, and P.~Frossard, ``Learning parametric dictionaries
  for signals on graphs,'' {\em IEEE Transactions Signal Processessing},
  vol.~62, no.~15, pp.~3849--3862, 2014.

\bibitem{camponogara2015models}
E.~Camponogara and L.~F. Nazari, ``Models and algorithms for optimal
  piecewise-linear function approximation,'' {\em Mathematical Problems in
  Engineering}, vol.~2015, 2015.

\bibitem{zhao2012huge}
T.~Zhao, H.~Liu, K.~Roeder, J.~Lafferty, and L.~Wasserman, ``The huge package
  for high-dimensional undirected graph estimation in {R},'' {\em Journal of
  Machine Learning Research}, vol.~13, no.~Apr, pp.~1059--1062, 2012.

\bibitem{yankelevsky2016}
Y.~Yankelevsky and M.~Elad, ``Dual graph regularized dictionary learning,''
  {\em IEEE Transactions on Signal and Information Processing over Networks},
  vol.~2, no.~4, pp.~611--624, 2016.

\bibitem{Zheng}
M.~Zheng, J.~Bu, C.~Chen, C.~Wang, L.~Zhang, G.~Qiu, and D.~Cai, ``Graph
  regularized sparse coding for image representation,'' {\em IEEE Transactions
  on Image Processing}, vol.~20, no.~5, pp.~1327--1336, 2011.

\bibitem{segarra2016blind}
S.~Segarra, A.~G. Marques, G.~Mateos, and A.~Ribeiro, ``Blind identification of
  graph filters with multiple sparse inputs.,'' in {\em Proc. IEEE
  International Conference on Acoustics, Speech and Signal Processing
  (ICASSP)}, pp.~4099--4103, 2016.

\bibitem{stankovic2017vertex}
L.~Stankovi{\' c}, M.~Dakovi{\' c}, and E.~Sejdi{\' c}, ``Vertex-frequency
  analysis: A way to localize graph spectral components [lecture notes],'' {\em
  IEEE Signal Processing Magazine}, vol.~34, no.~4, pp.~176--182, 2017.

\bibitem{stankovic2019vertexTEL}
L.~Stankovi{\'c} and E.~Sejdi{\'c}, {\em Vertex-Frequency Analysis of Graph
  Signals}.
\newblock Springer, 2019.

\bibitem{pasdeloup2019uncertainty}
B.~Pasdeloup, V.~Gripon, R.~Alami, and M.~G. Rabbat, ``Uncertainty principle on
  graphs,'' in {\em Vertex-Frequency Analysis of Graph Signals}, pp.~317--340,
  Springer, 2019.

\bibitem{dal2019wavelet}
A.~Dal~Col, P.~Valdivia, F.~Petronetto, F.~Dias, C.~T. Silva, and L.~G. Nonato,
  ``Wavelet-based visual data exploration,'' in {\em Vertex-Frequency Analysis
  of Graph Signals}, pp.~459--478, Springer, 2019.

\bibitem{tanaka2019oversampled}
Y.~Tanaka and A.~Sakiyama, ``Oversampled transforms for graph signals,'' in
  {\em Vertex-Frequency Analysis of Graph Signals}, pp.~223--254, Springer,
  2019.

\bibitem{bohannon2019filtering}
A.~W. Bohannon, B.~M. Sadler, and R.~V. Balan, ``A filtering framework for
  time-varying graph signals,'' in {\em Vertex-Frequency Analysis of Graph
  Signals}, pp.~341--376, Springer, 2019.

\bibitem{gu2019local}
Y.~Gu and X.~Wang, ``Local-set-based graph signal sampling and
  reconstruction,'' in {\em Vertex-Frequency Analysis of Graph Signals},
  pp.~255--292, Springer, 2019.

\bibitem{mao2019time}
X.~Mao and Y.~Gu, ``Time-varying graph signals reconstruction,'' in {\em
  Vertex-Frequency Analysis of Graph Signals}, pp.~293--316, Springer, 2019.

\bibitem{stankovic2001measure}
L.~Stankovi{\'c}, ``A measure of some time--frequency distributions
  concentration,'' {\em Signal Processing}, vol.~81, no.~3, pp.~621--631, 2001.

\bibitem{stankovictutorial}
L.~Stankovi{\'c}, E.~Sejdi{\'c}, S.~Stankovi{\'c}, M.~Dakovi{\'c}, and
  I.~Orovi{\'c}, ``A tutorial on sparse signal reconstruction and its
  applications in signal processing,'' {\em Circuits, Systems, and Signal
  Processing}, pp.~1--58, 2018.

\bibitem{stankovic2019understanding}
L.~Stankovic, D.~P. Mandic, M.~Dakovic, I.~Kisil, E.~Sejdic, and A.~G.
  Constantinides, ``Understanding the basis of graph signal processing via an
  intuitive example-driven approach [lecture notes],'' {\em IEEE Signal
  Processing Magazine}, vol.~36, no.~6, pp.~133--145, 2019.

\bibitem{candes2006robust}
E.~J. Cand{\`e}s, J.~Romberg, and T.~Tao, ``Robust uncertainty principles:
  Exact signal reconstruction from highly incomplete frequency information,''
  {\em IEEE Transactions on Information Theory}, vol.~52, no.~2, pp.~489--509,
  2006.

\bibitem{masuda2017random}
N.~Masuda, M.~A. Porter, and R.~Lambiotte, ``Random walks and diffusion on
  networks,'' {\em Physics reports}, vol.~716, pp.~1--58, 2017.

\bibitem{stankovic2015digital}
L.~Stankovi{\' c}, {\em Digital signal processing with selected topics}.
\newblock CreateSpace Independent Publishing Platform, An Amazon.com Company,
  2015.

\bibitem{dong2019learning}
X.~Dong, D.~Thanou, M.~Rabbat, and P.~Frossard, ``Learning graphs from data: A
  signal representation perspective,'' {\em IEEE Signal Processing Magazine},
  vol.~36, no.~3, pp.~44--63, 2019.

\bibitem{mateos2019connecting}
G.~Mateos, S.~Segarra, A.~G. Marques, and A.~Ribeiro, ``Connecting the dots:
  Identifying network structure via graph signal processing,'' {\em IEEE Signal
  Processing Magazine}, vol.~36, no.~3, pp.~16--43, 2019.

\bibitem{giannakis2018topology}
G.~B. Giannakis, Y.~Shen, and G.~V. Karanikolas, ``Topology identification and
  learning over graphs: Accounting for nonlinearities and dynamics,'' {\em
  Proceedings of the IEEE}, vol.~106, no.~5, pp.~787--807, 2018.

\bibitem{wai2019community}
H.-T. Wai, Y.~C. Eldar, A.~E. Ozdaglar, and A.~Scaglione, ``Community inference
  from graph signals with hidden nodes,'' in {\em ICASSP 2019-2019 IEEE
  International Conference on Acoustics, Speech and Signal Processing
  (ICASSP)}, pp.~4948--4952, IEEE, 2019.

\bibitem{grotas2019power}
S.~Grotas, Y.~Yakoby, I.~Gera, and T.~Routtenberg, ``Power systems topology and
  state estimation by graph blind source separation,'' {\em IEEE Transactions
  on Signal Processing}, vol.~67, no.~8, pp.~2036--2051, 2019.

\bibitem{kaplan2008structural}
D.~Kaplan, {\em Structural equation modeling: {F}oundations and extensions},
  vol.~10.
\newblock Sage Publications, 2008.

\bibitem{chen2011vector}
G.~Chen, D.~R. Glen, Z.~S. Saad, J.~P. Hamilton, M.~E. Thomason, I.~H. Gotlib,
  and R.~W. Cox, ``Vector autoregression, structural equation modeling, and
  their synthesis in neuroimaging data analysis,'' {\em Computers in Biology
  and Medicine}, vol.~41, no.~12, pp.~1142--1155, 2011.

\bibitem{ioannidis2019semi}
V.~N. Ioannidis, Y.~Shen, and G.~B. Giannakis, ``Semi-blind inference of
  topologies and dynamical processes over dynamic graphs,'' {\em IEEE
  Transactions on Signal Processing}, vol.~67, no.~9, pp.~2263--2274, 2019.

\bibitem{kolaczyk2014statistical}
E.~D. Kolaczyk, {\em Statistical Analysis of Network Data -- Methods and
  Models}.
\newblock Springer-Verlag New York, 2009.

\bibitem{baba2004partial}
K.~Baba, R.~Shibata, and M.~Sibuya, ``Partial correlation and conditional
  correlation as measures of conditional independence,'' {\em Australian \& New
  Zealand Journal of Statistics}, vol.~46, no.~4, pp.~657--664, 2004.

\bibitem{dempster1972covariance}
A.~P. Dempster, ``Covariance selection,'' {\em Biometrics}, pp.~157--175, 1972.

\bibitem{yuan2007model}
M.~Yuan and Y.~Lin, ``Model selection and estimation in the {G}aussian
  graphical model,'' {\em Biometrika}, vol.~94, no.~1, pp.~19--35, 2007.

\bibitem{banerjee2008model}
O.~Banerjee, L.~E. Ghaoui, and A.~d’Aspremont, ``Model selection through
  sparse maximum likelihood estimation for multivariate {G}aussian or binary
  data,'' {\em Journal of Machine Learning Research}, vol.~9, no.~Mar,
  pp.~485--516, 2008.

\bibitem{yuan2006model}
M.~Yuan and Y.~Lin, ``Model selection and estimation in regression with grouped
  variables,'' {\em Journal of the Royal Statistical Society: Series B
  (Statistical Methodology)}, vol.~68, no.~1, pp.~49--67, 2006.

\bibitem{kalofolias2016learn}
V.~Kalofolias, ``How to learn a graph from smooth signals,'' in {\em
  Proceedings of the Artificial Intelligence and Statistics}, pp.~920--929,
  2016.

\bibitem{chepuri2017learning}
S.~P. Chepuri, S.~Liu, G.~Leus, and A.~O. Hero, ``Learning sparse graphs under
  smoothness prior,'' in {\em Proceedings of the 2017 IEEE International
  Conference on Acoustics, Speech and Signal Processing (ICASSP)},
  pp.~6508--6512, IEEE, 2017.

\bibitem{segarra2017network}
S.~Segarra, A.~G. Marques, G.~Mateos, and A.~Ribeiro, ``Network topology
  inference from spectral templates,'' {\em IEEE Transactions on Signal and
  Information Processing over Networks}, vol.~3, no.~3, pp.~467--483, 2017.

\bibitem{thanou2017learning}
D.~Thanou, X.~Dong, D.~Kressner, and P.~Frossard, ``Learning heat diffusion
  graphs,'' {\em IEEE Transactions on Signal and Information Processing over
  Networks}, vol.~3, no.~3, pp.~484--499, 2017.

\bibitem{gori2005new}
M.~Gori, G.~Monfardini, and F.~Scarselli, ``A new model for learning in graph
  domains,'' in {\em Proceedings of the IEEE International Joint Conference on
  Neural Networks, 2005.}, vol.~2, pp.~729--734, IEEE, 2005.

\bibitem{scarselli2008graph}
F.~Scarselli, M.~Gori, A.~C. Tsoi, M.~Hagenbuchner, and G.~Monfardini, ``The
  graph neural network model,'' {\em IEEE Transactions on Neural Networks},
  vol.~20, no.~1, pp.~61--80, 2008.

\bibitem{micheli2009neural}
A.~Micheli, ``Neural network for graphs: A contextual constructive approach,''
  {\em IEEE Transactions on Neural Networks}, vol.~20, no.~3, pp.~498--511,
  2009.

\bibitem{zhou2018graph}
J.~Zhou, G.~Cui, Z.~Zhang, C.~Yang, Z.~Liu, and M.~Sun, ``Graph neural
  networks: A review of methods and applications,'' {\em arXiv preprint
  arXiv:1812.08434}, 2018.

\bibitem{wu2019comprehensive}
Z.~Wu, S.~Pan, F.~Chen, G.~Long, C.~Zhang, and P.~S. Yu, ``A comprehensive
  survey on graph neural networks,'' {\em arXiv preprint arXiv:1901.00596},
  2019.

\bibitem{zhu2005semi}
X.~J. Zhu, ``Semi-supervised learning literature survey,'' tech. rep.,
  University of Wisconsin-Madison Department of Computer Sciences, 2005.

\bibitem{lecun1998gradient}
Y.~LeCun, L.~Bottou, Y.~Bengio, P.~Haffner, {\em et~al.}, ``Gradient-based
  learning applied to document recognition,'' {\em Proceedings of the IEEE},
  vol.~86, no.~11, pp.~2278--2324, 1998.

\bibitem{wang2004image}
Z.~Wang, A.~C. Bovik, H.~R. Sheikh, E.~P. Simoncelli, {\em et~al.}, ``Image
  quality assessment: from error visibility to structural similarity,'' {\em
  IEEE Transactions on Image Processing}, vol.~13, no.~4, pp.~600--612, 2004.

\bibitem{mandic2007machine}
D.~Mandic, ``Machine learning and signal processing applications of fixed point
  theory,'' {\em Tutorial in IEEE ICASSP, 2007}, 2007.

\bibitem{li2015gated}
Y.~Li, D.~Tarlow, M.~Brockschmidt, and R.~Zemel, ``Gated graph sequence neural
  networks,'' {\em arXiv preprint arXiv:1511.05493}, 2015.

\bibitem{dai2018learning}
H.~Dai, Z.~Kozareva, B.~Dai, A.~Smola, and L.~Song, ``Learning steady-states of
  iterative algorithms over graphs,'' in {\em Proceedings of the International
  Conference on Machine Learning}, pp.~1114--1122, 2018.

\bibitem{atwood2016diffusion}
J.~Atwood and D.~Towsley, ``Diffusion-convolutional neural networks,'' in {\em
  Advances in Neural Information Processing Systems}, pp.~1993--2001, 2016.

\bibitem{bacciu2019non}
D.~Bacciu and L.~Di~Sotto, ``A non-negative factorization approach to node
  pooling in graph convolutional neural networks,'' in {\em Proceedings of the
  International Conference of the Italian Association for Artificial
  Intelligence}, pp.~294--306, Springer, 2019.

\bibitem{sakiyama2019eigendecomposition}
A.~Sakiyama, Y.~Tanaka, T.~Tanaka, and A.~Ortega, ``Eigendecomposition-free
  sampling set selection for graph signals,'' {\em IEEE Transactions on Signal
  Processing}, vol.~67, no.~10, pp.~2679--2692, 2019.

\bibitem{zhang2019star}
J.~Zhang, X.~Shi, S.~Zhao, and I.~King, ``S{TAR-GCN}: {S}tacked and
  reconstructed graph convolutional networks for recommender systems,'' {\em
  arXiv preprint arXiv:1905.13129}, 2019.

\bibitem{tanaka2019generalized}
Y.~Tanaka and Y.~C. Eldar, ``Generalized sampling on graphs with subspace and
  smoothness priors,'' {\em arXiv preprint arXiv:1905.04441}, 2019.

\bibitem{ioannidis2019graphsac}
V.~N. Ioannidis, D.~Berberidis, and G.~B. Giannakis, ``Graphsac: {D}etecting
  anomalies in large-scale graphs,'' {\em arXiv preprint arXiv:1910.09589},
  2019.

\bibitem{gilmer2017neural}
J.~Gilmer, S.~S. Schoenholz, P.~F. Riley, O.~Vinyals, and G.~E. Dahl, ``Neural
  message passing for quantum chemistry,'' in {\em Proceedings of the 34th
  International Conference on Machine Learning-Volume 70}, pp.~1263--1272,
  JMLR. org, 2017.

\bibitem{kipf2016semi}
T.~N. Kipf and M.~Welling, ``Semi-supervised classification with graph
  convolutional networks,'' {\em arXiv preprint arXiv:1609.02907}, 2016.

\bibitem{bacciu2018contextual}
D.~Bacciu, F.~Errica, and A.~Micheli, ``Contextual graph {M}arkov model: {A}
  deep and generative approach to graph processing,'' {\em arXiv preprint
  arXiv:1805.10636}, 2018.

\bibitem{hamilton2017inductive}
W.~Hamilton, Z.~Ying, and J.~Leskovec, ``Inductive representation learning on
  large graphs,'' in {\em Advances in Neural Information Processing Systems},
  pp.~1024--1034, 2017.

\bibitem{velivckovic2017graph}
P.~Veli{\v{c}}kovi{\'c}, G.~Cucurull, A.~Casanova, A.~Romero, P.~Lio, and
  Y.~Bengio, ``Graph attention networks,'' {\em arXiv preprint
  arXiv:1710.10903}, 2017.

\bibitem{monti2017geometric}
F.~Monti, D.~Boscaini, J.~Masci, E.~Rodola, J.~Svoboda, and M.~M. Bronstein,
  ``Geometric deep learning on graphs and manifolds using mixture model
  {CNN}s,'' in {\em Proceedings of the IEEE Conference on Computer Vision and
  Pattern Recognition}, pp.~5115--5124, 2017.

\bibitem{zhang2018gaan}
J.~Zhang, X.~Shi, J.~Xie, H.~Ma, I.~King, and D.-Y. Yeung, ``Gaan: {G}ated
  attention networks for learning on large and spatiotemporal graphs,'' {\em
  arXiv preprint arXiv:1803.07294}, 2018.

\bibitem{hammond2011wavelets}
D.~K. Hammond, P.~Vandergheynst, and R.~Gribonval, ``Wavelets on graphs via
  spectral graph theory,'' {\em Applied and Computational Harmonic Analysis},
  vol.~30, no.~2, pp.~129--150, 2011.

\bibitem{bruna2013spectral}
J.~Bruna, W.~Zaremba, A.~Szlam, and Y.~LeCun, ``Spectral networks and locally
  connected networks on graphs,'' {\em arXiv preprint arXiv:1312.6203}, 2013.

\bibitem{motl2015ctu}
J.~Motl and O.~Schulte, ``The {CTU} prague relational learning repository,''
  {\em arXiv preprint arXiv:1511.03086}, 2015.

\bibitem{defferrard2016convolutional}
M.~Defferrard, X.~Bresson, and P.~Vandergheynst, ``Convolutional neural
  networks on graphs with fast localized spectral filtering,'' in {\em Advances
  in Neural Information Processing Systems}, pp.~3844--3852, 2016.

\bibitem{Lin2008}
Y.~R. Lin, Y.~Chi, S.~Zhu, H.~Sundaram, and B.~L. Tseng, ``Facetnet: a
  framework for analyzing communities and their evolutions in dynamic
  networks,'' {\em In Proceedings of the International Conference on World Wide
  Web (WWW)}, pp.~685--694, 2008.

\bibitem{Lin2009}
Y.~Lin, J.~Sun, P.~Castro, R.~Konuru, H.~Sundaram, and A.~Kelliher, ``Metafac:
  community discovery via relational hypergraph factorization,'' {\em In
  Proceedings of the ACM KDD International Conference on Knowledge Discovery
  and Data Mining}, pp.~527--536, 2009.

\bibitem{Tan2009}
W.~Tang, Z.~Lu, and I.~S. Dhillon, ``Clustering with multiple graphs,'' {\em In
  Proceedings of Ninth IEEE International Conference on Data Mining},
  pp.~1016--1021, 2009.

\bibitem{Nickel2011}
M.~Nickel, V.~Tresp, and H.-P. Kriegel, ``A three-way model for collective
  learning on multi-relational data,'' {\em In Proceedings of the 28th
  International Conference on Machine Learning}, pp.~809--816, 2011.

\bibitem{Papalexakis2013}
E.~E. Papalexakis, L.~Akoglu, and D.~Lence, ``Do more views of a graph help?
  community detection and clustering in multi-graphs,'' {\em In Proceedings of
  the 16th International Conference on Information Fusion}, pp.~899--905, 2013.

\bibitem{Gauvin2014}
L.~Gauvin, A.~Panisson, and C.~Cattuto, ``Detecting the community structure and
  activity patterns of temporal networks: a non-negative tensor factorization
  approach,'' {\em PLOS ONE}, vol.~9, p.~e86028, 2014.

\bibitem{Verma2017}
A.~Verma and K.~K. Bharadwaj, ``A comparative study based on tensor
  factorization and clustering techniques for community mining in heterogeneous
  social network,'' {\em In Proceedings of the International Conference on
  Computing, Communication and Networking Technologies (ICCCNT)}, vol.~In
  Press, pp.~1--6, 2017.

\bibitem{Verma2017_2}
A.~Verma and K.~K. Bharadwaj, ``Identifying community structure in a
  multi-relational network employing non-negative tensor factorization and ga
  k-means clustering,'' {\em Wires: Data Mining and Knowledge Discovery},
  vol.~7, pp.~1--32, 2017.

\bibitem{Katsimpras2019}
G.~Katsimpras and G.~Paliouras, ``Class-aware tensor factorization for
  multi-relational classification,'' {\em Information Processing \&
  Management}, vol.~In Press, 2019.

\bibitem{Freeman1977}
L.~C. Freeman, ``A set of measures of centrality based on betweenness,'' {\em
  Sociometry}, vol.~40, pp.~35--41, 1977.

\bibitem{Brandes2005}
U.~Brandes, {\em Network Analysis: Methodological Foundations}.
\newblock Springer, 2005.

\bibitem{TFL}
``Transport for london.''

\bibitem{Markowitz1952}
H.~Markowitz, ``Portfolio selection,'' {\em Journal of Finance}, vol.~7, no.~1,
  pp.~77--91, 1952.

\bibitem{Kolm2014}
P.~N. Kolm, R.~Tutuncu, and F.~J. Fabozzi, ``60 years of portfolio
  optimization: Practical challenges and current trends,'' {\em European
  Journal of Operational Research}, vol.~234, no.~2, pp.~356--371, 2014.

\bibitem{Clarke2002}
R.~Clarke, H.~De~Silva, and S.~Thorley, ``Portfolio constraints and the
  fundamental law of active management,'' {\em Financial Analysts Journal},
  vol.~58, pp.~48--66, 2002.

\bibitem{Black1992}
F.~Black and R.~Litterman, ``Global portfolio optimization,'' {\em Financial
  Analysts Journal}, vol.~48, no.~5, pp.~280--291, 1992.

\bibitem{Ledoit2003}
O.~Ledoit and M.~Wolf, ``Improved estimation of the covariance matrix of stock
  returns with an application to portfolio selection,'' {\em Journal of
  Empirical Finance}, vol.~10, no.~5, pp.~603--621, 2003.

\bibitem{Boginski2003}
V.~Boginski, S.~Butenko, and P.~M. Pardalos, ``On structural properties of the
  market graph,'' in {\em Innovations in Financial and Economic Networks}
  (A.~Nagurney, ed.), pp.~29--45, Edward Elgar Publishers, 2003.

\bibitem{Simon1962}
H.~A. Simon, ``The architecture of complexity,'' {\em In Proceedings of the
  American Philosophical Society}, 1962.

\bibitem{LopezdePrado2014}
N.~J. Calkin and M.~Lopez~de Prado, ``Stochastic flow diagrams,'' {\em
  Algorithmic Finance}, vol.~3, no.~1--2, pp.~21--42, 2014.

\bibitem{LopezdePrado2014_2}
N.~J. Calkin and M.~Lopez~de Prado, ``The topology of macro financial flows: An
  application of stochastic flow diagrams,'' {\em Algorithmic Finance}, vol.~3,
  no.~1, pp.~43--85, 2014.

\bibitem{LopezdePrado2016}
N.~J. Calkin and M.~Lopez~de Prado, ``Building diversified portfolios that
  outperform out of sample,'' {\em The Journal of Portfolio Management},
  vol.~42, no.~4, pp.~59--69, 2016.

\bibitem{Peralta2016}
G.~Peralta and A.~Zareei, ``A network approach to portfolio selection,'' {\em
  Journal of Empirical Finance}, vol.~38, no.~A, pp.~157--180, 2016.

\bibitem{Li2019}
Y.~Li, X.~F. Jiang, Y.~Tian, S.~P. Li, and B.~Zheng, ``Portfolio optimization
  based on network topology,'' {\em Physica A}, vol.~515, pp.~671--681, 2019.

\bibitem{Raffinot2017}
T.~Raffinot, ``Hierarchical clustering-based asset allocation,'' {\em The
  Journal of Portfolio Management}, vol.~44, no.~2, pp.~89--99, 2017.

\bibitem{Boginski2005}
V.~Boginski, S.~Butenko, and P.~M. Pardalos, ``Statistical analysis of
  financial networks,'' {\em Computational Statistics \& Data Analysis},
  vol.~48, no.~2, pp.~431--443, 2005.

\bibitem{Boginski2006}
V.~Boginski, S.~Butenko, and P.~M. Pardalos, ``Mining market data: A network
  approach,'' {\em Computers \& Operations Research}, vol.~33, no.~11,
  pp.~3171--3184, 2006.

\bibitem{Gunawardena2012}
A.~A. Gunawardena, R.~R. Meyer, and W.~L. Dougan, ``Optimal selection of an
  independent set of cliques in a market graph,'' {\em In Proceedings of the
  International Conference on Economics, Business and Marketing Management},
  pp.~281--285, 2012.

\bibitem{Boginski2014}
V.~Boginski, S.~Butenko, S.~O., S.~Trunkhanov, and J.~Gil~Lafuente, ``A
  network-based data mining approach to portfolio selection via weighted clique
  relaxations,'' {\em Annals of Operations Research}, vol.~216, pp.~23--34,
  2014.

\bibitem{Kalyagin2014}
V.~Kalyagin, A.~Koldanov, P.~Koldanov, and V.~Zamaraev, ``{Market graph and
  Markowitz model},'' in {\em Optimization in Science and Engineering} (T.~M.
  Rassias, C.~A. Floudas, and S.~Butenko, eds.), pp.~293--306, Springer, 2014.

\bibitem{Schaeffer2007}
S.~E. Schaeffer, ``Graph clustering,'' {\em Computer Science Review}, vol.~1,
  no.~1, pp.~27--64, 2007.

\bibitem{Hagen1992}
L.~Hagen and A.~B. Kahng, ``New spectral methods for ratio cut partitioning and
  clustering,'' {\em IEEE Transactions on Computer-Aided Design of Integrated
  Circuits and Systems}, vol.~11, no.~9, pp.~1074--1085, 1992.

\bibitem{Shi2000}
J.~Shi and J.~Malik, ``Normalized cuts and image segmentation,'' {\em
  Departmental Papers (CIS)}, p.~107, 2000.

\bibitem{Fiedler1973}
M.~Fiedler, ``Algebraic connectivity of graphs,'' {\em Czechoslovak
  Mathematical Journal}, vol.~23, no.~2, pp.~298--305, 1973.

\bibitem{Ng2002}
A.~Y. Ng, M.~I. Jordan, and Y.~Weiss, ``On spectral clustering: Analysis and an
  algorithm,'' {\em In Proceedings of the Conference on Neural Information
  Processing Systems (NIPS)}, pp.~849--856, 2002.

\bibitem{Spielman2007}
D.~A. Spielman and S.~H. Teng, ``Spectral partitioning works: Planar graphs and
  finite element meshes,'' {\em Linear Algebra and its Applications}, vol.~421,
  no.~2-3, pp.~284--305, 2007.

\bibitem{Stankovic2019_1}
L.~Stankovi\'c, D.~P. Mandic, M.~Dakovi\'c, M.~Brajovi\'c, B.~Scalzo~Dees, and
  T.~Constantinides, ``Graph signal processing -- {P}art {I}: Graphs, graph
  spectra, and spectral clustering,'' {\em arXiv:1907.03467}, 2019.

\bibitem{stankovic2019graph}
L.~Stankovic, D.~Mandic, M.~Dakovic, M.~Brajovic, B.~Scalzo, and A.~G.
  Constantinides, ``Graph signal processing--{P}art {II}: Processing and
  analyzing signals on graphs,'' {\em arXiv preprint arXiv:1909.10325}, 2019.

\bibitem{DeMiguel2009}
L.~G. De~Miguel, V. and R.~R.~Uppal, ``Optimal versus naive diversification:
  How inefficient is the $1/n$ portfolio strategy?,'' {\em Review of Financial
  Studies}, vol.~22, pp.~1915--1953, 2009.

\end{thebibliography}

\end{document}